\documentclass[12pt]{article}

\usepackage{epsfig}

\renewcommand{\thesection}{\arabic{section}.}

\renewcommand{\theequation}{\arabic{section}.\arabic{equation}}

\begin{document}

\newcommand{\ez}{\epsilon_0}
\newcommand{\gz}{\gamma_0}
\newcommand{\smat}{{\cal{S}}}



\begin{center}
{{\Large{\bf Ab-initio methods for spin-transport at the nanoscale level}}}
\end{center}

\vspace{1.5cm}

\begin{center}
{\bf{Stefano Sanvito}}

\vspace{0.7cm}

Department of Physics, Trinity College Dublin, Ireland

\vspace{0.4cm}

email: sanvitos@tcd.ie
\end{center}

\date{today}

\begin{abstract}
Recent advances in atomic and nano-scale growth and characterization techniques 
have led to the production of modern magnetic materials and magneto-devices 
which reveal a range of new fascinating phenomena. The modeling of these is a
tough theoretical challenging since one has to describe accurately both 
electronic structure of the constituent materials, and their transport
properties. In this paper I review recent advances in modeling spin-transport 
at the atomic scale using first-principles methods, focusing both on the
methodological aspects and on the applications. The review, which is designed as
tutorial for students at postgraduate level, is structured in six main
sections: 1) Introduction, 2) General concepts in spin-transport, 3) Transport 
Theory: Linear Response, 4) Transport Theory: Non-equilibrium Transport, 
5) Results, 6) Conclusion. In addition an overview of the computational codes
available to date is also included.
\end{abstract}

\newpage

\tableofcontents

\newpage

\setcounter{equation}{0}
\setcounter{figure}{0}
\setcounter{table}{0}
\section{Introduction}

According to a recent issue of ``Physics Today'', the traditional study of magnetism 
and magnetic materials has an image of ``musty physics laboratories peopled by old 
codgers with iron filings under their nails'' \cite{Simonds}. However over the last 
few years new advances in atomic- and nano-scale growth and characterization techniques 
have led to the production of modern magnetic materials and magneto-devices which reveal 
a range of new fascinating phenomena. 

This new renaissance in the study of magnetism has been initiated by the
discovery of the Giant Magnetoresistance (GMR) effect in magnetic multilayers 
\cite{baibich_gmr,binasch_gmr}. GMR is the drastic change in the electrical resistance of 
a multilayer formed by alternating magnetic and non-magnetic materials when a magnetic 
field is applied. In absence of an external magnetic field the exchange coupling between
adjacent magnetic layers through the non-magnetic ones \cite{exchange} aligns the
magnetization vectors 
antiparallel to each other. Then when a magnetic field strong enough to overcome the 
antiferromagnetic coupling is applied, all the magnetization vectors align along the field 
direction. The new ``parallel'' configuration presents an electrical resistance considerably 
smaller than that of the antiparallel: this change in resistance is the GMR.

The main reason why GMR was such an important milestone is that not only the interplay between 
transport and magnetism was demonstrated, but also that the spin degree of freedom could be
engineered and exploited in a controlled way. In other words GMR established that the longly 
neglected electron spin could be used in the similar way that the electron charge in an electronic 
device. In the remarkable short time of about a decade, GMR evolved from an academic curiosity 
to a major commercial product.
Today GMR-based read/write heads for magnetic data storage devices (hard-drives) are in every 
computers with a huge impact over a multi-billion dollar industry, and magnetic random access 
memory (MRAM) based on metallic elements will soon impact another multi-billion industry.

Most recently, in particular after the advent of magnetic semiconductors \cite{DMS1,DMS2},
a new level of control of the spin-dynamics has been achieved. Very long spin lifetime \cite{Aws1} 
and coherence \cite{Aws2} in semiconductors, spin coherence transport across interfaces \cite{Aws3},
manipulation of both electronic and nuclear spins over fast time scales \cite{Aws4}, have been all 
already demonstrated. These phenomena, that collectively take the name of ``spintronics'' or ``spin 
electronics'' \cite{spintronics_wolf,spintronics_prinz1,spintronics_prinz2} open a new avenue to the 
next generation of electronic devices where the spin of an electron can be used both for logic and 
memory purposes. Quoting a recent review of the field ``the 
advantages of these new devices would be nonvolatility, increased data processing speed, decreased 
electrical power consumption, and increased integration densities compared with conventional 
semiconductor devices'' \cite{spintronics_wolf}.
The ultimate target is to go beyond ordinary 
binary logic and use the spin entanglement for new quantum computing strategies \cite{spin_computer}.
This will probably require the control of the spin dynamics on a single spin scale, a remote task that
will merge spintronics with the rapidly evolving field of molecular electronics \cite{moltronics}.

Interestingly, it is worth noting that most of the proposed spintronics implementations/ applications 
to date simply translate well-known concepts of conventional electronics to spin systems. 
The typical devices are mainly made by molecular beam epitaxy growth, and lithographic techniques; 
a bottom-up approach to spintronics devices has been only poorly explored. This is an area where the
convergence with molecular electronics may bring important new breakthroughs.
The idea of molecular electronics is to use molecular systems for electronic applications.  
This is indeed possible, and conventional electronic devices including, molecular transistors 
\cite{mol_trans}, negative differential resistance \cite{mol_ndr}, and rectifiers \cite{mol_rec} 
have been demonstrated at the molecular level. However in all these ``conventional'' molecular 
electronics applications the electron spin is neglected. 

Therefore it starts to be natural asking whether the fields of spin- and 
molecular electronics can be integrated. This basically means asking: ``how can we inject, manipulate 
and detect spins at the atomic and molecular level?'' It is worth noting that in addition to the 
fact that a molecular self-assembly approach can substitute expensive growing/processing technology 
with low-cost chemical methods, spintronics in low dimensional systems can offer genuine advantages 
over bulk metals and semiconductors. In fact, molecular systems are mainly made from light, 
non-magnetic atoms, and the conventional mechanisms for spin de-coherence (spin-orbit coupling, 
scattering to paramagnetic impurities) are strongly suppressed. Therefore the spin coherence time 
is expected to be several orders of magnitude larger in molecules than in semiconductors. Strong 
indications on possibility of integrating the two fields come from a few recent pioneering 
experiments demonstrating spin injection \cite{CNT_GMR} and magnetic proximity \cite{CNT_prox} into 
carbon nanotubes, molecular GMR \cite{mol_GMR}, huge GMR in ballistic point contacts 
\cite{BMR1,BMR2,BMR3}, spin injection into long polymeric materials \cite{GMR_poly} and spin
coherent transport through organic molecules \cite{Aws5}. 

It is therefore clear that a deep understanding of the spin-dynamics and of the spin transport at 
the molecular and atomic scale is fundamental for further advances in both spin- and molecular-
electronics. This is an area where we expect a large convergence from Physics, Chemistry, 
Materials Science, Electronical Engineering and in prospective Biology. The small lengths scale 
and the complexity of some of the systems studied put severe requirements to a quantitative 
theory.

Theory has been at the forefront of the spintronics ``revolution''
since the early days. For instance spin transport into magnetic multilayers has been very
successfully modeled by the widely known Valet-Fert theory \cite{valet_fert}. This is based 
on the Boltzmann's equations solved in the relaxation time approximation, which 
reduce to a resistor network model in the limit of long spin-flip length. 
The main aspect of this approach is that the details of the electronic 
structure of the constituent materials can be neglected in favor of some averaged
quantities such as the resistivity. Such methods are based on the idea that the scattering length scale is considerably 
shorter than the typical size of the entire device. Clearly the scheme breaks down when we 
consider spin transport at the molecular and atomic scale. In this situation individual scattering events 
may be responsible for the whole resistance of a device and an accurate description is needed in order to 
make quantitative predictions. In particular a theory of spin-transport will be predictive 
if it comprises the following two ingredients:
\begin{enumerate}
\item An accurate electronic structure calculation method which relies weakly on external parameters
\item A transport method able to describe charging effects
\end{enumerate}
At present there are several approaches to both electronic structures and transport methods, but very few 
algorithms that efficiently include both. The purpose of this paper is to present an organic review of spin 
transport methods at the nanoscale, in particular of approaches which are based on parameter-free 
{\it ab initio} electronic structure calculations. I will incrementally add details to the description with 
the idea to make the paper accessible to non-experts to the field. The prototypical device that I will 
consider is the spin-valve. This is a trilayer structure with a first magnetic
layer used as spin-polarizer, a non-magnetic spacer and a second magnetic layer used as analyzer.
I will consider as non-magnetic part either metals, or insulators or molecules. 

Since the field is rather large, this review does not pretend to be exhaustive, but only to provide a 
didactic  introduction to the fascinating world of the theory of both spin- and molecular-electronics. 
In doing that I will overlook several important aspects such as semiconductors spintronics, theory of optical 
spin excitation and femtosecond laser spectroscopy, and non-elastic spin transport in molecules, 
for which I remand to the appropriate literature. 

The paper is structured with a section describing the main ideas behind spin-transport at the nanoscale with 
reference to recent experimental advances, followed by a tutorial presentation of a transport algorithm based 
on the non-equilibrium Green's function approach. Here I will make a link with other approach and review the 
computer codes available to date. Finally I will review a selected number of problems where {\it ab initio} 
spin-transport theory has led to important new developments.

\setcounter{equation}{0}
\section{General concepts in spin-transport}

\subsection{Length scales}

\begin{figure}[ht]
\begin{center}
\includegraphics[width=7.5cm,clip=true]{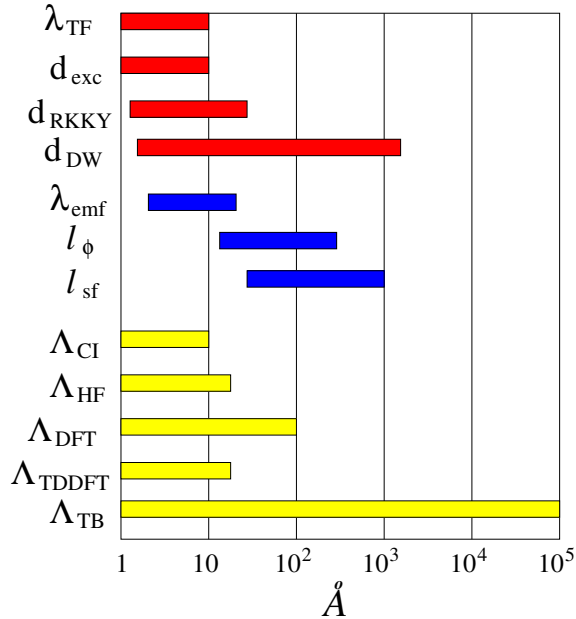}
\caption{\small{Relevant length scales for spin-transport theories. The magnetic and electronic length scales are in red, 
the transport length scales in blue and the computational are in yellow. Usually the transport scales 
change quite drastically from metals to semiconductors and here we report representative values
for metallic systems. The symbols are as following: $\lambda_\mathrm{TF}$ Thomas-Fermi screening length,
$d_\mathrm{exc}$ exchange length, $d_\mathrm{RKKY}$ exchange coupling (RKKY) length, 
$d_\mathrm{DW}$ domain wall thickness, $\lambda_\mathrm{emf}$ elastic mean free path,
$l_\phi$ phase coherence length $l_\mathrm{sf}$ spin diffusion length, 
$\Lambda_\mathrm{CI}$ range of configuration interaction (method), 
$\Lambda_\mathrm{HF}$ range of Hartree-Fock, $\Lambda_\mathrm{DFT}$ range of density functional theory,
$\Lambda_\mathrm{TDDFT}$  range of time-dependent density functional theory,
$\Lambda_\mathrm{TB}$ range of tight-binding.}}\label{fig1}
\end{center}
\end{figure}

The modeling of spin-transport at the nanoscale using {\it ab initio} methods brings together different
aspect of materials science such as magnetism, transport and electronic materials. Each one of those has
some characteristic length scales, and it is crucial to understand how these relate to each other. This 
of course allows us to identify the limit of validity of our theoretical description. Moreover it is 
important to compare these length scales with the range of applicability of modern quantum mechanical 
methods. Often we cannot describe a system just because it is ``too big'' for our computational 
capabilities. It is therefore crucial to have some understanding of the level of precision we want to achieve. 
The best method is the one offering the best tradeoff between accuracy and computational overheads for 
the problem under investigation. A summary of the length scales relevant for spin-transport 
in presented in figure \ref{fig1}.

\subsubsection{Electronic and Magnetic Lengths}

These are length scales connected with the range of the electrostatic and magnetic interactions. 

\vspace{0.3cm}
\noindent
{\it{\underline{Screening Length}}}

When a charge $Q$ is introduced in a neutral material (say a metal) this produces a perturbation 
in the existing electronic potential. In the absence of any screening effect this will add a 
coulombic-like potential to the existing potential generated by both electrons and ions. However when 
conduction electrons are present they will respond to the perturbation of their potential by
effectively screening the potential of the additional charge. Following the elementary 
Thomas-Fermi theory \cite{thomasfermi1,thomasfermi2,ashcroft,pettifor} the ``screened'' potential 
is of the form:
\begin{equation}
V(r)=\frac{Q}{r}\:\mathrm{e}^{-r/\lambda_\mathrm{TF}}\;,
\label{TFP}
\end{equation}
where $\lambda_\mathrm{TF}$ is the screening length. For a free electron gas this can be written as
\begin{equation}
\lambda_\mathrm{TF}=\left[\frac{e^2D(E_\mathrm{F})}{\epsilon_0}\right]^{-1/2}\:,
\label{TFSL}
\end{equation}
where $e$ is the electron charge, $D(E_\mathrm{F})$ the density of state (DOS) at the Fermi level 
$E_\mathrm{F}$ and $\epsilon_0$ the vacuum permittivity.
Given the large density of state at $E_\mathrm{F}$, for typical transition metals $\lambda_\mathrm{TF}$ is 
of the order of the lattice parameters 1-10~\AA. This means that the charge density of a transition 
metal is unchanged at about 10~\AA\ from an electrostatic perturbation such as a free surface
or an impurity.

\vspace{0.3cm}
\noindent
{\it{\underline{Exchange}}}

The coupling between spins in a magnetic material is governed by the exchange interaction, which 
ultimately is related to the exchange integral. This depends on the overlap between electronic 
orbitals and therefore is rather short range, usually of the order of the lattice parameter. 
Therefore the typical length scale for the exchange interaction $d_\mathrm{exc}$ is of the same order than the 
screening length. In addition in atomic scale junctions most of the atoms which are relevant for 
the transport reside close to free surfaces or form small clusters. These may present features 
at a length scale comparable with the lattice constant that are rather different from that of the bulk. 
For this reason specific calculations are needed for specific atomic arrangements and the 
extrapolation of the magnetic properties at the nanoscale from that of the bulk is often not correct.

\vspace{0.3cm}
\noindent
{\it{\underline{RKKY}}}

The spin polarization of conduction electrons near a magnetic impurity can act as an effective 
field to influence the polarization of nearby impurities. In the same way two magnetic layers 
can interact via the conduction electrons in a metallic spacer. This interaction, that is 
analogous to the well known Ruderman-Kittel-Kasuya-Yosida (RKKY) interaction 
\cite{RKKY1,RKKY2,RKKY3}, can be either ferromagnetic or antiferromagnetic depending on 
the thickness of the spacer and its chemical composition. Moreover it decay as a power law 
with the separation between the magnetic materials and it is usually negligible for length 
scale ($d_\mathrm{RKKY}$) larger that a few atomic planes $\sim10-30$\AA.

\vspace{0.3cm}
\noindent
{\it{\underline{Domain Wall}}}

The exchange interaction aligns the spins of a magnetic material. In a ferromagnet 
these are aligned parallel to each other giving rise to a net magnetization $\vec{M}$. 
In order to create this spontaneous magnetization the exchange energy must be larger that 
the magnetostatic energy $E_\mathrm{E_M}\propto\int {M}^2\:\mathrm{d}V$. Therefore it is 
energetically favorable for a ferromagnetic material to break the homogeneous magnetization 
in small regions (magnetic domains) where the magnetization is constant, but it is aligned 
in a different direction with respect to that of the neighboring regions. This reduces the 
magnetostatic energy without large energy costs against the exchange. The regions separating 
two domains, where the spins change their orientation, are called domain walls. The thickness of a 
domain walls $d_\mathrm{DW}$ depends critically on the material and its anisotropy \cite{nicola}. 
At the nanoscale the structure of a domain wall can be rather different from that in the bulk. 
In particular in a nanoconstriction the domain wall width is predicted to be of the same order
of the lateral size of the constriction itself \cite{brunodw}. This means that in an atomic point contact
domain walls as thick as a single atomic plane can form.

\subsubsection{Transport Lengths}

These are length scales connected to the motion of the electrons in a material. In contrast to the previous
scales that are due to static effects, these arise from the electron dynamics.

\vspace{0.3cm}
\noindent
{\it{\underline{Elastic mean free path}}}

It is the average distance traveled by an electron (or a hole) before changing its momentum.
The momentum change is given usually by scattering to impurities, to interfaces or generally is 
associated to any other scattering mechanism that does not change the electron energy. Since the 
electron energy is conserved the phase of the electron wave-function is also conserved. This means
that transport processes occurring at length scales shorter or comparable to the elastic mean free
$\lambda_\mathrm{emf}$ path are sensitive to quantum mechanical interference effects such as weak 
localization \cite{kramer_loc}. The elastic mean free path depends strongly on the metalicity of
a material (longer in semiconductors) and on its purity. In magnetic transition metals it is usually
of the order of a few atomic planes $\lambda_\mathrm{emf}\sim$10-20\AA.

\vspace{0.3cm}
\noindent
{\it{\underline{Phase coherence length}}}

This is the average distance that an electron (hole) travels before undergoing to a scattering 
event that change its energy. A few examples of these events are scattering to lattice
vibrations, electron-electron scattering or spin-wave scattering in the case of magnetic materials.
Note that all these processes allow energy exchange between the electrons and 
other degrees of freedom, therefore the phase coherent length $l_\phi$ is the length
that characterizes the bulk resistivity of a material $\rho$. According to the elementary Drude
theory this can be written as:
\begin{equation}
\rho=\frac{mv_\mathrm{F}}{ne^2l_\phi}\;,
\end{equation}
where $n$ is the electron density, $e$ and $m$ the electron charge and mass, and $v_\mathrm{F}$
the Fermi velocity \cite{ashcroft}.

Since energy non-conserving scattering also does not preserve the quantum-mechanical phase of 
the electron wave-function, one should not expect interference effects for lengths exceeding
$l_\phi$. Since not all the scattering event are inelastic generally $l_\phi>\lambda_\mathrm{emf}$ 
and varies largely with the sample composition, the metalicity and the temperature. 
In typical transition metal heterostructures at low temperature it is of the order
of several lattice spacing $l_\phi\sim$100--200~\AA\ \cite{in_mfp}.

\vspace{0.3cm}
\noindent
{\it{\underline{Spin diffusion length}}}

In contrast to the other quantities that are related to the current, the spin diffusion length
(sometime known as the spin-flip length) $l_\mathrm{sf}$ is related to the spin-current.
$l_\mathrm{sf}$ is defined as the average distance that an electron travel before losing 
``memory'' of its spin direction. Clearly if the typical length scale of a device is smaller
than the spin diffusion length, then each spin current can be treated as independent and
spin-mixing can be ignored. This approximation, introduced by Mott \cite{2current_mott}, is 
usually called the two-spin current model.

Many factors can affect the spin diffusion length, such as
spin-orbit interaction, scattering to magnetic impurities or spin-wave scattering in the case of
magnetic materials, and $l_\mathrm{sf}$ can change largely from material to material. In magnetic 
transition metals it can be of the order of 10$^3$~\AA\ \cite{sdl_Co}, but it reduces drastically in
magnetic permalloy $l_\mathrm{sf}\sim$50~\AA\ \cite{sdl_py}. Finally it is worth noting that 
one can observe extremely long spin diffusion length ($\sim$10$^6$\AA) in ordinary semiconductors
at low temperature \cite{Aws2}.

\subsubsection{Computational length scales}

Generally any solid state computational technique has a range of applicability. This is primarily
connected to the scaling properties of computational method. However other factors as
the particular numerical implementation, the availability of highly optimized numerical libraries,
or the possibility for parallelization are usually important. 
{\it Ab initio} transport methods interface numerical methods for electronic transport with
accurate electronic structure techniques. The most demanding part of a typical {\it ab initio}
transport algorithm is usually the electronic structure part, which sets the number of degrees of
freedom (number of basis functions) that the method can handle. Here we list the present and proposed 
electronic transport methods to be used in transport algorithms.

\vspace{0.3cm}
\noindent
{\it{\underline{Configuration interaction}}}

This is a method to calculate the excitation properties of material systems. The main idea is 
to expand the energy and the eigenfunction of a system of $N$  interacting particles over a finite 
number of $N$ particles non-interacting configurations. The scaling of this method with the number
of atoms is very severe (some high power law) and only small systems containing not more than 10
atoms can be tackled on ordinary computers. The characteristic length scale is therefore of the
order of the atomic spacing $\Lambda_\mathrm{CI}\sim$1--10~\AA. An attempt to apply this method 
to transport properties has been recently proposed \cite{pauljim}. 

\vspace{0.3cm}
\noindent
{\it{\underline{Hartree-Fock}}}

The Hartree-Fock approach is one of the numerous electronic structure methods based on the mean 
field approximation, hence electron-electron interaction is treated in an approximate way. 
It is a wavefunction-based method, where the only electron correlation enters through the
exchange interaction. For this reason the electronic gaps are largely overestimated. 
Usually the computational overheads of the Hartree-Fock scheme scale as $N^4$, where $N$ is the
number of atoms of the system under investigation. Therefore the typical length scale 
$\Lambda_\mathrm{HF}$ is again of the order of a few lattice constants 
$\Lambda_\mathrm{HF}\sim$5--20~\AA. A few Hartree-Fock-based quantum transport methods are available
at present \cite{HF_Ratner}.

\vspace{0.3cm}
\noindent
{\it{\underline{Density Functional Theory}}}

Density functional theory (DFT) is a non-wavefunction-based method for calculating electronic
structures. It is based on the Hohenberg-Kohn theorem stating that the ground state energy of a
system of $N$ interacting particles is a unique functional of the single particle charge density
\cite{HKohn}. The prescription to calculate both the charge density and the total energy is that to
map the exact functional problem onto a fictitious single-particle Hamiltonian problem, known as the
Kohn-Sham Hamiltonian \cite{KSham}. I will discuss extensively the use of this method for 
calculating transport in the next sections. 

A typical DFT calculation scales as $N^3$; order-$N$ methods are available \cite{pabloON} although
at present these are not implemented for spin-polarized systems. A large number of implementations 
exists at present and calculations involving between 100 and 1000 atoms are not uncommon. Therefore 
I assign to DFT a characteristic length scale of $\Lambda_\mathrm{DFT}\sim\:$100~\AA.

\vspace{0.3cm}
\noindent
{\it{\underline{Time Dependent Density Functional Theory (TDDFT)}}}

This is a time dependent generalization of DFT \cite{TDDFT_Gross}. 
It can be viewed as an alternative formulation of time-dependent quantum mechanics, where 
the basic quantity is the density matrix and not the wavefunction. The core
of the theory is the Runge-Gross theorem \cite{Runge_Gross}, which is the time-dependent
extension of the Hohenberg-Kohn theorem. The scaling of this method is similar to that of ordinary
DFT, although there is not a clear pathway to order-$N$ scaling yet. For this reason the typical
length scale if $\Lambda_\mathrm{TDDFT}\sim\:$20~\AA.

One of the benefit of TDDFT over static DFT is the ability to describe excitation spectra, hence it
appears very attractive for transport properties. At present a few schemes have been proposed
\cite{baer,burke_car}, although a robust TDDFT-based transport code is still not available.

\vspace{0.3cm}
\noindent
{\it{\underline{Tight-binding method}}}

These are semi-empirical methods design to handle large systems. The main idea is to expand the 
wavefunction over a linear combination of atomic orbitals and express the Hamiltonian in terms
of a small subset of parameters \cite{sutton}. These can then be calculated or simply fitted 
from experiments. For this reason the method usually is not self-consistent (although
self-consistent versions are available \cite{tbdft,Alex_tb}) and the scaling can be linear in the 
number of atoms.

There is a vast amount of literature over tight-binding methods for transport (for a review see 
\cite{todorov_tdtb}) and in the linear response limit the method can be used for infinitely long
systems with typical sub-linear running-time scaling \cite{ss_1}. For this reason I fix the length
scale for tight-binding methods to $\Lambda_\mathrm{TB}\sim\:$10$^6$~\AA\ (corresponding to
approximately 10$^5$ atoms).

\subsection{Spin-polarization of a device}
\label{spoad}

In this section I will discuss a few general concepts common to transport in magnetic
devices and how the spin-polarization of the materials forming the device affects the
magneto-transport properties. 

\subsubsection{Band structure of a magnetic transition metal}

Before discussing the main transport regimes in magnetic materials it is useful to recall the general
electronic properties of a transition metal and in particular of a magnetic transition metal (for a more 
complete review see for instance \cite{gehring}). 
\begin{figure}[ht]
\begin{center}
\includegraphics[width=4.0cm,clip=true]{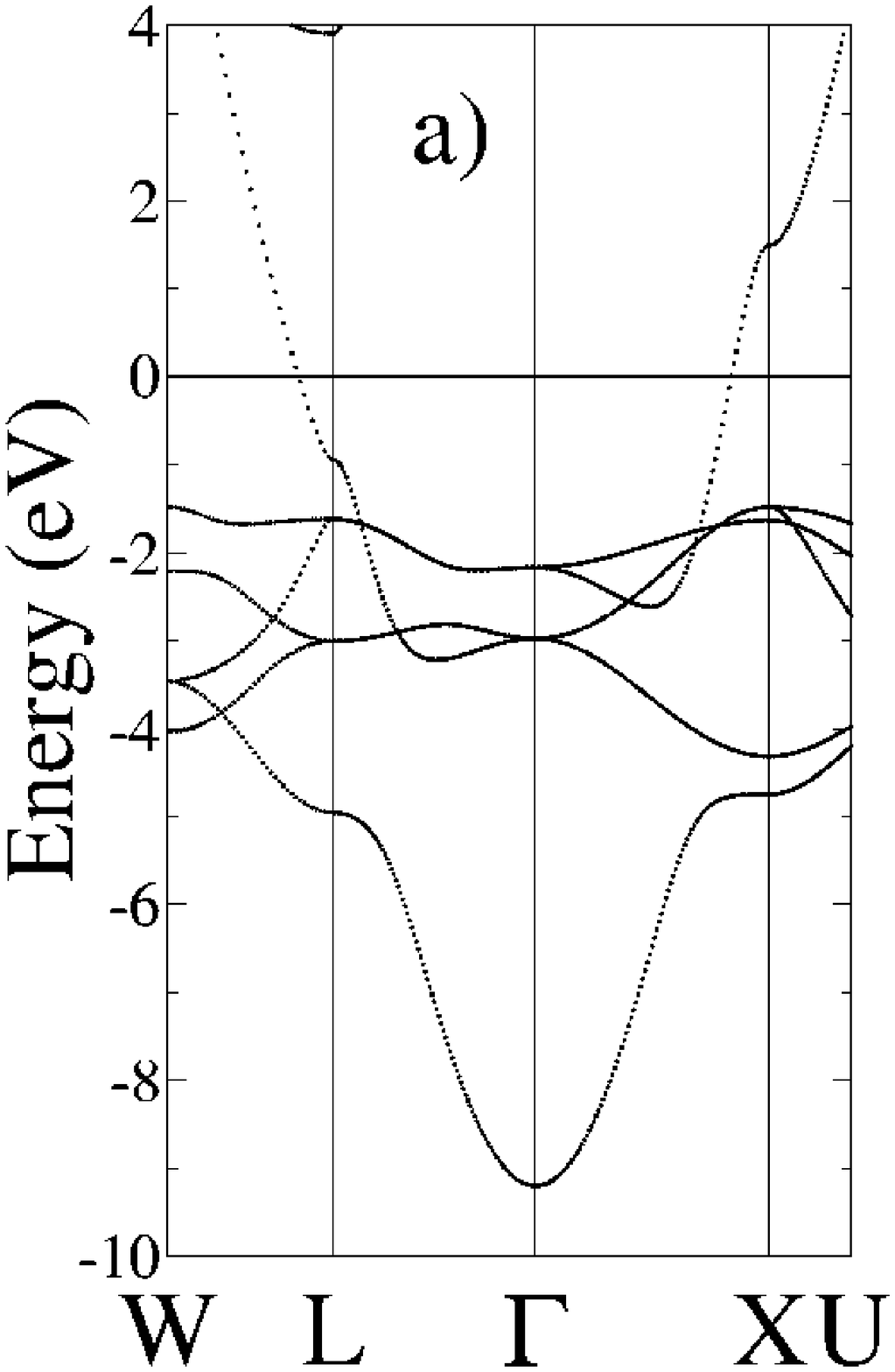}
\hspace{0.4cm}
\includegraphics[width=5.0cm,clip=true]{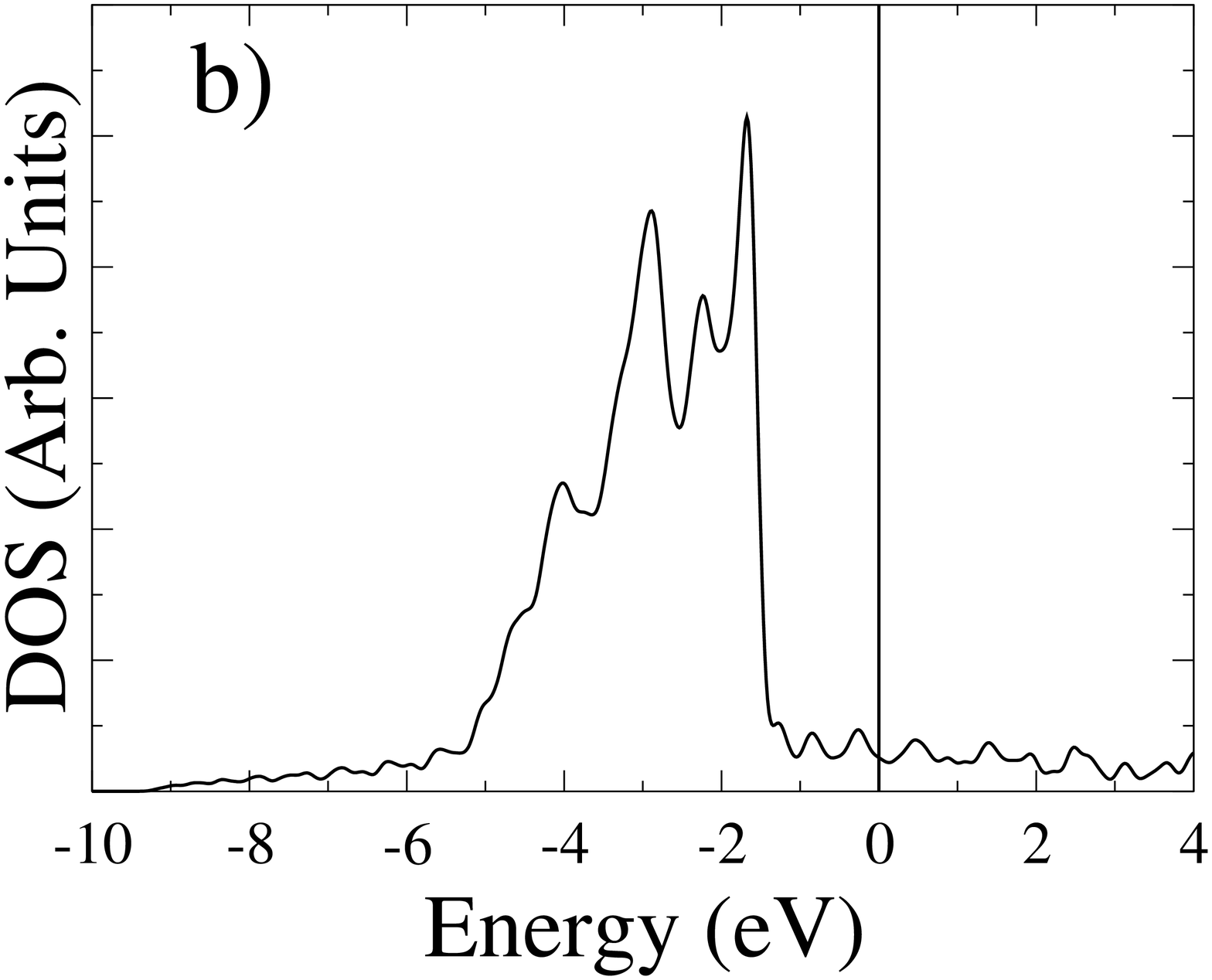}
\hspace{0.4cm}
\includegraphics[width=4.0cm,clip=true]{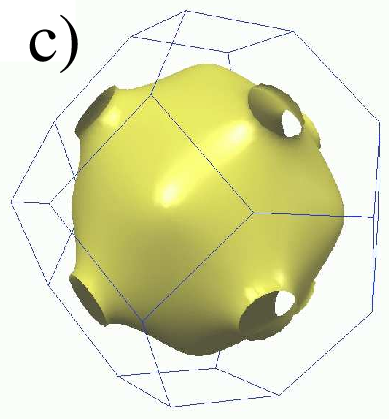}
\caption{a) Band structure, b) density of states, and c) Fermi surface for fcc Cu. a) and b) have been calculated
with density functional theory using the code SIESTA \cite{siesta}. c) has been obtained from
an $spd$ tight-binding Hamiltonian \cite{Fermi_per}.}\label{fig2a}
\end{center}
\end{figure}

The band structure, the corresponding DOS and the Fermi surface of fcc Cu is
presented in figure \ref{fig2a} (the Fermi surface is from reference \cite{Fermi_per}). 
The main feature of the dispersion of Cu is the presence 
of two rather different bands. The first is a high dispersion broad band with a minimum at 
the $\Gamma$ point $\sim9$~eV below the Fermi level, which then re-emerges at the edge of 
the Brillouin zone at the $L$ and $X$ points. The second is a 
rather narrow $\sim3$~eV wide band that extends through the entire Brillouin zone 1.5~eV below 
$E_\mathrm{F}$. By analyzing the orbital content of these two bands
within a tight-binding scheme \cite{slater_koster} one can attribute the broad band to 
$s$ and the narrow one to $d$ electrons. This is in agreement with the intuitive picture 
of the $d$ electrons much more tightly located around the atomic core and with an atomic-like 
$d^{10}$ configuration. Clearly there is hybridization for energies around the position of 
the $d$ band, which creates band distortion. However the hybridization occurs well below 
$E_\mathrm{F}$ and therefore the Fermi surface is entirely dominated by $s$ electrons and 
appears rather spherical (figure \ref{fig2a}c).

The situation is rather different in ferromagnetic metals. Since the nominal atomic configuration 
of the $d$ shell is $3d^8$, $3d^7$ and $3d^6$ respectively for Ni, Co and Fe the Fermi level 
of an hypothetical paramagnetic phase lays in a region of very high density of states. For this 
reason the material is Stoner unstable and develops a ferromagnetic ground state. In this case 
the electron energies for up and down spins are shifted with respect to each other by a constant 
splitting $\Delta=\epsilon_{\vec{k}\downarrow}-\epsilon_{\vec{k}\uparrow}$, where $\Delta$ is 
approximately 1.4~eV in Fe, 1.3~eV in Co and 1.0~eV in Ni. 
More sophisticate DFT calculations show that the simple Stoner picture is a good approximation 
of the real electronic structure of Ni, Co and Fe. 

\begin{figure}[ht]
\begin{center}
\includegraphics[width=5.5cm,clip=true]{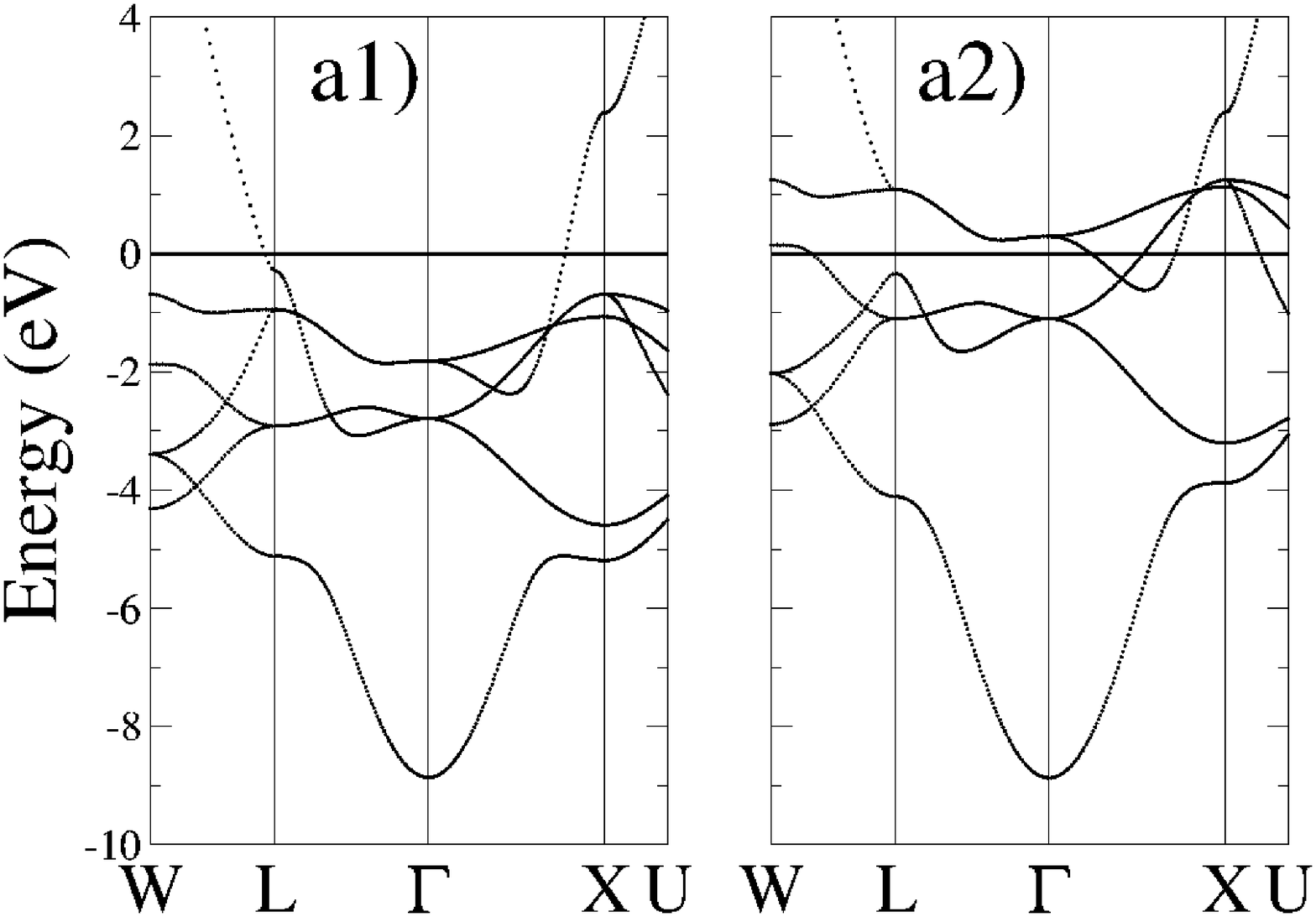}
\hspace{0.3cm}
\vspace{-0.3cm}
\includegraphics[width=5.0cm,clip=true]{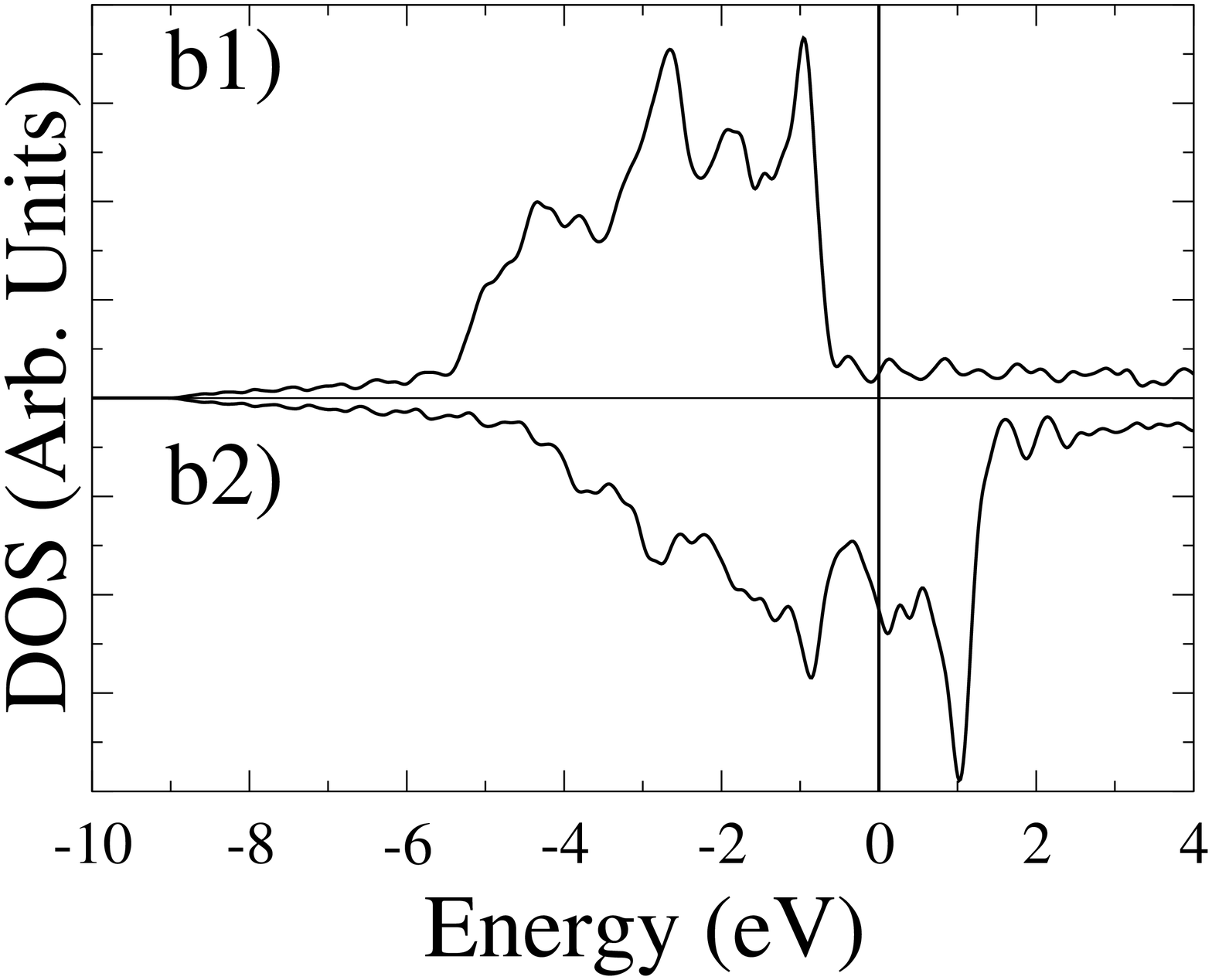}
\hspace{0.3cm}
\includegraphics[width=5.0cm,clip=true]{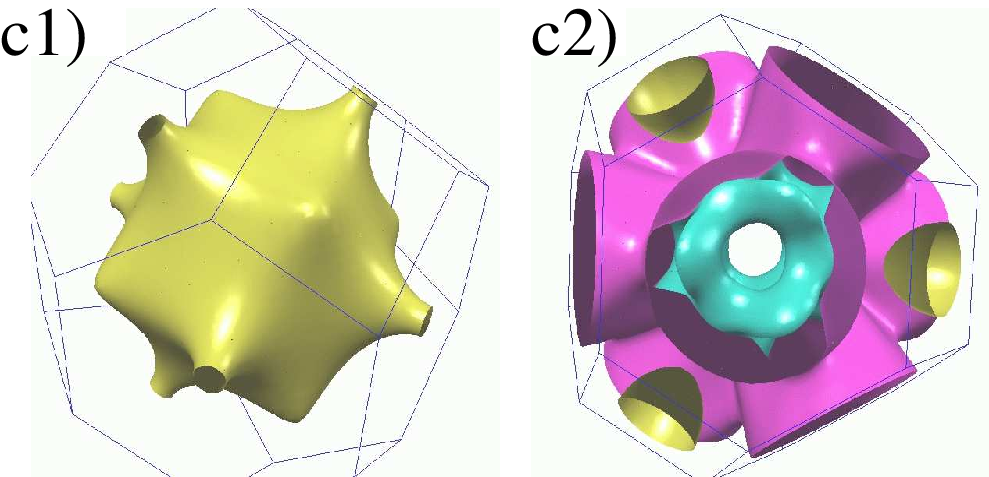}
\caption{a) Band structure, b) density of states, and c) Fermi surface for fcc Co. The figures a1), b1) and c1)
refers to the majority spin electrons, while a2), b2) and c2) to the minority. The pictures have been 
obtained with density functional theory using the code SIESTA \cite{siesta}, and from an
$spd$ tight-binding Hamiltonian \cite{Fermi_per}.}\label{fig2b}
\end{center}
\end{figure}

The main consequence of this electronic structure on the transport properties comes from the 
fact that the DOS at the Fermi level, and therefore the entire Fermi surface, for the up spins 
(usually called the majority spin band) is rather different from that of the down spins 
(minority spin band). This difference is more pronounced in the case of strong ferromagnet, 
where only one of the two spin bands is entirely occupied. An example of this situation is 
fcc Co (the high temperature phase), whose electronic structure is presented in figure \ref{fig2b}. 
It is important to observe that the majority spin band is dominated at the Fermi level by $s$ 
electrons, while the minority by $d$ electrons. With this respect the electronic structure (bands, 
DOS and Fermi surface) of the majority spin band looks remarkably similar to that of Cu. 

Finally it is worth mentioning that there are materials that at the Fermi level present a 
finite DOS for one spin specie and a gap for the other. These are known as half-metals 
\cite{ss_2} and are probably among the best candidates as materials for future magneto-electronics 
devices.

\subsubsection{Basic transport mechanism in a magnetic device}

\begin{figure}[ht]
\begin{center}
\includegraphics[width=6.5cm,clip=true]{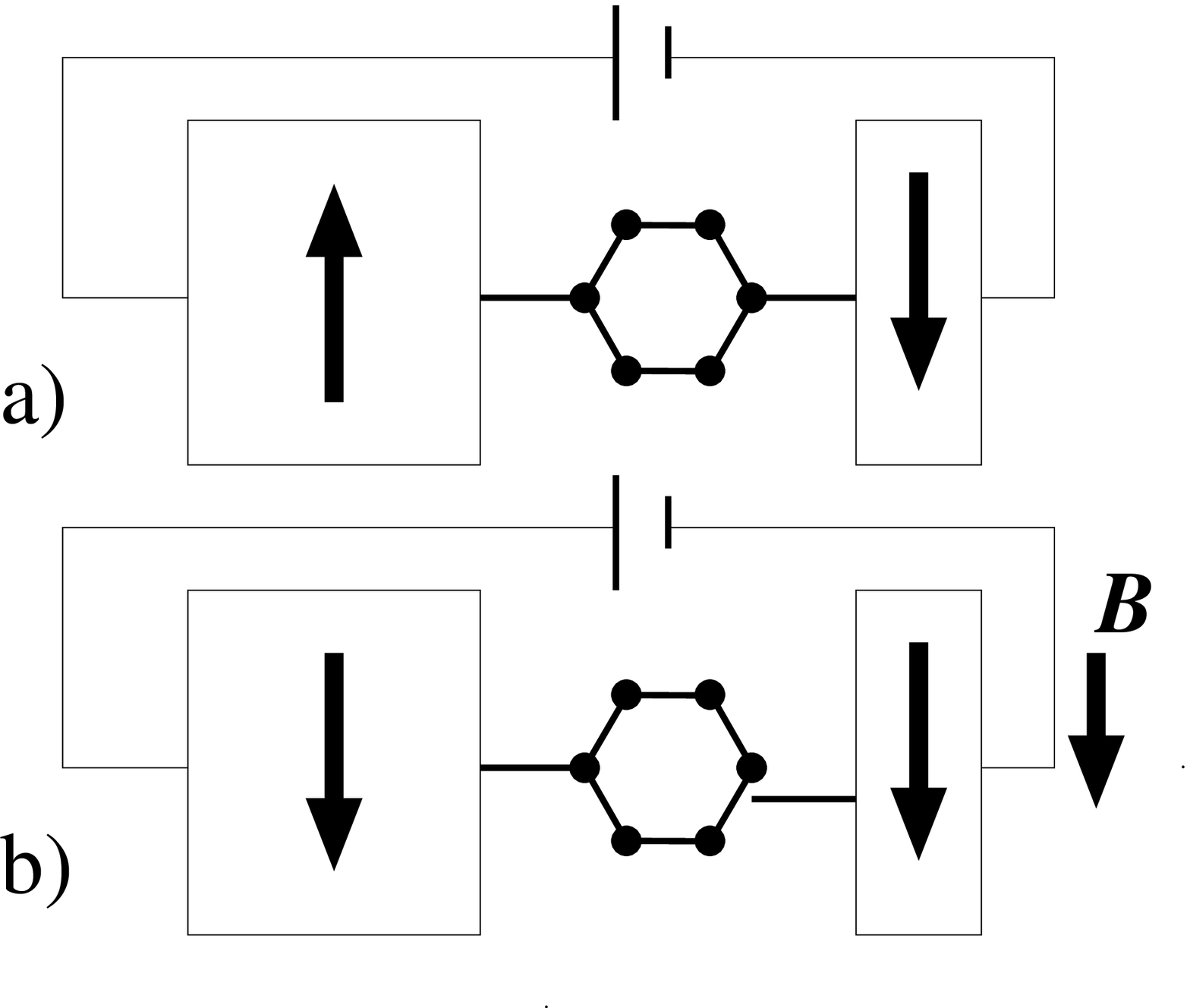}
\caption{Scheme of a spin-valve in the two resistance states: a) high resistance, b) low resistance. The arrows
indicate the direction of the magnetization vectors in the two magnetic layers. In this case we have
taken the spacer to be an organic molecule, a more conventional choice can be a transition metal 
or an insulator.}\label{fig2}
\end{center}
\end{figure}

Let us consider the prototypical GMR device: the spin-valve. A spin valve is formed by two magnetic
layers separated by a non-magnetic spacer. Usually the magnetic layers are metallic (typically Co,
Ni, Fe or some permalloy), while the spacer can be either a metal, a semiconductors, an insulator or
a nanoscale object such as a molecule or an atomic constriction. The typical operation of a
spin-valve is schematically illustrated in figure \ref{fig2}. Usually the two magnetic layers have a
rather different magnetic anisotropy with one layer being strongly pinned and the other free to
rotate along an external magnetic field. In this way the magneto-transport response of the device can be 
directly related to the direction of the magnetization of the free layer. In our discussion we consider 
only the two extreme cases in which the two magnetization vectors are either parallel (P) or antiparallel 
(AP) to each other. 

Here I will describe the current flowing perpendicular to the plane (CPP) since this is the relevant
current/voltage configuration for transport through a nanoscaled spacer. However in
the case of metallic spacer another possible setup is with the current flowing in the plane 
(CIP), as in the present generation of read/write heads used in magnetic data storage devices. 
As a further approximation I assume the two spin current model \cite{2current_mott}. This is justified 
by the fact that a typical CPP spin-valve is usually shorter that the spin-diffusion length. 

\begin{figure}[ht]
\begin{center}
\includegraphics[width=8.0cm,clip=true]{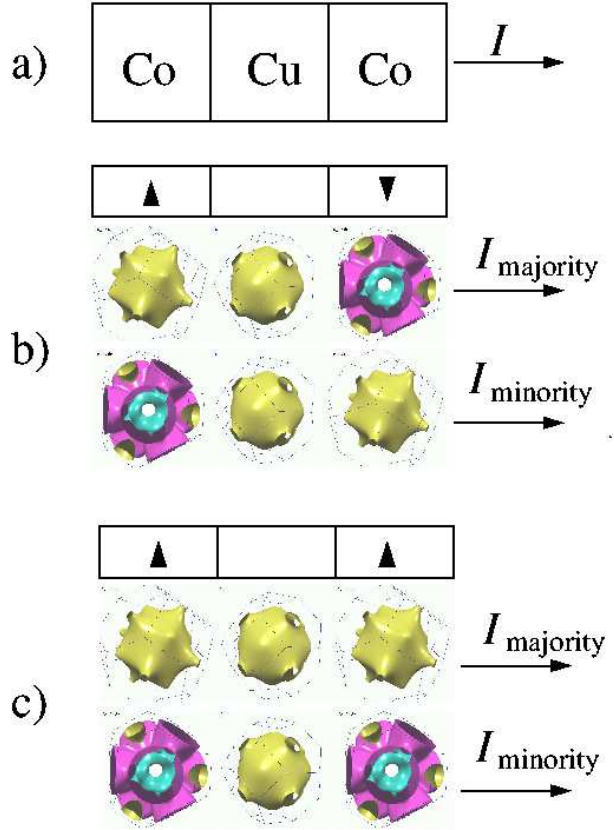}
\caption{Magnetoresistance mechanism in a Co/Cu/Co spin valve (panel a) in the two spin current approximation.
In the antiparallel state b) the alignment of the Fermi surface is such to have one high resistance interface
for both the spin channels. This is given by the interface between the minority spin band of Co and that of 
Cu. In the parallel case c) the resistance of the majority spin channel is considerably lowered since there
is good match of all the Fermi surfaces across the entire device. This high conductance channel is 
responsible for the GMR effect.}\label{fig3}
\end{center}
\end{figure}

To fix the idea consider a Co/Cu/Co spin-valve, and let us follow the path of both the electron 
spin species across the device. The Fermi surfaces line up for both the P and AP cases are 
presented in figure \ref{fig3}.
In the AP case the magnetization vector of the two magnetic layers points in opposite directions. 
This means that an electron belonging to the majority band in one layer, will belong to the 
minority in the other layer. Consequently in the AP case both the spin currents (usually called 
the spin channels) arise from electrons that have traveled within the Fermi surface of Cu and 
of both spins of Co. In contrast in the P case the two spin currents are rather different. The 
spin up current is made from electrons that have traveled within the Fermi surfaces of Cu and 
of the majority spin Co, while the down spin current 
from electrons that have traveled within the Fermi surfaces of Cu and of the minority spin Co. 

If we naively assume that the total resistance of the device can be obtained by adding in series 
the resistances of the materials forming the device (resistor network model) we obtain:
\begin{equation}
R_\mathrm{AP}=\frac{1}{2}\left(R^\mathrm{Co}_\uparrow+R^\mathrm{Co}_\downarrow+
R^\mathrm{Cu}\right)\;,
\end{equation}
\begin{equation}
R_\mathrm{P}=\left(\frac{1}{2R^\mathrm{Co}_\uparrow+R^\mathrm{Cu}}+
\frac{1}{2R^\mathrm{Co}_\downarrow+R^\mathrm{Cu}}\right)^{-1}
\;,
\end{equation}
where $R_\mathrm{P}$ and $R_\mathrm{AP}$ are the resistance for the parallel and antiparallel 
configuration respectively, $R^\mathrm{Cu}$ is the resistance of the Cu layer and 
$R^\mathrm{Co}_\uparrow$ and $R^\mathrm{Co}_\downarrow$ are the resistance of the Co layer for 
the majority ($\uparrow$) and the minority ($\downarrow$) spins. Usually 
$R^\mathrm{Co}_\uparrow\ll R^\mathrm{Co}_\downarrow$, hence $R_\mathrm{P}<R_\mathrm{AP}$. 
This produces the GMR effect.

Conventionally the magnitude of the effect is given by the GMR ratio $r_\mathrm{GMR}$ defined as:
\begin{equation}
r_\mathrm{GMR}=\frac{R_\mathrm{AP}-R_\mathrm{P}}{R_\mathrm{P}}\;.
\end{equation}
This is usually called the ``optimistic'' definition (since it gives large ratios). 
An alternative definition is obtained by normalizing the resistance difference by 
either $R_\mathrm{AP}$ or $R_\mathrm{P}+R_\mathrm{AP}$; in this last case
$r_\mathrm{GMR}$ is bounded between 0 and 1.

The discussion so far is based on the hypothesis of treating the spin-valve as a resistor network. 
This is strictly true only if $\lambda_\mathrm{emf}<l_\phi<L$, where $L$ is the typical 
thickness of the layers forming the spin-valve, but in general adding resistances in 
series may not be correct. However it is also clear that the magnitude of the magnetoresistance 
depends critically on the asymmetry of the two spin currents in the magnetic material, which ultimately 
depends on its electronic structure. It is therefore natural to introduce the concept of spin 
polarization $P$ of a magnetic metals as
\begin{equation}
P=\frac{I_\uparrow-I_\downarrow}{I_\uparrow+I_\downarrow}\:,
\label{Pdef}
\end{equation}
where $I_\sigma$ is the spin-$\sigma$ contribution to the current. Clearly $I_\sigma$ and $P$ are not 
directly observable and must be calculated or inferred from an indirect measurement. 
Unfortunately the way to relate the spin-current $I_\sigma$ to the electronic properties of a material 
is not uniquely defined and depends on the particular experiment. This is why it is 
crucial to analyze the equation (\ref{Pdef}) for different situations.

\subsubsection{Diffusive Transport}

As brilliantly pointed out by Mazin \cite{Mazin_PRL}, the relation between the 
spin-polarization of a magnetic material and its electronic structure depends critically 
on the transport regime that one considers. Let us start by looking at the diffusive 
transport. Here the phase coherence length is rather short and quantum
interference is averaged out. The transport is then described by the Boltzmann's equations, 
which govern the evolution of the electron momentum distribution function
\cite{BOEqs}. Within the relaxation time approximation \cite{ashcroft}, assuming that the relaxation times
does not depend on the electron spin the current is simply proportional to $N_\mathrm{F}
v_\mathrm{F}^2$, where $N_\mathrm{F}$  and $v_\mathrm{F}$ are the density of states at the Fermi
level and the Fermi velocity respectively. 

This leads us to the ``$Nv^2$'' definition of the spin-polarization:
\begin{equation}
P_{Nv^2}=\frac{N_\mathrm{F}^\uparrow v_\mathrm{F}^{\uparrow^2}-
N_\mathrm{F}^\downarrow v_\mathrm{F}^{\downarrow^2}}
{N_\mathrm{F}^\uparrow v_\mathrm{F}^{\uparrow^2}+
N_\mathrm{F}^\downarrow v_\mathrm{F}^{\downarrow^2}}\:.
\label{Pnv2}
\end{equation}
Values of $P_{Nv^2}$ for typical transition metals are reported in table \ref{Table1} \cite{CrO,
LCMO,TMO}.
\begin{table}[ht]
\begin{center}
\begin{tabular}{lccc}
\hline
\hline \\[-0.2cm]
  & $P_{Nv^2}$ (\%) & $P_{Nv}$ (\%) & $P_{N}$ (\%)  \\[0.1cm]
\hline \\[-0.2cm]
Fe      			& 20 & 30 & 60 \\
Ni      			& 0 & -49 & -82 \\
CrO$_2$ 			& 100 & 100 & 100 \\
La$_{0.67}$Ca$_{0.33}$MnO$_3$ 	& 92 & 76 & 36 \\
Tl$_2$Mn$_2$O$_7$		& -71 & -5 & 66 \\[0.2cm]
\hline
\hline
\end{tabular}
\end{center}
\caption{Spin-polarization of typical magnetic metals according to the various definitions given
in the text. The data are taken from literature as follows: Ni and Fe \cite{Mazin_PRL}, 
CrO$_2$ \cite{CrO}, La$_{0.67}$Ca$_{0.33}$MnO$_3$ \cite{LCMO} and Tl$_2$Mn$_2$O$_7$
\cite{TMO}.}
\label{Table1}
\end{table}

\subsubsection{Ballistic Transport}
\label{224}

In this case $l_\phi$ is much longer than the size of the magnetic device. The 
energy is not dissipated as resistance in the device and the current can be calculated using 
the Landauer formalism \cite{landauer_original1,landauer_original2,multi_channel}. 
I will talk extensively about this approach in the next sections while here I just wish 
to mention that in the ballistic limit the
conductance, and hence the current, are simply proportional to $N_\mathrm{F}v_\mathrm{F}$. 
Moreover in the Landauer approach, as we will see, the electron velocity and the density of states 
exactly cancel. This means that $N_\mathrm{F}v_\mathrm{F}$ is just an integer proportional to
the number of bands crossing the Fermi level in the direction of the transport, or alternatively to
the projection of the Fermi surface on the plane perpendicular to the direction of the transport
(for a rigorous derivation see reference \cite{sharvin,sharvin2}).

This leads to the ``$Nv$'' definition of spin-polarization
\begin{equation}
P_{Nv}=\frac{N_\mathrm{F}^\uparrow v_\mathrm{F}^{\uparrow}-
N_\mathrm{F}^\downarrow v_\mathrm{F}^{\downarrow}}
{N_\mathrm{F}^\uparrow v_\mathrm{F}^{\uparrow}+
N_\mathrm{F}^\downarrow v_\mathrm{F}^{\downarrow}}\:.
\label{Pnv}
\end{equation}

\subsubsection{Tunneling}

It is generally acknowledged that in tunneling experiments the GMR ratio of the specific device
is given by some density of states. This was firstly observed by Jullier almost three decades ago 
\cite{jullier} and it is based on the fact that typical tunneling times are much faster then $L/v_\mathrm{F}$,
with $L$ the length of the tunneling barrier. This means that the electron velocity in the metal
is irrelevant in the tunneling process. Although it is now clear that the relevant density of states 
for magneto-tunneling processes is not necessarily that of the bulk magnetic metal, but it must
take into account of the structure of the tunneling barrier and of the bonding between the barrier
and the metal \cite{tsymbal}, we can still introduce the ``$N$'' definition of polarization:
\begin{equation}
P_{N}=\frac{N_\mathrm{F}^\uparrow-
N_\mathrm{F}^\downarrow}
{N_\mathrm{F}^\uparrow+
N_\mathrm{F}^\downarrow}\:.
\label{Pn}
\end{equation}

Clearly the three definitions may give rise to different spin-polarizations, since the relative
weight of $N$ and $v$ is different. In particular $P_N$ favors electrons
with high density, while $P_{Nv^2}$ electrons with high mobility. In magnetic transition
metals, where high mobility low density $s$ electrons coexist with low mobility high density
$d$ electrons, these differences can be largely amplified. In principle one can speculate around 
materials that are normal metals according to one definition and half-metals according to another.
This is for instance the case of La$_{0.7}$A$_{0.3}$MnO$_3$ with A=Ca, Sr, .. 
\cite{ss_2}, in which the majority band is dominated by delocalized e$_g$ states
and the minority by localized t$_{2g}$ electrons. Therefore La$_{0.7}$A$_{0.3}$MnO$_3$ is a conventional ferromagnet 
according to the definitions $P_{Nv}$ and $P_{N}$ and it is an half-metal according to $P_{Nv^2}$.

\subsubsection{Andreev Reflection}

Andreev reflection is the relevant scattering mechanism for sub-gap transport across
an interface between a normal metal and a superconductor \cite{Andreev}. The main idea is that
an electron incising the interface from the metal side, can be reflected as a hole with opposite
spin leaving a net charge of $2e$ inside the superconductor. In the case of normal metal the efficiency of
this mechanism is solely given by the transparency of the interface. In contrast for magnetic metals,
the Fermi surface of the incident electron and that of the reflected hole are different since their
spins are opposite. This leads to a suppression of the Andreev reflection and therefore
to a way for measuring the spin polarization of a magnetic metal. 

Unfortunately also in the case of Andreev reflection it is not easy to relate the measured polarization
with the electronic structures. In the case of a tunneling barrier between the magnetic metal and the
superconductor, then the spin polarization is that given by $P_{Nv^2}$ and indeed the values
obtained from Andreev measurements \cite{TedrovM} agree rather well with those estimated from
diffusive GMR experiments \cite{sdl_Co}.

In contrast in the case of ballistic junctions the situation is more complicated. Intuitively one would guess 
that the degree of suppression of the Andreev reflection in a magnetic metal is proportional to the
overlap between the Fermi surfaces of the two spin-species \cite{Beenaker}. In reality the transmittivity 
of the interface and the nature of the bonding at the interface enters in the problem \cite{Mazin_PRL}
and an estimate requires an accurate knowledge of the details of both the materials and the interface
\cite{ss_3}

\subsubsection{The spin-injection problem}

Spin injection is one of the central concepts of spintronics in semiconductors \cite{spintronics_wolf}.
The main idea is to produce some spin-polarization of the current flowing in a normal semiconductor by
injecting it from a magnetic material. This is of course very desirable, since the spin-lifetime in 
semiconductors is extremely long, and coherent spin manipulation is envisioned. 

From a fundamental point of view one has the problem to transfer spins across systems with very 
different electronic properties. In particular the huge difference in DOS at the Fermi level
between an ordinary semiconductor and a magnetic transition metal, has the important consequence
that the two materials show very different conductivities. This sets a fundamental limitation to 
spin-injection, or at least to the production of semiconductor-based GMR-like devices. In fact, if one describes 
the magnetic response of such devices in terms of the resistor model, it is easy to see the large spin 
{\it independent} resistance of the semiconductor will dominate over the small spin {\it dependent}
resistances of the magnetic metals. Clearly the total resistance will be very weakly spin-dependent
(for a more formal demonstration see reference \cite{Schmit}). 

Although several schemes have been proposed to overcome this problem including spin-dependent
barriers at the metal/semiconductor interface \cite{Rashba1}, and ballistic devices
\cite{Rashba2}, spin-injection from magnetic metals remains to date rather elusive \cite{Roukes}. 
A significant breakthrough comes with the advent of diluted magnetic semiconductors since all 
semiconductor devices can be made, avoiding any resistance mismatch. Indeed spin-injection
in all-semiconductor devices has been demonstrated \cite{Aws6,Molenkamp}.

\subsubsection{Crossover between different transport regimes}

The classification of the different transport regimes that I have provided so far
should not be taken as completely rigorous and one may imagine situations where the
electrons in a device behave ballistically at low temperature and diffusively at
high temperature. In this cases, unless a microscopic theory is available, it becomes rather 
difficult to correlate the electronic structure of the device with its transport properties. 
For instance consider a magnetic multilayer at a temperature such that the phase
coherent length $l_\phi$ is longer than the individual layer thickness. In this case 
the transport will be ballistic up to $l_\phi$ and therefore it will depend on the electronic 
structure of as many layers as those comprised in one phase coherent length. These may
include interfaces between different materials. Thus the transport will not
depend on the electronic properties of the constituent materials individually, nor on the
electronic structure of the whole device. 

These are among the most difficult situations that a theory can address, since some hybrid 
methods crossing different transport regimes are needed.
At present an efficient {\it ab initio} theory for spin-transport, and in general for quantum
transport, capable to span across different length scales is not available.

\subsection{Spin-transport at the atomic level}

Most of the concepts introduced in the previous sections are usually a good starting point for
describing spin-transport at the atomic scale. The main idea is now to shrink the dimensions of
the device in such a way that its sensitive part will be of a size comparable with the Fermi 
wave-length. In this case the transport is ballistic and depends critically on the 
entire device. Therefore it can be hardly inferred from the properties of its components, such as 
the spin-polarization of the current/voltage electrodes. 

Let us use again the spin-valve as a prototypical example, and consider two magnetic bulk 
contacts separated by an atomic scale object. This can be a point contact or for instance a 
molecule. There are two main differences with respect to the bulk case: 1) the Fermi surface
of the spacer can be highly degenerate, in the extreme limit collapsing into a single point, 
2) the coupling between the magnetic surfaces and the spacer can be strongly orbital
dependent. The crucial point is that in both cases the transport characteristics will be
given by {\it local} properties of the Fermi surfaces of the magnetic material, which means
either from a particular region in $k$-space, or a particular orbital manifold.

\begin{figure}[ht]
\begin{center}
\includegraphics[width=7.5cm,clip=true]{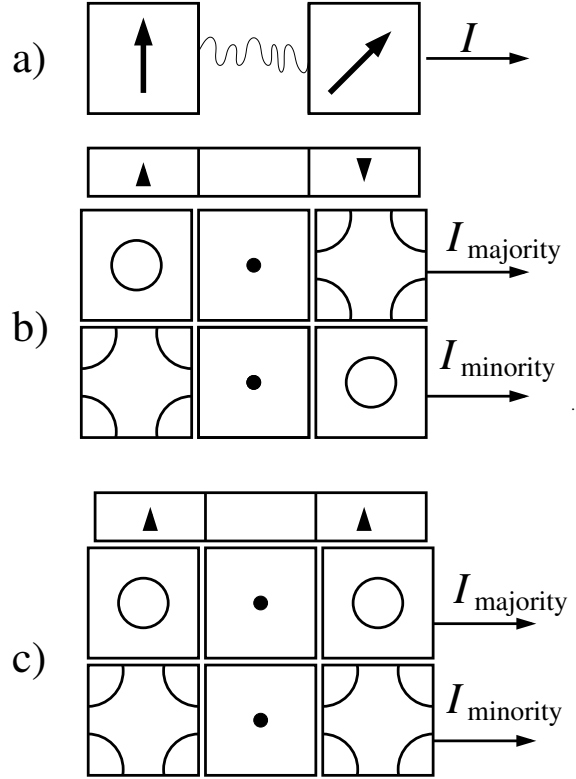}
\caption{Magnetoresistance mechanism in a spin valve constructed with a atomic scaled spacer (panel a)
in the two spin current approximation. In the antiparallel state b) the alignment of the Fermi 
surfaces generates ``infinite'' resistance for both the spin channels. This is given 
by the interface between the minority spin band and that of the atomic scale object. The two surfaces
infact has no overlap and no electrons can be transmitted in the ballistic regime.
In the parallel case c) there is transport in the majority channel giving rise to the
GMR effect.}\label{fig4}
\end{center}
\end{figure}

Here I will illustrate only the first concept, since I will discuss extensively the second later on.
Consider figure \ref{fig4} where I present an hypothetical device
formed by two metallic surfaces sandwiching a spacer whose Fermi surface is a single point 
(the $\Gamma$ point). For the sake of simplicity I consider a model ferromagnet, namely 
a single orbital two-dimensional simple cubic lattice, with Fermi surfaces centered at the band
center and at the band edges respectively for majority and minority spins. In this case the
Fermi surface of the spacer overlaps only with the majority Fermi surface of the magnetic material.
For this reason we expect zero transmission for the minority spins and for the antiparallel
configuration, leading to an infinite GMR ratio. Note that this is the same result that 
one would expect in the case of half-metallic contacts \cite{AlexB_1}, although none of 
the materials here is an half-metal.

\subsubsection{Magnetic point contacts}

Point contacts are usually obtained by gently breaking a metallic contact thus forming
a tiny junction comprising only a few atoms \cite{PR_Agrait}. In the extreme limit of 
single atom contact the junction remains metallic but its resistance is given by
the electronic structure of that particular atom. A vast amount of experimental data is
nowaday available for point contacts constructed from noble metals like Au (for a comprehensive
review see \cite{PR_Agrait}) since their large malleability and low reactivity.

Recently there was a considerable interest in magnetic point contacts, since they
offer the unique chance to study spin transport at atomic scale, and they may be used
to construct ultra-small and ultra-sensitive magnetic field sensors. Indeed very large
GMR ratios have been measured \cite{BMR1,BMR2,BMR3}, although there is at present a large
debate on whether the effect is of mechanical or electronical origin.
As a general feature transport in magnetic point contact is expected to be ballistic
with both $s$ and $d$ electron contributing to the current. Moreover since the metallic nature of
the contacts and the small screening length local charge neutrality is expected.

\subsubsection{Molecular Spin transport}

The growing interest in interfacing conventional electronic devices with organic
compounds has brought to the construction of spin-valves using molecules as a 
spacer. These include carbon nanotubes \cite{CNT_GMR,CNT_prox}, elementary
molecules \cite{mol_GMR,Aws5} and polymers \cite{GMR_poly}. Spin-transport through
these objects can be highly non-conventional and vary from metallic-like,
to Coulomb-blockade like, to tunneling-like. Moreover the molecule can be either 
chemisorbed of physisorbed depending on the molecular end groups, and
the same spacer can give rise to different transport regimes.

In this case the simple requirement of local charge neutrality is not enough to
describe the physics of the spacer and an accurate description of the drop of
the electrostatic potential across the device is needed. Note that the transport
can still be completely ballistic, in the sense that the electrons do not change their
energy while crossing the spacer.

A more complicate situation arises for polymer-like spacers. In most polymers in fact
the transport is not band transport but it is due to hopping and it is associated
with the formation and propagation of polarons \cite{ssh}. Clearly this adds additional 
complication to the problem since now the electronic and ionic degrees of freedom cannot
be decoupled in the usual Born-Oppenheimer approximation. At present there is very
little theoretical work on spin-transport in polymers.

\subsubsection{Dynamical effects: Magnetization reversal and Domain Wall motion}

The GMR effect demonstrates that the electrical current depends on the magnetic state
of a device. Of course it is interesting to ask the opposite question: ``can a spin current
change the magnetic state of a junction?''. The answer is indeed positive and experimental
demonstrations of current-induced magnetization switching \cite{torque_exp} and current induced 
domain-wall motion \cite{dw_motion_exp} are now available.

The main idea, contained in the seminal works of Berger, is that a spin polarized current
can exert both a force \cite{dw_motion_the} and a torque \cite{torque_the} on the magnetization
of a magnetic material. The force is produced by the momentum transfer between the
conduction electrons and the local magnetic moment, and it is due to the electron reflection.
This can be understood by imagining an electron at the Fermi energy completely reflected
by a domain wall. In this case the electron transfers a momentum $2k_\mathrm{F}$ ($k_\mathrm{F}$
is the Fermi wave vector) to the domain wall, therefore producing a force. Thus the force
is proportional to both the current density (number of electrons scattered) and the domain
wall resistance (momentum transferred per electron). Since the domain wall resistance is usually
rather small, unless the domain wall is rather sharp, this effect is not negligible only for 
very thin walls.

In contrast the torque is due to the spin transfer between the spin-current and the
magnetization and it is proportional to $\vec{S}\times\vec{s}$, where $\vec{S}$ is
the local magnetic moment and $\vec{s}$ is the spin-density of the current currying 
electrons. This effect is dominant for thick walls, where the spin of the electron
follow the magnetization adiabatically.

Although several semi-phenomenological theories are available \cite{tatara} an {\it ab initio}
formulation of these dynamical problems has not been produced yet. A first reason for this 
is that do date very few algorithms for spin-transport at microscopic level have been produced; 
a second and perhaps the most important one is that to date there is not a clear formulation for 
current-induced forces, and specifically for current-induced magnetic forces. 
The problem of calculating the domain wall motion or the magnetization switching from a 
microscopic point of view is rather analogous to the problem of calculating electromigration 
transition rates. This is particularly demanding since it is not clear whether or not current
induced forces are conservative \cite{TTod_Max}.

\setcounter{equation}{0}
\section{Transport Theory: Linear Response}

\subsection{Introduction: Tight-Binding Method}

In this section I will develop the formalism needed for computing spin-transport using
first principles electronic structures. Throughout the derivation of the different methods
I will always make the assumption that the Hamiltonian of the system can be written
in a tight-binding-like form, or equivalently that the wavefunction can be expanded
over a finite set of atomic orbitals. This is a rather general request that in principle
does not set any limitations on the origin of such Hamiltonian nor on the level of accuracy
of the calculation.

The main idea behind the tight-binding method (see for instance \cite{pettifor,sutton,
slater_koster}) is that the wave-function of
an electronic system can be written as a linear combination of localized atomic
orbitals $|\vec{R}\:\alpha\rangle$, where $\vec{R}$ labels the position of the atoms
and $\alpha$ is a collective variable describing all the relevant quantum numbers 
(i.e. $\alpha=n,l,m ...$). The specific choice of the basis set depends on the particular 
problem. A typical choice is to consider a linear combination of atomic orbitals (LCAO)
\begin{equation}
\langle\vec{r}|\vec{0}\:\alpha\rangle=\Psi_\alpha(\vec{r})=R_{nl}(r)Y_{lm}(\theta,\varphi)\:,
\end{equation}
where $R_{nl}(r)$ is the radial component depending on the principal quantum $n$ and the 
angular momentum $l$, and $Y_{lm}(\theta,\varphi)$ is a spherical harmonic describing the 
angular component. This latter depends also on the magnetic quantum number $m$.

For a periodic system the wave-function is then constructed as a Bloch function from the 
localized basis set $|\vec{R}\:\alpha\rangle$
\begin{equation}
|\psi_{\vec{k}}\rangle=\frac{1}{\sqrt{N}}\sum_{\vec{R},\alpha}
\mathrm{e}^{i\vec{k}\cdot\vec{R}}\phi_\alpha^{\vec{k}}|\vec{R}\:\alpha\rangle\:,
\label{tb_wf}
\end{equation}
where $\phi_\alpha^{\vec{k}}$  are expansion coefficients, the sum runs over all the lattice 
sites and $N=N_{\mathrm{site}}\times N_{\alpha}$, with $N_{\mathrm{site}}$ the number 
of atomic sites and $N_{\alpha}$ the number of degrees of freedom per site. If one now 
substitutes $|\psi_{\vec{k}}\rangle$ in the Schr\"odinger equation and then projects over 
$|\vec{R}^\prime\:\alpha^\prime\rangle$, he will find the following matricial equation
for the coefficients $\phi_\alpha^{\vec{k}}$ (secular equation)
\begin{equation}
E(\vec{k})\sum_{\vec{R},\alpha}\phi_\alpha^{\vec{k}}\mathrm{e}^{i\vec{k}\cdot\vec{R}}
\langle\vec{R}^\prime\:\alpha^\prime|\vec{R}\:\alpha\rangle=
\sum_{\vec{R},\alpha}\phi_\alpha^{\vec{k}}\mathrm{e}^{i\vec{k}\cdot\vec{R}}
\langle\vec{R}^\prime\:\alpha^\prime|H|\vec{R}\:\alpha\rangle\:,
\end{equation}
where $E(\vec{k})$ is the energy and $H$ the Hamiltonian. This is usually written in the 
compact form 
\begin{equation}
E(\vec{k})\sum_{\vec{R},\alpha}\phi_\alpha^{\vec{k}}\mathrm{e}^{i\vec{k}\cdot\vec{R}}
S_{\vec{R}^\prime\alpha^\prime\:,\:\vec{R}\:\alpha}=
\sum_{\vec{R},\alpha}\phi_\alpha^{\vec{k}}\mathrm{e}^{i\vec{k}\cdot\vec{R}}
H_{\vec{R}^\prime\alpha^\prime\:,\:\vec{R}\:\alpha}\:,
\label{tb_bs}
\end{equation}
where we have now introduced the Hamiltonian and overlap matrices $H$ and $S$.

Up to this point the formalism is rather general. The specific Hamiltonian to use in the 
equation (\ref{tb_bs}) depends on the problem one wishes to tackle and on the level of accuracy needed. 
In general there are
two main strategies: 1) non-self-consistent Hamiltonian, and 2) self-consistent Hamiltonian.
In the first case one assumes that the matrix elements of $H$ and $S$ can be written in terms
of a small subset of parameters either to calculate or to fit from experiments \cite{slater_koster}.
In addition one usually assumes that the matrix elements of both $S$ and $H$ vanish if the
atoms are not in nearest neighboring positions (nearest neighbors approximation).
The Hamiltonian is then set at the beginning of the calculation and no additional iterations are
needed. This approach is rather powerful for bulk systems where good parameterizations are 
available \cite{papacon,papa_web}, and it is computationally attractive since the size of
the calculation scales linearly with the system size (sub-linearly in the case of some transport
applications \cite{ss_1}).

In contrast in self-consistent methods the Hamiltonian has some functional 
dependence on the electronic structure (typically on the charge density), and needs to be
calculated self-consistently. These methods are intrinsically more demanding since 
several iterations are needed before the energy spectrum can be calculated, although the
final computational costs can vary massively depending on the specific method used
\cite{pabloON}.

\begin{figure}[ht]
\begin{center}
\includegraphics[width=12.5cm,clip=true]{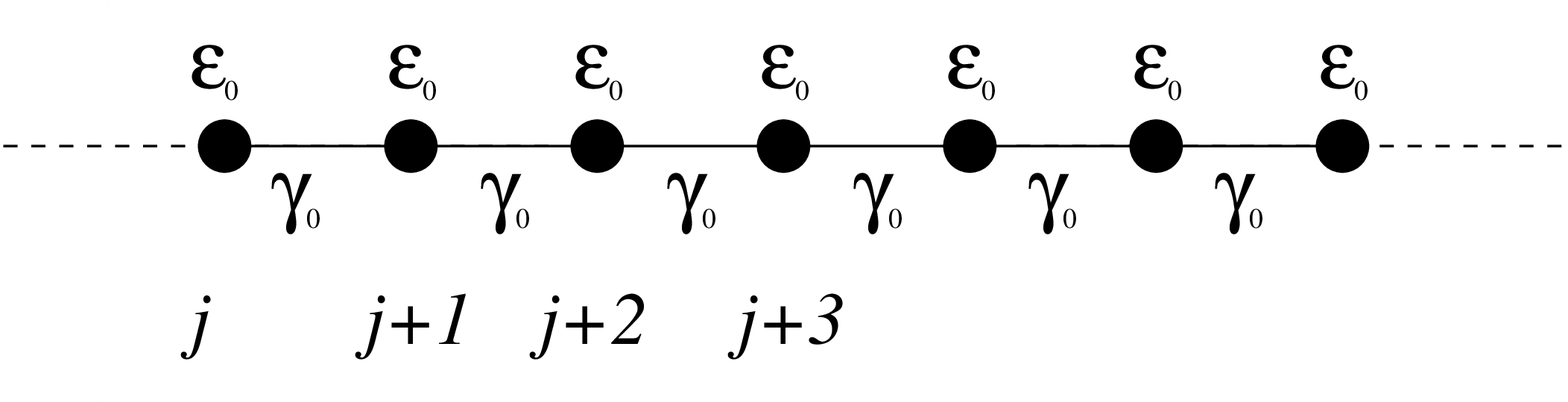}
\vspace{0.6cm}
\includegraphics[width=7.5cm,clip=true]{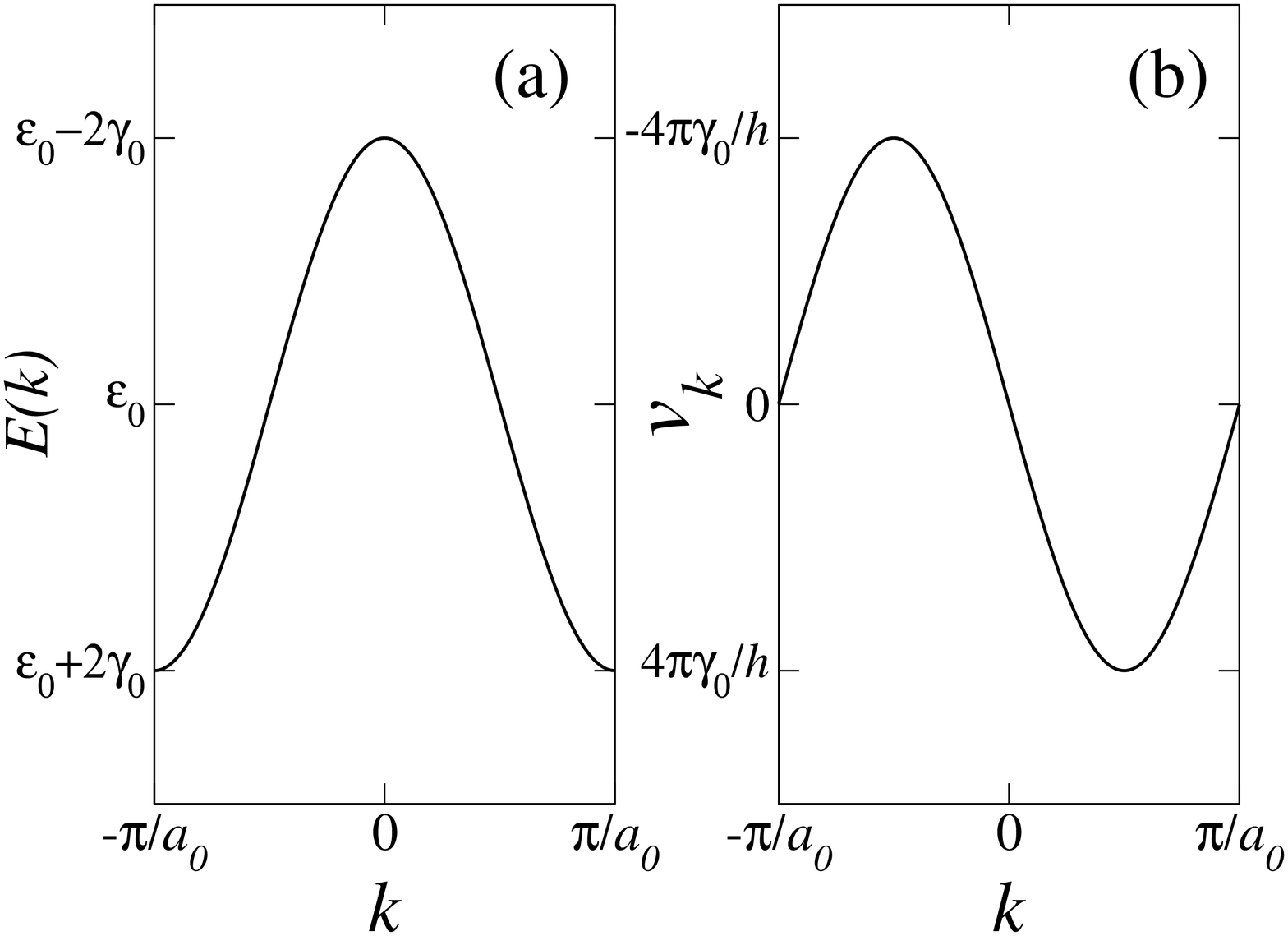}
\caption{Linear chain of H atoms. $\epsilon_0$ is the on-site energy and $\gamma_0$ the nearest neighbour 
hopping integral ($\gamma_0<0$). $j$ labels the atomic position in the chain $R=la_0$ with 
$a_0$ the lattice constant. In (a) I present the the dispersion relation $E(k)$ and in (b)
the corresponding group velocity $v_k$.}\label{fig5}
\end{center}
\end{figure}

Throughout this section I will consider non-self-consistent methods, while 
all the next sections will be devoted to the self-consistent ones. One important
approximation, to the equation (\ref{tb_bs}) is to assume that basis functions 
located at different sites are orthogonal. In this case 
$S_{\vec{R}^\prime\alpha^\prime\:,\:\vec{R}\:\alpha}=
\delta_{\vec{R}^\prime\alpha^\prime\:,\:\vec{R}\:\alpha}$ and the
secular equation reads
\begin{equation}
E(\vec{k})\phi_{\alpha^\prime}^{\vec{k}}=
\sum_{\vec{R},\alpha}\phi_\alpha^{\vec{k}}\mathrm{e}^{i\vec{k}\cdot(\vec{R}-\vec{R}^\prime)}
H_{\vec{R}^\prime\alpha^\prime\:,\:\vec{R}\:\alpha}\:.
\label{tb_bsorth}
\end{equation}
As an example consider an infinite linear chain of hydrogen atoms (see figure \ref{fig5})
described by a single-orbital orthogonal nearest neighbor tight binding model. 
In this case the basis set is simply given by the H 1$s$ orbital $|j\rangle$ where 
$j$ is an integer spanning the positions in chain ($R=ja_0$ with $a_0$ the lattice constant) 
and the only not vanishing matrix elements are:
\begin{equation}
\langle j|H|j\rangle=\ez\:,\;\;\;\;\;
\langle j|H|j\pm1\rangle=\gz\:,\;\;\;\;\;
\langle j\pm1|H|j\rangle=\gz\:,
\label{matrix_element}
\end{equation}
which are called respectively the on-site energy and the hopping integral. 
Finally from the equations (\ref{tb_wf}) and (\ref{tb_bsorth}) it is easy to see that
($\phi_\alpha^k=1$)
\begin{equation}
|\psi_k\rangle=\frac{1}{\sqrt{N}}\sum_{j}
\mathrm{e}^{ikja_0}|j\rangle
\end{equation}
and
\begin{equation}
E(k)=\ez+2\gz\cos(ka_0)\:,
\end{equation}
with $k$ to be taken in the first Brillouin zone $-\pi/a_0<k<\pi/a_0$, and $N\rightarrow\infty$.
A picture of the dispersion relation is presented in figure \ref{fig5}.

\subsection{The Landauer formula}

Here I will derive a general relation between the conductance of a ballistic conductor
and its scattering properties. This relation is known as the Landauer formula
and it is at the core of modern transport theory \cite{landauer_original1,landauer_original2}.

\subsubsection{Current and orbital current}
\label{Cur_orb_cur}

Let us consider again the case of an infinite hydrogen chain discussed before 
and ask ourselves the question ``what is the current carried by one Bloch function?''. 
In order to answer this question I will calculate the time evolution of the density matrix 
$\rho_t=|\psi_t\rangle\langle\psi_t|$ associated to a particular time-dependent quantum 
state $|\psi_t\rangle$. This simply reads
\begin{equation}
\frac{\mathrm{d}}{\mathrm{d}t}|\psi_t\rangle\langle\psi_t|=
\frac{1}{i\hbar}\left[H|\psi_t\rangle\langle\psi_t|-|\psi_t\rangle
\langle\psi_t|H\right]\:,
\end{equation}
where on the left hand side I have used the time-dependent Schr\"odinger equation
$i\hbar\frac{\mathrm{d}}{\mathrm{d}t}|\psi_t\rangle=H|\psi_t\rangle$. Without any loss
of generality one can expand the generic state $|\psi_t\rangle$ over the LCAO
basis $|j\rangle$
\begin{equation}
|\psi_t\rangle=\sum_{j}^N\psi_j(t)|j\rangle\:,
\label{tb_wf2}
\end{equation}
and obtain
\begin{equation}
\frac{\mathrm{d}\rho_t}{\mathrm{d}t}=
\frac{1}{i\hbar}\left[\sum_{jj^\prime}H|j\rangle
\langle j^\prime|\:\psi_j\psi_{j^\prime}^*-
\sum_{jj^\prime}|j\rangle\langle j^\prime|H\:
\psi_j\psi_{j^\prime}^*
\right]
\:.
\label{tddm}
\end{equation}
This is the fundamental equation for the time evolution of the density matrix written
in a tight-binding-like form, where $\psi_j=\psi_j(t)$ are the time dependent expansion coefficients.

In order to work out the current through the H chain we need to consider the
evolution of the charge density at a specific site $x=la_0$. This is obtained by taking the
expectation value of both sides of equation (\ref{tddm}) over the state $|l\rangle$
\begin{equation}
\frac{\mathrm{d}(\rho_t)_{ll}}{\mathrm{d}t}=
\frac{1}{i\hbar}\left[\sum_{jj^\prime}\langle l|H|j\rangle
\langle j^\prime|l \rangle\:\psi_j\psi_{j^\prime}^*-
\sum_{jj^\prime}\langle l|j\rangle\langle j^\prime|H|l \rangle\:
\psi_j\psi_{j^\prime}^*
\right]
\:.
\end{equation}
Finally recalling that I am considering the first nearest neighbor orthogonal 
tight-binding approximation, I obtain
\begin{equation}
\frac{\mathrm{d}(\rho_t)_{ll}}{\mathrm{d}t}=
{\cal J}_{l+1\:\rightarrow\:l}+{\cal J}_{l-1\:\rightarrow\:l}
\:,
\label{tb_cur}
\end{equation}
with
\begin{equation}
{\cal J}_{l+1\:\rightarrow\:l}=\:-\:\frac{i}{\hbar}
[\langle l|H|l+1\rangle\psi_{l+1}\psi_{l}^*-\langle l+1|H|l\rangle\psi_{l}\psi_{l+1}^*]
\label{occ_lr}
\:,
\end{equation}
and
\begin{equation}
{\cal J}_{l-1\:\rightarrow\:l}=\:-\:\frac{i}{\hbar}
[\langle l|H|l-1\rangle\psi_{l-1}\psi_{l}^*-\langle l-1|H|l\rangle\psi_{l}\psi_{l-1}^*]
\label{occ_rl}
\:.
\end{equation}
We can now interpret the results of equation (\ref{tb_cur}) in the following way. The change in
the charge density at the site $x=la_0$ is the result of two currents, one given by electrons
moving from right to left ${\cal J}_{l+1\:\rightarrow\:l}$ and one given by electrons
moving from left to right ${\cal J}_{l-1\:\rightarrow\:l}$. The net current through the chain
is then given by ${\cal J}={\cal J}_{l-1\:\rightarrow\:l}+{\cal J}_{l+1\:\rightarrow\:l}$
(note that here I am considering the current in units of the $e$).

Finally let us calculate the total current carried by a Bloch state. In this case it is simple
to see that the expansion coefficients of the wave-function (\ref{tb_wf2}) are
\begin{equation}
\psi_j(t)=\mathrm{e}^{-iE(k)t/\hbar}\frac{\mathrm{e}^{ikja_0}}{\sqrt{N}}\:,
\end{equation}
and the corresponding currents
\begin{equation}
{\cal J}_{l+1\:\rightarrow\:l}=\frac{2\gamma_0}{\hbar N}\:\sin(ka_0)=
-\frac{1}{L}\:v_k\:,
\label{oc_lr}
\end{equation}
and
\begin{equation}
{\cal J}_{l-1\:\rightarrow\:l}=-\frac{2\gamma_0}{\hbar N}\:\sin(ka_0)=
\frac{1}{L}\:v_k\:,
\label{oc_rl}
\end{equation}
where $L=Na_0$ is the chain length ($L\rightarrow\infty$) and we have introduced the group 
velocity $v_k=\frac{1}{\hbar}\:\frac{\partial E(k)}{\partial k}$. Therefore in the case of a
pure Bloch state the current is exactly zero. This is the result of an exact
balance between left- and right-going currents. However it is worth noting that
the individual currents ${\cal J}_{l+1\:\rightarrow\:l}$ and ${\cal J}_{l-1\:\rightarrow\:l}$ 
are not zero and they are indeed proportional to the 
group velocity associated to the Bloch state. This suggests that, although there is no
net current, it may be possible to associate a conductance to this single quantum state.
However the notion of conductance needs the introduction of the notion of bias voltage, or
more generally of chemical potential difference. This will be introduced in the next section through
the Landauer formula.

Before finishing this section I would like to make a final remark on the actual derivation
of the equation for the current. Here I have projected the time evolution of the
density matrix over a specific basis set, representing an atomic orbital located
at an arbitrary site. In this case the choice of the basis function to project on is
immaterial since we have only one degree of freedom per atom, and one atom per unit cell. 
In the case of more than one degree of freedom 
per unit cell (the cell may contain more that one atom, and each atom can be described by more 
than one orbital), this choice becomes more critical. In general the current calculated from 
the equations (\ref{oc_lr}) and (\ref{oc_rl}) depends on the specific orbital used in the
projection, and it is usually defined as ``orbital current'' or ``bond current'' \cite{todorov_tdtb}. 
Orbital currents associated to different orbitals are usually different.
This may lead to the erroneous conclusion that the current in not locally conserved.
Indeed this feature originates from the incompleteness of the LCAO basis set. To overcome the 
problem a working definition is that of ``current per cell'', where the total current 
is obtained by integrating all the orbital currents of those orbitals belonging to a unit 
cell (special care should be taken in the case of non-orthogonal tight-binding). 
In this way the current is conserved ``locally'' only over a unit cell.
This basically means that the ``most local'' measurement of the current 
allowed by our basis set is that over the whole unit cell.

\subsubsection{Landauer Formula}

\begin{figure}[ht]
\begin{center}
\includegraphics[width=12.5cm,clip=true]{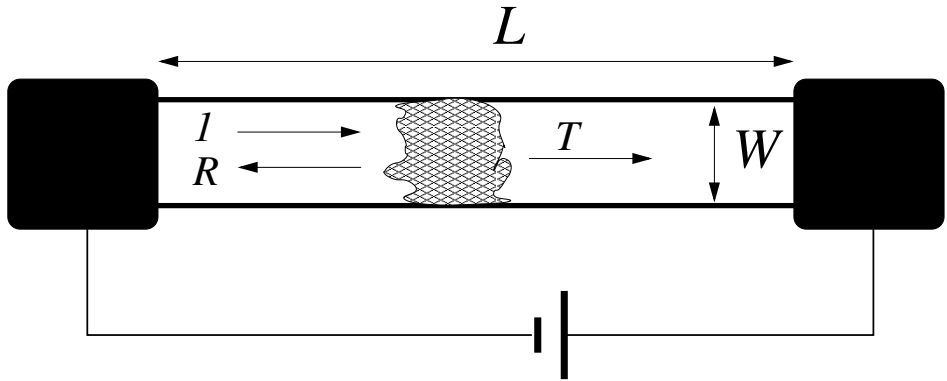}
\caption{Schematic description of a ballistic device. The two black squares represent two bulk 
current/voltage contacts setting the external chemical potential. The scattering region
is long $L$ and $W$ wide with both $L$ and $W$ shorter than the phase coherent length.
The dashed area represents a scattering potential with the corresponding transmission $T$
and reflection $R$ probability.}\label{fig6}
\end{center}
\end{figure}

The crucial aspect of the scattering theory of electronic transport is to relate the scattering
properties of a device, and therefore its electronic structure, to the current flowing through the
device. This is the main result contained in the Landauer formula. Let us follow the original
Landauer's idea \cite{landauer_original1,landauer_original2}. Consider a device formed by two bulk
contacts that act as current/voltage electrodes, connecting through a scattering region (see figure
\ref{fig6}). The main assumptions behind the Landauer formula are the following: 1) the two 
current/voltage probes act as electron reservoirs feeding uncorrelated electrons to the central region
at their own chemical potentials, 2) the difference between the chemical potential of the left $\mu_1$ 
and right $\mu_2$ lead is such that $\mu_1-\mu_2\rightarrow0^+$, 3) electrons can be fed to and 
absorbed from the scattering region without any scattering. Under these assumptions it is easy to
calculate the current between the two leads.

Let us further assume that the scattering region, if extended to infinite, supports only one Bloch
state. In this state the current is given by the product of the current associated to a single Bloch state 
$\frac{1}{L}\:v_k$ times the number of states contained in the energy window between the two 
chemical potentials $\left(\frac{\mathrm{d}n}{\mathrm{d}E}\right)(\mu_1-\mu_2)$
\begin{equation}
I=e\:\frac{v_k}{L}\left(\frac{\mathrm{d}n}{\mathrm{d}E}\right)(\mu_1-\mu_2)\:,
\end{equation}
where $\frac{\mathrm{d}n}{\mathrm{d}E}$ is the DOS. It is now important to note that the DOS can
be easily written in terms of group velocity
\begin{equation}
\frac{\mathrm{d}n}{\mathrm{d}E}=\frac{\partial n}{\partial k}\:\frac{\mathrm{d}k}{\mathrm{d}E}=
\frac{L}{hv_k}\:,
\end{equation}
and therefore the current becomes
\begin{equation}
I=\frac{e}{h}(\mu_1-\mu_2)\:.
\end{equation}
Finally if one introduced the bias voltage through the chemical potential difference
$\Delta V=e\Delta\mu$, it is possible to define the conductance $G$ for such a system
\begin{equation}
G=\frac{2e^2}{h}=G_0\:,
\end{equation}
where the factor 2 takes into account the spin. $G_0=\frac{2e^2}{h}$
is known as ``quantum conductance''. 

At this point I wish to make a few comments. The
result just derived says that in the case of reflectionless electrodes the conductance through
a scattering-free object is quantized in units of $G_0$ independently on the nature of the
conductor itself. This result alone means that the whole definition of conductivity (and therefore
resistivity) becomes meaningless and the only well define quantities are the current and the
conductance. I also with to stress that this result arises from the exact cancellation
between the group velocity and the DOS. This basically means that one should expect conductance 
quantum independently on the DOS or the band dispersion of the conducting electron. 
If we want to apply this concept to a ballistic contact formed 
from a magnetic transition metals, it is then clear that both $s$ and $d$ electrons can give 
the same contribution to the current, independently on their own dispersions.

Finally we assume that the central region is not completely scattering free, and we define
$T$ and $R$ respectively as the total transmission and reflection {\it probabilities} across the
scattering region ($T+R=1$). Then the current will be given by $I=\frac{e}{h}T(\mu_1-\mu_2)$
and the conductance
\begin{equation}
G=\frac{2e^2}{h}T=G_0T\:.
\label{lb_t}
\end{equation}
Note that in general $T=T(E)$ is energy dependent and since the condition sustaining the 
equation (\ref{lb_t}) is $\mu_1-\mu_2\rightarrow 0^+$, then the relevant $T$ is that
calculated at the Fermi level $T(E_\mathrm{F})$.

In the case of spin-dependent transport the total transmission probability is spin dependent 
and the total current must be written in term of the two individual spin-currents
\begin{equation}
G=\frac{e^2}{h}\sum_{\sigma=\uparrow,\downarrow}T^\sigma=\frac{G_0}{2}(T^\uparrow+T^\downarrow)\:.
\end{equation}
where $T^\uparrow$ ($T^\downarrow$) is the transmission probability for majority (minority)
spins.

\subsubsection{Multichannel formalism}
\label{sec_mcf}

The formalism developed in the previous section is based on the assumption that in the scattering
region there is only one Bloch state for a given energy. This is true in the
case of most linear chains, but in general several Bloch states can be available at the same energy. 
The situation can be understood by considering 
a two dimensional regular square lattice of H atoms with lattice constant $a_0$. 
Again we describe the system using the H 1$s$ orbitals $|lj\rangle$, where now $l$ and $j$ 
run respectively on the $x$ and $y$ direction ($\vec{R}=(l,j)a_0$). 
In this case it is simple to see that the wave-functions are 
\begin{equation}
|\psi_{\vec{k}}\rangle=\sum_{lj}^{N_xN_y}\left(\frac{1}{N_xN_y}\right)^{1/2}
\mathrm{e}^{i(k_xa_0l+k_ya_0j)}|lj\rangle\:,
\end{equation}
and the band dispersion is
\begin{equation}
E(k)=\epsilon_0+2\gamma_0\left[\cos(k_xa_0)+\cos(k_ya_0)\right]\:,
\end{equation}
where $\epsilon_0$ and $\gamma_0$ are respectively the on-site energy and the hopping integral,
and $N_x$ and $N_y$ are the number of atomic sites in the $x$ and $y$ directions.

\begin{figure}[ht]
\begin{center}
\includegraphics[width=10.5cm,clip=true]{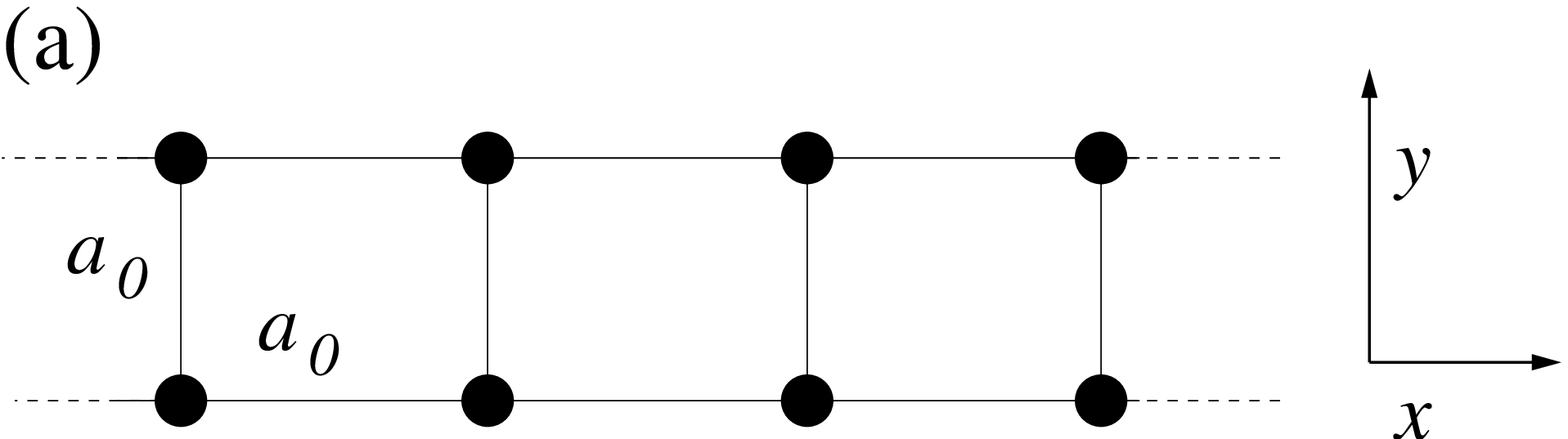}
\vspace{0.6cm}
\includegraphics[width=10.5cm,clip=true]{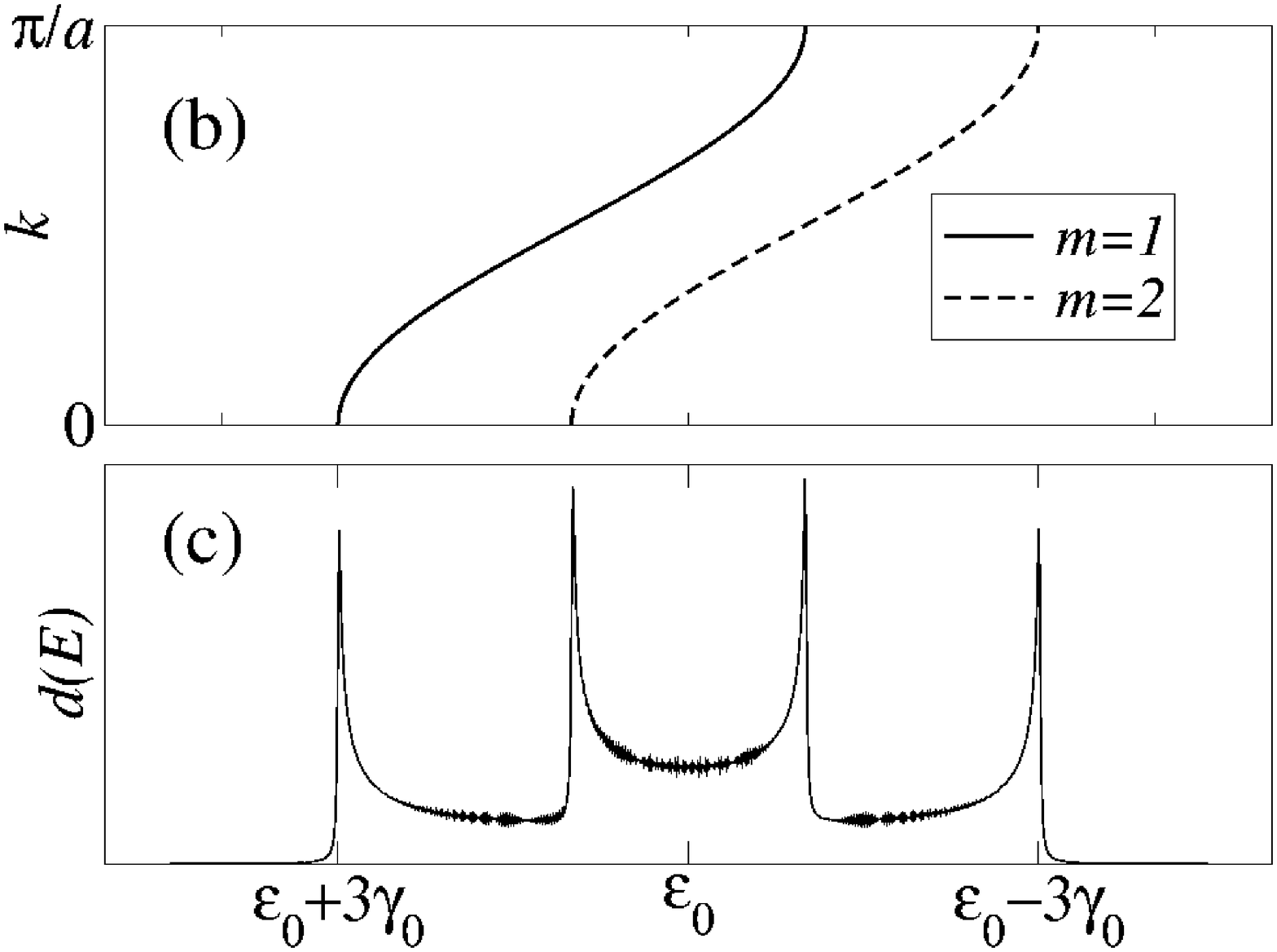}
\caption{(a) Infinite slab made from a two dimensional square lattice of H atoms: (b) band
structure for the two Bloch states $|\psi_{\vec{k}}^m\rangle$, and (c) corresponding
density of states.}\label{fig7}
\end{center}
\end{figure}

Let us now cut a slab along $x$ direction forming an infinite stripe containing 
$N_y$ sites in the cross section (see figure \ref{fig7}). In this case vanishing 
boundary conditions along the $y$ direction must be satisfied and the new wave-functions
and dispersion now read
\begin{equation}
|\psi_{\vec{k}}^m\rangle=\left[\left(\frac{1}{N_x}\right)^{1/2}\left(\frac{2}{N_y+1}\right)^{1/2}
\mathrm{e}^{ik_xa_0l}\sin\left(\frac{m\pi}{N_y+1}j\right)\right]
|lj\rangle\:,
\end{equation}
and
\begin{equation}
E(k)=\epsilon_0+\epsilon_m+2\gamma_0\cos(k_xa_0)\:,
\end{equation}
with
\begin{equation}
\epsilon_m=2\gamma_0\cos\left(\frac{m\pi}{N_y+1}\right)\:.
\end{equation}
In this case the wave function is the product of a running wave along $x$ and a standing
wave along $y$ and the system is usually denoted as quasi-1D. The band structure
(see figure \ref{fig7}) is made from a set of $N_y$ one-dimensional bands along $k_x$
centered at energies $\epsilon_0+\epsilon_m$. These are usually called mini-bands.

Since every mini-band corresponds to a Bloch state, therefore to a pure eigenvalue
of the periodic system, it is possible to associate to each $|\psi_{\vec{k}}^m\rangle$
a current, exactly as we did for the true one dimensional case. The states 
$|\psi_{\vec{k}}^m\rangle$ are called {\it channels}.
This leads us to the generalization of the Landauer formula due to B\"uttiker 
\cite{multi_channel}. 
In this case we assume that the leads inject into each individual channel
un-correlated electrons without any scattering. This means that, if there is no 
scattering also in the conductor, the conductance will now be
\begin{equation}
G=\frac{2e^2}{h}M=G_0M\:,
\end{equation}
where $M=M(E_\mathrm{F})$ is the number of channels at the Fermi level. Note that the 
conductance is simply obtained by counting the number of channels at the Fermi level, and
that the contribution of each channel to the conductance is $2e^2/h$ independently
from the band dispersion of the specific channel. This is again the result of the
cancellation between the group velocity and the density of state in the definition
of the current per channel.

Finally, consider the case where some scattering is present in the conductor. In general the 
$i$-channel traveling through the conductor from left to right has a probability $T_{ij}$
of being transmitted through the scattering region in the $j$-channel.
Therefore the conductance $G_i$ associated to the $i$-th channel will be
\begin{equation}
G_i=\frac{2e^2}{h}\sum_jT_{ij}\:,
\end{equation}
where the sum runs over all the final states.
The total conductance of the whole system is then
\begin{equation}
G=\sum_iG_i=\frac{2e^2}{h}\sum_{ij}T_{ij}\:,
\label{lb_fcm}
\end{equation}
where $T_{ij}=T_{ij}(E_\mathrm{F})$. 
This is the multi-channel generalization of the Landauer formula, that defines a 
complete mapping between transport and scattering properties of a device. 

Finally if we define $R_{ij}$ as the total probability for the $i$-th channel to be
reflected into the $j$-th channel, from the particle conservation requirement we obtain
the following relations
\begin{equation}
\sum_j(R_{ij}+T_{ij})=1\;\;\;\;\;\;\mathrm{and}\;\;\;\;\;\;\sum_{ij}(R_{ij}+T_{ij})=M\:.
\end{equation}

\subsubsection{Finite Bias}

In the previous sections I have established a relation between the zero-bias conductance 
and the scattering properties of an electronic system, here I will generalized the formulation 
to the case of finite bias. The treatment is not going to be rigorous, 
however it gives a very transparent picture of electron transport under bias.
\begin{figure}[ht]
\begin{center}
\includegraphics[width=9.5cm,clip=true]{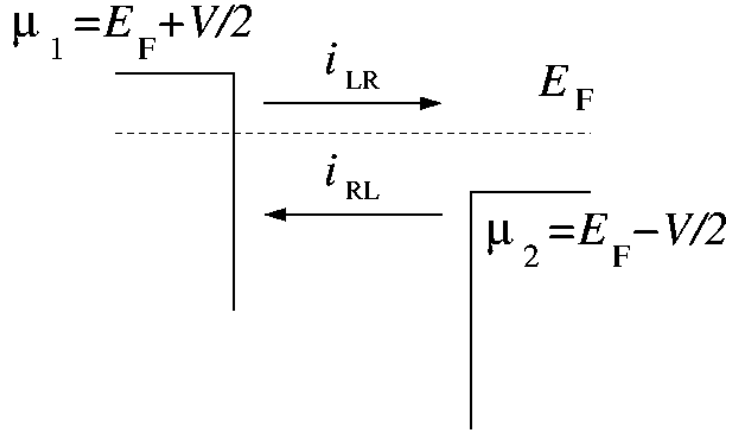}
\caption{Scheme for two terminals electron transport under bias. The left and the right leads are kept at
the chemical potentials $\mu_1=E_\mathrm{F}+V/2$  and $\mu_2=E_\mathrm{F}-V/2$ respectively,
with a potential difference $V$. The total current is then given by the difference between the
right-going current $i_\mathrm{LR}$ and the left-going current $i_\mathrm{RL}$.}\label{fig8}
\end{center}
\end{figure}

Let us consider the situation of figure \ref{fig8} where two current/voltage leads
are connected through a conductor. A bias voltage $V$ is applied by shifting
the two chemical potentials of the leads respectively of $\pm eV/2$, in such a way that 
$\mu_1-\mu_2=V$. At a given energy $E$ the electron flux flowing from the left to the 
right of the device $i_{\mathrm{L}\rightarrow\mathrm{R}}$ is simply given by the 
Landauer-B\"uttiker formula (\ref{lb_fcm})
\begin{equation}
i_{\mathrm{L}\rightarrow\mathrm{R}}(E)=\frac{2e}{h}
\left[\sum_{ij}T_{ij}(E)\right]F_\mathrm{L}(E)\:,
\label{c_lr}
\end{equation}
where $F_\mathrm{L}(E)$ is the Fermi distribution in the left-hand side lead
($k_B$ is the Boltzmann constant)
\begin{equation}
F_{\mathrm{L}}(E)=\frac{1}{1+\mathrm{e}^{(E-\mu_{1})/k_BT}}=
\frac{1}{1+\mathrm{e}^{(E-E_\mathrm{F}- eV/2)/k_BT}}\:.
\end{equation}
The meaning of the 
equation (\ref{c_lr}) is rather transparent. It says that the flux (in units of $2e/h$) 
from left to right is given by the total transmission probability at the energy $E$, multiplied 
by the filling probability $F_\mathrm{L}$. Note that in Landauer's
spirit, when only elastic scattering is considered, backscattered electrons do not compete
for final states and therefore the term $(1-F_\mathrm{R})$ needed to ensure the Pauli's principle 
should not be introduced ($F_\mathrm{R}$ is the Fermi distribution of the right lead).
In the same way the electron flux from right to left is
\begin{equation}
i_{\mathrm{R}\rightarrow\mathrm{L}}(E)=\frac{2e}{h}
\left[\sum_{ij}T^\prime_{ij}(E)\right]F_\mathrm{R}(E)\:,
\label{c_rl}
\end{equation}
where $\sum_{ij}T^\prime_{ij}(E)$ is the total transmission probability at the energy $E$
for electrons moving from right to left. The total flux $i(E)$ at the energy $E$ is then given by 
$i=i_{\mathrm{L}\rightarrow\mathrm{R}}-i_{\mathrm{R}\rightarrow\mathrm{L}}$
\begin{equation}
i(E)=\frac{2e}{h}
\left[\sum_{ij}T_{ij}(E)F_\mathrm{L}(E)
-\sum_{ij}T^\prime_{ij}(E)F_\mathrm{R}(E)\right]\:.
\end{equation}
Finally the total current is obtained by integrating $i(E)$ over the energy. In doing so
I consider the fact that when the system has time reversal symmetry (when there are no magnetic
field or inelastic processes), then $T_{ij}(E)=T^\prime_{ij}(E)$ \cite{But_IBM}. This leads to the
expression
\begin{equation}
I=\frac{2e}{h}\int\sum_{ij}T_{ij}(E)
[F_\mathrm{L}(E)-F_\mathrm{R}(E)]\mathrm{d}E\:.
\label{lb_cub}
\end{equation}

The equation (\ref{lb_cub}) allows us to evaluate $I-V$ characteristics of
ballistic devices, and it is rather useful in many situations. However it should be used with
caution. First, in general the transmission coefficient $T(E)$ is not
only a function of the energy $E$ but also of the bias voltage $V$, $T=T(E,V)$. 
This means that the scattering potential creating the quantum mechanical reflection and 
transmission depends on the bias applied. This dependence is weak in the case of good
metals, where there is local charge neutrality, however it becomes important in
non-metallic cases, like in molecules, where the electronic structure of the conductor may
change substantially under bias. 

The second reason is more fundamental and it has to do with the derivation of equation
(\ref{lb_cub}). In fact this has been derived assuming the Landauer formula to be valid
away from $\mu_1-\mu_2\rightarrow 0^+$ and $T=0$, which are the limits in which the Landauer 
formula is valid. Therefore the arguments sustaining equation (\ref{lb_cub}) are compelling, but a formal 
derivation is still lacking \cite{Kohn_transport}.

\subsection{Green's Functions scattering technique}
\label{greensection}

Here I will present a complete scheme to calculate ballistic transport in the linear (Landauer)
limit. The technique is based on Green's functions. These are usually preferred to the simpler
wave-functions since they posses richer properties and therefore they are more versatile for
transport calculations. Similar approaches based on wave-functions can be find in the
literature \cite{wave_function_method1}.

In the linear response limit the fundamental elements of a scattering technique
are the asymptotic wave-functions (``channels'') and the scattering potential. Information 
regarding the detailed shape of the wave-function and of the charge density inside the 
scattering region are not important, since the current can be determined solely from the asymptotic 
states.
Therefore it is natural to divide the calculation into
three fundamental steps: 1) the calculation of the asymptotic states, 2) the construction 
of an effective coupling matrix between the surfaces of the leads (the scattering potential), 
3) the evaluation of the scattering probabilities $T$ and $R$. From a numerical point of view 
it is also convenient to decouple the first and the second part, because the same leads 
can be used with different scatterers, saving considerable computation time. 

\subsubsection{Elementary scattering theory and ${\cal{S}}$ matrix}

The theory developed so far has been written in term of total transmission
and reflection probabilities, here I will relate those to the quantum mechanical
scattering amplitudes.
\begin{figure}[ht]
\begin{center}
\includegraphics[width=12.5cm,clip=true]{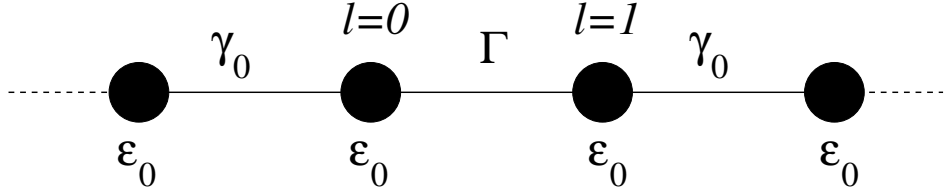}
\caption{Linear tight-binding chains connected through the hopping $\Gamma$.
$\epsilon_0=0$ and $\gamma_1$ are the on-site energy and 
hopping integral.}\label{fig9}
\end{center}
\end{figure}

Consider two semi-infinite linear chains of lattice constant $a_0$ described by a 
tight-binding model with one degree of freedom per atomic site (see figure \ref{fig9}). 
The on-site energy is set to zero ($\epsilon_0=0$) and the hopping integral is $\gamma_0$ 
($\gamma_0<0$). The left-hand side chain is terminated at the atomic position $l=0$ 
and the right-hand side chain starts at the position $l=1$. The chains are coupled 
through the hopping integral $\Gamma$. 

A general quantum state associated with such a system can be written as
\begin{equation}
|\psi_k\rangle=\sum_{l=-\infty}^{+\infty}\phi_l|l\rangle\;,
\label{3.40}
\end{equation}
with
\begin{equation}
\phi_l=\frac{a_l}{\sqrt{v_k}}\mathrm{e}^{ikla_0}+\frac{b_l}{\sqrt{v_k}}
\mathrm{e}^{-ikla_0}\:.
\label{gen_wf}
\end{equation}
In the equation (\ref{gen_wf}) the normalization has been chosen in order to give unit
flux (see section \ref{Cur_orb_cur}). The state $|\psi_k\rangle$ is a linear
combination of right- and left-moving plane waves, and a scattering state corresponds to the
following particular choice of $a_j$ and $b_j$
\begin{equation}
\phi_l=\left\{
\begin{array}{ll}
\frac{\mathrm{e}^{ikla_0}}{\sqrt{v_k}}+r\:\frac{\mathrm{e}^{-ikla_0}}{\sqrt{v_k}} & l\le0 \\
\\
t\:\frac{\mathrm{e}^{ikl}}{\sqrt{v_k}} & l\ge1
\end{array}
\right.\:.
\label{sc_state}
\end{equation}

This means that an electron incising the scattering region placed at $l=0$ from the 
left-hand side, has a probability $T=|t|^2$ of being found at any positions $l>0$. 
Similarly the probability of finding the same electron at $l<0$ is $R=|r|^2$. 
$r$ and $t$ are called respectively the reflection and transmission 
{\it coefficients} and from particle conservation it follows the condition $|t|^2+|r|^2=1$.

A convenient way to collect all the information regarding transmission and
reflection is via the scattering matrix $\smat$. In general this is defined as
\begin{equation}
\psi_{\mathrm{OUT}}=\smat\:\psi_{\mathrm{IN}}\:,
\end{equation}
where $\psi_{\mathrm{IN}}$ ($\psi_{\mathrm{OUT}}$) is a vector containing the amplitudes
of the quantum states approaching to (emerging from) the scattering region.
Using the definition of scattering states given in equation (\ref{sc_state}) the $\smat$ matrix reads
\begin{equation}
\smat=
\left(
\begin{array}{cc}
r & t^\prime  \\
t & r^\prime  
\end{array}
\right)\:,
\label{Smatr}
\end{equation}
where the quantity $t^\prime$ and $r^\prime$ are respectively the transmission and
reflection coefficients for states approaching the scattering region from the right.

Clearly the expressions given above can be generalized to the case of multi-channels.
A generic scattering channel can be written as a linear combination of all possible
channels and the scattering amplitudes define the $\smat$ matrix.
If $k$, ($k^\prime$) is a real incoming (outgoing) wave-vector of energy $E$, then an 
incident plane-wave in one of the leads, with longitudinal wave-vector $k$, will scatter 
into outgoing plane-waves $k^\prime$ with amplitudes $s_{k'k}(E,H)$. If all plane-waves 
are normalized to unit flux, (by dividing by the square-root of their group velocities) then 
provided the plane-wave basis diagonalizes the current operator in the leads, the outgoing 
flux along channel $k^\prime$ is $\vert s_{k'k}(E,H)\vert^2$ and ${\cal S}$ will be unitary.
If $H$ is real, then ${\cal S}$ will be symmetric, but more generally
time reversal symmetry implies $s_{k'k}(E,H)=s_{kk'}(E,H^*)$.
Hence, if $k,k^\prime$ belong to the left (right) lead,  then I define reflection coefficients via $r_{k'k}=s_{k'k}$ ($r'_{k'k}=s_{k'k}$), whereas if $k,k^\prime$ belong to left and right leads 
respectively (right and left leads respectively) I define transmission coefficients
$t_{k'k}=s_{k'k}$ ($t'_{k'k}=s_{k'k}$).

In this way the scattering matrix $\smat$ preserves the form (\ref{Smatr}) and 
$t$ and $r$ are matrices describing all possible reflection and transmission 
coefficients between the different channels. These are called respectively the
transmission and reflection matrix. Finally the total transmission and reflection
probabilities can be written as
\begin{equation}
T=\sum_{ij}t_{ij}t_{ij}^*=\mathrm{Tr}\:tt^\dagger\:,
\end{equation}
and
\begin{equation}
R=\sum_{ij}r_{ij}r_{ij}^*=\mathrm{Tr}\:rr^\dagger\:.
\end{equation}

\subsubsection{A simple 1D example}
\label{s1dc}

Before going into a detailed analysis of the general scattering technique, I will
present a simple example in which the main ideas are introduced. Let us consider
the two semi-infinite linear chains connected through the sites $l=0$ and $l=1$.

For an infinite chain with on-site energy $\epsilon_0$ and hopping integral $\gamma_0$
($\gamma_0<0$) the retarded Green function (RGF) $\hat{g}$ is simply (see for example 
\cite{econbook})
\begin{equation}
\hat{g}=\sum_{jl}g_{jl}|j\rangle\langle l|=
\sum_{jl}\frac{\mathrm{e}^{ik|j-l|}}{i\hbar v_k}|j\rangle\langle l| \; ,
\label{eq2.2}
\end{equation}
where $v_k$ is the group velocity. In what follows and for the remaining of this
section I will always use the adimensional
$k$-vector $k=ka_0$ and I will consider only the matrix of the coefficients
$g_{jl}$ (the Green's function matrix). This satisfies the Green's equation 
$\sum_j(E+i\xi-H_{ij})g_{jl}=\delta_{il}$ with $\xi\rightarrow 0^+$.

The equation (\ref{eq2.2}) describes the RGF of an infinite system. 
However our main goal is to describe an electron approaching the scattering region,
not one propagating through an infinite periodic chain, therefore the relevant 
RGF is that of a semi-infinite system. This can be computed starting 
from the one of the double-infinite system (equation (\ref{eq2.2})) by imposing the 
appropriate boundary conditions.
Suppose the infinite system (extending from $-\infty$)
is terminated at the position $l=i_0-1$ in such a way that 
there is no potential for $l=i_0$. Then the Green function $g_{jl}$ 
with source at $l=i<i_0$ must vanish for $j=i_0$
(scattering channels approaching the boundaries from the left). 
This is achieved by adding to the expression (\ref{eq2.2}) the following wave-function
\begin{equation}
\phi_j(l,i_0)=-\frac{\mathrm{e}^{-ik(j-2i_0+l)}}{i\hbar v_k} \; 
\label{eq2.4}
\end{equation}
and noting that adding a wave-function to a Green function results in a new Green function 
with the same causality. The final RGF for the semi-infinite system is then 
${\tilde{g}}=g+\phi$. Note that its value at the boundary of the scattering region $j=l=i_0-1$
is
\begin{equation}
\tilde{g}_{i_0-1,i_0-1}=\frac{\mathrm{e}^{ik}}{\gamma_0} \; .
\label{eq2.5}
\end{equation}
This is usually called ``surface Green's function''. The surface Green function is independent 
from the position of the boundary, as expected from the translational invariance of the problem.
An identical expression can be derived for the surface Green function of a
semi-infinite linear chain starting at $l=i_0$ and extending to $+\infty$. 

Going back to the initial problem, the surface Green function for two decoupled ($\Gamma=0$)
chains facing through the sites $i=0$ and $i=1$ is an infinite matrix of the form 
\begin{equation}
g=\left(
\begin{array}[h]{rr}
\tilde{g}_L & 0 \;\; \\
0 \;\; & \tilde{g}_R \\
\end{array}
\right){\;},
\label{eq2.6}
\end{equation}
where $\tilde{g}_L$ ($\tilde{g}_R$) is the RGF for the left (right) semi-infinite
chain. Note that $g$ has vanishing off-diagonal terms, reflecting the fact that 
the two chains are decoupled.

Let us now switch on the coupling $\Gamma$ between the two chains. The total Green's function
$G$ for the two chains connected through the sites $l=0$ and $l=1$ can be found
by solving the Dyson's equation 
\begin{equation}
G= (g^{-1}-W)^{-1}\; ,
\label{eq2.7}
\end{equation}
where $W$ is an infinite matrix, whose only non-zero elements are $W_{10}=W_{01}=\Gamma$. 

At this point it is important to observe that current conservation gives us the freedom to 
decide the ``most convenient'' surface to use for computing the current (in fact 
is is identical across any surface). Since the current across a generic surface 
is calculated from those matrix elements of the total Green function $G$ corresponding to the 
atoms facing each other through the surface, the criterion behind our choice
is to select a surface for which the calculation of $G$ is easier. A particularly
convenient choice is to use the surface between the atoms placed at $l=0$ and $l=1$. 
In this case it is possible to show that the matrix elements $G_{00}$,  $G_{11}$,
$G_{10}$, and $G_{01}$ can be obtained from the ``reduced'' Dyson's equation (\ref{eq2.7})
for the finite matrices $g$ and $W$ which contain only the matrix elements between
the position $l=0$ and $l=1$. In the present case these are
\begin{equation}
g=\left(
\begin{array}[h]{rr}
\frac{\mathrm{e}^{ik}}{\gamma_0} & 0 \;\;\; \\
0 & \frac{\mathrm{e}^{ik}}{\gamma_0} \\
\end{array}
\right){\;},
\label{red_gf}
\end{equation}
and
\begin{equation}
W=\left(
\begin{array}[h]{rr}
0 & \Gamma  \\
\Gamma  & 0\\
\end{array}
\right){\;}.
\label{red_w}
\end{equation}

This basically establishes that the surface Green's functions
(\ref{red_gf}) and the coupling potential $W$ (equation (\ref{red_w}))
are sufficient to compute the total Green's function across the scattering region,
and therefore the current.
In this way the total Green's function across the plane $l=0$ and $l=1$ is simply
\begin{equation}
G=\frac{1}{\gamma_0^2\mathrm{e}^{-i2k}-\Gamma^2}
\left(
\begin{array}[h]{rr}
{\mathrm{e}^{-ik}}\gamma_0 & \Gamma \;\;\; \\
\Gamma \;\;\; & {\mathrm{e}^{-ik}}{\gamma_0}\\
\end{array}
\right){\;}.
\label{eq2.7bis}
\end{equation}

The remaining task is to extract from the total Green function $G$ the ${\cal S}$ matrix.
First note that the general wave-function $|\psi_k\rangle$ for an electron approaching 
the scattering region from the left has the form (equations (\ref{3.40}) through (\ref{sc_state}))
\begin{equation}
|\psi_k\rangle=\sum\phi_l|l\rangle \;\;\;\;\;\;\mathrm{with}\;\;\;\;\;\;
\phi_l=\left\{
\begin{array}[4]{r}
\frac{\mathrm{e}^{ikl}}{\sqrt{v_k}}+
\frac{r}{\sqrt{v_k}}
\mathrm{e}^{-i{kl}}
\;\;\;\;\;l\leq{0} \\
\\
\; \frac{t}{\sqrt{v_k}}
\mathrm{e}^{i{k}l}
\;\;\;\;\;l\geq{1} \\
\end{array}
\right.\;
\;{,}
\label{eq2.8}
\end{equation}
where the transmission $t$ and reflection $r$ coefficients are introduced and
the incoming wave-function $\frac{\mathrm{e}^{ikl}}{\sqrt{v_k}}$ is
normalized to unit flux (see previous section).
This normalization guarantees the unitarity of the {${\cal{S}}$ matrix $|t|^2+|r|^2=1$. 
Since all the information contained in $|\psi_k\rangle$ are also contained in the RGF
(away from the source a Green function is identical to a wave-function),
the final step is to project the total Green function $G$ over the wave-function $|\psi_k\rangle$.
It is possible to show (see Appendix \ref{AppD}) that the projector
that projects the retarded Green function for an infinite system over the
unitary flux Bloch-function 
$|\psi_k\rangle=\sum_j\frac{\mathrm{e}^{ikl}}{\sqrt{v_k}}|j\rangle$, 
also projects the total
Green function $G$ over the (\ref{eq2.8}). Such projector is easily calculated
through the relation
\begin{equation}
g_{lj} P(j)=\frac{\mathrm{e}^{ikl}}{\sqrt{v_k}}\;\;\;\;\;{\mathrm{for}}\;\;\;\;l\geq j \; ,
\label{eq2.9}
\end{equation}
and it is simply
\begin{equation}
P(j)=i\mathrm{e}^{ikj}\sqrt{v_k} \; .
\label{eq2.10}
\end{equation}
Now I can now use $P(j)$ to extract $t$ and $r$. In fact by applying $P(j)$ to
$G_{lj}$ and by taking the limit $l\rightarrow 0 \;\;$ I obtain
\begin{equation}
G_{00}P(0)=\frac{1}{\sqrt{v_k}}+\frac{r}{\sqrt{v_k}}\; ,
\label{eq2.11}
\end{equation}
from which the reflection coefficient is easily calculated
\begin{equation}
r=G_{00}P(0){\sqrt{v_k}}-1\; .
\label{eq2.12-r}
\end{equation}
In the same way the transmission coefficient is simply
\begin{equation}
t=G_{10}P(0)\sqrt{v_k}\mathrm{e}^{-ik}\; .
\label{eq2.12-t}
\end{equation}
Note that the same technique can be used to calculate $t^\prime$ and $r^\prime$ for electrons
incoming the scattering region from the right.

To conclude this section I want to summarize the calculation scheme presented in
this example. First I calculated the Green function for an infinite
system. From this I have derived the surface Green function for the corresponding 
semi-infinite leads by using the appropriate boundary conditions.
Secondly I switched on the interaction between the leads by solving the Dyson
equation with a given coupling matrix between the two lead surfaces. 
Finally I calculated the ${\cal S}$ matrix by introducing a projector that maps the total
Green function onto the total scattered wave-function. 

The advantage of this technique is twofold. On the one hand the calculation of
the RGF for the infinite system enables us to obtain useful information
regarding the leads (density of state, conductance) and on the other hand the scattering
region is treated separately and added to the leads only before evaluating the
${\cal S}$ matrix. As noted above this latter aspect is particularly
useful in the case in which a large number of computations of different
scatterers with the same leads are needed.

\subsubsection{Structure of the Green Functions}

In this section I will discuss the general structure of the surface Green function
for a quasi one dimensional system. I will start from a simple case, namely the 
two-dimensional simple-cubic lattice of H atoms discussed in section \ref{sec_mcf}. 

Be $x$ the direction of transport and $y$ the transverse direction comprising $N_y$ 
atomic sites. As usual $\epsilon_0$ is the on-site energy and $\gamma_0$ the 
hopping integral ($\gamma_0<0$). By solving the Green's equation one can find that the 
Green's function for such a system is \cite{econbook}
\begin{equation}
g_{lj,l^{\prime}j^{\prime}}=\sum_{m=1}^{N_y}\left(\frac{2}{N_y+1}\right)
\sin\left(\frac{m\pi}{N_y+1}j\right)
\sin\left(\frac{m\pi}{N_y+1}j^{\prime}\right)
\frac{e^{ik_x^m(E)|l-l^{\prime}|}}{i\hbar{v_x^m(E)}}{\;},
\label{eq2.13}
\end{equation}
where $l$ and $j$ label respectively the $x$ and $y$ coordinates and $k_z^m(E)$ is the longitudinal 
momentum. This satisfies the dispersion relation
\begin{equation}
E=\epsilon_0+2\gamma_0\left[\cos\left(\frac{m\pi}{N_y+1}\right)+\cos{k_z^m(E)}\right]\;{,}
\label{eq2.14}
\end{equation}
and $v_z^m(E)$ is the group velocity associated to the $m$-th channel
\begin{equation}
\frac{\partial{E}(k_z^m)}{\partial{k_z^m}}=\hbar{v_z^m(E)}{\;}.
\label{eq2.15}
\end{equation}
Let us look at the structure of $g$.
The equation (\ref{eq2.13}) can be written as
\begin{equation}
g_{lj,l^{\prime}j^{\prime}}=\sum_{m=1}^{N_y}\phi^m_j
\frac{e^{ik_x^m(E)|l-l^{\prime}|}}{i\hbar{v_x^m(E)}}\phi^{m*}_{j^\prime}{\;}.
\label{eq2.16}
\end{equation}
$g_{lj,l^{\prime}j^{\prime}}$ consists of the sum of all the allowed
longitudinal mode $e^{ik_x^m(E)l}$ (with $k_x^m(E)$'s both real and 
imaginary) weighted by the corresponding transverse component
$\phi^m_j$ that in this case is simply
\begin{equation}
\phi^m_j=\left(\frac{2}{N_y+1}\right)^{1/2}\sin\left(\frac{m\pi}{N_y+1}j\right)\:.
\label{eq2.17a}
\end{equation}

It is then easy to identify the plane-waves $\frac{e^{ik_x^ml}}{\sqrt{v_z^m}}$ 
with the scattering channels defined previously. Note that in the case of a one-dimensional
linear chain the equation (\ref{eq2.16}) reduces to the expression (\ref{eq2.2}), 
where the $\phi^m=1$. 

The possible scattering channels can be then divided into four classes.
The left-moving scattering channels {\it lm} (right-moving
scattering channels {\it rm}) are propagating states 
($k_x^m$ is a real number) having negative (positive) group velocity.
Similarly the left-decaying scattering channels {\it ld} 
(right-decaying scattering channels {\it rd}) are states whose 
wave-functions have a real exponential decay, with $k_x^m$ possessing
a negative (positive) imaginary part. Note that in the case in which
time-reversal symmetry is valid, the number of left- and right-moving scattering
channels must be the same, as well as the number of left- and right-decaying
scattering channels. Moving and decaying channels are conventionally called respectively
``open'' and ``closed'' channels. A schematic picture of all the scattering channels is 
given in figure \ref{fig10}.
\begin{figure}[ht]
\begin{center}
\includegraphics[width=9.5cm,clip=true]{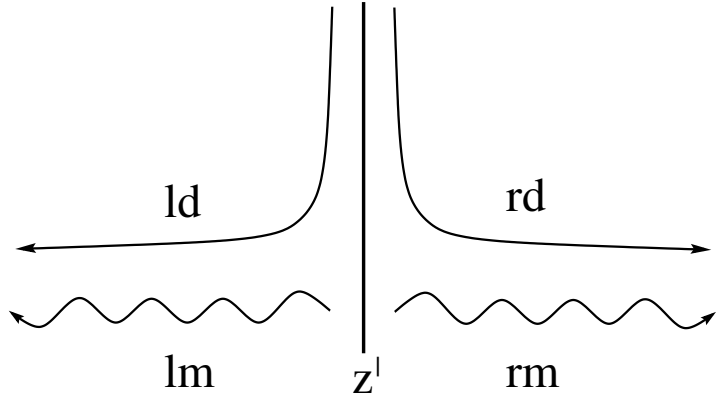}
\caption{Green function structure. {\it lm} ({\it rm}) denotes the left- (right-) moving channels, {\it ld} ({\it rd}
the left- (right-) decaying channels. $z^\prime$ is the position of the source.
}\label{fig10}
\end{center}
\end{figure}

Clearly there are $N_y$ scattering channels and the retarded Green function of equation 
(\ref{eq2.13}) is obtained by summing up all open and closed channels, with their relative
transverse wave-components. This structure is the starting point for our general approach.

\subsubsection{General surface Green function}
\label{gsgf}

In this section I will present a general technique to construct the surface Green 
functions of an arbitrary crystalline lead. This is the first step toward a general Green's 
function method for ballistic transport. An important feature of this section is that the Green 
function will be defined by a semi-analytic formula, which can be applied to any crystalline 
structure. As explained for the simple 1D case, the computation of the Green function for a 
semi-infinite crystalline lead starts from calculating the Green function of the associated 
doubly infinite system. Then the semi-infinite case is derived by applying vanishing boundary 
conditions at the end of the lead. To this goal, consider the doubly infinite system shown 
in figure \ref{fig11}.
\begin{figure}[ht]
\begin{center}
\includegraphics[width=10.5cm,clip=true]{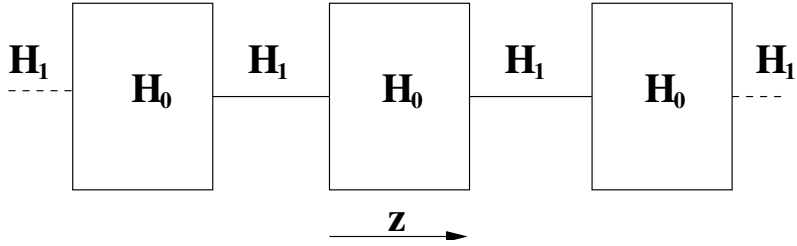}
\caption{Infinite system formed from periodically repeated slices. $H_0$
describes the interaction within a slice and $H_1$ describes the coupling
between adjacent slices.}\label{fig11}
\end{center}
\end{figure}

Let us define $z$ as the direction of transport, and write the Hamiltonian in a block-like
form along such direction. We can break down the system as a periodic sequence of ``slices'', 
described by an intra-slice tight-binding matrix $H_0$, coupled only to the nearest neighbor 
slices via the inter-slice tight-binding hopping matrix $H_1$ \cite{Joann1,Joann2}. 
This procedure is always possible
if the lattice has some periodicity along the direction of the transport, since the 
interaction range of the tight-binding Hamiltonian is finite. Note that the same is true
when considering non-orthogonal tight-binding models. In fact the overlap matrix $S$ has
an interaction range, which is always shorter than that of the Hamiltonian matrix $H$. This means
that if $H$ is written in a block-like form, also $S$ will. In all the remaining I will consider
for simplicity only orthogonal tight-binding, although the theory can be easily generalized
to the non-orthogonal case. Note also that the nature of the slices does not need to be specified at this 
stage. They can describe a single atom in an atomic chain, an atomic plane or a more complex cell.

For such a general system, the total Hamiltonian $H$ can be written as an 
infinite matrix of the form
\begin{equation}
H=\left(
\begin{array}[h]{rrrrrrrr}
... & ... \; & ... \; & ... \; & ... \; & ... \; & ... \; & ... \\
... & H_0 \: & H_1 \: & 0 \;\: & ... \; & ... \; & ... \; & ... \\
... & H_{-1} & H_0 \: & H_1 \: & 0 \;\: & ... \; & ... \; & ... \\
... & 0 \;\: & H_{-1} & H_0 \: & H_1 \: & 0 \;\: & ... \; & ... \\
... & 0 \;\: & 0 \;\: & H_{-1} & H_0 \: & H_1 \: & 0 \;\: & ... \\
... & ... \; & ... \; & ... \; & ... \; & ... \; & ... \; & ... \\
... & ... \; & ... \; & ... \; & ... \; & ... \; & ... \; & ... \\
\end{array}
\right){\;},
\label{eq2.18}
\end{equation}
where $H_0$ is Hermitian and $H_{-1}=H_1^\dagger$. To fix the ideas let us assume that the 
number of atomic orbitals describing each slice is $M$, and therefore $H_0$ and $H_1$ are 
$M\times M$ matrices. From the equation (\ref{tb_bs}), the Schr\"odinger 
equation for this system is of the form
\begin{equation}
H_0\psi_z+H_1\psi_{z+1}+H_{-1}\psi_{z-1}=E\psi_z{\;},
\label{eq2.19}
\end{equation}
where $\psi_z$ is a column vector corresponding to the slice at the position $z=ja_0$ with 
$j$ an integer and $a_0$ the inter-slice distance. Let the quantum numbers 
corresponding to the degrees of freedom within a slice be $\mu=1,2,\ldots,M$ and the 
corresponding components of $\psi_z$ be $\psi_z^\mu$. The Schr\"odinger equation may then be 
solved by introducing the Bloch state, 
\begin{equation}
\psi_z=n_{k_\perp}^{1/2}e^{ik_\perp{z}}\phi^{k_\perp}{\;},
\label{eq2.20}
\end{equation}
where $\phi^{k_\perp}$ is a normalized $M$-component column vector and $n_{k_\perp}^{1/2}$ an 
arbitrary constant. Note that throughout this paper I will use the symbol ``$\perp$'' to 
indicate the direction of the current and the symbol ``$\parallel$'' to label the transverse 
plane.
Substituting (\ref{eq2.20}) into the equation (\ref{eq2.19}) gives
\begin{equation}
\left(H_0+H_1e^{ik_\perp}+H_{-1}e^{-ik_\perp}-E\right)
{\phi}^{k_\perp}=0{\;}.
\label{eq2.21}
\end{equation}
The task is now to compute the Green function $g$ of such a system, for all real energies, 
using the general Green function structure discussed in the previous section.
For a given energy $E$, the first goal is to determine all possible values (both real and complex)
of the wave-vectors $k_{\perp}$ by solving the secular equation
\begin{equation}
\det(H_0+H_1\chi+H_{-1}/\chi-E)=0\;\;\;\;\;\mathrm{with}\;\;\;\;\;\chi=e^{ik_\perp}\:.
\label{eq2.22}
\end{equation}

In contrast to conventional band-theory, where the problem is to compute the $M$ values of 
$E$ for a given (real) choice of $k_\perp$, my aim is to compute the complex roots $\chi$
of the polynomial (\ref{eq2.22}) for a given (real) choice of $E$ (remember that both open 
and closed scattering channels enter in the definition of the Green function).
Consider first the case where $H_1$ is not singular. Note that for real $k_\perp$,
conventional band-theory yields $M$ energy bands $E_n(k_\perp)$, $n=1,\ldots,M$, with 
$E_n(k_\perp+2\pi)=E_n(k_\perp)$. As a consequence, for a given choice of $E$, to each real 
solution $k_\perp = k$, for which the group velocity
\begin{equation}
{v_k}=\frac{1}{\hbar}
\frac{\partial{E}(k)}{\partial{k}}
{\;}
\label{eq2.23}
\end{equation}
is positive (right-moving channel), there exists a second solution $k_\perp = \bar{k}$ for 
which the group velocity
\begin{equation}
{v_{\bar k}}=\frac{1}{\hbar}
\frac{\partial{E}(\bar{k})}{\partial{\bar{k}}}
{\;}
\label{eq2.24}
\end{equation}
is negative (left-moving channel). In the simplest case, where $H_1 = H_{-1}$, one finds 
$k=-\bar k$. Also note that to each solution $k_\perp$ the hermitian conjugate of
(\ref{eq2.21}) shows that $k_\perp^*$ is also a solution. Hence to each right-decaying 
solution $k$ possessing a positive imaginary part, there is a left-decaying solution $\bar k$
with a negative imaginary part. For the purpose of constructing the Green function,
this suggests dividing the roots of (\ref{eq2.21}) into two sets: the first set
of $M$ wave-vectors labeled $k_m$ ($m= 1,...,M$) correspond to right-moving
and right-decaying plane-waves and the second set labeled $\bar k_m$ ($m= 1,...,M$) correspond 
to left-moving and left-decaying plane-waves.

Although the solutions to (\ref{eq2.22}) can be found using a root tracking algorithm, for 
numerical purposes it is more convenient to map (\ref{eq2.21}) onto an equivalent eigenvalue 
problem by introducing the matrix $\cal{H}$ 
\begin{equation}
\cal{H}=\left(
\begin{array}[4]{rr}
-H_1^{-1}({H_0}-E) & -H_1^{-1}H_{-1} \\
{\cal I}\;\;\;\;\;\;\;\;\; & 0\;\;\;\;\;\;\;\;\; \\
\end{array}
\right){\;},
\label{eq2.25}
\end{equation}
where ${\cal I}$ is the $M$ dimensional identity matrix. The eigenvalues of $\cal{H}$ are the 
$2M$ roots $e^{ik_m}$, $e^{i\bar k_m}$  and the upper $M$ components of the eigenvectors of 
$\cal{H}$ are the  corresponding eigenvectors ${\phi}^{k_m}=\phi^m$, 
${\phi}^{\bar{k}_m}=\phi^{\bar{m}}$.
It is clear that in the case in which $H_1$ is singular, the matrix $\cal{H}$ is not defined.
However a few methods to regularize $H_1$ are available \cite{smeagol1,smeagol2} and in what
follows I will assume $H_1$ not to be singular.

To construct the retarded Green function $g_{zz^\prime}$ of the doubly infinite system, 
note that except at the source $z=z^\prime$, $g$ is simply a wave-function and hence must have 
the form
\begin{equation}
g_{zz^\prime}=\left\{
\begin{array}[4]{r}
\sum_{m=1}^M{\phi}^me^{ik_m(z-z^\prime)}
{\mathsf w}^{m\dag}\;\;\;\;\;z\geq{z}^\prime \\
\\
\sum_{m=1}^M{\phi}^{\bar{m}}e^{i\bar{k}_m(z-z^\prime)}
{\mathsf w}^{\bar{m}^\dag}
\;\;\;\;\;z\leq{z}^\prime \\
\end{array}
\right.\;,
\label{eq2.26}
\end{equation}
where the $M$-component vectors ${\mathsf w}^{k_m}={\mathsf w}^{m}$ 
and ${\mathsf w}^{\bar{k}_m}={\mathsf w}^{\bar{m}}$ are  to be
determined. At this point it is important to observe that the structure of the Green
function of equation (\ref{eq2.26}) is identical to the one discussed in the
previous section, and that the vectors ${\phi}^{m}$ and ${\mathsf w}^{m}$ 
(equivalent to the $\phi^m$'s of the previous section) include all the degrees of freedom of 
the transverse direction. Note that in representing $g$ I have chosen a slightly different notation
with respect to the previous section, and I do not consider explicitly the orbital dependence on the 
transverse degree of freedom. It is then understood that the Green's function between
the orbital $\mu$ placed in the slice $z$ and the orbital $\mu^\prime$ placed in the 
slice $z^\prime$ is
\begin{equation}
g_{z\mu\:z^\prime\mu^\prime}=\left\{
\begin{array}[4]{r}
\sum_{m=1}^M{\phi}^{m}_\mu\:e^{ik_m(z-z^\prime)}\:
{\mathsf w}^{m\dag}_{\mu^\prime}\;\;\;\;\;z\geq{z}^\prime \\
\\
\sum_{m=1}^M{\phi}^{\bar{m}}_\mu\:e^{i\bar{k}_m(z-z^\prime)}\:
{\mathsf w}^{\bar{m}\dag}_{\mu^\prime}
\;\;\;\;\;z\leq{z}^\prime \\
\end{array}
\right.\;.
\label{eq2.26explic}
\end{equation}

Since $g_{zz^\prime}$ is  retarded both in $z$ and $z^\prime$,
it satisfies the Green function equation corresponding to (\ref{eq2.19})
and is continuous at the point $z=z^\prime$ (see Appendix \ref{AppB} for a detailed
calculation), one obtains 
\begin{equation}
g_{zz^\prime}=\left\{
\begin{array}[4]{r}
\sum_{m=1}^M{\phi}^{m}e^{ik_m(z-z^\prime)}
\tilde{{\phi}}^{m\dag}{\cal{V}}^{-1}\;\;\;\;\;z\geq{z}^\prime \\
\\
\sum_{m=1}^M{\phi}^{\bar{m}}e^{i\bar{k}_m(z-z^\prime)}
\tilde{{\phi}}^{\bar{m}\dag}{\cal{V}}^{-1}
\;\;\;\;\;z\leq{z}^\prime \\
\end{array}
\right.\;{.}
\label{eq2.27}
\end{equation}
The matrix ${\cal{V}}$ is defined by
\begin{equation}
{\cal{V}}=\sum^{M}_{l=1}H_{-1}\left[
{\phi}^{m}e^{-ik_m}\tilde{{\phi}}^{m\dag}-
{\phi}^{\bar{m}}e^{-i\bar{k}_m}\tilde{{\phi}}^{\bar{m}\dag}
\right]\;{,}
\label{eq2.28}
\end{equation}
and the set of vectors $\tilde{{\phi}}^{m\dag}$ 
($\tilde{{\phi}}^{\bar{m}\dag}$) are the duals of the set
${\phi}^{m}$ (${\phi}^{\bar{m}}$), defined by
\begin{equation}
\tilde{{\phi}}^{m\dag}{\phi}^{n}=
\tilde{{\phi}}^{\bar{m}\dag}{\phi}^{\bar{n}}=
\delta_{mn}\;{,}
\label{eq2.29}
\end{equation}
from which follows the completeness conditions
\begin{equation}
\sum^{M}_{m=1}
{\phi}^{m}\tilde{{\phi}}^{m\dag}=
\sum^{M}_{m=1}{\phi}^{\bar{m}}\tilde{{\phi}}^{\bar{m}\dag}=
{\cal I}\;{.}
\label{eq2.30}
\end{equation}

Equation (\ref{eq2.27}) is the RGF for a doubly infinite system.
For the semi-infinite one, this must vanish at the end of the lead.
In complete analogy with the one dimensional case of the previous section, consider first the 
left lead, which extends to $z=-\infty$ and terminates at $ z=z_0-1$, such that the position 
of the first missing slice is $z=z_0$.
In order to satisfy the vanishing boundary condition at $z=z_0$, one must subtract from the right 
hand side of (\ref{eq2.27}) a wave-function of the form 
\begin{equation}
\Delta_z(z',z_0)=\sum^{M}_{mn}{\phi}^{\bar{n}}
e^{i\bar{k}_nz}
\Delta_{nm}(z',z_0)\;{,}
\label{eq2.31}
\end{equation}
where $\Delta_{nm}(z',z_0)$ is a complex matrix, determined from the
condition that the Green function vanishes at $z_0$, which
yields
\begin{eqnarray}
\nonumber
\Delta_z(z^\prime,z_0)=\Delta_{z^\prime}(z,z_0)=\;\;\;\;\;\;\;\;\;\;\;\;\;\; \\
\nonumber \\
\sum^{M}_{m,n=1}{\phi}^{\bar{n}}e^{i\bar{k}_{n}(z-z_0)}
\tilde{{\phi}}^{\bar{n}\dag}{\phi}^{m}
e^{ik_m(z_0-z^\prime)}\tilde{{\phi}}^{m\dag}{\cal{V}}^{-1}
\;{.}\nonumber \\
\label{eq2.32}
\end{eqnarray}
For the purpose of computing the scattering matrix, I will require the
Green function of the semi-infinite left-lead
$\tilde{g}_{zz^\prime}(z_0)=g_{zz^\prime}-\Delta_z(z^\prime,z_0)$ evaluated on
the surface of the lead, namely at $z=z'=z_0-1$.
Note that in contrast with the Green's function of a doubly infinite lead, which
depends only on the difference between $z$ and $z^\prime$, the Green's
function $\tilde{g}$ of a semi-infinite lead for arbitrary $z, z^\prime$
is also a function of the position $z_0$ of the first missing slice beyond the
termination point of the lead.
Writing $g_{\rm{L}}=g_{(z_0-1)(z_0-1)}(z_0)$ yields for this surface
Green function
\begin{equation}
g_{\rm{L}}=\left[{\cal I}-\sum_{m,n}
{\phi}^{\bar{n}}e^{-i\bar{k}_n}
\tilde{{\phi}}^{\bar{n}\dag}{\phi}^{m}
e^{ik_m}\tilde{{\phi}}^{m\dag}
\right]{\cal{V}}^{-1}\;{.}
\label{eq2.33}
\end{equation}
Similarly on the surface of the right lead, which extends to $z=+\infty$,
the corresponding surface Green function is
\begin{equation}
g_{\rm{R}}=\left[{\cal I}-\sum_{m,n}
{\phi}^{n}e^{ik_n}
\tilde{{\phi}}^{n\dag}{\phi}^{\bar{m}}
e^{-i\bar{k}_m}
\tilde{{\phi}}^{\bar{m}\dag}
\right]{\cal{V}}^{-1}\;{,}
\label{eq2.34}
\end{equation}
which has been obtained by subtracting from $g$ the following wave-function 
\begin{eqnarray}
\nonumber
\Delta_z(z^\prime,z_0)=\Delta_{z^\prime}(z,z_0)=\;\;\;\;\;\;\;\;\;\;\;\;\;\; \\
\nonumber \\
\sum^{M}_{m,n=1}{\phi}^{n}e^{i{k}_n(z-z_0)}
\tilde{{\phi}}^{n\dag}{\phi}^{\bar{m}}
e^{i\bar{k}_m(z_0-z^\prime)}\tilde{{\phi}}^{\bar{m}\dag}{\cal{V}}^{-1}
\;{,}\nonumber \\
\label{eq2.34bis}
\end{eqnarray}
and
considering $g_{\rm{R}}=g_{(z_0+1)(z_0+1)}(z_0)$ ($z_0+1$ is the position of the
first slice of the right lead).

The expressions (\ref{eq2.33}) and (\ref{eq2.34}), when used in
conjunction with (\ref{eq2.25}) form a versatile method of determining
lead Green functions, without the need to perform $k$-space integrals
or a contour integration.
As a consequence of translational invariance of the doubly infinite system,
the surface Green functions are independent of the position of the surface
$z_0$. 

\subsubsection{The effective Hamiltonian of the scattering region: {\it decimation} technique}

I have already shown that, given the coupling matrix between the external surfaces of the leads, 
the Green function of the scatterer plus leads can be computed via Dyson's equation.
Generally the scattering region is not simply described by a coupling matrix
between surfaces, but is a complex Hamiltonian involving all the degrees 
of freedom of the scatterer. 
Therefore it is useful to develop a method that transforms such a detailed
Hamiltonian into an effective coupling matrix between the two surfaces.
For structures, which possess a quasi-one dimensional
geometry and a Hamiltonian which is block tri-diagonal, this can be
achieved by projecting out the internal
degrees of freedom of the scatterer. In the literature, depending
on the context or details of implementation, this
procedure is sometimes referred to as ``the recursive Green function
technique'' or ``the decimation method'', but is no more than
an efficient implementation of Gaussian elimination.

Consider a scatterer composed on $N-2M$ degrees of freedom. Then the
Hamiltonian for the scatterer plus semi-infinite leads is of the form
$H=H_L+H_R+\tilde H$, where $H_L$, $H_R$ are the Hamiltonians of the
left and right isolated leads and $\tilde H$ a $N\times N$ Hamiltonian
describing the scattering region and any additional couplings involving
surface sites of the leads induced by the presence of the scatterer.
The aim of the decimation method is to successively eliminate 
the internal degrees of freedom of the scatterer, $i= 1,2,...,N-2M$, to yield a 
$(2M) {\times} (2M)$ effective Hamiltonian $H_{\rm eff}$. 
After eliminating the degree of freedom $i=1$, $\tilde H$ is reduced to
a $(N-1) {\times} (N-1)$ matrix with elements
\begin{equation}
H_{ij}^{(1)}=\tilde H_{ij}+\frac{\tilde H_{i1}\tilde H_{1j}}{E-\tilde H_{11}}
\; . 
\label{eq2.35}
\end{equation}
Repeating this procedure $l$ times one obtains the ``decimated''
Hamiltonian at $l$-th order
\begin{equation}
H_{ij}^{(l)}=H_{ij}^{(l-1)}+\frac{H_{il}^{(l-1)}H_{lj}^{(l-1)}}{E-H_{ll}^{(l-1)}}
\;{,}
\label{eq2.36}
\end{equation}
and after $N-2M$ such steps, an effective Hamiltonian $H_{\rm eff}=H^{N-2M}$
of the form
\begin{equation}
H_{\rm{eff}}(E)=\left(
\begin{array}[4]{rr}
H_{\rm{L}}^*(E) & H_{\rm{LR}}^*(E)\\
H_{\rm{RL}}^*(E) & H_{\rm{R}}^*(E)\\
\end{array}
\right){\;}.
\label{eq2.37}
\end{equation}
In this expression,
$H_{\rm{L}}^*(E)$ ($H_{\rm{R}}^*(E)$) describes intra-surface
couplings involving degrees of freedom belonging to the surface of
the left- (right-) hand side
lead and $H_{\rm{LR}}^*(E)=H_{\rm{RL}}^*(E)^{\dagger}$
describes the effective coupling
between the surfaces of the left-hand side and the right-hand side leads.

Since the effective Hamiltonian is energy dependent, this procedure
is particularly useful when one wishes to compute the Green function
at a given energy. It is also very efficient in the presence of short range
interactions, because only matrix elements involving degrees of freedom
coupled to the decimated one, are redefined.
This latter aspect is very useful in the case that the scatterer has 
some periodicity and allows clever numerical optimizations \cite{ss_1}. 

Since the problem now involves only  $(2M) {\times} (2M)$ matrices,
it is straightforward to obtain the surface Green function for the whole system
(i.e. the two surfaces coupled through the scattering region)
by solving  Dyson's equation
\begin{equation}
G(E)=[g(E)^{-1}-{{H}}_{\rm{eff}}(E)]^{-1}
\;{,}
\label{eq2.38}
\end{equation}
where
\begin{equation}
g(E)=\left(
\begin{array}[4]{rr}
g_{\rm L}(E) & 0\;\;\;\;\; \\
0\;\;\;\;\; & g_{\rm R}(E) \\
\end{array}
\right){\;},
\label{eq2.39}
\end{equation}
with $g_L$ and $g_R$ given by equations (\ref{eq2.33}) and (\ref{eq2.34}).

\subsubsection{The ${\cal S}$ matrix and the scattering coefficients}

To extract the transport coefficients from the Green function, I will use a generalization 
\cite{sanvito} of the method described in reference \cite{lhr93} (in particular see A.26 of \cite{lhr93})
to the case of non-orthogonal scattering channels. The same method has been used in the one
dimensional case of section \ref{s1dc}.

Consider first the probability current for an electron in the Bloch state (\ref{eq2.20})
\begin{equation}
J_k=n_{k_\perp}v_{k_\perp}
\;{,}
\label{eq2.40}
\end{equation}
where $n_{k_\perp}$ is the probability of finding an electron 
in a slice and $v_{k_\perp}$ is the corresponding group velocity.
It follows that the vector 
\begin{equation}
\psi_{z}=\frac{1}{\sqrt{v_k}}e^{ikz}{\phi}^k
\;{,}
\label{eq2.41}
\end{equation}
is normalized to unit flux.
To compute the group velocity note that if $|\psi_k\rangle$
is an eigenstate (\ref{eq2.18}), whose projection onto slice $z$ is
$\psi_z$, then
\begin{eqnarray}
\nonumber
v_k=\frac{1}{\hbar}\frac{\partial}{\partial k}\langle\psi_k|H|\psi_k\rangle=
\;\;\;\;\;\;\;\;\;\;\; \\
\nonumber \\
=\frac{1}{\hbar}\frac{\partial}{\partial k}
\left[{\phi}^{k\dag}\left(H_0+H_1e^{ik}+H_{-1}e^{-ik}
\right){\phi}^k\right]=
\;\; \\
\nonumber \\
=\frac{i}{\hbar}
{\phi}^{k\dag}\left(H_1e^{ik}-H_{-1}e^{-ik}\right)
{\phi}^k
\;\;\;\;\;\;\;\;{,}\nonumber \\
\label{eq2.42}
\end{eqnarray} 
where the last step follows from equation (\ref{eq2.21})
and normalization of $\phi^k$.

It can be shown that the states (\ref{eq2.41})
diagonalize the current operator only if they correspond to distinct 
$k$ values.  In the case of degenerate $k$'s, the current is in general non-diagonal.
Nevertheless  it is always possible to define a rotation in the degenerate subspace
for which the current operator is diagonal
and in what follows, when a degeneracy
is encountered, I assume that such a rotation has been performed
(see Appendix \ref{AppC}).
With this convention, the current carried by a state of the form
\begin{equation}
\psi_z=\sum_m a_m\frac{e^{ik_mz}}{\sqrt{v_m}}{\phi}^{m}
\;{,}
\label{eq2.43}
\end{equation}
is simply $\sum_m \vert a_m\vert^2$.

It is now straightforward to generalize the analysis of \cite{lhr93} 
(end of paragraph 2.1) to
the case of non-orthogonal scattering channels.
Consider first a doubly infinite periodic structure, whose
Green function is given by
equation (\ref{eq2.27}). For $z\geq z'$,
acting on $g_{zz^\prime}$ from the right with
the following projector
\begin{equation}
P_m(z^\prime)=\frac{e^{ik_mz^\prime}}
{\sqrt{v_m}}{\cal{V}}{\phi}^{m}
\;{,}
\label{eq2.44}
\end{equation}
yields the normalized plane-wave (\ref{eq2.41}). Similarly by acting
on the Green function $g_{zz^\prime}(z_0)$ of a semi-infinite left-lead 
terminating
at $z_0$, one obtains for $z\geq z'$, $z_0\geq z$, an eigenstate of a
semi-infinite lead arising from a normalized incident wave along channel
$k_l$. Note that the projector introduced through the (\ref{eq2.44}) is
the generalization of the one defined for the one dimensional case.
In Appendix \ref{AppD} I will formally show that the projector that
projects the Green function for a double infinite system onto its 
corresponding wave-function, projects also the total Green function.

Thus the operator $P_m(z')$ and its left-moving counterpart
$\bar{P}_m(z')$ allow one to project-out wave-functions from the Green
function of a given structure.
For example, following the same procedure of the introduction,
if $G_{zz^\prime}$ is the retarded
Green function for a scattering region
sandwiched between two perfect leads whose surfaces are located
at the points $z=0$ and $z=L$,
then for $z'\leq 0$, the projected wave-function is of the form
\begin{equation}
\psi_z=\left\{
\begin{array}[4]{r}
\frac{e^{ik_mz}}{\sqrt{v_m}}{\phi}^{m}+
\sum_{n}\frac{r_{nm}}{\sqrt{\bar{v}_{n}}}
e^{i\bar{k}_{n}z}{\phi}^{\bar{n}}
\;\;\;\;\;z\leq{0} \\
\\
\;\;\;\;\ \sum_{n}\frac{t_{nm}}{\sqrt{{v}_{n}}}
e^{i{k}_{n}z}
{\phi}^{n}\;\;\;\;\;z\geq{L} \\
\end{array}
\right.\;
\;{,}
\label{eq2.45}
\end{equation}
where $r_{nm}=r_{\bar k_{n}, k_{m}}$, $t_{nm}=t_{k_{n},k_{m}}$ are
reflection  and transmission coefficients associated with an incoming state
from the left. In particular for $z=L$, $z'=0$, one obtains
\begin{equation}
\sum_{n}\frac{t_{nm}}{\sqrt{{v}_{n}}}
e^{i{k}_{n}L}{\phi}^{n}=G_{L0}P_m(0)
\;{,}
\label{eq2.46}
\end{equation}
and hence
\begin{equation}
t_{nm}=\tilde{{\phi}}^{n\dag}
G_{L0}{\cal{V}}{\phi}^{m}
\sqrt{\frac{{v}_{n}}{{v}_{m}}}e^{-ik_{n}L}
\;{,}
\label{eq2.47}
\end{equation}
where I used the definition of the dual vector $\tilde{{\phi}}$ given in 
equation (\ref{eq2.29}).
With the same method one evaluates all the other elements of the ${\cal S}$ matrix
\begin{equation}
t_{nm}^\prime=\tilde{{\phi}}^{\bar{n}\dag}
G_{0L}{\cal{V}}{\phi}^{\bar{m}}
\sqrt{\frac{{v}_{n}}{{v}_{m}}}e^{i\bar{k}_{n}L}
\;{,}
\label{eq2.48}
\end{equation}
\begin{equation}
r_{nm}=\tilde{{\phi}}^{\bar{n\dag}}
(G_{00}{\cal{V}}-{\cal I}){\phi}^{m}
\sqrt{\frac{v_n}{v_m}}
\;{,}
\label{eq2.49}
\end{equation}
\begin{equation}
r^\prime_{nm}=\tilde{{\phi}}^{n\dag}
(G_{LL}{\cal{V}}-{\cal I}){\phi}^{\bar{m}}
\sqrt{\frac{v_n}{v_m}}
\;{.}
\label{eq2.50}
\end{equation}
Since the right-hand sides of (\ref{eq2.47}-\ref{eq2.50}) involve only
the surface Green function of equation (\ref{eq2.38})
the transport coefficients are determined.
Moreover, since the above analysis is valid for any choice of the Hamiltonians 
$H_0$ and $H_1$, this approach is completely general.

\subsection{Alternative approach to linear response transport}

In this section I will review several alternative approaches to linear response transport. They are
all based on the Landauer formula, and the main differences concern the method for 
calculating the surface Green's functions, and for extracting the transport coefficients. 
Since the field is rather vast, here I will review only those methods that have been largely 
used for spin-transport calculations.

\subsubsection{Sharvin Conductance}
\label{sharsection}

A simple way to calculate the linear conductance of a nanoscaled system with a cross section
smaller than $\lambda_\mathrm{emf}$ and larger than the Fermi wave-length, is the calculation
of the Sharvin conductance \cite{sharvin,sharvin2}. The main idea is that of assuming the conductor
to be scattering free and evaluating the conductance by counting the number of open channels
at the Fermi level, for a system of finite cross section. These are proportional to the area of the 
cross section $A$ and the projections $S_\nu$ of the Fermi surface for different bands $\nu$
onto the plane normal to the transport direction $\hat{n}$. In this way the conductance for spin-$\sigma$
electrons can be written as
\begin{equation}
G^\sigma(\hat{n})=\frac{e^2}{h}\frac{A}{8\pi^2}\:{\sum_\nu}\:S_\nu^\sigma(\hat{n})=
\frac{e^2}{h}\frac{A}{8\pi^2}\:{\sum_\nu}\:\int\mathrm{d}\vec{k}|\hat{n}\cdot\vec{\nabla}_{\vec{k}}
\epsilon_{\nu\sigma}(\vec{k})|\delta(\epsilon_{\nu\sigma}(\vec{k})-E_\mathrm{F})\:,
\label{schar_cd}
\end{equation}
where $\epsilon_{\nu\sigma}(\vec{k})$ are the energy bands. The formula above has been
derived from both the Boltzmann equation \cite{wexler} and the Landauer formalism \cite{bauer1}.

The method is clearly rather crude, since the scattering problem is not solved. However, when combined with
accurate band structure calculations, it may provide useful preliminary insights. In addition it is worth pointing
out that the expression (\ref{schar_cd}) does not need a tight-binding-like Hamiltonian and it can be combined
with any band structure method, such as plain-wave based DFT.

For instance, this approach has been very successful in understanding the r\^ole of the electronic structure 
in the GMR effect \cite{kelly1,kelly2} and the transport through digital ferromagnetic heterostructures 
\cite{ss_4,ss_5}.

\subsubsection{Recursive methods}

The main idea behind recursive methods is to calculate either the wave- function or the 
Green's functions of an infinite system, by ``propagating'' the asymptotic solution 
(channel) across the scattering region. Here I will explain the basic concept by using 
wave-functions and a simple one dimensional hydrogen chain, and I will refer to the literature 
for more details.
\begin{figure}[ht]
\begin{center}
\includegraphics[width=15.0cm,clip=true,angle=0]{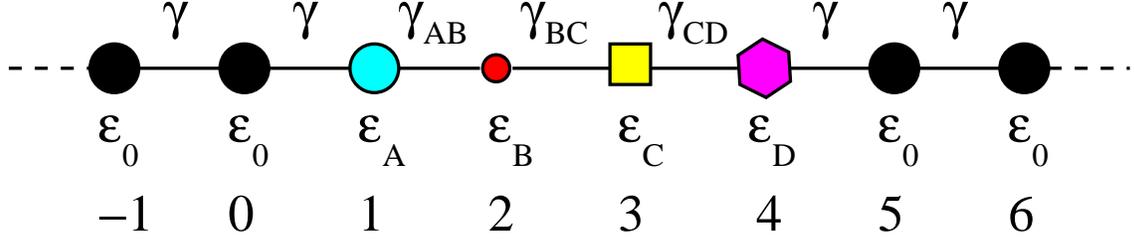}
\caption{One dimensional chain made from two semi-infinite linear chains with on-site energy $\epsilon_0$
and hopping integral $\gamma_0$ extending respectively between $5<j<+\infty$ and $-\infty<j<0$.
The two chains are connected through four different sites with varying on-site energies $\epsilon_j$
($j=1,...,4$) and hopping integrals $\gamma_{ij}$.}\label{fig12}
\end{center}
\end{figure}

Consider two semi-infinite linear chains described by a nearest neighbors single-orbital
tight-binding model with on-site energy $\epsilon_0$ and hopping integral $\gamma_0$
($\gamma_0<0$). The two chains are terminated respectively at the sites $l=0$ and $l=5$,
and are connected through four atoms with on-site energies $\epsilon_l$ ($l=1,..,4$) and
hopping integrals $\gamma_{ij}$ (see figure \ref{fig12}). As usual the general wave function 
can be written as
\begin{equation}
|\psi\rangle=\sum_l\psi_l|l\rangle\:,
\end{equation}
where $|l\rangle$ is the atomic orbital corresponding to the site $l$. In complete analogy to what
developed in the previous sections, it is clear that the knowledge of the coefficients $\psi_l$ over the
scattering region $l=1,..,4$ is not necessary in order to extract the transport coefficients.
The main idea is then to write the coefficients for $l<0$ in terms of those for $l>5$. This can be achieved 
by ``propagating'' the wave function is real-space over the scattering region.

Consider the Schr\"odinger for the coefficient $\psi_l$ at site $l$ equation associated with 
the linear chain under consideration
\begin{equation}
(\epsilon_l-E)\psi_l+\gamma_{l,\:l+1}\psi_{l+1}+\gamma_{l-1,\:l}\psi_{l-1}=0\:.
\end{equation}
This can be re-written as 
\begin{equation}
\left\{
\begin{array}{l}
(\epsilon_l-E)\psi_l+\gamma_{l,\:l+1}\psi_{l+1}+\gamma_{l-1,\:l}\psi_{l-1}=0 \\
\\
\psi_l=\psi_l
\end{array}
\right.\:,
\end{equation}
or in the convenient matrix form
\begin{equation}
\left(
\begin{array}{c}
\psi_{l+1} \\
\psi_{l}
\end{array}
\right)=
\left(
\begin{array}{cc}
-\gamma_{l,\:l+1}^{-1}(\epsilon_l-E) & -\gamma_{l,\:l+1}^{-1}\gamma_{l-1,\:l} \\
1 & 0
\end{array}
\right)
\left(
\begin{array}{c}
\psi_{l} \\
\psi_{l-1}
\end{array}
\right)=T_l(E)
\left(
\begin{array}{c}
\psi_{l} \\
\psi_{l-1}
\end{array}
\right)\:.
\label{tmeq}
\end{equation}

The matrix $T_l(E)$ connects the coefficients of the wave function at the sites $l+1$ and $l$ to
those at $l$ and $l-1$. This is called {\it transfer} matrix and allow us to express the coefficients of
a wave-function at a certain position in terms of those at another position. Thus the coefficient
at $l+n+1$ and $l+n$ are connected to those at $l$ and $l-1$ by
\begin{equation}
\left(
\begin{array}{c}
\psi_{l+n+1} \\
\psi_{l+n}
\end{array}
\right)=
T_{l+n}T_{l+n-1}...T_{l+1}T_l
\left(
\begin{array}{c}
\psi_{l} \\
\psi_{l-1}
\end{array}
\right)\:.
\end{equation}

Note that in the case of a infinite periodic system ($\epsilon_l=\epsilon_0$ and 
$\gamma_{ij}=\gamma_0$) the coefficients of the wave function are simple plane waves
$\psi_l=\phi^k\mathrm{e}^{ikl}$ and the equation (\ref{tmeq}) simply reduces to
\begin{equation}
\left(
\begin{array}{cc}
-\gamma_{0}^{-1}(\epsilon_0-E) & -1 \\
1 & 0
\end{array}
\right)
\left(
\begin{array}{c}
\psi_l \\
\psi_{l-1}
\end{array}
\right)=
\mathrm{e}^{ik}
\left(
\begin{array}{c}
\psi_l \\
\psi_{l-1}
\end{array}
\right)
\:.
\end{equation}
This is entirely equivalent to the equation (\ref{eq2.25}) for a generic quasi-one dimensional
system, once we have replaced $\epsilon_0$ with $H_0$ and $\gamma_0$ with $H_1$ (note that
in this case $H_{-1}=H_1^\dagger$ and therfore $H_1^{-1}H_{-1}={\cal I}$). 
Indeed the equation (\ref{eq2.25}) can be derived from the generalization of the 
transfer matrix method to multi-orbital tight-binding Hamiltonian. Hence the inverse band
structure problem $k=k(E)$ is entirely equivalent to finding a diagonal form for the 
transfer matrix $T_l$.

Going back to our original problem let us suppose to have calculated the transfer matrices connecting
the sites $l=-1,0$ to the sites $l=5,6$
\begin{equation}
\left(
\begin{array}{c}
\psi_{6} \\
\psi_{5}
\end{array}
\right)=
T_{5}T_{4}T_{3}T_2T_1T_0
\left(
\begin{array}{c}
\psi_{0} \\
\psi_{-1}
\end{array}
\right)={\cal T}
\left(
\begin{array}{c}
\psi_{0} \\
\psi_{-1}
\end{array}
\right)
\label{tm_lc}
\end{equation}
The crucial point is that for $l>5$ and $l<0$ the coefficients of the wave function are just plane waves
(channels)
\begin{equation}
\psi_l=\left\{
\begin{array}{ll}
a_1\mathrm{e}^{ikl}+b_1\:\mathrm{e}^{-ikl} & l\le0 \\
\\
a_2\mathrm{e}^{-ikl}+b_2\mathrm{e}^{ikl} & l\ge5
\end{array}
\right.
\end{equation}
This allows us to write a matrix equation for the amplitudes $a$ and $b$ 
\begin{equation}
\left(
\begin{array}{c}
a_2 \\
b_2
\end{array}
\right)
=
{\cal T}_k\left(
\begin{array}{c}
a_1 \\
b_1
\end{array}
\right)\:,
\end{equation}
with
\begin{equation}
{\cal T}_k=
\left(
\begin{array}{cc}
\mathrm{e}^{-i6k} & \mathrm{e}^{i6k} \\
\mathrm{e}^{-i5k} & \mathrm{e}^{i5k} 
\end{array}
\right)^{-1}
{\cal T}
\left(
\begin{array}{cc}
1 & 1 \\
\mathrm{e}^{-ik} & \mathrm{e}^{ik}
\end{array}
\right)\;.
\end{equation}

Finally, recalling the definition of ${\cal S}$ matrix
\begin{equation}
\left(
\begin{array}{c}
b_1 \\
b_2
\end{array}
\right)
=
{\cal S}\left(
\begin{array}{c}
a_1 \\
a_2
\end{array}
\right)
\end{equation}
one finds
\begin{equation}
{\cal S}=
\left(
\begin{array}{cc}
-{\cal T}_{21}/{\cal T}_{22} & 1/{\cal T}_{22} \\
1/{\cal T}^\star_{11} & {\cal T}_{12}/{\cal T}_{22}
\end{array}
\right)\:.
\end{equation}

It is then demonstrated that the transfer matrix method can fully describe a quantum mechanical
scattering problem. Moreover it can be generalized to both multi-orbitals tight-binding model, in the
spirit of the quasi-one dimensional systems described in section \ref{gsgf}, and to the 
calculation of the surface Green's functions (instead of the wave-functions).
In particular the use of the transfer matrix method to calculate the RGF for scattering problem
was introduced two decades ago by Lopez-Sancho {\it et al.} \cite{lopez1,lopez2} and then
used in a large range of scattering problems \cite{marcoBN}.

An interesting use of the transfer matrix method combined with Green's functions, is the ``adlayer''
method proposed by Mathon \cite{mat1,mat2}. Here the idea is to construct the RGF across the scattering
region by propagating the surface Green function for one of the leads. In this way the scatter is add
layer by layer to the lead by solving recursively the Dyson's equation.

\subsubsection{Layer KKR Methods}

The Korringa, Kohn and Rostoker (KKR) method is an electronic structures method
originally designed for calculating band structures of solids \cite{KKR1,KKR2} and then 
optimized to problems of defects and interfaces. The essential idea is to perform a 
muffin-tip approximation of the real ionic potential \cite{ashcroft} and to expand the
wave-function over a set of spherical harmonics. The KKR approximation consists
in truncating this expansion.

Recently the method has been combined with DFT and quantum transport with good results in
describing ballistic transport and tunneling. In particular large emphasis has been
given to the so called layer KKR method (LKKR) \cite{LKKR1}. Here the idea is to
divide the system into layers and to ``propagate'' the wave-function from layer to layer
with a recursive method resembling the transfer matrix technique. The final output is the 
electronic structure of a solid. The crucial point is that, since the algorithm is
essentially recursive, there is no need of periodicity in three dimensions. This is
why LKKR is suitable for transport problems. In this case one starts from a layer
deep the two leads and propagates the wave function across the scattering
region. Once the electronic structure is calculated,
the ballistic (linear response) conductance can be extracted. Several implementations have
been proposed within this scheme. These include a wave-function approach with Landauer
conductance \cite{LKKR2} and a Green's function scheme \cite{LKKR3,LKKR4} with current
extracted from the Kubo formula \cite{kubo}.

\subsubsection{Wannier Functions}

A Wannier function $w_{n\vec{R}}(\vec{r})$, labeled by the Bravais lattice vector 
$\vec{R}$, is obtained as a unitary transformation of a Bloch function 
$\psi_{n\vec{k}}(\vec{r})$ of the $n$-th band
\begin{equation}
w_{n\vec{R}}(\vec{r})=\frac{V}{(2\pi)^3}\int_\mathrm{BZ}
\psi_{n\vec{k}}(\vec{r})\:\mathrm{e}^{-i\vec{k}\cdot\vec{R}}\:d^3k\:,
\end{equation}
with $V$ the volume of the unit cell and the integration extending over the entire
Brillouin zone. The resulting function $w_{n\vec{R}}(\vec{r})$ is an atomic
orbital like function and it can be used to construct a tight-binding like Hamiltonian.

Recently a novel scheme for calculating ballistic transport using Wannier
functions have been proposed \cite{wan1}. The method consists first in performing
a DFT calculation and then to transform the Kohn-Sham eigenvectors into
a Wannier function basis set. Particular care is taken to construct rather localized
Wannier functions, and the maximally localized scheme is used \cite{marzari} . 
In this way the self-consistent Hamiltonian can be written
in a tight-binding form and the ballistic transport calculated using standard
Green's function techniques.
Although to my knowledge the method has never been used for spin-transport it appears
as rather versatile post-processing step in standard electronic structure calculations,
and it may become popular in the future.

\subsubsection{Ring Geometries}

One of the fundamental problems of the Landauer theory of transport is that although
it is based on the scattering states and therefore on a purely quantum mechanical system,
it needs the definition of two distinct chemical potentials in order to derive 
a relation between current and voltage. Of course the two pictures reconciliate in the 
zero bias limit, but a conceptual problem remains in the case of finite bias.

In recent years there have been several attempts to recast the Landauer theory
in a completely Hamiltonian form. The basic idea is to map the problem of two electronic
current/voltage probes coupled through a scattering region (as in the case of a two 
terminal transport measurement), on a macroscopically large ring with bias
applied as a time-dependent vector potential (see figure \ref{fig13}). In this case
the problem becomes completely Hamiltonian and it can be in principle solved
exactly, even in the presence of strongly interacting systems \cite{Kohn_transport}.
Although it is not clear whether the method will be ever extended to realistic materials, 
several fundamental results have been obtained. 
\begin{figure}[ht]
\begin{center}
\includegraphics[width=6.5cm,clip=true,angle=-90]{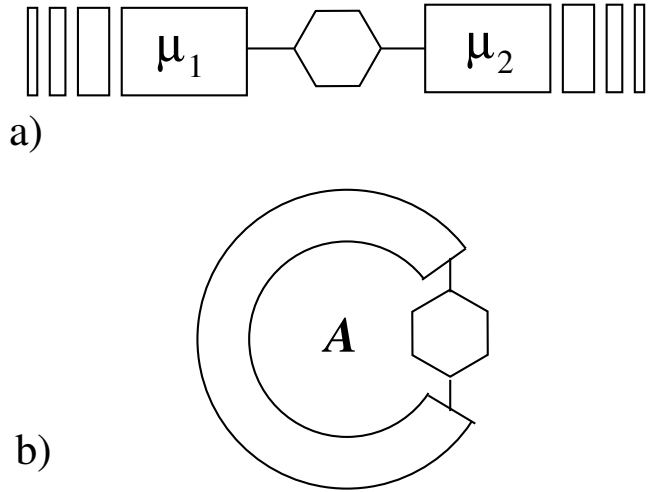}
\caption{Schematic diagram of the ring geometry. A conventional two terminal device a)
with leads carrying two different chemical potentials $\mu_1$ and $\mu_2$ is
mapped onto a large but finite ring b) with a magnetic flux threaded through the
center. The flux is expressed in terms of the vector potential $\vec{A}$.}\label{fig13}
\end{center}
\end{figure}

In particular Ram\v{s}ak and Rejec demonstrated that for non-interacting leads connected
to an interacting Fermi liquid region the zero-temperature conductance can be
calculated from the variation of the ground state energy with respect to the magnetic flux
\cite{ramsak1,ramsak2,ramsak3}. Moreover Burke et al. in the DFT framework showed that
the Landauer formula at zero bias can be recovered if the exchange correlation potential 
is local or semi-local \cite{kieron}. Finally Gebauer and Car proposed a method
in which the the density matrix of a many-electron system is propagated in time under 
the action of an external electric field \cite{burke_car,kieron2}. The method uses TDDFT in the 
master equation formulation.

\subsubsection{Superconductivity}
\label{supsection}

One interesting application of the linear response transport theory is in the study of
sub-gap transport across superconductor interfaces. This is particularly relevant for 
spin-transport since it was proposed as a method for measuring the spin-polarization of 
magnetic metals \cite{Upd1,Upd2,Soul}.
As explained in the previous section the relevant scattering mechanism for sub-gap 
conductance across a superconductor interface is the Andreev reflection \cite{Andreev}, 
where an electron is back-scattered from a superconducting interface into a normal metal 
leaving two electronic charges in the superconductor. This mechanism competes with normal
reflection and the conductance must be calculated by taking into account the two possible
scattering events. 

The basis for computing the scattering coefficients is formed by the Bogoliubov de 
Gennes equation \cite{BdG}, which are obtained by diagonalizing the mean-field BCS
Hamiltonian. For a non-magnetic, spin-singlet superconductor within a one dimensional 
single orbital tight-binding model they read
\begin{equation}
\begin{array}{c}
E\psi_i=\epsilon_i\psi_i-\gamma_0\sum_\delta\psi_{i+\delta}+\Delta_i\phi_i \\
\\
E\phi_i=-\epsilon_i\phi_i+\gamma_0^*\sum_\delta\phi_{i+\delta}+\Delta_i^*\psi_i\:,
\end{array}
\end{equation}
where $\psi_i$ ($\phi_i$) are the amplitudes of the electron (hole) wave function,
$\Delta_i$ is the superconducting gap and the sums run over nearest neighbors.
The model is formally identical to two one-dimensional chains with on-site energies and
hopping integrals respectively $\epsilon_i$, $\gamma_0$ and $-\epsilon_i$,
$-\gamma_0^*$, and coupled through the superconducting gap energy $\Delta_i$.
Clearly in the normal (non-superconducting) region electrons and holes are
not coupled and $\Delta_i=0$. Since the formal analogy with a normal problem, all the
scattering techniques developed in the previous sections can be used also for the 
superconducting problem. For an extensive review I refer the reader to more
specialized literature \cite{Lambert_Raimondi}.

Finally the same formalism can be applied to the case of ferromagnetic materials
in contact to superconductors. In this case the explicit spin-dependence
of the single-particle Hamiltonian (in the Bogoliubov de Gennes spirit) must be
introduced giving
\begin{equation}
H_{\rm BG}=
\left(
\begin{array}[h]{rrrr}
H_0^\uparrow & 0 \;\; & \underline{\Delta} \;\; & 0 \;\;\; \\
0 \;\; & H_0^\downarrow & 0 \;\;\; & -\underline{\Delta} \;\; \\
\underline{\Delta}^* & 0 \;\;  & -H_0^{\downarrow *} & 0 \;\;\; \\
0 \;\; & -\underline{\Delta}^* & 0 \;\;\; & -H_0^{\uparrow *} \\
\end{array}
\right)
\;{,}
\label{eq6.8}
\end{equation}
where $H_0^\sigma$ is the spin-dependent Hamiltonian describing the
normal state and $\underline{\Delta}=\Delta\times\cal{I}$ with $\cal{I}$ the
unit matrix. Note that I have now generalized the formalism to arbitrary dimensions
and multiple orbitals tight-binding model, being $H_0$ a general $N\times N$ matrix.
In the case of a ferromagnet we have $\Delta=0$, while for a superconductor
$H_0^\uparrow=H_0^\downarrow$. Again all the scattering techniques developed so far can be
used for calculating the linear response conductance. Finally it is important to note that 
the exchange coupling in magnetic transition metals is typically three order of 
magnitudes larger than the superconducting gap in ordinary superconductors. For this 
reason one does not expect the spin-polarization of the magnetic metal to change with 
the onset of the superconductivity, and therefore the same tight-binding parameters can 
be used in the normal and superconducting case. Usually one considers abrupt interfaces
and both the magnetization and the superconducting parameter are taken as step
functions across the interface.

\setcounter{equation}{0}
\section{Transport Theory: Non-equilibrium Transport}

\subsection{Introduction}

The scattering theory outlined in the previous section is based on the simple idea of measuring 
the probability of an electron leaving one current/voltage probe to be absorbed by another probe. 
In all the discussion the details of the electronic structure of the scattering region was 
secondary and we were only interested in the resulting scattering potential. In actual fact I 
have deliberately eliminated all the information about the scattering region with the decimation 
scheme. Clearly this procedure compromises our knowledge of how the charge density distributes 
in the scattering region, and of the detailed shape of the scattering potential. At zero bias, 
in the limit where the Landauer approach is strictly applicable, this information is somehow 
immaterial, but it becomes crucial at finite bias.

In this section I will develop an alternative method to evaluate the current between two probes, 
which fully preserves the knowledge of the electronic charge density and potential. This method 
is way less intuitive and transparent than the previous one, but it has the advantage of allowing 
a self-consistent evaluation of the potential, and in principle to include inelastic scattering. 
The development will not be formal and aims to give a consistent intuitive picture of the method. 
More formal derivations can be find a in rather vast literature \cite{caroli,ferrer,datta,Haug,Xue}. 

\subsubsection{A simple model}

Let me start with a simple example designed by Datta \cite{datta2}. Consider two current/voltage probes
having two different chemical potentials $\mu_1$ and $\mu_2$ and connected through an atom described by
a single energy level $\epsilon$ (see Figure \ref{fig14}). If the energy level is in equilibrium with the first contact
its occupation will be given by 
\begin{equation}
f_1(\epsilon)=\frac{1}{1+\mathrm{e}^{(\epsilon-\mu_1)/k_\mathrm{B}T}}\:.
\end{equation}
In the same way the condition of equilibrium with the second contact gives an occupation 
$f_2(\epsilon)$. However if $\mu_1\ne\mu_2$ equilibrium cannot be established with either 
the first and the second probe and the average number $N$ of electrons occupying the state 
will be a value intermediate between $f_1$ and $f_2$. This imbalance between the equilibrium 
occupation and the real occupation produces the current to flow. In fact the net flux $I_1$ 
across the left contact is proportional to $f_1-N$:
\begin{equation}
I_1=e\frac{\gamma_1}{\hbar}(f_1-N)\;,
\label{cur1}
\end{equation}
while the flux $I_2$ across the right contact is
\begin{equation}
I_2=e\frac{\gamma_2}{\hbar}(f_2-N)\;.
\label{cur2}
\end{equation}
where $\gamma_\alpha/\hbar$ are the transmission rate between the contact $\alpha$ and the 
single energy level $\epsilon$.
\begin{figure}[ht]
\begin{center}
\includegraphics[width=4.5cm,clip=true,angle=-90]{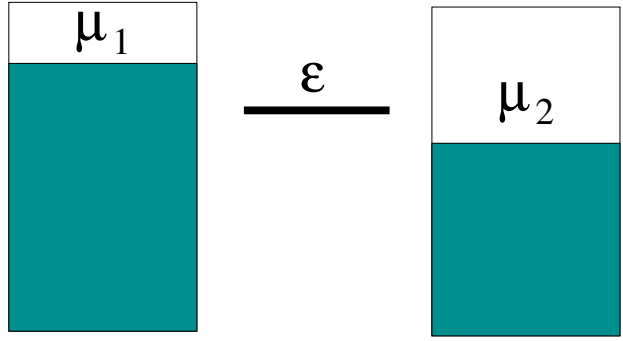}
\caption{Schematic diagram of two current/voltage probes attached through a single level system
of energy $\epsilon$. The two probes are kept at the two chemical potentials
$\mu_1$ and $\mu_2$.}\label{fig14}
\end{center}
\end{figure}

At steady state there is no net flux into or out of the energy level, 
thus $I_1+I_2=0$, and one can simply derive an equation for $N$
\begin{equation}
N=\frac{\gamma_1f_1+\gamma_2f_2}{\gamma_1+\gamma_2}\;.
\label{Nsm}
\end{equation}
This allow us to finally obtain the current, by substituting equation (\ref{Nsm}) into one of the 
equations for the flux
\begin{equation}
I=-I_1=I_2=\frac{e}{\hbar}\frac{\gamma_1\gamma_2}{\gamma_1+\gamma_2}[f_1(\epsilon)-f_2(\epsilon)]\;.
\label{Nsc}
\end{equation}
Unless one considers very high temperatures the current will flow only if the energy level $\epsilon$ 
is located between the two chemical potentials. In fact if $\epsilon$ is below both $\mu_1$ and $\mu_2$ 
then, $f_1\sim f_2\sim 1$, while if it is above $f_1\sim f_2\sim 0$.

Note that the model developed so far provides information on both the occupation of the energy level 
(equation (\ref{Nsm})) and the steady state current (equation (\ref{Nsc})). This is a major advantage with
respect to the methodology developed in the previous section, since it naturally allow us to introduce 
self-consistent procedures. For instance if one assumes that the position of the energy level is a
function of its occupation $\epsilon=\epsilon(N)$, then one can calculate self-consistently both N and
$\epsilon$ by looping over the equation (\ref{Nsm}), as suggested in figure \ref{fig15}. 
\begin{figure}[ht]
\begin{center}
\includegraphics[width=8.5cm,clip=true,angle=-90]{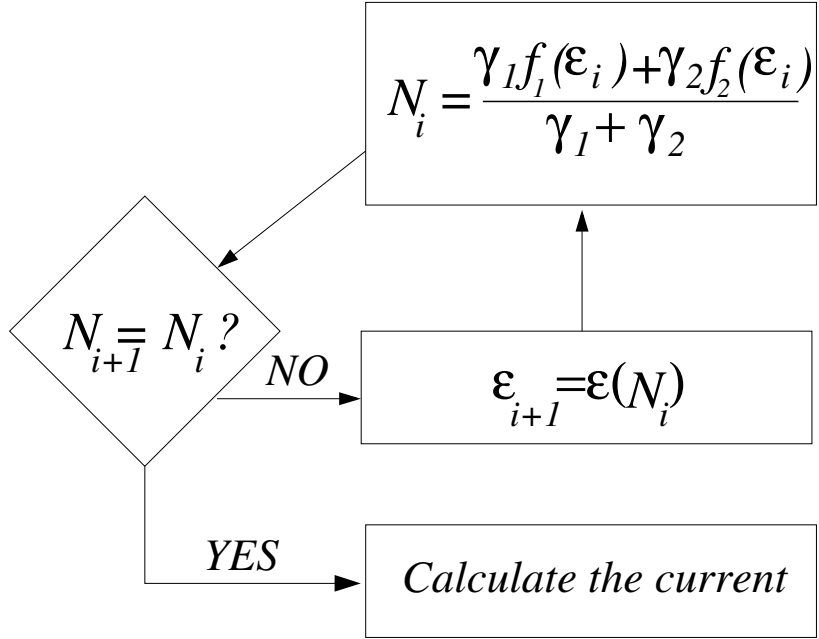}
\caption{Flow diagram of the self-consistent procedure for calculating $\epsilon$ and $N$.}\label{fig15}
\end{center}
\end{figure}

\subsubsection{Broadening and Green's function}

Our simple model contains already most of the aspects of a more general theory based on the 
non-equilibrium Green's function method. To complete the analogy I will introduce an additional 
element in the description, this is the broadening of the energy level. In fact, since the energy level 
$\epsilon$
is coupled to the leads by the transmission rates $\gamma_1/\hbar$ and $\gamma_2/\hbar$, its 
lifetime is finite and given by $\hbar/\gamma$ ($\gamma=\gamma_1+\gamma_2$). The uncertainty principle 
then requires the state to have an energy spread of $\gamma$. This is introduced by describing the
DOS of the energy level $D_\epsilon(E)$ as a Lorentzian function centered around $E=\epsilon$
\begin{equation}
D_\epsilon(E)=\frac{\gamma/2\pi}{(E-\epsilon)^2+(\gamma/2)^2}\;.
\label{Lor}
\end{equation}
Clearly $D_\epsilon$ should have the property to hold only one electron, hence its energy integral is
equal to one. It is then quite intuitive to re-define the expressions for the occupation and the current across
the two terminals in terms of the DOS associated to $\epsilon$ as follows
\begin{equation}
N=\int_{-\infty}^{+\infty}\mathrm{d}E\:D_\epsilon(E)
\frac{\gamma_1f_1(E)+\gamma_2f_2(E)}{\gamma}\;,
\label{Nsmb}
\end{equation}
\begin{equation}
I=\frac{e}{\hbar}\int_{-\infty}^{+\infty}\mathrm{d}E\:D_\epsilon(E)
\frac{\gamma_1\gamma_2}{\gamma}[f_1(E)-f_2(E)]\;,
\label{Nscb}
\end{equation}
where I have deliberately left the $\gamma$'s inside the integral since in general the broadening 
can be a function of the energy $\gamma=\gamma(E)$. Note that the argument of the integral in equation
(\ref{Nsmb}) is nothing but the electron charge density at the steady state $n(E)=D_\epsilon(E)
[\gamma_1f_1(E)+\gamma_2f_2(E)]/\gamma$, and that the expression for the two terminals current can be 
written as
\begin{equation}
I=\frac{e}{\hbar}\int_{-\infty}^{+\infty}T(E)[f_1(E)-f_2(E)]\;,
\label{Nsct}
\end{equation}
with
\begin{equation}
T(E)=D_\epsilon(E)\frac{\gamma_1\gamma_2}{\gamma}\;.
\label{Tra}
\end{equation}
$T(E)$ is simply the total transmission probability and the equation (\ref{Nsct}) reproduces the result
obtained when starting from the two terminal Landauer theory of equation (\ref{lb_cub}).
It is also useful to rewrite the two terminal currents (\ref{cur1}) and (\ref{cur2}) in terms 
of $n(E)$ and $D(E)$, since this gives us a powerful tool to generalized our formalism to 
multi-terminal devices. For a generic lead $\alpha$ this can be written in a compact form as
\begin{equation}
I_\alpha=\frac{e}{\hbar}\int_{-\infty}^{+\infty}\mathrm{d}E[\gamma_\alpha^\mathrm{in}D_\epsilon(E)-
\gamma_\alpha n(E)]\:,
\end{equation}
where we have introduced the in-scattering function $\gamma_\alpha^\mathrm{in}=\gamma_\alpha
f_\alpha$.

In the expression developed so far I have written both the current and the occupation of
the energy level only in terms of the DOS, the electron charge density and transmission probabilities.
Furthermore it is important to note that I did not introduce any notion of quantum mechanics, except for
the uncertainty principle, and that all my arguments are essentially based on a system of rate
equations. It is now useful to recast the previous equations in terms of a more powerful analytical 
tool, the single particle Green's function $G$
\begin{equation}
G(E)=\frac{1}{E-\epsilon+i\gamma/2}\:.
\label{elspgf}
\end{equation}
This will offer us the opportunity to generalized the simple concepts introduced so far to more complicated 
electronic structure and self-consistent methods. Both the DOS and the charge density
can be written as a function of $G$
\begin{equation}
D_\epsilon(E)=\frac{1}{2\pi}G(E)\gamma(E)G^*(E)=i[G(E)-G^*(E)]\:,
\end{equation}
\begin{equation}
n(E)=\frac{1}{2\pi}G(E)\gamma^\mathrm{in}(E)G^*(E)\:.
\label{cdens}
\end{equation}
where $\gamma^\mathrm{in}=\gamma_1^\mathrm{in}+\gamma_2^\mathrm{in}$. Similarly
the two terminal current is
\begin{equation}
I=\frac{e}{h}\int_{-\infty}^{+\infty}\mathrm{d}E\:G(E)\gamma_1(E)G^*(E)\gamma_2(E)[f_1(E)-f_2(E)]\;.
\label{CurrG}
\end{equation}
The non-equilibrium Green's functions (NEGF) method consists in generalizing the equation
written above to more complicated electronic structures.

Before leaving this section let me summarize the philosophy adopted throughout this derivation. 
The main difference from scattering theory is that now I focus my attention on the
energy level (the actual conductor), instead then on the leads. Hence I define the conditions
for the steady state, where the current is obtained as a balance of the currents entering 
and leaving the conductor. Finally I recast all the formalism in terms of single particle Green's 
function, which is the proper theoretical tool for bridging this simple model with the more 
powerful NEGF theory. 

\subsubsection{NEGF Method}
\label{NEGFM}

The simple model introduced in the previous section does not describe two important aspects 
of a real device: the electronic structure of the contacts and the details of the scattering 
region Hamiltonian. These are replaced respectively by the Fermi distribution functions 
and by a single energy level. It is the goal of this section to bring these details back in our
formalism.

Let us assume again that we can express the Hamiltonian in an
orthogonal tight-binding form. In the last part of the previous section I have re-written
all the fundamental quantities (energy level occupation and current) in terms of the 
energy level Green's function $G(E)$ (see equation (\ref{elspgf})), and the coupling with
the leads ($\gamma$'s). It is therefore natural to generalize $G(E)$
to a more detailed Hamiltonian $H_\mathrm{S}$, describing all the degrees of freedom of the
scattering region
\begin{equation}
G(E)=\lim_{\eta\rightarrow 0}[(E+i\eta)-H_\mathrm{S}-\Sigma_\mathrm{1}-\Sigma_2]^{-1}\:,
\label{negfmx}
\end{equation}
where $\eta$ has been introduced to respect causality. In this general form the coupling of the
scattering region with the left and right lead is described respectively by the left and right
{\it self-energy} $\Sigma_\mathrm{1}$ and $\Sigma_2$. These are energy dependent 
matrices containing all the details of the electronic structures of the leads and their
coupling with the scattering region. Therefore $G(E)$ is formally the Green's function
associated to
an effective Hamiltonian $H_\mathrm{eff}=H_\mathrm{S}+\Sigma_\mathrm{1}+\Sigma_2$,
with the self-energies taking the r\^ole of external potentials. It is also important to note
that the self-energy matrices are in general complex quantities, with the expected result
that the total number of electrons in the scattering region is not a conserved quantity.
The self-energies naturally extend the idea of transmission rates $\gamma$ introduced in 
the previous section and their pure complex part $\Gamma_\alpha$ ($\alpha=$1, 2) is 
usually called the broadening matrix
\begin{equation}
\Gamma_\alpha=i[\Sigma_\alpha-\Sigma_\alpha^\dagger]\:.
\label{broadmat}
\end{equation}
In addition the $\Sigma$'s possess a non-vanishing real part. These shift the energy levels 
of the scattering region, an effect that was neglected in the simple model.

We can then define the two terminal current as
\begin{equation}
I=\frac{e}{h}\int_{-\infty}^{+\infty}\mathrm{d}E\:\mathrm{Tr}[G(E)\Gamma_\mathrm{1}(E)
G^\dagger(E)\Gamma_2(E)][f_\mathrm{1}(E)-f_2(E)]\;.
\label{CurrNEGF}
\end{equation}
This is similar to that of equation (\ref{CurrG}), where now the total transmission coefficient
is obtained as the trace of $G\Gamma_\mathrm{1}G^\dagger\Gamma_2$. The derivation of 
equation (\ref{CurrNEGF}) follows the same arguments of {\it in going} and {\it out-going}
currents used for the derivation of its single energy counterpart (\ref{CurrG}), and quantities
such as the {\it in scattering} matrix $\Sigma^\mathrm{in}_\alpha=\Gamma_\alpha\:f_\alpha$
can be defined.

The equations (\ref{negfmx}) and (\ref{CurrNEGF}) allow us to calculate the current once the
Hamiltonian for the scattering region and the self-energies are known. While I will give an 
expression for the self-energies in the next section, here I would like to point out that
the Hamiltonian $H_\mathrm{S}$ is not always known exactly. In general its functional dependence
on the electronic structure is known but its exact value needs to be calculated 
self-consistently. In general I will assume that $H_\mathrm{S}$ depends on the 
single particle density matrix $\rho$ associated to the scattering region 
$H_\mathrm{S}=H_\mathrm{S}(\rho)$. This is a rather general assumption underpinning 
the fact that $H_\mathrm{S}$ can be constructed with a single particle
electronic structure method (DFT, Hartree Fock etc.). It can be viewed as a natural
generalization of the dependence of the energy level $\epsilon$ on its occupation
$\epsilon=\epsilon(N)$ discussed in the previous section.

It is therefore important to evaluate the density matrix $\rho$ from
the Green's function $G(E)$ of the scattering region. Following the analogy with
our simple model the {\it energy dependent} density matrix $n(E)$ can be 
written as (see equation (\ref{cdens}))
\begin{equation}
n(E)=\frac{1}{2\pi}G(E)\Gamma^\mathrm{in}(E)G^\dagger(E)=
\frac{1}{2\pi}G(E)[\Gamma_\mathrm{1}(E)f_\mathrm{1}(E)+\Gamma_2(E)f_2(E)]
G^\dagger(E)\:,
\label{cdensnegf}
\end{equation}
while the density matrix is obtained by integrating $n(E)$
\begin{equation}
\rho=\frac{1}{2\pi}\int dE\:
G(E)[\Gamma_\mathrm{1}(E)f_\mathrm{1}(E)+\Gamma_2(E)f_2(E)]
G^\dagger(E)\:.
\label{denmatrix}
\end{equation}

As in the case of the simple model we have now a clear prescription on how to construct 
a self-consistent calculation. Let us suppose to have calculated the self-energies and that
these do not change during our self-consistent procedure (see section \ref{hcse}).
We first compute the scattering region Green's function
(equation (\ref{negfmx})) for $H_{\mathrm{S}}=H_{\mathrm{S}}(\rho_0)$, where the Hamiltonian 
$H_{\mathrm{S}}$ is evaluated at a given initial charge density $\rho_0$. Then from the
Green's function $G$ we calculate the new charge density $\rho_1$
(equation (\ref{denmatrix})), which is then used to 
construct the new Hamiltonian $H_{\mathrm{S}}\left(\rho_1\right)$. This procedure is iterated 
until reaching self-consistency, that is when $\rho_{\mathrm{n+1}}=\rho_\mathrm{n}$.
At this point the scattering region Hamiltonian is known exactly and finally the current can be 
extracted by using the expression (\ref{CurrNEGF}).

\subsubsection{How to calculate the Self-energies}
\label{hcse}

In this section I will present a simple argument which will enable us to evaluate the
leads self-energies. Let us consider a scattering region attached
to only one lead as shown in figure \ref{fig16}. In this case the Hamiltonian $H$ for 
the whole system (scattering region plus lead) is a semi-infinite matrix of the form
\begin{equation}
H=
\left(
\begin{array}{cc}
H_1 & H_\mathrm{1S}  \\
H_\mathrm{1S}^\dagger & H_\mathrm{S}  
\end{array}
\right)\:.
\label{Htot}
\end{equation}

The corresponding RGF $G=[E+i\eta-H]^{-1}$ can be partitioned
in a block structure and obtained by formal inversion of the Hamiltonian
\begin{equation}
\left(
\begin{array}{cc}
G_1 & G_\mathrm{1S}  \\
G_\mathrm{S1} & G_\mathrm{S}  
\end{array}
\right)=
\left(
\begin{array}{cc}
(E+i\eta){\cal{I}}+H_1 & H_\mathrm{1S}  \\
H_\mathrm{1S}^\dagger & E{\cal{I}}+H_\mathrm{S}  
\end{array}
\right)^{-1}\:.
\label{Gtot}
\end{equation}
Our task is to extract an expression for the portion $G_\mathrm{S}$, which represents
the Green's function for the scattering region attached to the lead 1. From the equation
(\ref{Gtot}) one can write
\begin{equation}
[(E+i\eta){\cal{I}}-H_1]G_\mathrm{1S}+H_\mathrm{1S}G_\mathrm{S}=0\:,
\label{gfeq1}
\end{equation}
and
\begin{equation}
[E{\cal{I}}-H_\mathrm{S}]G_\mathrm{S}+H_\mathrm{1S}^\dagger G_\mathrm{1S}=0\:.
\label{gfeq2}
\end{equation}
From the first equation we can extract
\begin{equation}
G_\mathrm{1S}=-g_1H_\mathrm{1S}G_\mathrm{S}\:,
\label{gfeq3}
\end{equation}
with
\begin{equation}
g_1=[(E+i\eta){\cal{I}}-H_1]^{-1}
\:.
\label{gfeq4}
\end{equation}
Note that $g_1$ is the retarded Green's function for the semi-infinite lead 1,
a quantity that we have already calculated in section \ref{gsgf}. Finally substituting 
Eq.(\ref{gfeq3}) into Eq.(\ref{gfeq2}) we derive the expression for the Green's function
of the scattering region
\begin{equation}
G_\mathrm{S}=[E{\cal{I}}-H_S-H_\mathrm{1S}^\dagger g_1H_\mathrm{1S}]^{-1}
\:.
\label{gfeq5}
\end{equation}
The expression (\ref{gfeq5}) is identical to that of equations (\ref{negfmx}),
which allows us to identify $H_\mathrm{1S}^\dagger g_1H_\mathrm{1S}$ as the
self-energy of lead 1. 
\begin{figure}[ht]
\begin{center}
\includegraphics[width=4.5cm,clip=true,angle=-90]{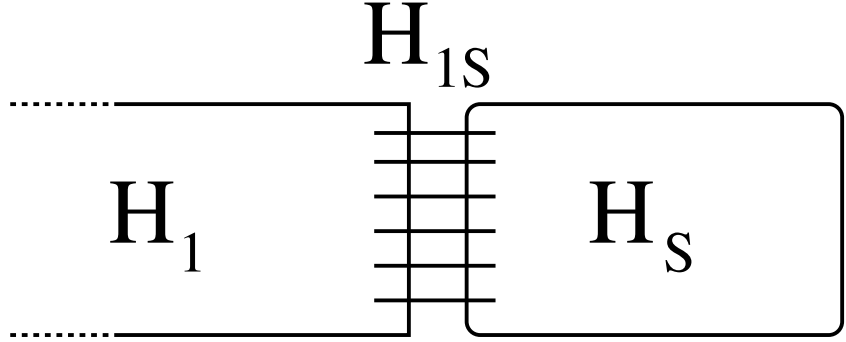}
\caption{A scattering region described by the Hamiltonian $H_\mathrm{S}$ is attached
to the lead 1. This is described by the semi-infinite Hamiltonian
matrix $H_1$. The coupling between the scattering region and the lead
is described by the matrix $H_\mathrm{1S}$}\label{fig16}
\end{center}
\end{figure}

Although $H_\mathrm{1S}$ is in principle a $N\times N$ matrix with $N$
the number of degrees of freedom in the scattering region, in practice
it will couple only a limited number of atoms (that we generally called 
surface atoms). This means that only surface atoms on the leads are relevant 
for the product $H_\mathrm{1S}^\dagger g_1H_\mathrm{1S}$,
allowing us to identify $g_1$ with the surface Green's function.

It is important to point out that a similar result could have been obtained by decimating
layer by layer all the degrees of freedom of the lead (see for example \cite{marcoBN}). This procedure,
introduced for the first time by Lopez-Sancho et al. \cite{lopez1,lopez2} does not require
the decimation of the whole semi-infinite lead and a few iterations of a recursive set of 
equations are sufficient. However, the recursive method nicely highlights the differences
between the NEGF method and the scattering theory. With the first
method one decimates the leads in order to extract the self-energies (effects of the leads
on the scattering region). In the second, one decimates the scattering region in order
to evaluate an effective scattering potential between the leads. Clearly the two methods are 
equivalent and one has to decide which is the best method for the specific problem.

\subsubsection{Introducing the bias}

The expression for the current given by the equation (\ref{CurrNEGF}) is very similar to its
Landauer-B\"uttiker counterpart (\ref{lb_cub}) and essentially it involves the integration
of the transmission coefficient $T(E)=\mathrm{Tr}[\Gamma_\mathrm{1}G^\dagger\Gamma_2]$ over 
the bias window.
However in the present case the availability of information about the charge density in the
scattering region (and thus about the Hamiltonian), gives us the chance to introduce a 
self-consistent evaluation of the potential drop, and therefore to replace $T(E)$ with
$T(E,V)$ in the expression for the current.

There are several ways of introducing finite bias in the self-consistent calculation
and the details usually depend on the particular numerical implementation chosen. Here 
I will only provide a few general concepts.

\vspace{0.3cm}
\noindent
{\it{\underline{The ``effective'' scattering region}}}

The key consideration in setting up a self-consistent calculation is to observe that 
the potential drop will only affect a small portion of the entire device (defined as
leads plus scattering region). In general the leads are metallic and one expects 
the electrical potential to relax rapidly to its bulk value when moving away from the 
surface. This is simply due to the strong electron screening properties of ordinary metals.
For this reason one usually performs a self-consistent calculation of the potential
drop by defining an ``effective'' scattering region comprising the scatterer and a few
layers of the leads. The number of layers introduced in this new scattering region
should be large enough to reproduce the bulk potential at the edges. 

\vspace{0.3cm}
\noindent
{\it{\underline{Displacing the potential in the leads}}}

The effect of applying a finite bias over the leads is that of shifting their relative chemical
potentials. This, in addition to the necessary condition of local charge neutrality, results
in a relative displacement of the whole band structure of the leads. In practice one displaces
the bottom of the left-hand side lead by $+V/2$ and that of the right-hand side lead by
$-V/2$ (and the opposite for reverse bias), with $V$ the bias applied. In a tight-binding language
this means shifting rigidly the Hamiltonian $H_\alpha$ ($\alpha$=1, 2) of the leads 
\begin{equation}
H_{1/2}\rightarrow H_{1/2}\pm\frac{V}{2}{\cal{I}}\:,
\label{VHam}
\end{equation}
and in terms of self-energies it translates into the following expression
\begin{equation}
\Sigma_{1/2}(E)\rightarrow\Sigma_{1/2}(E\mp V/2)\:.
\label{VselfEn}
\end{equation}

\vspace{0.3cm}
\noindent
{\it{\underline{Trial potential drop and self-consistent procedure}}}

In principle the rigid shift of the leads potential and the appropriate choice of the
effective scattering region should be enough to determine the potential drop. In fact
the first sets the boundary conditions and the second ensures that those boundary 
conditions can be met by an appropriate electrostatic calculation. At this point one
needs an efficient way to solve the Poisson equation for the effective scattering
region under the given boundary conditions. A conventional way is to add a linear potential
drop to the Hamiltonian describing the scattering region
\begin{equation}
H_\mathrm{S}(V)=H_\mathrm{S}+\beta V\:{\cal{I}}\:,
\label{linear}
\end{equation}
and then to iterate the self-consistent procedure described in section \ref{NEGFM}
once the shifted self-energies of equation (\ref{VselfEn}) are used.

\subsection{DFT method}

Density functional theory (DFT) is nowadays the most popular electronic
structure method among several communities. It is based on the famous Hohenberg-Kohn 
theorem stating that the ground state energy of a system of $N$ interacting particles 
is a unique functional of the single particle charge density \cite{HKohn}.
The prescription for calculating both the charge density and the total energy is that to
map the exact functional problem onto a fictitious single-particle Hamiltonian problem, 
known as the Kohn-Sham Hamiltonian \cite{KSham}. This comprises a kinetic part,
a part describing the interaction between electrons and nuclei, a classical
electrostatic part and the exchange and correlation potential. This latter in principle
includes all the many body effects, although its actual form in unknown.

\vspace{0.3cm}
\noindent
{\it{\underline{DFT for magnetic materials}}}

Density functional calculations for magnetic materials became
widespread in the late 1970s, with a number of studies of third
and fourth row transition metals \cite{Janak_Williams,Wang_Callaway,Moruzzi_77}.
These studies established that the local density
approximation gives results that are in reasonable agreement with
experiments for quantities such as cohesive energy, bulk modulus and
magnetic moments, provided that spin polarization is included explicitly,
by extending the LDA to the local spin density approximation (LSDA) \cite{vbh}.
They also noted, however, that the calculated properties are very sensitive to
details of the structure and magnetic ordering, which can lead to
discrepancies between the LSDA results and experiment. The
most notorious of these is the well known prediction of the
incorrect ground state of iron (face centered cubic and antiferromagnetic,
rather than the correct body centered cubic and ferromagnetic)
by the LSDA.
 
A number of technical developments have facilitated the study of
magnetic materials, perhaps the most important being the
introduction of the fixed spin moment (FSM)
method \cite{Schwarz_Mohn,Moruzzi_FSM}. In the FSM method the ground
state of a constrained system with a fixed magnetic moment is
calculated. Not only this does speed convergence, but the total energy
{\it surface} in magnetic moment/volume space can be determined,
giving additional information particularly  about metastable magnetic
phases.  Also, the implementation of Gaussian smearing \cite{Fu_Ho} and
related schemes have helped to speed convergence of calculations for magnetic
metals with partially filled $d$ bands and complex Fermi surfaces, in
which it is difficult to carry out integrals over the occupied part of
the Brillouin Zone.
 
In parallel with these technical developments, extensions and improvements
to the LSDA have also been explored. The usual generalized gradient approximation (GGA)
\cite{PBE96} that gives improved results for
non-magnetic systems do not tend to give systematic improvement for magnetic
materials, although the GGA does at least predict the correct ground state
for iron. For more information about these approximations see reference
\cite{David_Singhs_book} and references therein. 
Finally it is worth mentioning that methods such as the
LDA+U \cite{anis1,anis2}, and self-interaction-correction \cite{pzu} are 
specifically tailored to treat strongly correlated systems, and therefore 
are more appropriate for magnetic systems with narrow $d$ or $f$ bands.  

\vspace{0.3cm}
\noindent
{\it{\underline{DFT for transport}}}

The use of DFT in quantum transport algorithms has certainly a recent history.
Most of the early activity has been focused on using DFT for extracting parameters
for simpler models (tight-binding) and in conjunction with 
Landauer theory. Only in the last five years the connection between DFT and 
self-consistent finite-bias transport has become strong. The general idea is
to use DFT as a single-particle theory. In this way the Kohn-Sham eigenstates
are interpreted as single particle excitations and non-equilibrium transport methods
can be applied. This is clearly not completely satisfactory, since formally 
the Kohn-Sham eigenstates do not represent an effective single particle theory, but only 
the eigenstates of the fictitious Kohn-Sham Hamiltonian. Nevertheless they often 
reproduce rather well the single particle excitations and one expects that in the same limit
also the transport will be well described. The discussion on this and related
issues is at present very active.

To date there are a few codes available for performing DFT-transport. Most of them
are based on the NEGF method and they essentially differ
for the numerical implementations considered. To our knowledge there is only one 
fully self-consistent DFT code, which can deal with spin-polarized systems.
This is {\it Smeagol}, a code developed by the present author and collaborators 
and entirely designed 
for magnetic materials. In what follows I will provide a short description of
{\it Smeagol} and the other DFT codes available, and finally of a few non-DFT based 
transport code.

\subsubsection{{\it Smeagol}}

{\it Smeagol} (Spin and Molecular Electronics on Atomically Generated Orbital
Landscape) \cite{smeagol1,smeagol2} is a fully spin-polarized code, which combines the 
NEGF method with the DFT code SIESTA \cite{siesta}.
SIESTA is a numerical implementation of DFT constructed on a non-orthogonal localized atomic
orbital basis set \cite{PAO}. This, in conjunction with the use of efficient
scalar relativistic Troullier-Martins pseudopotentials \cite{TM} with non linear core 
corrections \cite{Lou82}, makes SIESTA particularly suitable for handling systems
with a rather large number of degrees of freedom. It is therefore ideal for typical
magneto-transport problems.

Since the final product of SIESTA is a tight-binding like Hamiltonian, SIESTA can be
easily interfaced to the NEGF method discussed previously. 
This is implemented into {\it Smeagol}. To my knowledge {\it Smeagol} is the only fully 
spin-polarized code available at present. This goes beyond the use spin polarized exchange 
correlation potentials, including the non-collinear spin option, since magnetic materials 
carry additional difficulties with respect to non-magnetic systems. In fact the localized 
$d$ electrons co-exist in the valence with the extended $s$ electrons making the Hamiltonian 
matrix rather sparse. In this situation conventional recursive methods have problems to
converge the surface Green's functions of the leads, which are essential to construct
the self-energies. In {\it Smeagol} surface Green's functions are calculated using the 
semi-analytical expression \cite{sanvito} discussed in section \ref{gsgf}, after having
regularized $H_1$. 

Finally it is worth mentioning the way in which {\it Smeagol} calculates the potential
drop. This is obtained by solving the Poisson equation for the scattering region. However,
instead of using a real-space method (perhaps more desirable) {\it Smeagol} does that
in $k$-space, making use of the fast Fourier transform algorithm. For this purpose a supercell 
containing the scattering region is constructed and a saw-like external potential is
superimposed at any iteration.

\subsubsection{Other DFT codes}

Several DFT-based non-equilibrium transport codes are available, although at present 
none of them has the ability to deal with spin-polarized systems. In what
follows we list a few of them with a short description of their main technical details.

\vspace{0.3cm}
\noindent
{\it{\underline{The McDCal Program}}}

McDCal is a package created in Guo's group at the McGill University \cite{mcdcal}.
It combines NEGF method with a numerical implementation of DFT based
on pseudopotentials and the Fireball linear combination of atomic orbitals basis set 
\cite{fireball}. The self-energies of the leads are calculated by using a technique very 
similar to that of section \ref{gsgf}, while the electrostatic problem is addressed by
solving the Poisson equation for the Hartree potential using a real-space multigrid
approach \cite{multigrid}.
McDCal has been very successful in describing transport through molecular objects
\cite{mcdcal1,mcdcal2} and metallic nanowires \cite{mcdcal3}.

\vspace{0.3cm}
\noindent
{\it{\underline{TranSIESTA}}}

TranSIESTA \cite{transiesta} is a combination of NEGF with the DFT code SIESTA \cite{siesta}.
In many respects TranSIESTA is similar to both {\it Smeagol} and McDCal. The electrostatic potential
is calculated by solving the Poisson equation in $k$-space superimposing to the 
Hartree potential a linear potential ramp, while the surface Green's functions are calculated
by direct integration. Although it does not have spin-polarized functionalities,
TranSIESTA allows the calculation of current-induced forces \cite{transiesta1} and therefore the
study the mechanical stability of current-carrying nanodevices.

\vspace{0.3cm}
\noindent
{\it{\underline{GECM transport}}}

Recently the NEGF method has been implemented in conjunction with the quantum chemistry package
Gaussian98 \cite{gaussian98} to give the Gaussian embedded-cluster scheme (GECM) 
for quantum transport \cite{palacios1,palacios2}. One of the main difference of this methods with other
existing algorithms is the use of tight-binding Bethe lattice \cite{bethe} when constructing the
self-energies. The self-energy of a Bethe lattice is essentially analytical and somehow it averages out
effects like disorder of thermal smearing. For this reason usually both the transmission coefficients
and the $I$-$V$ characteristics are rather smooth.

\vspace{0.3cm}
\noindent
{\it{\underline{Lippmann-Schwinger method}}}

One of the early approach to DFT-based transport consists in the solution of the
Lippmann-Schwinger equation involving the Green's function for a biased metallic
junction \cite{Lang2}. The method is implemented in a plane-wave DFT framework and 
also in this case there is a self-consistent procedure for evaluating the potential drop.
The method has been extensively used for calculating fundamental transport properties
of elementary molecules \cite{diVentra}.

\vspace{0.3cm}
\noindent
{\it{\underline{Nanohub}}}

One of the world-leading group in theory of quantum transport is the Purdue
University group. Over the last few years they have developed a series of
packages for calculating realistic $I$-$V$ characteristics at different
level of complexity, including DFT methods \cite{Xue}. Part of this work
can be found at the web-site www.nanohub.org.

\subsection{Transport using non DFT electronic structure methods}

In addition to DFT-based methods recently a few novel computational tools
have been produced. The purposes of these tools are various, from introducing strong
correlation effects in the calculation, to lightweight semi-empirical algorithms
capable to handle a large number of degrees of freedom. 
A quick summary of these codes is reported here.

\subsubsection{Semi-empirical TB}

Semi-empirical tight-binding methods are half-way between fully self-consistent localized
basis DFT algorithms and non self-consistent Hamiltonians. The main idea is to expand the
Kohn-Sham Hamiltonian written on a localized atomic orbital basis set about a reference
charge density $\rho_0$ \cite{tbdft,tbdft2}. Thus one obtains a second order expansion 
in a tight-binding like form. This depends on a series of parameters that can be either
calculated directly or fitted from LDA calculations. In addition, usually the on-site 
Coulomb potential is replaced by Hubbard $U$ terms, in order to correct the LDA inability
to evaluate correctly the on-site Coulomb energy.
Since the Hamiltonian is written in a tight-binding like form 
this method can be readily implemented in a NEGF scheme for transport. A code
developed by the group of Di Carlo is currently free available  
(http://icode.eln.uniroma2.it) \cite{dicarlo}.

\subsubsection{Configuration Interaction method}

In an attempt to go beyond DFT in describing the electronic transport a novel method based on
the configuration interaction scheme has been recently proposed \cite{pauljim}. Here the 
many-body wave-function is expanded as spin-dependent Slater determinants
and the scattering boundary conditions are obtained in terms of Wigner function. From the results
available to date it is clear that many-body effects may be relevant for transport through small molecules,
although it is not obvious whether the method (quite computational intensive) may be
upscaled to more complicated systems.

\subsubsection{TDDFT method}

Despite the success of DFT-based quantum transport codes, several important questions
at the foundation of the method still remain. In particular the use of the Landauer formalism out
of equilibrium and the interpretation of the Kohn-Sham eigenstates as single particle states,
are somehow not completely under control. Recently a new DFT theorem 
encompassing transport at finite bias has been proved \cite{burke_car,kieron2}. This is based on the
reformulation of TDDFT including dissipation to phonons using an associated
Kohn-Sham master equation. The method is a radical departure from conventional
DFT-based methods for quantum transport. However at present it is difficult to forecast 
how easily efficient numerical implementations handling a large number of degrees of
freedom may be constructed.

\setcounter{equation}{0}
\section{Results}

In this section I will review the main successes of {\it ab initio}
transport theory in describing magnetic devices. I will start with the
more conventional GMR effect in magnetic multilayers, which is at the basis of the ``spintronics era'', 
and its tunneling counterpart, the tunneling magnetoresistance (TMR). Devices
based on these two effects are already in production (GMR-based read heads for
magnetic data storage) or are at a prototype level (TMR-based 
Magnetic Random Access Memories: MRAM). Then I will discuss atomic scaled
magnetic junctions, made either from metals such as magnetic point contacts or from
organic molecules. These will probably drive the next revolution in the field and
appear very attractive to theory since their reduced dimensions offer the unique possibility 
to compare directly predictions with experiments.
Finally I will briefly discuss a very recent development, the magnetic
proximity effect. This is the induction of a net magnetization in a non-magnetic 
material when brought into contact with a magnetic one. 

\subsection{GMR}

As already mentioned the giant magnetoresistance (GMR) effect
is the drastic change in resistance of a magnetic multilayer when a magnetic field
is applied. This is related to the change of the mutual orientation of the magnetic
moments of the magnetic layers. 

In metallic systems adjacent magnetic layers are magnetically coupled to each other, through the
non-magnetic ones. The sign of this exchange coupling, discovered by Stuart Parkin 
in the early nineties \cite{parkin_ex}, is an oscillatory function of the separation 
between the magnetic layers, whose details depend on the Fermi surface of the
non-magnetic one \cite{Bruno_ex}. In practice one can tune the thickness of the 
non-magnetic layers to obtain an overall antiferromagnetic (AF) state of the multilayer. 
In this situation the multilayer is in a high resistance state. 
When a magnetic field strong enough to align the magnetic layer along 
the same direction is applied, thus overcoming the antiferromagnetic exchange coupling, 
the multilayer resistance drops. Now the system is in a ferromagnetic (FM) configuration
corresponding to a low resistance state. The relative change in resistance is the GMR effect.

Early GMR experiments \cite{baibich_gmr,binasch_gmr} have been conducted 
with the so-called current in the plane configuration (CIP) (see figure 
\ref{fig17}) in which the current flows in the plane of the layers. In these 
experiments the typical cross sections are of the order of 1~mm$^2$ and the transport 
is mainly diffusive. This is the favorite configuration for devices, since the
resistances are rather large and they can be measured with conventional 
four-probe technique.
\begin{figure}[ht]
\begin{center}
\includegraphics[width=6.5cm,clip=true,angle=-90]{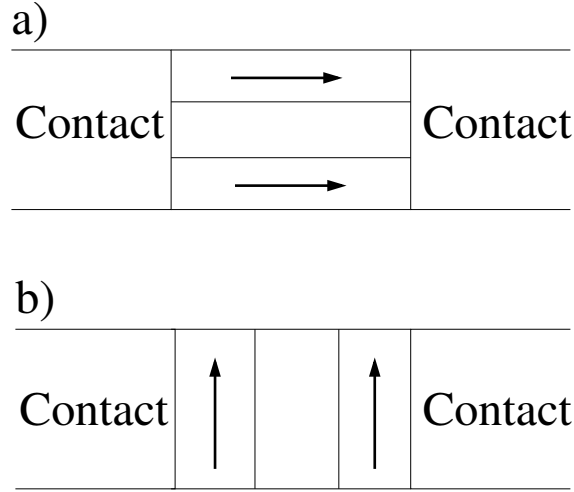}
\caption{Schematic representation of a typical GMR experiment: a) Current
In Plane (CIP) configuration, b) Current Perpendicular to the Planes
(CPP) configuration.}\label{fig17}
\end{center}
\end{figure}

An important breakthrough was the possibility to study the transport of a
multilayer with the current flowing perpendicular to the planes (CPP GMR). 
In this case the resistances are rather small and difficult to measure, and one
must either use superconducting contacts \cite{MSU91} or
shape the samples to very small cross sections \cite{Gijs93}. In these
experiments the electrons have to cross the entire multilayer over distances
smaller than 1 $\mu$m. The spin filtering is more effective and the transport can be
phase-coherent (for a comprehensive review on CPP GMR see 
\cite{Gijs97,Ansermet98}). The CPP arrangement is preferred by
theoreticians since {\it ab initio} calculations can be carried out.

\subsubsection{Electronic structure calculations for CPP GMR}

There are two main differences between CIP and CPP GMR.
First, in the CPP configuration an electron must cross the whole structure before 
being collected by the leads, while in the CIP channeling into individual
layers is possible. Secondly, in contrast to CIP GMR, where the dimensions
are certainly macroscopic, in the CPP the typical device dimensions are 
mesoscopic. This means that in the CPP configuration, particularly at low 
temperature, the transport may be largely phase coherent.
The early theory of CPP GMR, based on the Boltzmann's equations
neglected completely any quantum effects \cite{valet_fert}. All the
details of the electronic structure of the materials forming the multilayer were
``averaged'' out, and the only information needed to describe a material were the
layer resistivity, the interface resistivity between two materials, and two
parameters quantifying the spin-polarization of the current and of the interfaces. 

In 1995 Schep et al. \cite{kelly1,kelly2} challenged this conventional 
picture and showed that large values of GMR (of order 
120\%) could be obtained in Co/Cu multilayers, only as a results of the
Co and Cu band mismatch. The calculation consists in evaluating the 
Sharvin resistance (see section \ref{sharsection}) of an infinite Co/Cu multilayer,
whose electronic structure is obtained by DFT-LDA.
In figure \ref{fig18} I present the projection of the Fermi surface for the
a) majority and b) minority spin electrons in the parallel case and that for both
the spins in the antiparallel one (c), for a Co$_5$/Cu$_5$ multilayer (the index
labels the number of atomic planes in the unit cell along the multilayer stacking). 
The projection for the majority spin resembles that of free electrons, 
indicating that the transport is mainly given by $s$ electrons 
(see figure \ref{fig2b}). In contrast the projection for the minority spin 
is certainly not free-electron-like, and it is determined by the complicated Fermi 
surface of the Co minority spin. A similar argument can be applied for both
the spin channels in the antiparallel case. From the picture it is clear that
the conductance (proportional to area of the surface) of the majority spin electrons 
in the ferromagnetic state
is by far the largest one, and it is the suppression of this channel that
gives the GMR effect. The GMR ratio obtained in this way ranged between 30\% and
120\% depending on the layer thickness.
\begin{figure}[ht]
\begin{center}
\includegraphics[width=10.5cm,clip=true,angle=0.0]{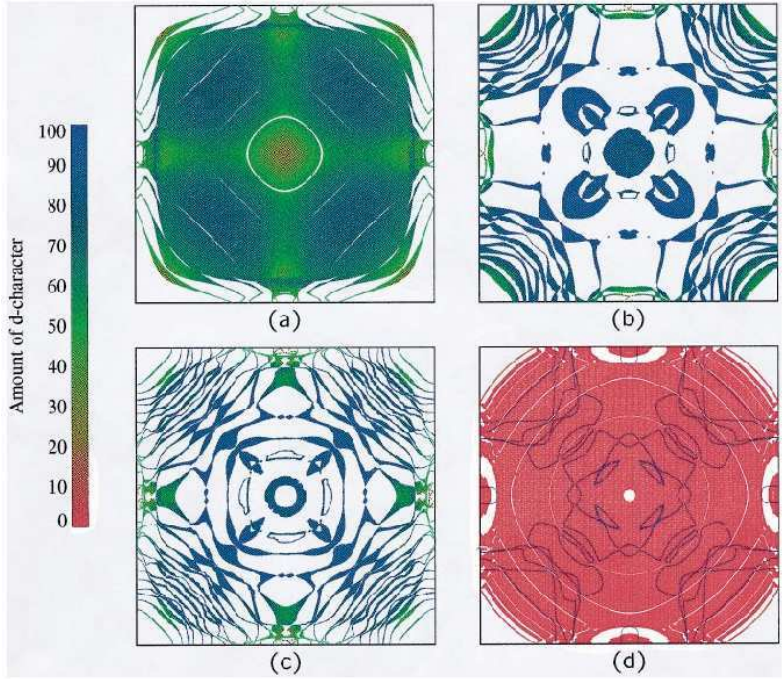}
\caption{Projections inside the first Brillouin zone of different Fermi surfaces 
for a (100) oriented Co$_5$Cu$_5$ multilayer on a plane parallel to the 
interfaces. The amount of $d$ character is given by the color code on the 
left-hand side of the figure. The $\Gamma$ point is at the center of each 
panel. (a) Majority spin and (b) minority spin in the parallel configuration. 
(c) Either spin in the antiparallel configuration, (d) minority spin in the 
parallel configuration where the sp-d hybridization is omitted. 
Reprinted with permission from \cite{kelly1}. Copyright (1995) American
Physical Society.}\label{fig18}
\end{center}
\end{figure}

One crucial result of this seminal work was to point out the
r\^ole of the $d$ electrons in the conduction (in particular in the case of
the AF configuration), and that of the band structure mismatch
in the determination of the GMR. In this spirit we have systematically
studied magnetic multilayers made from different materials \cite{sanvito},
using a $s$-$p$-$d$ tight-binding Hamiltonian and the Green's function method 
for ballistic transport presented in section \ref{greensection}. From this study we 
have identified two main mechanisms affecting the spin-transport: 1) a strong band
mismatch between the magnetic and the non-magnetic material, and 2) a
strong inter-band scattering. 

Let me explain this second concept with the example of a Co/Ag multilayer. 
First it is worth noting that the ballistic current in Ag is mainly carried by 
$s$ and $p$ electrons. These have a rather small DOS at the Fermi level, but a 
rather large group velocity, so they can be characterized as ``a few fast electrons''.
In contrast in Co the majority spins have a mixture of $s$, $p$ and $d$ character,
while the minority are mainly $d$ electrons. 
This means that an electron in Ag, whose spin is in the same direction of 
the magnetization, can enter Co as an $sp$-electron without the need for strong 
inter-band scattering (i.e. without changing much its orbital character). 
On the other hand, if its spin points in the opposite 
direction, it will undergo inter-band scattering because
in the minority band the electron must propagate as a $d$-electron.
This additional source of scattering is primarily related to the very
different dispersion relations of the $sp$-electrons with respect to the $d$-
electrons, suggesting that the minimal model to describe spin-transport
in transition metals is a two-band tight-binding model \cite{ss_1}.

Therefore the best GMR multilayer must be able to maximize the 
electron propagation in one of the two spin-bands and to minimize it in
the another. To achieve this result, the high conduction spin-band
should have a small band mismatch and weak inter-band scattering at
the interfaces, while the low conduction band should have a large
band mismatch and strong inter-band scattering.
Note that at this stage there are no general predictions on the total
polarization of a multilayer, being dependent on the band structure details of
both the magnetic and non-magnetic materials.
\begin{table}[htbp]
\begin{center}
\begin{tabular}{cccccccc} \hline\hline
\ Multilayer \ & \ $G^\uparrow_\mathrm{FM}$ \ & \  $G^\downarrow_\mathrm{FM}$ \ &
 \ \  $G_\mathrm{AF}$ \ \ & \ Multilayer \ & \ $G^\uparrow_\mathrm{FM}$ \ & \ $G^\downarrow_\mathrm{FM}$ \ &
 \ \  $G_\mathrm{AF}$ \ \\ \hline\hline
{\bf Co/Cu} & 0.59 &  0.32  & 0.18               & {\bf Ni/Cu}  &   0.66  &  0.47  &   0.44 	      \\ 
{\bf Co/Ag} & 0.63 &  0.32  & 0.21               & {\bf Ni/Ag}  &   0.69  &  0.43  &   0.42 	      \\ 
{\bf Co/Pd} & 0.33 &  0.16  & $9.41\cdot10^{-2}$ & {\bf Ni/Pd}  &   0.26  &  0.54  &   0.20 	      \\ 
{\bf Co/Pt} & 0.37 &  0.19  & 0.10               & {\bf Ni/Pt}  &   0.28  &  0.52  &   0.20 	      \\ 
{\bf Co/Au} & 0.24 &  0.16  & $8.40\cdot10^{-2}$ & {\bf Ni/Au}  &   0.64  &  0.49  &   0.44 	      \\ 
{\bf Co/Rh} & 0.17 &  0.25  & $7.81\cdot10^{-2}$ & {\bf Ni/Rh}  &   0.18  &  0.51  &   0.15 	      \\ 
{\bf Co/Ir} & 0.17 &  0.31  & $8.73\cdot10^{-2}$ & {\bf Ni/Ir}  &   0.19  &  0.49  &   0.17 	      \\ 
{\bf Co/Pb} & 0.15 &  0.17  & $6.47\cdot10^{-2}$ & {\bf Ni/Pb}  &  0.14  &  0.16  &  $7.74\cdot10^{-2}$ \\ 
{\bf Co/Al} & 0.12 &  0.16  & $5.28\cdot10^{-2}$ & {\bf Ni/Al}  &  0.13  &  0.11  &  $5.83\cdot10^{-2}$ \\ \hline\hline
\end{tabular}
\caption{Average conductance for different Co-based and Ni-based [Co(Ni)$_{10}$/NM$_x$]$_{\times 10}$
multilayers. The Co (Ni) thickness is fixed to 10 atomic planes, while that of the 
non-magnetic material varies from 1 to 40. The number of superlattice 
supercell is 10. The values reported are obtained averaging over the different 
non-magnetic materials thicknesses considered. The conductance of each 
multilayer is normalized to the conductance of the corresponding non-magnetic 
metal, which composes the leads.}
\label{Table2}
\end{center} 
\end{table}

In tables \ref{Table2} I report the average conductance 
(normalized to the conductance in the leads) for the majority and minority
spins in the ferromagnetic configuration ($G_\mathrm{FM}^\uparrow$ and 
$G_\mathrm{FM}^\downarrow$)
and for both spins in the antiferromagnetic ($G_\mathrm{AF}$),
for various Co- and Ni-based multilayers. The thickness of the Co (or Ni)
layers was fixed to ten atomic planes and that of the non-magnetic materials
varies between 1 and 40 atomic planes.
In table \ref{Table4} we report the corresponding GMR ratios.
\begin{table}[htbp]
\begin{center}
\begin{tabular}{cccccccc} \hline\hline
\ Multilayer \ & \ \ GMR (\%) \ \ & \ $\Delta$ (\%) \ &
 \ $\Delta$/GMR (\%) \ \\ \hline\hline
{\bf Co/Cu} & 150.7 & 9.2 &  6.1   & {\bf Ni/Cu} & 29.1 & 2.9 & 10.1	 \\ 
{\bf Co/Ag} & 131.0 & 7.6 &  5.8   & {\bf Ni/Ag} & 35.8 & 2.8 & 7.9	 \\ 
{\bf Co/Pd} & 165.2 & 31.1 &  18.8 & {\bf Ni/Pd} & 100.2 & 10.8 & 10.8   \\ 
{\bf Co/Pt} & 175.7 & 14.8 &  8.4  & {\bf Ni/Pt} & 94.3 & 10.6 & 11.2	 \\ 
{\bf Co/Au} & 138.8 & 20.1 &  14.5 & {\bf Ni/Au} & 26.9 & 3.3 & 12.37	 \\ 
{\bf Co/Rh} & 171.9 & 15.1 &  8.7  & {\bf Ni/Rh} & 131.3 & 6.4 & 4.9	 \\ 
{\bf Co/Ir} & 175.4 & 13.6 &  7.7  & {\bf Ni/Ir} & 107.3 & 6.0 & 5.6	 \\ 
{\bf Co/Pb} & 154.7 & 25.2 &  16.3 & {\bf Ni/Pb} & 97.8 & 12.1  & 12.4   \\ 
{\bf Co/Al} & 169.6 & 35.7 &  21.1 & {\bf Ni/Al} & 107.7 & 19.7  & 18.3  \\ \hline\hline
\end{tabular}
\caption{GMR ratio and GMR oscillations for different Co-based and Ni-based metallic multilayers.
The Co (Ni) thickness is fixed to 10 atomic planes, while that of the non-magnetic
material varies from 1 to 40. The number of superlattice supercell is 10. 
The values reported are obtained averaging over the different non-magnetic
materials thicknesses considered.}
\label{Table4}
\end{center}
\end{table}

Generally speaking the GMR for Co-based multilayers is considerably larger
than that of Ni-based. This is a direct consequence of the large exchange 
splitting in Co (1.3~eV to compare with 1.0~eV in Ni). Therefore the majority 
and minority bands in Ni are more similar to each other
than in Co and both the scattering and the GMR are reduced.

Let us now focus our attention on the dependence of the GMR and the conductances 
on the nature of the non-magnetic material. Here I consider only Ni-based
multilayers, which show a more clear behavior.
From the tables one can identify three typical situations depending on the value of 
the spin-conductance in the FM configuration: 1) the conductance is rather large for the majority 
spin and rather small for the minority spin (Cu, Ag, Au), 2) the conductance is rather
large for the minority spin and rather small for the majority (Pd, Pt, Rh. Ir), 
3) both the conductances are small (Pb, Al).
In the first case the conductance at the Fermi level of the non-magnetic material 
in the bulk is a mixture of $s$-$p$-$d$ orbitals. This means that the the match
with the majority spin-band of Ni is rather good, and that with the minority
rather poor. In contrast in Pt, Pd, Rh and Ir the Fermi energy cuts through $d$-like 
region, and the band mismatch is strong in the majority band and weak in the
minority. Finally Pb and Al behave like free-electron metals and the band match is poor
for both spin-bands.

An interesting aspect is that of the dependence of the spin-polarization of the multilayer
on the non-magnetic material. Since the
calculation is fully phase coherent (ballistic transport) the definition to use for
the spin-polarization is $P_{Nv}$ (see section \ref{224}). 
If we analyze $P_{Nv}$ for the various
multilayers of table \ref{Table2} we find that the spin-polarization
of the whole structure (obtained from the conductances in the parallel case)
varies widely depending on the non-magnetic layer. For example it is 
$P_\mathrm{Ni/Ag}$=0.23 for Ni/Ag multilayers and it becomes 
$P_\mathrm{Ni/Pd}$=-0.34 for Ni/Pd multilayers. This means that in these phase
coherent structures the spin-polarization of the magnetic metal says little about
the spin-polarization of the whole nanostructure, that in turns depends on
the details of the band mismatch. 
Similar ballistic calculations obtained either with DFT \cite{mertig1,mertig2} or 
tight-binding \cite{tsy1,mat3}
methods are available in the literature.

\subsubsection{R\^ole of disorder and breakdown of the resistor model}

All the calculations in the previous section assume disorder-free systems with
translational invariance in the direction orthogonal to the current. 
Here I will present studies on how disorder may affect the spin transport. This is
particularly relevant for magnetic systems not produced with Molecular Beam
Epitaxy (MBE) techniques, where structural defects or impurities may be abundant,
or for finite temperature measurements.

Two fundamentally different approaches have been used to describe CPP GMR in disordered 
systems. The first is the Valet and Fert (VF) model \cite{valet_fert}. This assumes that 
the transport is diffusive, and it is based on the semi-classical Boltzmann's equation within the 
relaxation time approximation. It has the great advantage that the same formalism can 
describe both CIP and CPP experiments. 
Disorder is included in the definition of the spin $\sigma$ dependent mean 
free path $\lambda_{\sigma}$ and the spin diffusion length $l_{\mathrm{sf}}$ and
finite temperature can be considered \cite{el3}. 
In the limit that the spin diffusion length is much larger 
than the layer thickness (infinite spin diffusion length limit), this model
reduces to a classical two current resistor network, where additional 
spin-dependent scattering at the interfaces is considered. 

The limitation of the VF model is of neglecting the band structure of the system: 
all the parameters are phenomenological.
An extension of the model to include band structure has been made recently 
\cite{mertig1,mertig2}, implementing the Boltzmann transport theory 
within DFT-LDA. In this calculation, the scattering due to
impurities is treated quantum mechanically, while transport is described
semi-classically. The same method was
previously used to describe the spin-polarization of the current in diluted
Ni- \cite{mertig3} and Co- \cite{mertig4} alloys. The polarization is generally
reproduced correctly for light impurities, while the absence of
spin-orbit interaction seems to be a severe limitation in the case of heavy 
impurities. 

The main assumption behind the use of a Boltzmann-like equation is that
interference effects are neglected and that transport is completely local.
As a consequence both the spin-polarization of the current and the GMR do not 
change with the length of the systems. This picture
is generally consistent with experiments. Nevertheless
it has been shown \cite{leed99,MSU97,Didier99} that in magnetic multilayers 
the GMR increases with the number of magnetic/non-magnetic layers, and
depends critically on the order of the layers. 
These results suggest that the relevant length scale for CPP GMR 
is not only the spin-diffusion length but also the elastic mean free path, and
that non-local contributions to the conductance are important. 
For these reasons a strictly local description of the transport 
may breakdown and a quantum approach is needed.

Quantum transport theory offers the possibility of studying phase-coherent
transport in disorder systems, and therefore can bridge across different
transport regimes. In this case full 
{\it ab initio} DFT calculations \cite{kelly1,kelly2} are not practical since the massive 
computer overheads, and tight-binding methods are more promising.
The only calculations carried out to date with accurate $s$-$p$-$d$ Hamiltonian
are still computational extensive, and for this reason they either involve infinite
superlattices in the diffusive regime \cite{tsy1} where small
unit cells can be used, or finite superlattices in which
disorder is introduced without breaking transverse translational symmetry \cite{ss_7,mat2}.
In the latter case the system is an effective quasi 1D system, whereas
real multilayers are 3D systems with roughness at the interfaces which breaks 
translational invariance.

To address systematically the issue of disorder a simpler tight-binding model with two degrees 
of freedom ($s$-$d$) per atomic site \cite{ss_1,ss_6} arranged over a 3D simple cubic lattice
is a better choice. Usually one considers two $s$ orbitals with hopping integrals chosen in such 
a way to give one dispersive and one flat band, with the understanding that these mimic respectively 
the $s$ and the $d$ band of a strong ferromagnet (Ni or Co). As an example In figure \ref{fig19} 
I present the DOS and the ballistic conductance for a set of parameters corresponding to copper.
\begin{figure}[ht]
\begin{center}
\includegraphics[width=12.5cm,clip=true,angle=0.0]{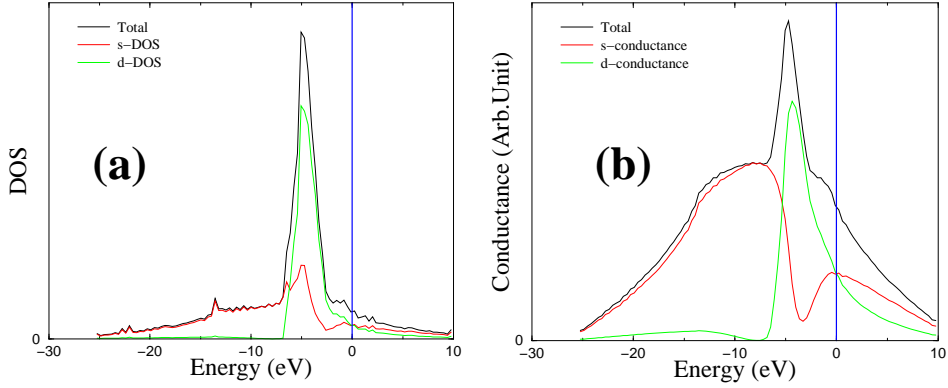}
\caption{DOS (a) and ballistic conductance (b) obtained from the two-band model, and their
projection over the $s$ and $d$ orbitals. The parameters used are the ones corresponding 
to Cu. The vertical line denotes the position of the Fermi energy
used in the calculation.}\label{fig19}
\end{center}
\end{figure}

Although several models for disorder may be considered, here I focus only on
the on-site Anderson-like model \cite{And80}, which consists in adding a random 
potential $V$ to each on-site energy, with a uniform distribution of width
$W$ ($-W/2\leq V \leq W/2$).
In order to investigate the different conductance regimes that may occur and
their dependence on the magnetic state of the system it is convenient to
consider as a scaling quantity the  ``reduced'' conductance $g$
\begin{equation}
g^\sigma=\frac{h}{e^2}\frac{\langle G^\sigma\rangle}
{N_{\mathrm{open}}}\: L
\;{,}
\label{eq5.7}
\end{equation}
where $\langle G^\sigma\rangle$ is the average conductance, $N_{\mathrm{open}}$
the number of open channels in the leads and $L$ the multilayer length.
In the ballistic limit $g$ increases linearly with length, in the diffusive (metallic) limit 
$g$ is constant, and in the localized regime $g$ decays as 
$\exp(-L/\xi)$ with $\xi$ the localization length \cite{Been97,Kram93}.
\begin{figure}[ht]
\begin{center}
\includegraphics[width=8.5cm,clip=true,angle=0.0]{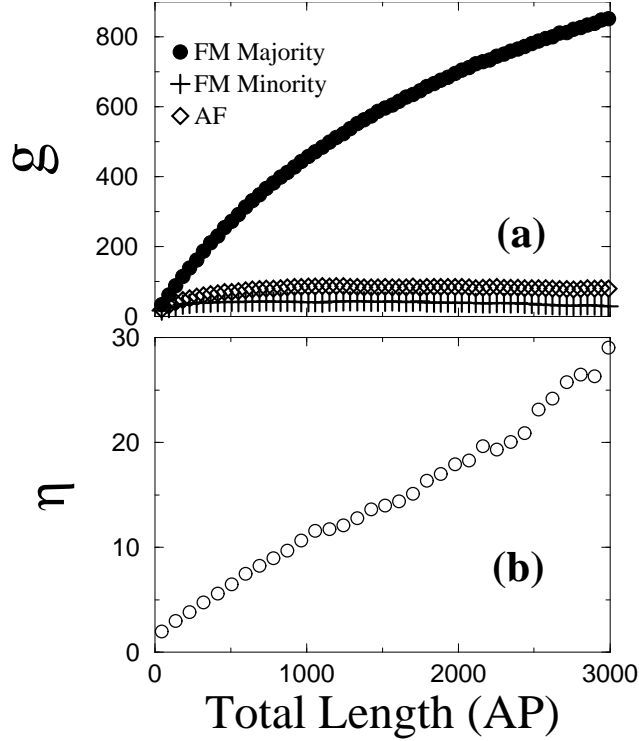}
\caption{Reduced conductance $g^\sigma$ and spin asymmetry
$\eta=g^\uparrow_{\mathrm FM}/g^\downarrow_{\mathrm FM}$
as a function of the multilayer length for Cu/Co multilayers
with random on-site potential. The random potential has a normal distribution 
of width 0.6eV, and the layer thicknesses are $t_{\mathrm Cu}=8$AP 
and $t_{\mathrm Co}=15$AP. Each point corresponds to a cell
Co/Cu/Co/Cu of total thickness 46AP.}\label{fig20}
\end{center}
\end{figure}

In figure \ref{fig20} I present the quantity $g$ for the two spin sub-bands in the FM 
and AF configurations along with the ratio 
$\eta=g^\uparrow_{\mathrm{FM}}/g^\downarrow_{\mathrm{FM}}$.
These results are obtained for Co/Cu multilayers (parameters from reference \cite{ss_1})
and a disorder strength of $W=$0.6~eV. The standard deviation of the mean is negligible on the 
scale of the symbols, and each point corresponds to an additional Cu/Co double bilayer.
From the figure it is immediately clear that the spin-asymmetry of $g$ (ie of 
the conductance) is increased by the disorder, which as a consequence of the 
band structure, turns out to be more effective in the minority band and in the 
AF configuration. This is because the mobility edge is at the Fermi level for
$d$ electrons while it is far from it for the $s$.
Since the current in the majority band of the FM configuration is carried mostly 
by $s$-electrons the majority spin sub-band will not be strongly affected by the disorder. 
In contrast in the minority band and in both bands in the AF configuration, the 
current is carried by $d$-electrons, for which the scattering is strong.

A second remarkable result is that in the FM configuration the almost ballistic 
majority electrons can co-exist with diffusive minority carriers. In the regime 
of phase coherent transport the definition of spin-dependent mean free paths 
for individual materials within the multilayer is not meaningful, and one must
consider the spin-dependent mean free path for the whole multilayered structure.
At finite temperature, when the phase breaking length $l_\phi$ is 
shorter than the elastic mean free path, $l_\phi$ becomes
the relevant length scale, and the scattering properties of such a structure are solely 
determined by elastic transport up to a length $l_\phi$.
Turning now the attention to GMR, it is clear from figure
\ref{fig20} and the definition of the GMR ratio that enhanced 
spin asymmetry will increase the GMR ratio because of the high transmission 
in the majority band. 

Finally I wish to discuss the breakdown of the resistor network model in case the
mean free path becomes longer than the typical layer length.
To illustrate this breakdown, consider a multilayer consisting of two independent 
building blocks, namely a (N/M) and a (N/M$^\prime$) bilayer, where M and M$^\prime$ 
represent magnetic layers either made from different materials or of  
different thicknesses and N represents normal metal `spacer' layers.
This is the experimental setup of references \cite{leed99,MSU97,Didier99}.
Two kinds of multilayer can be created. The first (type I or ``interleaved'' 
\cite{Didier99}), consists of a (N/M/N/M$^\prime$)$\times \mu$ sequence 
where the species M and M$^\prime$ are separated by an N layer and the group of four 
layers is repeated $\mu$ times. The second (type II or ``separated''), consists of a
(N/M)$\times \mu$(N/M$^\prime$)$\times \mu$ sequence, where the multilayers 
(N/M)$\times \mu$ and (N/M$^\prime$)$\times \mu$ are arranged in series.
From the point of view of a resistor network description of transport, the two 
types of multilayers are equivalent, because they possess the same number of 
magnetic and non-magnetic layers, and the same number of N/M and N/M$^\prime$ 
interfaces. Hence the GMR ratio must be the same.
In contrast the GMR ratio of type I multilayers is found experimentally 
to be larger  than that of type II multilayers \cite{leed99,MSU97,Didier99}, 
and the difference between the two GMR ratios increases with the number of 
bilayers. 

Also in this case one can use the simple $s$-$d$ Hamiltonian considered previously \cite{ss_6}. 
Here the parameters are again for Co and Cu with layer thicknesses 
$t_{\mathrm Cu}=10$AP, $t_{\mathrm Co}=10$AP, $t_{\mathrm Co}^\prime=40$AP.
In figure \ref{fig21} I present the mean GMR ratio for type I (type II) 
multilayers GMR$_{\mathrm I}$ (GMR$_{\mathrm{II}}$) and the difference
between the GMR ratios of type I and type II multilayers 
$\Delta$GMR=GMR$_{\mathrm I}$-GMR$_{\mathrm{II}}$, 
as a function of $\mu$ for different values of the on-site random potential. 
In the figure I display the standard deviation of the mean only
for $\Delta$GMR because for GMR$_{\mathrm I}$ and GMR$_{\mathrm{II}}$
it is negligible on the scale of the symbols.
It is clear that type I multilayers possess a larger GMR ratio than type II 
multilayers, and that both the GMR ratios and their difference increase
for large $\mu$.
These features are in agreement with experiments \cite{leed99,MSU97,Didier99} 
and cannot be explained within the resistor network model of CPP GMR.
\begin{figure}[ht]
\begin{center}
\includegraphics[width=8.5cm,clip=true,angle=0.0]{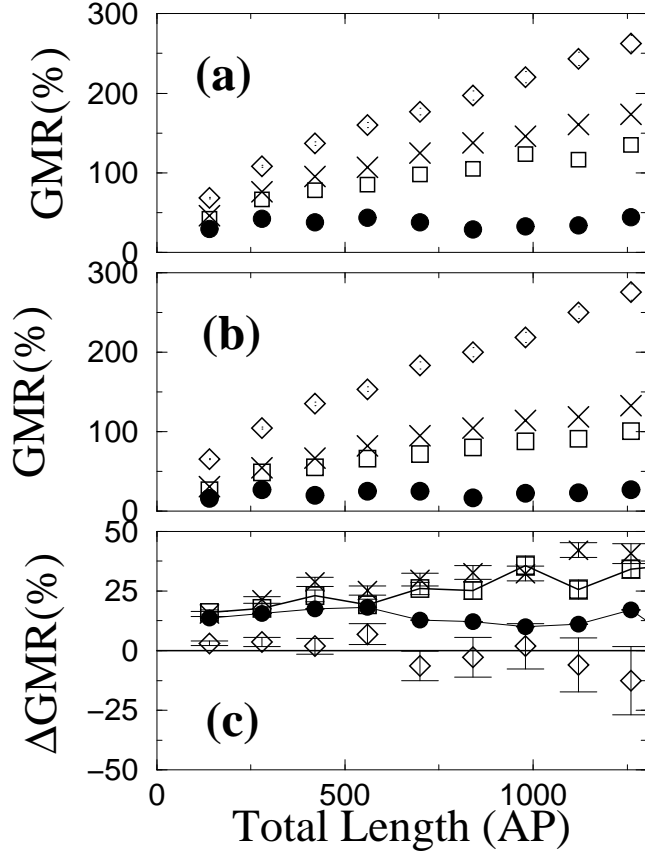}
\caption{GMR for type I (a) and type II (b) multilayers,
and $\Delta$GMR (c) in the case of thin (10AP) 
and thick (40AP) Co layers, as a function of the number
of double bilayers Co/Cu/Co/Cu for different values
of disorder. The symbols represent respectively
$W=0$ (circle), $W=0.3$eV (square),
$W=0.6$eV (star), $W=1.5$eV (diamond).
As an example the calculated mean free paths for $W=0.6$eV are
$\lambda_{\mathrm FM}^{{\mathrm (I)}\:\uparrow}\ge 4000$AP, 
$\lambda_{\mathrm FM}^{{\mathrm (I)}\:\downarrow}= 1300$AP,
$\lambda_{\mathrm AF}^{{\mathrm  (I)}\:\uparrow\downarrow}= 1800$AP,
$\lambda_{\mathrm FM}^{{\mathrm (II)}\:\uparrow}\ge 4000$AP, 
$l_{\mathrm FM}^{{\mathrm (II)}\:\downarrow}= 1700$AP,
$\lambda_{\mathrm AF}^{{\mathrm (II)}\:\uparrow\downarrow}= 2300$AP.
}\label{fig21}
\end{center}
\end{figure}

The increase of the GMR ratio as a function of the number of bilayers is a 
consequence of enhancement of the spin asymmetry of the current due to disorder. 
The different GMR ratios of type I and type II multilayers
are a consequence of the inter-band scattering, which occurs whenever
an electron phase-coherently crosses a region where two magnetic layers have AF
magnetizations.
This occurs in each (N/M/N/M$^\prime$) cell for type I multilayers,
while only in the central cell for type II multilayer.
Hence the contribution to the conductance in the AF alignment due to inter-band 
scattering is smaller in type I than in type II multilayers.
Eventually when the elastic mean free path is comparable with a single Co/Cu 
cell one expects the resistor model to become valid as illustrated
in reference \cite{ss_6}.

\subsubsection{The effects of using superconducting contacts}

Let us now turn our attention to the case of GMR measurements using superconducting 
contacts. The interest of these systems is twofold. On the one hand superconducting
contacts have been always employed \cite{MSU91} 
to achieved a uniform distribution of the current across the cross-section of the
multilayer, and to perform squid measurements of the resistance. 
On the other hand, at a fundamental level, new physics associated 
with such structures arises from the proximity of two electronic ground states with 
different correlations (ferromagnetism and superconductivity), which can reveal 
novel scattering processes not apparent in the separate materials. 
The basic feature of the transport in ferromagnetic/superconductor and 
ferromagnetic-multilayer/superconductor systems is that the current 
is spin-polarized in the magnetic material, but it is not spin-polarized in the superconductor. 

Below the superconducting gap the current is solely determined by Andreev reflection \cite{Andreev},
which involves electrons and holes with different spin orientations. 
This means that the Andreev reflection is a non spin-conserving
process and the two spin-bands are coupled, reflecting the fact that the supercurrent in the
superconductor is not spin-polarized. Therefore when a superconductor is brought into contact with
a material in which the current is spin-polarized, one expects extra resistance
at the interface \cite{Beenaker,Falko99,Falko99b} and the presence of depolarizing effects.
Since the GMR in magnetic multilayers is an effect which arises from the
spin-polarization of the current, it is reasonable to expect strong modifications
by adding superconducting contacts.

In order to understand the effects of superconducting contacts on the GMR it is interesting to
consider a magnetic multilayer in which one contact is a superconductor
and the other is a normal metal \cite{ss_8}. This situation corresponds to a
phase coherent length shorter than the entire multilayer. As pointed out
in section \ref{supsection} in the case of subgap conductivity we can simply consider the
linear response limit and neglect the effect of finite bias. Here I consider
realistic $spd$ tight-binding band structure with parameters corresponding
to Cu, Co and Pb, and with a superconducting gap $\Delta$ 
equal that of bulk Pb ($\Delta_{\rm Pb}=$1.331$\cdot 10^{-3}$ eV) \cite{ss_8}.

Figure~\ref{fig22}a shows results for the GMR ratio in the normal and
superconducting states as a function of the Cu thickness for a 
Cu/[Co$_{7}$/Cu$_{10}$]$_{\times 10}$/Pb multilayer, and clearly demonstrates a 
dramatic superconductivity-induced suppression of GMR. Figure \ref{fig22}b and \ref{fig22}c show 
results for the individual conductances per open channel and demonstrate that 
the GMR ratio suppression arises because the conductance in the FM configuration
in presence of superconductivity $G_{\mathrm{NS}}^{\mathrm{FM}}$ 
is drastically reduced compared with that in absence of superconductivity 
$G_{\mathrm{NN}}^{\mathrm{FM}}$ and equals that of the AF configuration
$G_{\mathrm{NS}}^{\mathrm{AF}}$.
\begin{figure}[ht]
\begin{center}
\includegraphics[width=8.5cm,clip=true,angle=0.0]{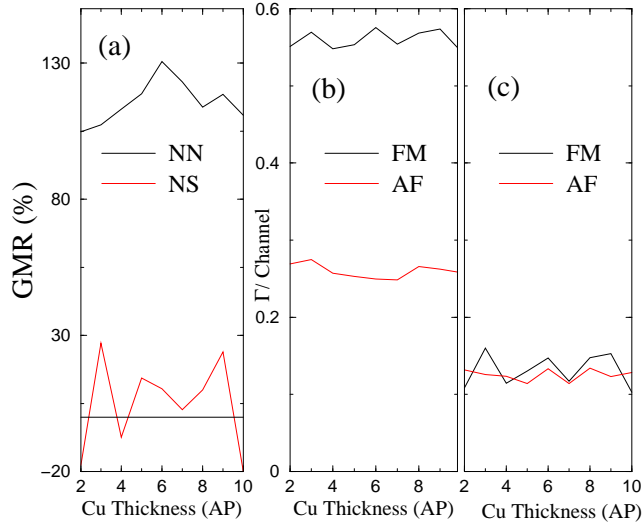}
\caption{GMR ratio (a) and conductance in the FM and AF configurations (b and c) for the disorder-free
system Cu/[Co$_{7}$/Cu$_{10}$]$_{\times 10}$/Pb. NN refers to the case in which Pb is in the 
normal state, NS to the case in which Pb is in the superconducting state. (b) shows the 
conductances in the NN case and (c) in the NS case. Note the dramatic suppression of the 
spin-polarization of the current when superconductivity is introduced.}\label{fig22}
\end{center}
\end{figure}

One can understand this effect by considering the simple model of
spin-dependent boundary scattering shown in figure \ref{fig23}.
This is a Kr\"onig-Penney potential where high barriers correspond to scattering of
minority spin electrons and small barrier to that of majority.
Since the minority spins see the higher barrier, one expects 
a higher transmission for majority electrons $T^\uparrow_\mathrm{FM}$
than for minority $T^\downarrow_\mathrm{FM}$.
Figures \ref{fig23}c and \ref{fig23}d show the scattering potentials for 
anti-ferromagnetically aligned layers. In this case there is an equal 
number of high and low barriers and the transmission $T_\mathrm{AF}$ is
in between that of the majority and minority band in the FM case.
For such an ideal structure, GMR arises from the fact that 
$T_{\mathrm{FM}}^{\uparrow}\gg T_{\mathrm{FM}}^{\downarrow}$ and
$T_{\mathrm{AF}}^{\uparrow}$.
In the presence of a single superconducting contact this picture is drastically 
changed. For ferromagnetically aligned layers, figure \ref{fig23}e shows an 
incident majority electron scattering from a series of low barriers, 
which Andreev reflects as a minority hole and then scatters from a series of 
high barriers (figure \ref{fig23}f). 
The reverse process occurs for an incident minority electron, 
illustrating the rigorous result that the Andreev reflection coefficient is 
spin-independent. Figures \ref{fig23}g and \ref{fig23}h show
Andreev reflection in the anti-aligned state. The crucial point 
is that in presence of a S contact for both the aligned 
(figures \ref{fig23}e and \ref{fig23}f) and anti-aligned (figures 
\ref{fig23}g and \ref{fig23}h) states the quasi-particle scatters from N 
(=4 in the figures) high barriers and N (=4) low barriers and therefore, 
one expects $G_{\mathrm NS}^{\mathrm{FM}}\sim G_{\mathrm{NS}}^{\mathrm{AF}}$.
This suppresses completely the GMR. Note that
a similar result has been demonstrated for diffusive multilayers by using the
simple two band $s$-$d$ model \cite{ss_8}.
\begin{figure}[ht]
\begin{center}
\includegraphics[width=8.0cm,clip=true,angle=0.0]{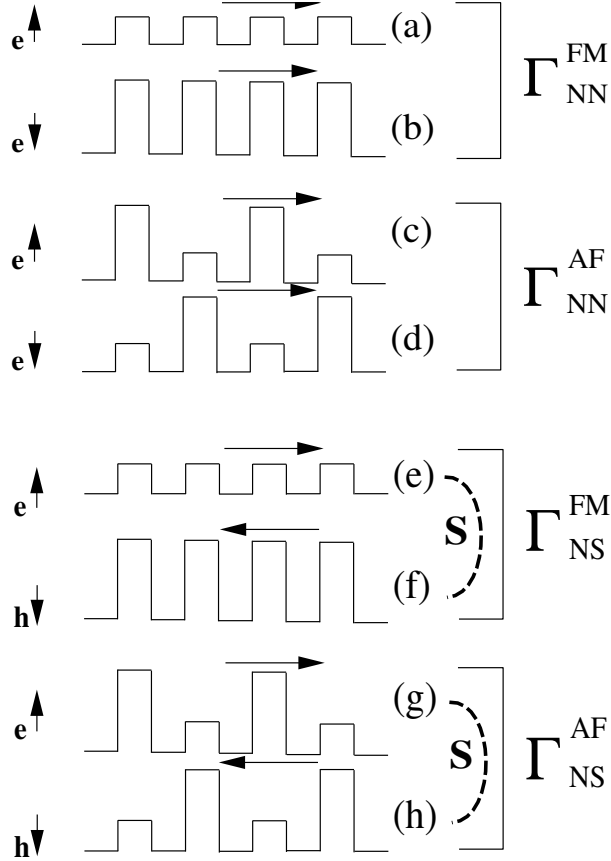}
\caption{Cartoon of the different scattering processes. Figures (a), (b), (c) 
and (d) describes the transmission of spin electrons $e^{\uparrow(\downarrow)}$
in a NN system. Figures (e), (f), (g) and (h) describe the NS case. Note that 
in the FM case a majority (minority) spin electron $e^{\uparrow}$ 
($e^{\downarrow}$) is Andreev reflected as a minority 
(majority) hole $h^{\downarrow}$ ($h^{\uparrow}$). In the 
antiferromagnetic (AF) case the path of the incoming electrons and out-coming 
holes is identical for both spins. The total number of large barriers
is the same in the AF and FM case, and this produces GMR suppression.
}\label{fig23}
\end{center}
\end{figure}

These results set a puzzle since almost all the experiments carried out with
superconducting contacts show a large GMR. This means that some
extra mechanism at the interface between the multilayer and the superconductor
must occur. One possibility is that spin-flip at the interfaces can account for such
a discrepancy \cite{ss_9}. Consider in fact the cartoon of figure \ref{fig24}, where now I
describe the Andreev reflection in presence of spin-flip at the interface. 
If a majority electron is Andreev reflected and spin-flipped, the corresponding
outgoing hole will possess an up spin, and therefore propagate in the majority
band. In this way the high transmission majority band is restored and the GMR
ratio will not be suppressed \cite{ss_9}. It is important to note that in this case the
electrons responsible for the GMR signal are the ones which undergo to spin-flip
at the interface. This situation is exactly opposite to the case in which no
superconductors are present. Unfortunately these predictions have not been 
verified yet and new experiments in which the superconductivity of the contacts can be 
switched on and off arbitrarily are welcome. 
\begin{figure}[ht]
\begin{center}
\includegraphics[width=8.0cm,clip=true,angle=0.0]{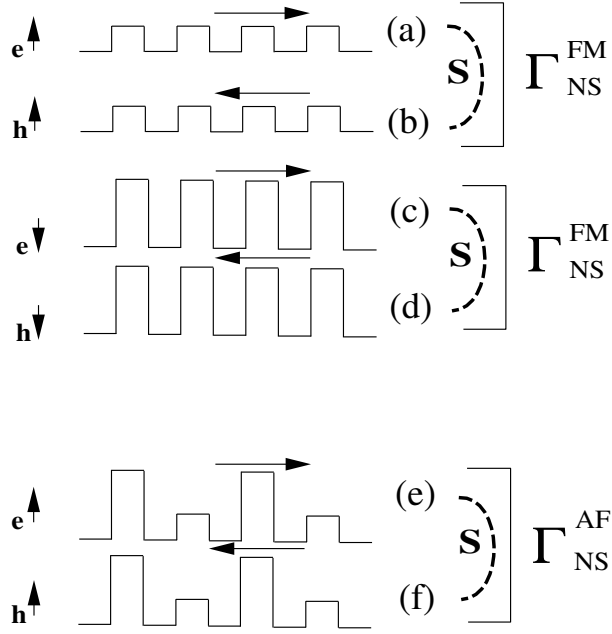}
\caption{Cartoon of Andreev reflection in presence of spin-flip at the N/S interface.
Figures (a-d) describe the FM configuration and figures (e-f) the AF configuration.
Note that a majority spin electron is reflected like a majority spin hole, if
spin-flip occurs at the interface (figures a and b). This produces high transmission in
the majority spin-channel and therefore large GMR.}\label{fig24}
\end{center}
\end{figure}

\subsubsection{Superconductor/Ferromagnet ballistic junctions}

One recent and successful way to obtain information about the spin polarization of 
a system is by using ballistic F/S junctions and measuring the change of the
conductance due to the onset of the superconductivity 
\cite{Upd1,Upd2,Soul}. In the typical experimental setup a small
constriction (usually 30nm long and 3-10nm thick) is made between a 
superconducting metal and another metal that can be either ferromagnetic or
normal. The system is then cooled below the critical temperature for the
superconductor and the $I$-$V$ curve at small biases is measured. As a
reference usually also the $I$-$V$ curve for the equivalent F/N junction is
measured at the same temperature. This is achieved by applying a magnetic field
higher than the critical field of the superconductor. 
The quantity which is of interest is the normalized conductance $g(V)$ 
as a function of the bias voltage $V$
\begin{equation}
g(V)=\frac{G_{\rm FS}(V)-G_{\rm FN}(V)}{G_{\rm FN}(0)}
\;{,}
\label{eq6.12}
\end{equation}
where $G_{\rm FS}(V)$ ($G_{\rm FN}(V)$) is the measured differential conductance 
for the F/S (F/N) junction. Experimentally, although the individual conductances
fluctuate from sample to sample by up to one order of magnitude, the quantity $g(V)$
is constant. This is a demonstration that the transport is ballistic and that
the fluctuations of the conductance depend only on the size of the constriction
(which can vary from sample to sample).
Finally a fit of $g(V)$ is performed by using a modification of the
Blonder-Tinkham-Klapwijk theory \cite{BTK} with spin-dependent $\delta$-like
scattering potential at the interface. Thus the polarization is evaluated. 
Usually a remarkable good agreement with the experimental data is achieved
particularly in the low bias region.

Again realistic $spd$ tight-binding models in conjunction with
the linear response scattering technique can give important insights.
The system in this case is assumed to be perfectly translational invariant across the 
entire structure (which means perfect lattice match at the interface) and
the differential conductance is calculated by integrating the zero bias
transmission coefficient.
\begin{figure}[ht]
\begin{center}
\includegraphics[width=6.5cm,clip=true,angle=0.0]{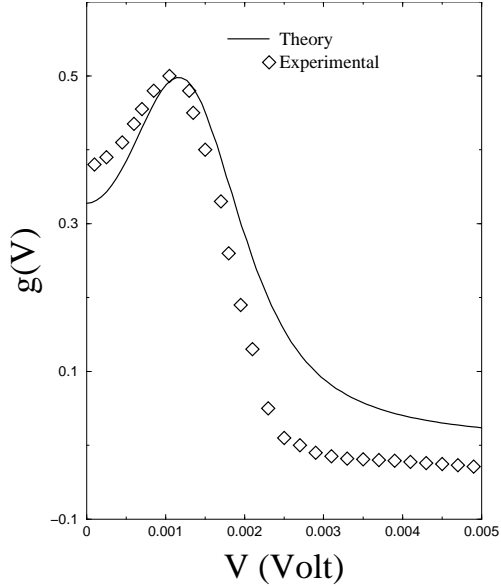}
\caption{$g(V)$ curve for a Cu/Pb ballistic junction at 4.2$^o$K. The solid line
represents the calculated curve and the squares the experimental data from
reference \cite{Upd1}. Note that the agreement is remarkably good
particularly at low bias.}\label{fig25}
\end{center}
\end{figure}

The calculated $g(V)$ curve for Cu/Pb \cite{ss_3} is shown in figure \ref{fig25}
together with the experimental data from reference \cite{Upd2}.
The agreement is surprisingly good particularly for low bias. Note that the
experimental data show a negative $g(V)$ for large biases which is in
contradiction with the elementary expectation of $g(V) \sim 0$ for 
$eV > \Delta$. Nevertheless this seems to be consistent with the 
experimental error on the determination of
$g(V)$ and therefore the agreement of the theoretical curve may be
considered almost perfect over the whole voltage range.
Better agreement can be obtained by reducing the superconducting gap $\Delta$
below the bulk value for Pb. This is reasonable if one considers that in the
constriction region size effects can suppress the superconductivity.
Similar results have been obtained with an analogous transport technique and 
DFT-LDA electronic structure \cite{KellySup}.
\begin{figure}[ht]
\begin{center}
\includegraphics[width=8.5cm,clip=true,angle=0.0]{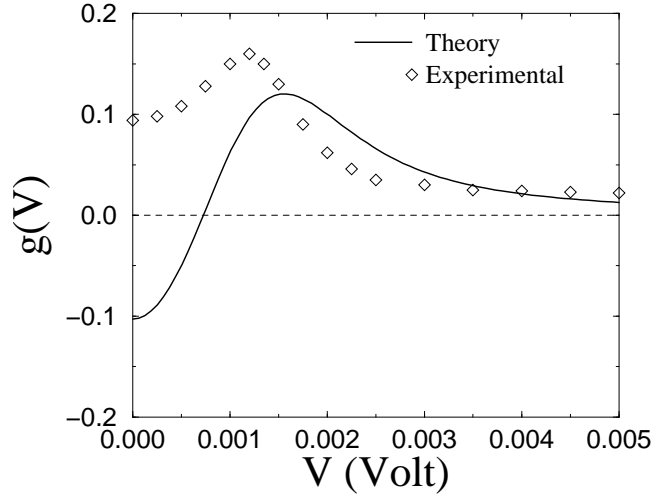}
\caption{$g(V)$ curve for a Co/Pb ballistic junction at 4.2$^o$K. The solid line
represents the calculated curve and the squares the experimental data from
reference \cite{Upd1}. Note that at low bias the calculated curve presents
a $g(V)$ value with an opposite sign with respect to what found in experiments.}\label{fig26}
\end{center}
\end{figure}

Figure \ref{fig26} shows the same curve for a Co/Pb
junction, where the poor agreement between theory and experiment is evident. 
In particular at zero bias the normalized conductance $g(V)$ is
negative, which is the result of a strong under-estimation of $\Gamma_{\rm FS}$
with respect to the experiments. Also in this case DFT does agree with the tight-binding 
model \cite{KellySup} and the reason for this disagreement remains
an open question. Disorder at the
interface producing additional scattering (both in the normal and the superconducting case)
seems not to play a decisive r\^ole \cite{KellySup,ss_9}, and a few speculations have been
made including spin-flip and enhanced exchange at the interface \cite{ss_3,ss_10}.

\subsection{TMR}

Despite the indisputable success of the CPP GMR either as scientific tool and as
building block for devices, it presents some disadvantages in practical applications. 
Firstly, since the layer thicknesses are rather small there is the need of measuring 
the resistance with sophisticated techniques such as superconducting contacts, which clearly
are not practical for applications. Secondly it is generally difficult to magnetically 
decouple the layers, large magnetic fields are needed and complex micromagnetic effects 
are unavoidable. 

In order to overcome these difficulties a much simpler structure has been proposed. This
is a tunneling junction, formed by two magnetic layers sandwiching an insulating material
and connected to two current/voltage probes. The two layers are now magnetically decoupled 
and engineered to have different coercive fields, hence their mutual orientation can be changed
by applying a tiny magnetic field. Also in this case the high current state is the
ferromagnetic and the low current state the antiferromagnetic. The quality of the device
is measured by the tunneling magnetoresistance ratio (TMR) using the same definition of that
for GMR.

The main difference between GMR and TMR is that in TMR the current is a tunneling current
and there is no conductance associated with the insulating barrier. From
the point of view of the scattering theory this means that not only the match between the
asymptotic wave-functions through the scattering region is important, but also how
these wave-functions decay within the tunneling barrier. 

\subsubsection{Ballistic Tunneling}

Early theoretical work on magneto-tunneling was based on the famous Julliere model
\cite{Julliere75}. Here the degree of spin-polarization of a tunneling junction 
and therefore its TMR was attributed to the degree of spin-polarization of the
magnetic layers. Unfortunately in the original paper there was no clear indication 
on what type of spin-polarization to consider (see the discussion of section \ref{spoad})
and the common understanding was to use that of the DOS. A similar argument based
on Fermi wave-vector mismatch was also considered \cite{Slon89}. An important consequence
of these models is the fact that the TMR signal does not depend on the nature of the 
insulating barrier nor on its thickness.

This vision appeared to be at least incomplete when it was experimentally demonstrated
\cite{deTeresa} that the TMR signal could be changed and also reversed only 
by replacing one insulating material with another one. Clearly the insulator
or at least the interface between the insulator and the magnetic layers plays
an important r\^ole in determining the TMR signal. For this reason realistic 
electronic structure calculations can give important insights.

Realistic band structures have been introduced in the calculation of the tunneling current 
either through {\it ab initio} DFT \cite{Mac99,Wang98} or through tight-binding models 
\cite{Tsy97t,Math97t}. In all the cases the system is assumed translational 
invariant in the direction perpendicular to the current and the transport in the 
linear response limit. This implicitly assumes that the device is perfectly
epitaxial. Although these early calculations gave rise to a controversy regarding 
the actual polarization of a tunneling junction and on the relevant factors which 
affect the tunneling, they also showed two common features: 1) the spin-polarization of the 
junction and therefore the TMR increases as the thickness of the barrier increases, 2) the 
transmission is resonant over the Brillouin zone associated to the plane perpendicular
to the current. 

In order to illustrate these two aspects in figure \ref{fig27} I present the transmission
coefficients for majority and minority spins and the corresponding junction polarization 
for a Co/INS/Co tunneling junctions with Cu probes. The calculation is carried out using 
$spd$ tight-binding model with parameters for Cu and Co extracted from reference \cite{papacon},
and with a model insulator INS whose DOS is presented in figure \ref{fig28}.
In addition in figures \ref{fig29} and \ref{fig30} the decomposition of the transmission
coefficient $T$ over the two-dimensional transverse Brillouin zone $T(k_x,k_y)$ 
is presented. For instance in this specific case the spin-polarization of the device
changes sign when the insulator thickness is increased.
\begin{figure}[ht]
\begin{center}
\includegraphics[width=10.5cm,clip=true,angle=0.0]{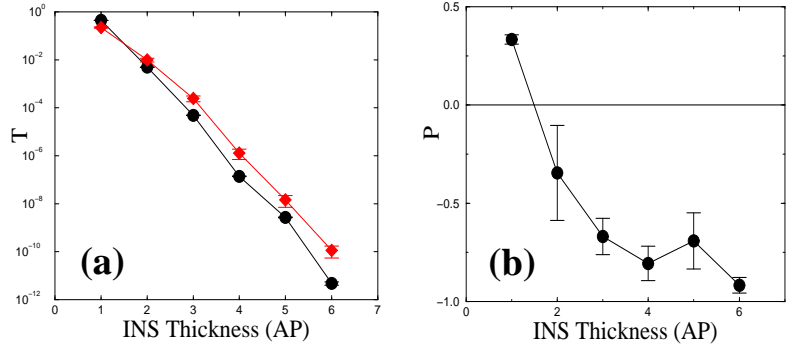}
\caption{Transmission coefficient (a) and polarization (b) of Cu/Co/INS/Co/Cu 
tunneling junctions as a function of INS thickness. The thickness of the
right-hand side Co layer is varied from 1AP to 55AP and each point corresponds
to the average value over these thicknesses.}\label{fig27}
\end{center}
\end{figure}
\begin{figure}[ht]
\begin{center}
\includegraphics[width=10.5cm,clip=true,angle=0.0]{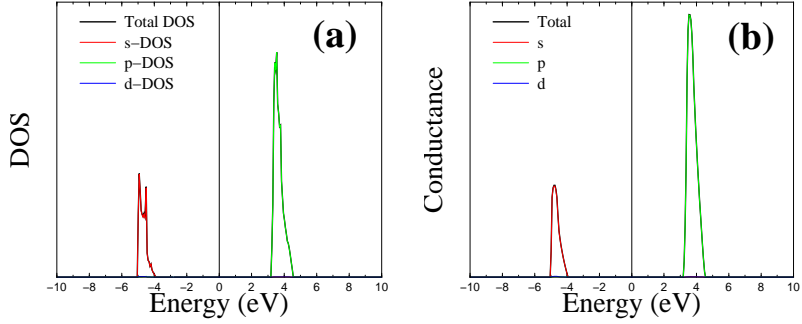}
\caption{Partial DOS (a) and partial conductance (b) for the fcc insulator used in the calculation. 
The vertical line denotes the position of the Fermi energy.
}\label{fig28}
\end{center}
\end{figure}

A simple rationale of these effects starts from noting that in Co the 
spin-polarization of the $s$-electrons at the Fermi energy is positive, 
while that of the $d$-electrons is negative. Therefore, if the barrier acts 
selectively on the $s$- and $d$-electrons, different polarizations of the 
tunneling junction are expected depending on the details of the insulator.
Along this line Tsymbal et al. \cite{Tsy97t} showed that in a Co-based tunneling 
junction with an $s$-insulator, the polarization is positive if one considers 
only ss$\sigma$ coupling at the interface and becomes negative if sd$\sigma$ is 
also included. Similar calculations based on multi-orbital models have also
been presented \cite{wang98bis}.
\begin{figure}[ht]
\begin{center}
\includegraphics[width=8.5cm,clip=true,angle=0.0]{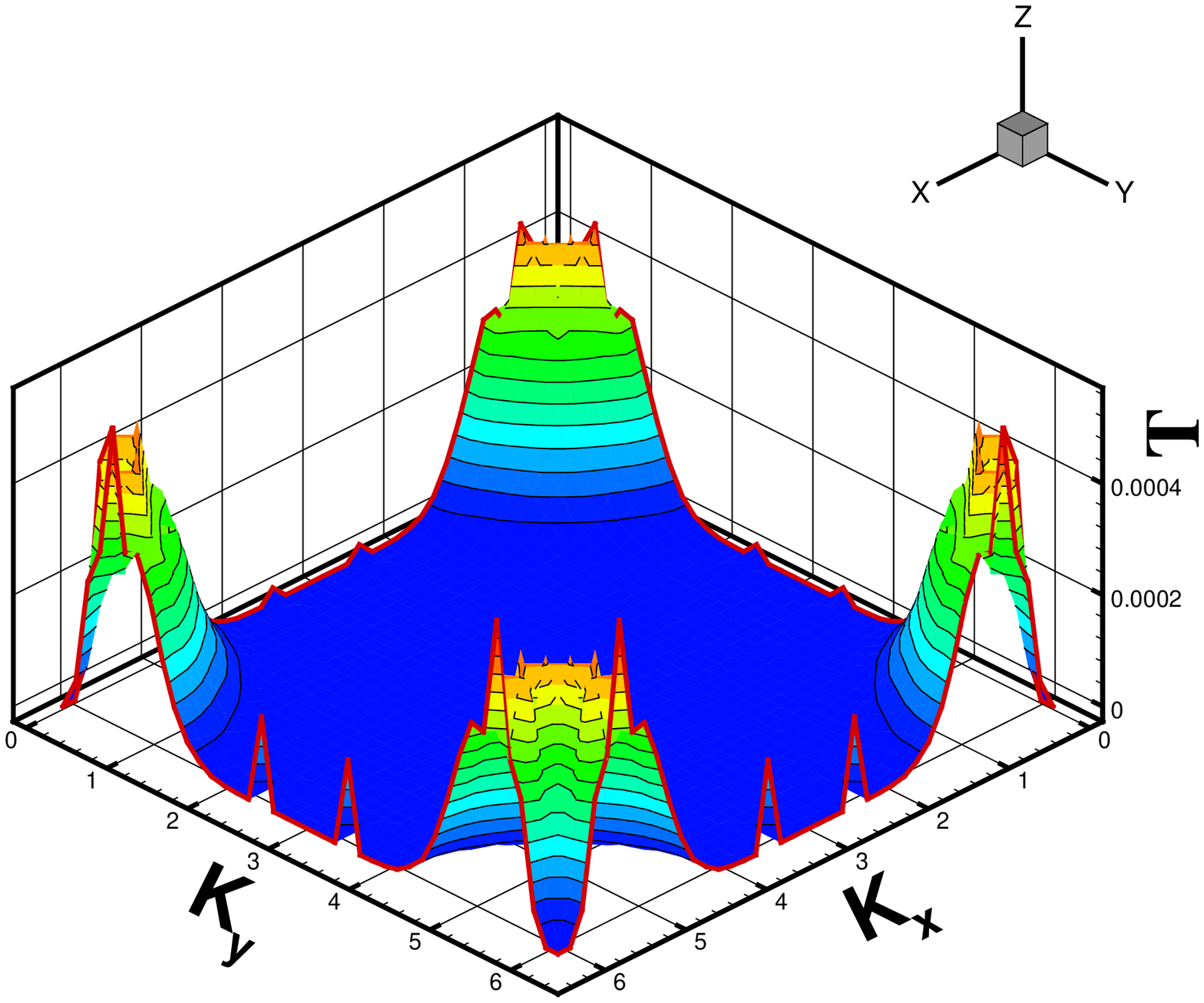}
\caption{$T(k_x,k_y)$ for Cu/Co/INS/Co/Cu tunneling junction in the parallel configuration: 
majority spins. The center of 
the Brillouin zone ($\Gamma$ point) is at the four corners of the picture.}\label{fig29}
\end{center}
\end{figure}
\begin{figure}[ht]
\begin{center}
\includegraphics[width=8.5cm,clip=true,angle=0.0]{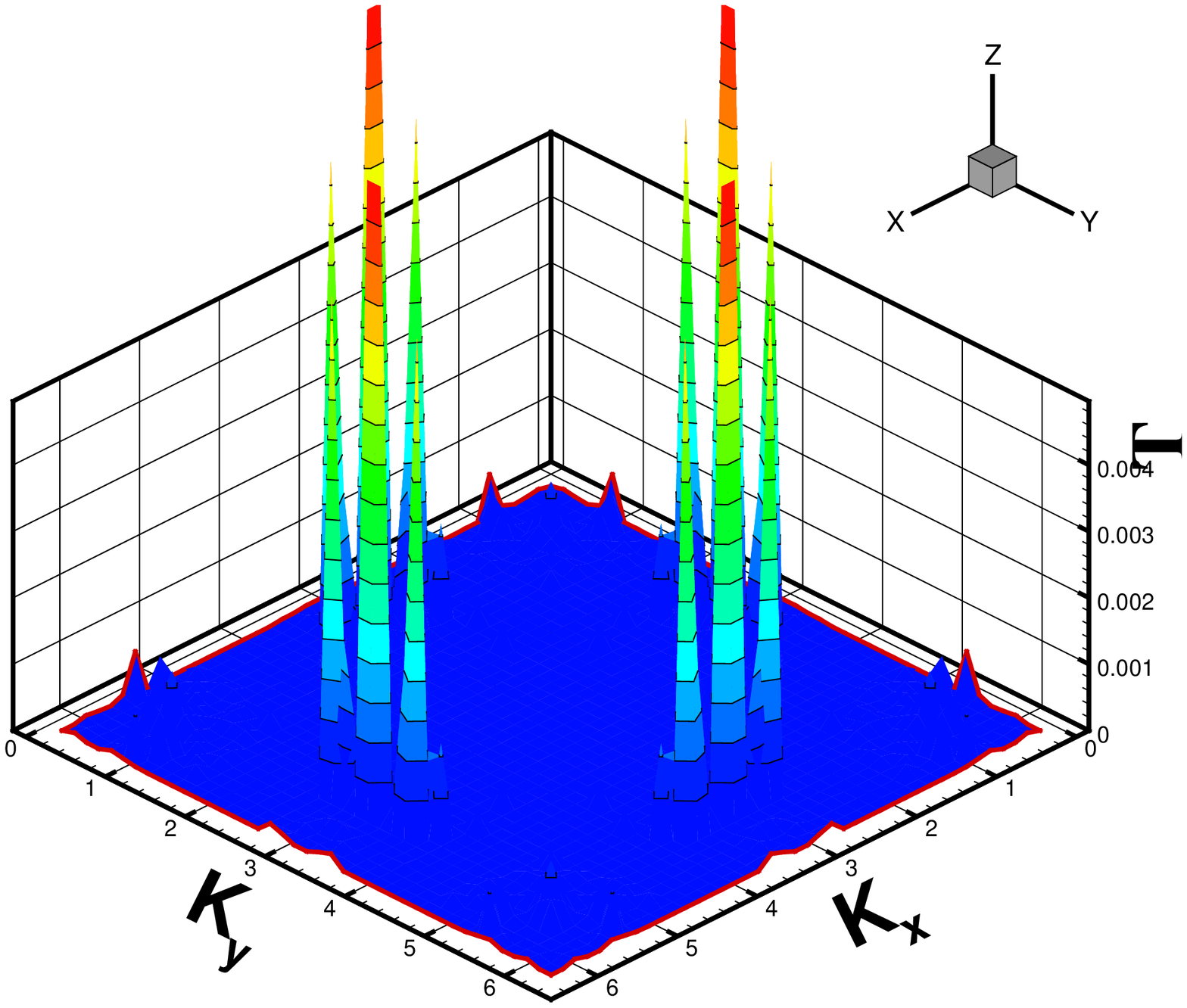}
\caption{$T(k_x,k_y)$ for Cu/Co/INS/Co/Cu tunneling junction in the parallel configuration: 
minority spins. The center of 
the Brillouin zone ($\Gamma$ point) is at the four corners of the picture.
}\label{fig30}
\end{center}
\end{figure}

Modern theory of magnetic tunneling junctions is essentially based on the concept
of complex band structure \cite{But01,wenn02}. Let us consider a Bloch function in the
leads, propagating toward the tunneling barrier. In the ``layer'' notation used in the 
previous section this reads
\begin{equation}
\psi=n_{k_\perp}^{1/2}\sum_z\mathrm{e}^{ik_\perp{z}}\phi^{k_\perp}{\;},
\label{bw1}
\end{equation}
where the sum runs over all the possible ``layers'' forming the lead, 
$n_{k_\perp}$ is a normalization constant, $k_\perp$ is the 
$k$-vector parallel to the transport direction and $\phi^{k_\perp}$
is a $N$ dimensional column vector describing the degrees of freedom of a layer
(for an infinite system in the transverse direction $N\rightarrow\infty$).
In the case of a perfectly crystalline system in the direction orthogonal to
the transport we can apply the Bloch Theorem in such direction (the $x$-$y$ plane in the
present notation) and re-write our Bloch function as
\begin{equation}
\psi=n_{k_\perp}^{1/2}\sum_z\sum_{\vec{x}}\mathrm{e}^{ik_\perp{z}}
\mathrm{e}^{i\vec{k}_\parallel\cdot\vec{x}_\parallel}\phi^{k_\perp}(k_\parallel){\;},
\end{equation}
where the $\vec{x}$ sum runs over all the cells in the two dimensional
plane orthogonal to the current, and $\phi^{k_\perp}(k_\parallel)$ is now
a $k_\parallel$-dependent $M$-dimensional column vector where $M$ is
the number of degrees of freedom (orbitals) in the unit cell.
Essentially we have reduced the problem of calculating
the transport coefficients of the Bloch wave (\ref{bw1}) to the problem of calculating 
the transmission coefficients for every Fourier component $k_\parallel$. 
This means that at the interface between the leads and the insulating barrier
each Fourier component $k_\parallel$ can scatter in the $M$ possible channels
with the same $k_\parallel$. However since there are no energy states in the
insulating region, all the possible scattering channels are closed channels, i.e.
$k_\perp$ is imaginary ($k_\perp=i\kappa_\perp$).
The extension of the band structure to the complex plane in the $z$ direction for the 
insulator is called ``complex band structure''.

If we now consider a particular Fourier component $k_\parallel$ the transmission
coefficient $T_{k_1,k_2}$ for an open channel $k_1$ in the left hand-side lead to be 
transmitted in the $k_2$ channel in the right hand-side lead is simply
\begin{equation}
T_{k_1,k_2}=\sum_m t_{k_1,\kappa_m}\mathrm{e}^{-\kappa_m z}t_{\kappa_m, k_2}\:,
\end{equation}
where $t_{k_1,\kappa_m}$ is the scattering amplitude from an open channel $k_1$ in the
leads to a closed channel $\kappa_m$ in the insulator. Finally the total transmission 
coefficient $T(k_\parallel)$ for the specific $k_\parallel$ is simply
\begin{equation}
T(k_\parallel)=\sum_{k_1}\sum_{k_2}\sum_m 
t_{k_1,\kappa_m}\mathrm{e}^{-\kappa_m z}t_{\kappa_m, k_2}=
\sum_{m}\mathrm{e}^{-\kappa_m z}\:{\cal{T}}_m\:.
\label{ftcbarrier}
\end{equation}
The equation (\ref{ftcbarrier}) essentially shows that the total transmission coefficient
for a particular Fourier component is simply given by the sum over all the 
decay states in the barrier of an exponential term times a prefactor. The exponential
decay is solely a property of the insulator while the prefactor takes into account
the matching of the wave-function at the interface between the barrier and the
leads. For short barrier the current, and therefore the spin polarization of the
junction, will depend mostly on the prefactor and therefore on the fine
details of the structure. However for large barriers the transmission is completely
dominated by the leading term (the one with the slowest decay) among all
the allowed $\kappa_m$.

It is therefore not surprising that both the TMR ratio and the spin-polarization
of the junction change when the barrier thickness increases. Moreover since the leading
term in the sum of equation (\ref{ftcbarrier}) is in principle different for different $k_\parallel$, 
the transmission coefficient will be strongly resonating over $k_\parallel$. 
Clearly this argument is strictly valid in the case of perfect translational invariance
both in the barrier and in the leads. When this hypothesis is relaxed and scattering between
$k_\parallel$ components is allowed the picture changes. In general the effect of lack
of translational invariance is that of smearing the Fermi surface of the leads. This essentially
means that the total transmission coefficient will be some average of the $T(k_\parallel)$
of equation (\ref{ftcbarrier}). In particular Tsymbal and Pettifor \cite{TsyPet} demonstrated with a 
tight-binding model that, if the barrier is rich of impurities a simple exponential decay of 
the wave-function governed by the complex band structure of the insulator is not an appropriate. 
In this situation the electron transport is via hopping between randomly placed trap states in the 
barrier and the decay of the transmission coefficient with the barrier width is much slower
than that suggested by the simple exponential decay of the leading term of equation
(\ref{ftcbarrier}). Interestingly in the large disorder limit the polarization of the junction is 
well described by the Julliere model \cite{olenik}.

\subsubsection{Inelastic effects and bias}

To date there are several first principles calculations of tunneling conductance 
of trilayers \cite{Mac99,Wang98,Tsy97t,Math97t,mertigtun,mathontun} obtained in the 
linear response limit.  These generally report quite large TMR ratios, often much larger 
than those found in experiments, and it is therefore important to investigate the main
factors leading to a reduction of the TMR. 

In broad terms inelastic scattering may reduce the TMR ratio, 
although differences are expected depending on whether or not the scattering is
spin conserving. These effects have been studied by Bratkovsky using a simple
free-electron like model \cite{Alex97tris} and considering scattering either with 
impurities, phonons or spin-waves. These three sources of scattering all decrease
the TMR ratio although they produce different temperature and bias dependence of
the junction current and hence of the spin-polarization. In particular spin-waves 
absorption/emission results in a drastic reduction of the TMR signal even in the
presence of half-metallic contacts, for which the simple density of state argument 
leads to the expectation of an infinite TMR ratio \cite{Alex97,Alex97bis}.
Unfortunately a fully {\it ab initio} theory of inelastic transport in the quantum limit is 
still missing and I am not aware of any calculation in that direction.

A somehow easier problem is that connected to the study of finite bias. One of the 
differences between GMR and TMR junctions is that, while the former operate always
at very small bias (then in the linear response limit), in the latter usually a bias of the order
of 1~Volt is applied across the insulating layer. Experimentally it was observed that a
bias voltage across the device can greatly reduce the TMR signal \cite{TMRexpBIAS}
and this was explained in terms impurity scattering at the interface. Clearly an applied bias
can change the equilibrium distribution of the inelastic elementary excitations
(phonons and spin-waves) and therefore give rise to a general reduction of the TMR
signal \cite{Alex97tris}. 

Recently Zhang et al. \cite{TMRbias} conducted a first principles study of the conductance
and the TMR of a Fe/FeO/MgO/Fe tunneling junction under bias. They used a bias-dependent 
version the KKR method, where a perfect translational invariance of the system is assumed.
For this type of trilayer the translational invariance perhaps is not a bad approximation since 
epitaxial growth of MgO on Fe has been experimentally demonstrated \cite{MgO}.
Their main finding is that although the changes in the electronic structure due to the bias
are minimal and the effective capacitance is consistent with the dielectric constant of MgO,
the $I$-$V$ characteristic is highly non-linear and the TMR ratio increases as a function of bias.
This behavior is essentially due to a reduction of the current associated to the minority
spin and in the antiparallel alignment as the bias is applied, and it is explained as a suppression
of the resonant contribution to the transmission coefficient coming from surface states.
These results, which are expected to be rather general, are in stark contradiction with
the experimental data and more investigation in this direction is certainly necessary. 

\subsection{Domain wall scattering and GMR in atomic point contacts}

The typical macroscopic configuration of magnetic materials is that of a collection
of regions, the magnetic domains, where the magnetic moments of the individual atoms 
point in the same direction. In normal conditions the macroscopic magnetization vectors
of these regions are randomly aligned in order to minimize the magnetostatic energy. 
This means that in the interstitial areas between two domains the magnetization
vector change direction. It is then natural to ask whether or not 
these domain walls (DW) bring additional contribution to the electrical resistance.
Intuitively one would expect some sort of contribution since the exchange potential
depends on the magnetic state.
\begin{figure}[ht]
\begin{center}
\includegraphics[width=9.5cm,clip=true,angle=0.0]{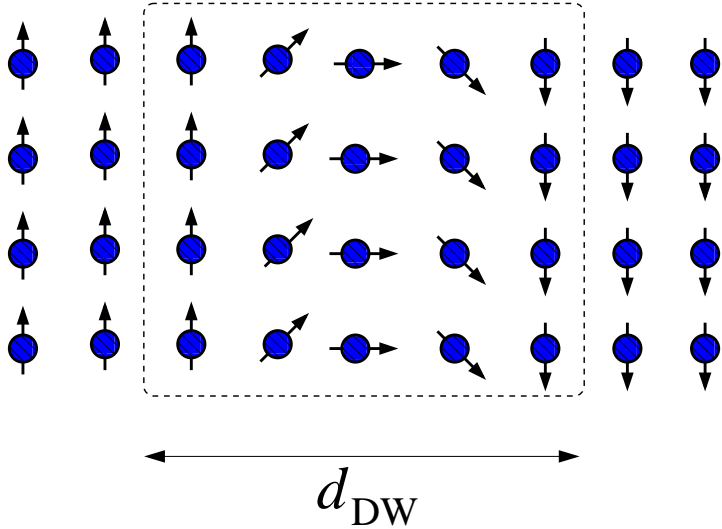}
\caption{Schematic representation of a typical domain wall magnetoresistance 
experiment. $d_\mathrm{DW}$ is the domain wall thickness.}\label{fig31}
\end{center}
\end{figure}

The typical configuration for a domain wall scattering experiment is that presented 
in figure \ref{fig31}. Essentially the idea is to measure the resistance of a domain 
wall using a GMR spin-valve experiment, where the non-magnetic spacer is now 
replaced by the wall itself. The current measured for the configuration of figure
\ref{fig31} is then compared with that obtained for the parallel alignment of the 
magnetization vectors of the contacts, i.e. when the domain wall is removed. Again the
quality factor of the device is given by the GMR ratio defined before.

\subsubsection{Infinite domain wall: early theory}

The first studies on domain wall scattering were focused on domain walls
with infinite cross section. This situation is a good approximation for bulk
transition metals, where the magnetic domains can be macroscopic objects.
The first calculations on domain wall scattering is due to Cabrera and 
Falicov \cite{cabrera}, who calculated the resistance associated to a domain
wall in terms of a tunneling process across the wall. The calculation was carried
out for a simple free-electron model and resulted in an exponentially small 
contribution to the resistance. A few years later Tatara and Fukuyama \cite{tatara1} 
calculated the contribution of a DW to the resistivity of a magnetic material.
They found that in general the presence of a DW enhances the Boltzmann electrical
resistivity, and that the DW contribution scales as $d_\mathrm{DW}^{-1}$. However, in 
situations where the Boltzmann resistivity is small (clean limit at low temperature), 
the presence of a wall contributes to the decoherence of the electrons, hence
decreasing the total electrical resistivity. Therefore the magnetoresistance
associated to a domain wall can be either positive or negative, depending on the
transport regime encountered.

This last result triggered a considerable interest and several experiments 
were performed in order to measure the DW resistance. Unfortunately these were 
often inconclusive and difficult to relate to each other, and both positive 
\cite{otani,ruediger} and negative magnetoresistance \cite{gregg,viret} were measured.
The main difficulty in these experiments was in separating the tiny change in resistance 
due to a DW wall from the ordinary contribution of the anisotropic magnetoresistance (AMR).

The reason for the calculated, and indeed measured, small contribution of
DW scattering to the resistivity is essentially that the energy associated
to the magnetic moment rotation is tiny with respect to the exchange energy. This
means that the presence of a DW does not considerably modify the Fermi surface.
However early theories were based on simplified descriptions of the electronic structure
of the transition metals neglecting the fact that the separation of the bands in the 
$d$ manifold is smaller than the typical exchange. Therefore it becomes increasingly 
important to study the effect of a DW scattering using realistic bands calculations. 
The first work of this type was performed by van Hoof et al. using realistic DFT-derived 
bandstructure and the Landauer formula \cite{vanHoof}. In their calculations a local spin 
rotation was introduced in the direction of the transport, while the system was considered 
translational invariant in the transverse direction. This reproduces the case of a DW of
infinite cross section but finite width. A summary of their results is reported in table \ref{Table6}.
\begin{table}[htbp]
\begin{center}
\begin{tabular}{cccc} \hline
\ Property \ & \ \ Fe \ \ & \ Ni \ & \ Co \ \\ \hline\hline
Crystal structure & bcc & fcc & fcc    \\ 
Layer direction & (100) & (100) & (111)     \\ 
$d_\mathrm{DW}$ (nm) & 40 & 100 & 15  \\ 
$d_\mathrm{DW}$ (monolayer) & 276 & 570 & 72   \\ 
&  &  &     \\ 
DW-MR &  &  &     \\ 
Bulk DW & -0.39\% & -0.16\% & -0.46\%    \\ 
Abrupt DW & -71\% & -58\%  & -67\%  \\ \hline\hline
\end{tabular}
\caption{DW magnetoresistance for Fe, Ni and Co. Bulk-DW indicates a DW whose
thickness is the one reported in the table, while the abroupt DW is a 
DW formed from one monolayer only. Adapted from reference \cite{vanHoof}.}
\label{Table6}
\end{center} 
\end{table}

The main result from this work is that a DW always decreases the electrical conductance,
which scales approximately as $d_\mathrm{DW}^{-1}$. As a consequence the MR associated
to a bulk DW is always very small, never exceeding 1\% in transition metals. These results
were later confirmed by other {\it ab initio} calculations \cite{kudDW,kudDW2}.

Interestingly the original van Hoof's calculation showed the possibility of remarkably large
GMR in the case of abrupt DW (see table \ref{Table6}), which is a domain wall where a 
180$^o$ degree spin-rotation is achieved in one atomic layer only. For instance in the case of
Co DW the GMR ratio is of the same order that the one calculated with similar methods for
Co/Cu multilayers \cite{kelly1}. This is not surprising. In fact a majority (minority) electron
traveling on the left hand side of the DW, must travel as a minority (majority) on the right
hand side. This means that in absence of spin-flip both spin electrons undergo to strong scattering
in crossing the DW. Moreover since the bandstructure of the majority band of Co (figure \ref{fig2a})
is very similar to that of Cu (figure \ref{fig2b}), one expects that the scattering for either spins 
when crossing a DW is similar to that of the minority electrons in crossing a Co/Cu interface. For 
this reason it does not appear strange that an abrupt DW presents large GMR. 
Unfortunately the exchange energy of an abrupt DW is rather large and therefore these are
impossible to form in the bulk. A possible way out is to consider very narrow constriction. 
In this case the energy costs are associated to the rotation of only a small number of magnetic 
moments, and DW as thick as the constriction lateral dimensions have been predicted \cite{brunodw}. 
This prediction opened a new avenue to DW GMR and several investigations on the electrical
properties of magnetic point contacts began.

\subsubsection{Magnetic point contacts and ballistic MR}

The experimental situation on DW GMR in magnetic point contact is still controversial. On the one hand
there are a number of reports of extremely large GMR \cite{BMR1,BMR2,BMR3} reaching up
1,000,000\% \cite{BMR4} in either transition metals or other magnetic materials. On the other hand there 
is an equally long list of works in which either very small \cite{viret1,viret2} or a non-existing GMR
was found \cite{egel1,egel2}. The experiments are usually GMR-type of experiments where an
external magnetic field is applied to a point contact, and the resistance is recorded. Hence
the essential point of controversy is whether or not the large GMR has electronic origin or is just 
a consequence of the rearrangement of the point contact when the field is applied. Numerous reasons
for the mechanical hypothesis have been given including magnetostriction, dipole-dipole interaction
between the contact apexes \cite{egel3} or magnetically induced stress relief \cite{JEW}. In all these 
situations the point contact becomes compressed upon the application of a magnetic field, and
its cross section increases with a consequent reduction of the two terminal resistance.

In contrast if one wants to follow the electronic-only hypothesis, the fundamental point to bare in 
mind is that in atomic point contacts the last atom in the constriction is the one that determines the
conductance \cite{sheer,agrait}. Hence this cannot exceed the number of valence 
electrons. Under this assumption we can immediately set an upper bound for the GMR in
a point contact made from a strong ferromagnet (Co or Ni). When the magnetization vectors of both 
sides of the constriction point in the same direction (parallel configuration) the conductance
can be as large as 7~$e^2/h$, with a contribution of 2~$e^2/h$ coming from the $s$ electrons 
(both spins contribute) and a contribution of 5~$e^2/h$ coming from the minority $d$ electrons. In this
case the majority $d$ states are completely filled and they do not contribute to the zero-bias current.
In contrast in the antiparallel configuration the conductance of the $d$ electrons is blocked and one is
left with the 2~$e^2/h$ contribution from the $s$ electrons only. This leads to a GMR ratio of
250\%. 

Unfortunately this intuitive picture is hardly found in reality, and with the exception of noble metals
like gold, the simple one on one mapping of conductance over the number of valence electrons does
not hold. The reality is that hybridization mixes the orbital states and usually one has several
scattering channels with transmission coefficients smaller than unity. For instance in the case of Co 
and Ni point contacts conductance histograms with peaks positioned everywhere from $e^2/h$ to 
4~$e^2/h$ have been reported \cite{his1,his2,his3,his4}. 

To date several calculations on point contact transport through magnetic transition metals have
been published. These are all based on the linear response theory of transport using either
DFT \cite{dalcorso,mertigpc,palaciospc} or tight-binding \cite{butlerpc} Hamiltonian. Despite differences
due to the details of the methods and the geometrical configuration of the junctions investigated, all
these calculations agree on a rather small value of GMR. The common feature is that there are
always two channels with $s$-like character (one per spin direction), which are almost perfectly
transmitted across the junction. These are quite robust against the contact geometry \cite{mertigpc}
and clearly do not give rise to any GMR. In contrast the transmission of $d$-type channels is rather
sensitive on the details of the junction, although their contribution to the conductance appear to be
always smaller than the upper bound of 5~$e^2/h$ set by the valence. Clearly a precise prediction 
should take into account the correct experimental configuration of the contact \cite{palaciospc}, 
however from this analysis it appears unlikely that a large GMR of purely electronic origin can be found.

A possible additional element to include into this picture is given by the presence of impurities in the 
point contact. In particular one can speculate that strongly electronegative impurities (for example
oxygen) can bind the $s$ electrons of the transition metal, therefore eliminating their
contributions to the conductance. If this is the case, the conductance will be given by $d$ electrons
only and one will expect very large GMR ratios. Calculations on these type of systems are particularly
problematic since the conventional LSDA (or indeed GGA) to DFT fails in describing the physics
of a Mott insulator, which characterizes transition metal mono-oxides such as NiO. A seminal calculation
of GMR across an oxygen monolayer sandwiched between (001) Ni electrodes obtained with DFT-LSDA demonstrated a large MR \cite{papapc}. However the investigation of realistic geometries and of
different DFT functionals is still lacking.

\subsubsection{Point contacts under bias}

In all the theoretical works cited in the previous section the calculations
were carried out in the linear response limit. This is appropriate since in
actual experiments very small biases are applied in order to prevent the point contact
fracture. However in the case of electrodeposited constrictions, in which
the typical cross section is much larger than that of single atom contacts,
biases in excess of 2~Volt can be applied resulting in interesting diode-like
$I$-$V$ characteristics \cite{oscar1,oscar2}. 

Charging and bias effects in magnetic point contacts were studied recently
by Rocha and the present author \cite{alexpc}. For the calculations we used
a self-consistent tight-binding model, with two orbitals per atomic site representing
respectively the $s$ and $d$ electrons. In this case the on-site energies present both 
a Stoner-type term giving rise to the ferromagnetic ground state, and a charging term
ensuring local charge neutrality. Finite bias $I$-$V$ characteristics were calculated
using the NEGF method. 

The main result of this calculation is to show that non-symmetric $I$-$V$ characteristics
can arise from a non-symmetric magnetic configuration of the junction, although the 
structure itself is perfectly symmetric. In order to demonstrate this concept we model a
point contact as a $2\times 2$ atomic chain comprising four atomic planes and
sandwiched between two semi-infinite simple cubic leads 
with a $3\times 3$ atom cross section. We investigate two possible situations
respectively when the DW is positioned symmetrically or asymmetrically in the junction
(see figure \ref{fig32}). The $I$-$V$ curves for these two configurations are presented in
figure \ref{fig33}, where a clear asymmetry for the asymmetric DW appears.
\begin{figure}[ht]
\begin{center}
\includegraphics[width=9.5cm,clip=true]{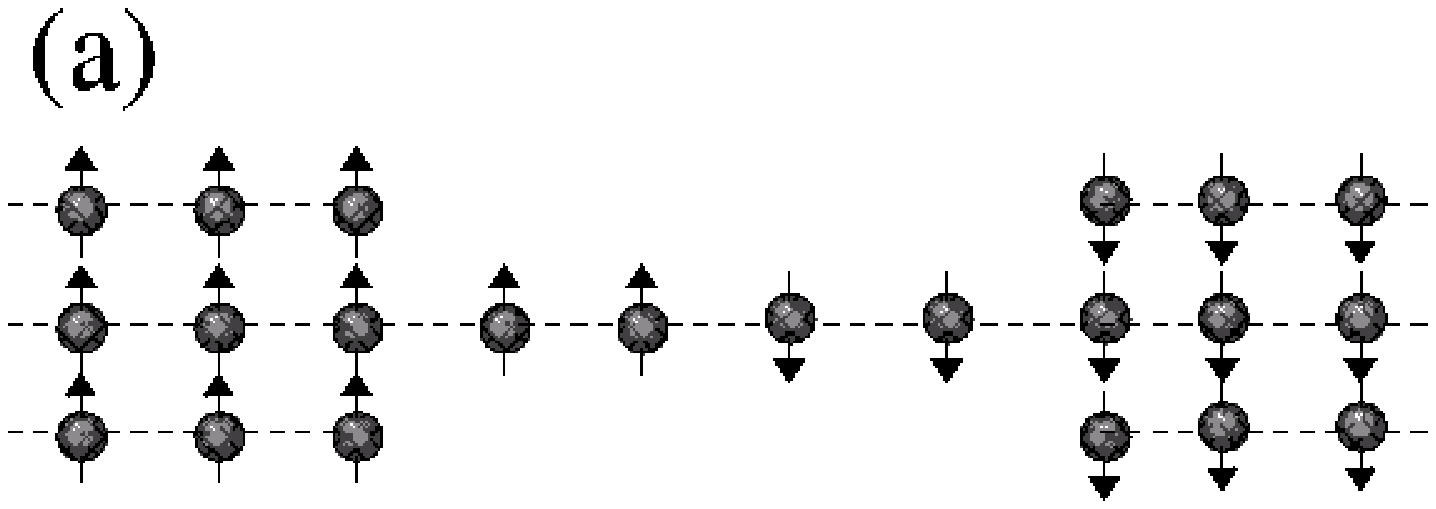}
\vspace{2.2cm}
\includegraphics[width=9.5cm,clip=true]{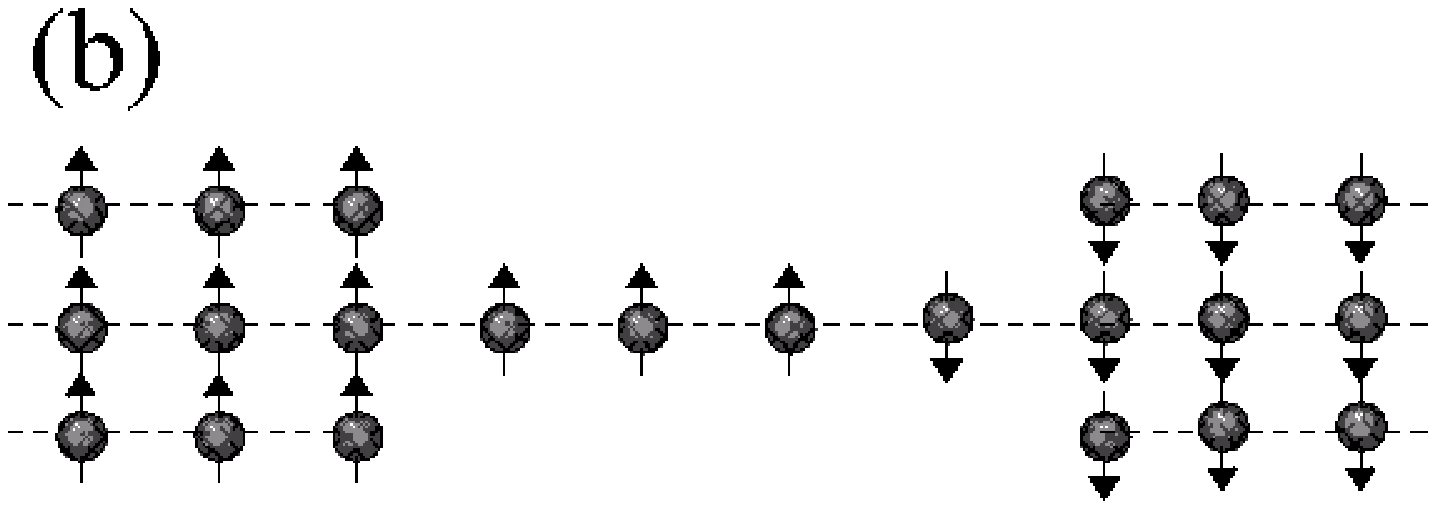}
\caption{Schematic representation of the point contact studied in reference \cite{alexpc}. An atomically 
sharp domain wall is positioned either in a symmetric (a) or an asymmetric (b) position. }\label{fig32}
\end{center}
\end{figure}
\begin{figure}[ht]
\begin{center}
\includegraphics[width=8.5cm,clip=true,angle=0.0]{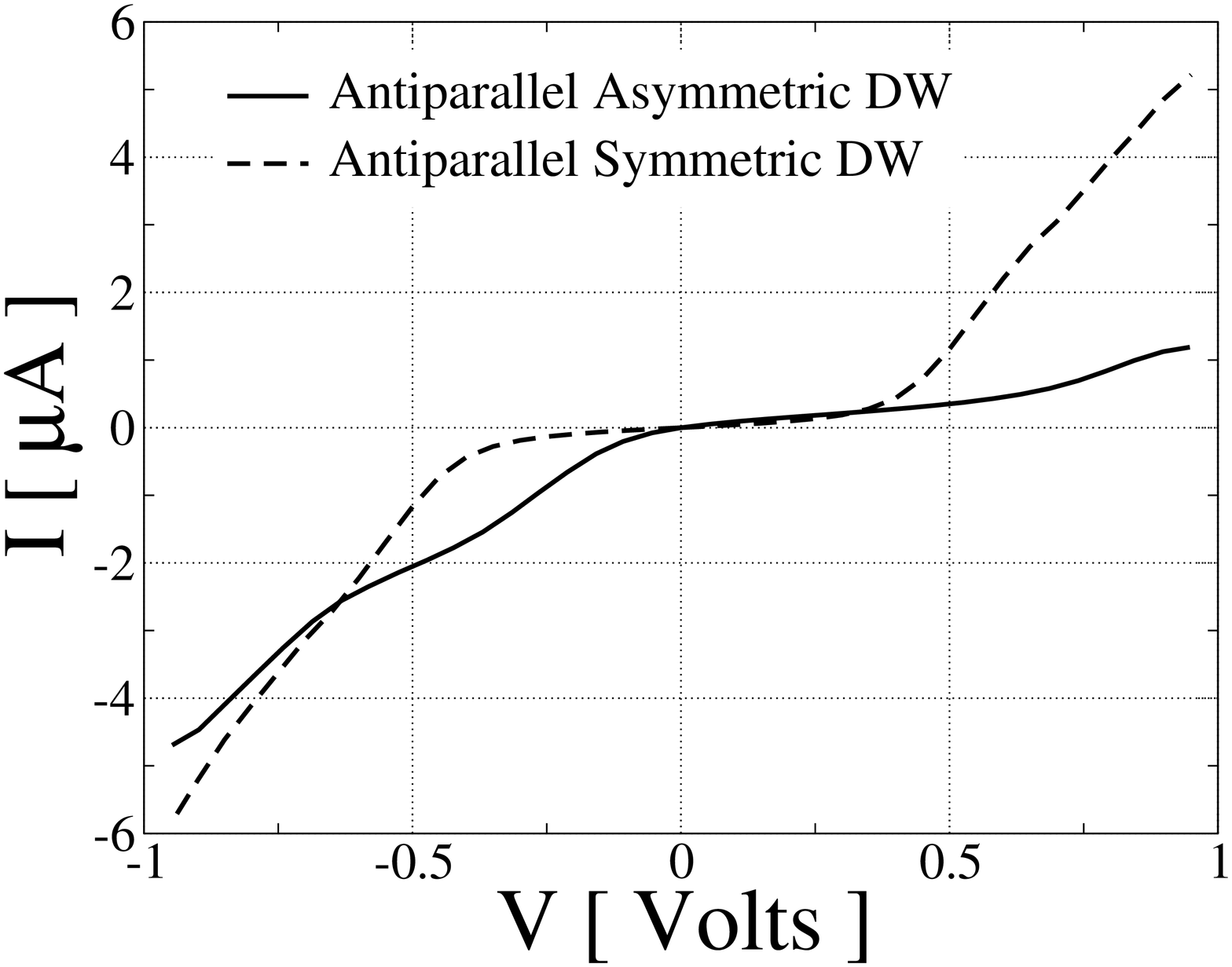}
\caption{Current as a function of the bias for an atomic point contact in which a DW is placed in either
a symmetric (configuration (a) of figure \ref{fig32}) or an asymmetric (configuration (b) of figure 
\ref{fig32}) position. Note the large asymmetry of the $I$-$V$ characteristic for the asymmetric DW.}\label{fig33}
\end{center}
\end{figure}

In order to understand this feature we model our magnetic point contact as a magnetic 
molecule. When the magnetization of the two leads are aligned antiparallel to each other 
the DW formed inside the molecule splits the HOMO and LUMO states. Hence our system 
appears as two molecules strongly attached respectively to the left and the right lead, 
but weakly coupled to each other. This gives rise to the level scheme presented in 
figure \ref{fig34}a, which is strictly valid only in the case the DW resistance is infinite. 
Within this scheme the left hand-side part of the PC couples strongly with the left lead 
therefore presenting a majority spin HOMO and a minority spin LUMO.
The situation is opposite on the right hand-side since the magnetization of the right lead is
rotated. Recalling the fact that we do not consider the possibility of spin mixing, this 
configuration presents a large resistance since there are no resonant states with the same spin extending through the entire point contact.
If we now apply a bias there will be level shifting. This aligns the majority spins HOMO on the left
with the LUMO on the right for positive bias (figure \ref{fig34}b) and the minority LUMO on the left
with the HOMO on the right for negative bias (figure \ref{fig34}c).
\begin{figure}[ht]
\begin{center}
\includegraphics[width=8.5cm,clip=true,angle=-90.0]{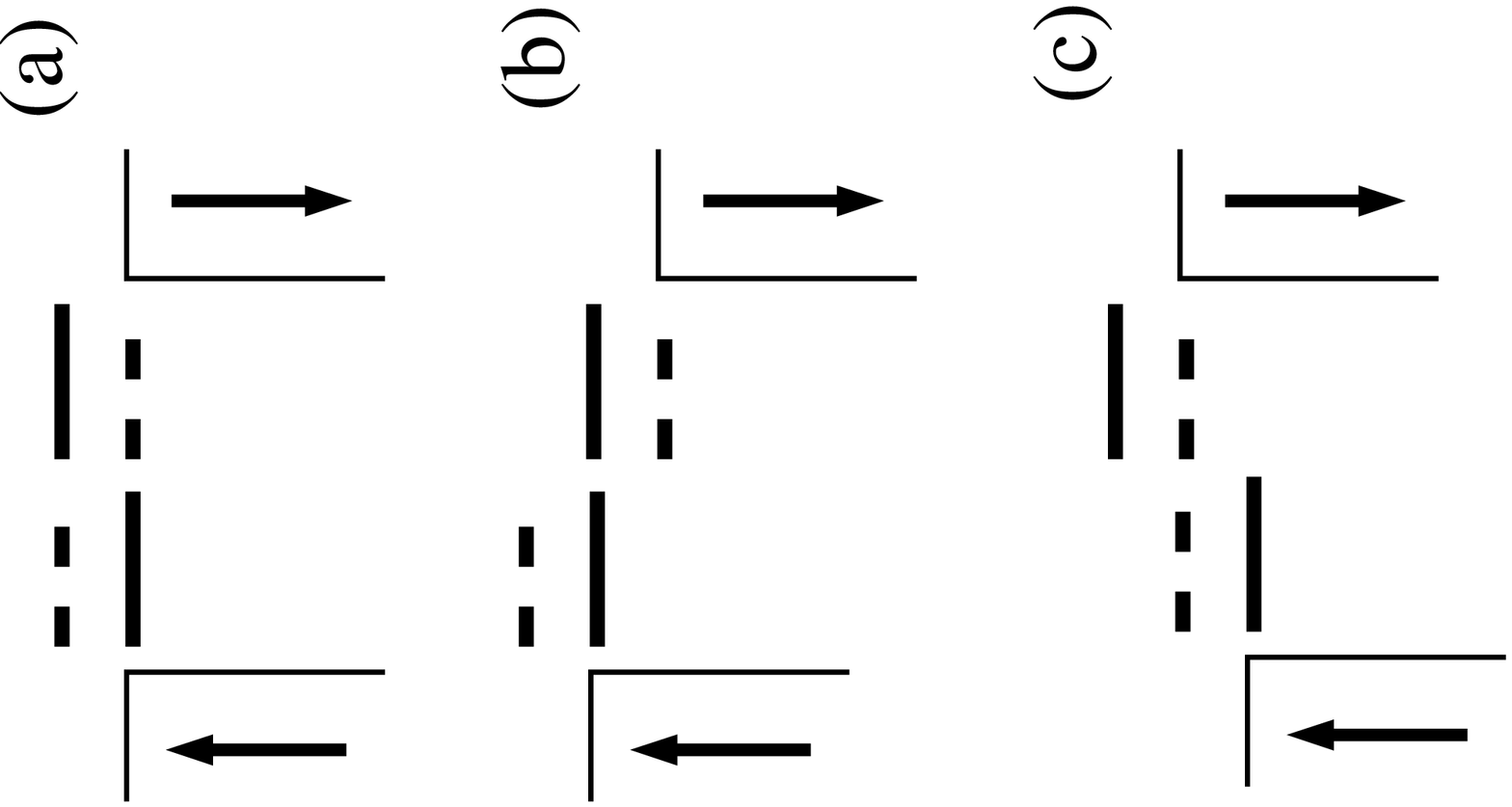}
\caption{Cartoon showing the levels alignment in the magnetic point contact. The solid 
(dashed) line denotes a majority (minority) spin molecular state. a)
symmetric case at zero bias, b) symmetric case at positive bias,
and c) symmetric case at negative bias. Note that for the antiparallel case 
the spin of the resonant level is opposite for opposite bias direction.}\label{fig34}
\end{center}
\end{figure}

In addition for positive bias DW charge is accumulated in the first two atomics 
planes to the left of the DW and it is depleted in the two atomic planes to the right. In the case of a symmetric DW the molecular states to the left and to the right of the Bloch wall charge in a 
symmetric way with respect to the bias direction. This means that the level alignment responsible 
for the large current will occur for the same bias difference independently from the bias polarity.
In contrast, when the DW is between the third and the fourth plane, more charge can be
accumulated (depleted) to the left of the DW with respect to the right,
since more states (atomic planes) are available. This means that the alignment of the energy 
level now depends on the bias polarity, leading to an asymmetric $I$-$V$ curve. 

The study of magnetic point contact under bias is indeed in its infancy and for a more 
quantitative analysis {\it ab initio} transport methods will be welcome. These probably should
include some sort of strong correlation corrections, although perhaps only at the mean field 
level \cite{anis1,anis2}, for describing the ferromagnetism in such reduced dimensions. Moreover
we believe that the study of the interplay between magnetic and mechanical effects will
be crucial for understanding issues connected with the stability of the contacts and of their
magnetic state.

\subsection{Spin-transport through carbon nanotubes and nanowires}

As I have anticipated in the introduction recently there was a considerable effort in combining
the field of molecular- and spin-electronics. The typical device consists in a spin-valve that uses
a molecular object as spacer. Potentially this has several advantages compared to more conventional
materials since molecules are free of strong spin-flip scattering mechanisms such as
spin-orbit and hyperfine interaction or scattering to magnetic impurities. In this race to organic
spin-devices the use of carbon nanotubes occupies an important place. 

Carbon nanotubes are almost defect-free graphene sheets rolled up in forming 1D 
molecules with enormous aspect ratios \cite{Dress}. Their conducting state (metalicity) depends on
their chirality, however in the metallic configuration they are ideal conductors with a remarkably long
phase coherent length \cite{Heer,todorovCN}. An important aspect is that the relevant physics at
the Fermi level is entirely dominated by the $p_z$ orbitals, which are radially aligned with respect 
to the tube axis. These include the bonding properties with other materials and between tubes. 
Therefore carbon nanotubes appear as an ideal playground for investigating both GMR and TMR
through molecules. In fact one can expect that two tubes with different chirality will bond to a magnetic
surface in a similar way, allowing us to isolate the effects of the molecule from that of the contacts.
Indeed TMR-like transport through carbon nanotubes has been experimentally reported by several 
groups \cite{CNT_GMR,CNT1,CNT2,CNT4,CNT5}.

Now the fundamental question is: why will one expect a large GMR from a carbon nanotube? In
order to answer this question I will use an argument derived by Tersoff \cite{Tersoff} and then 
subsequently refined \cite{MaxJTersoff} for explaining the contact resistance of a C nanotube
on an ordinary metal. In order to fix the ideas let me consider an armchair nanotube (metallic).
Its Fermi surface consists only of two points symmetric with respect to the
$\Gamma$ point (see figure \ref{fig35}). The Fermi wave-vector is then
$k_{\mathrm F}=2\pi/3z_0$ with $z_0=d_0\sqrt{3}/2$ and $d_0$ the C-C bond 
distance ($d_0$=1.42~\AA). 
\begin{figure}[ht]
\begin{center}
\includegraphics[width=9.5cm,clip=true,angle=0.0]{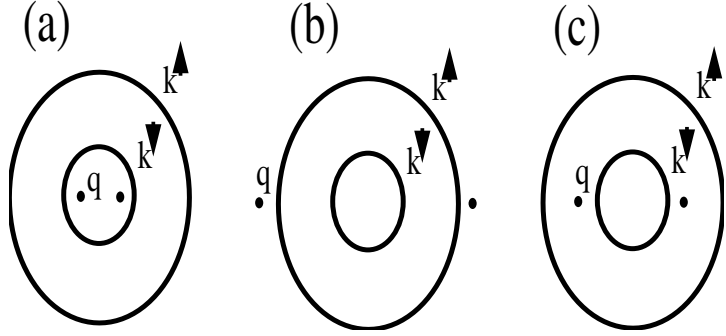}
\caption{Fermi surfaces of an armchair carbon nanotube and of a magnetic
transition metal. The Fermi surface of the nanotube consists in two points
$k_{\mathrm F}^{\mathrm N}=q$,
symmetric with respect to the $\Gamma$ point. The Fermi surface of a
transition magnetic metal consists of two spheres (for up and down spins) whose
different diameters depend on the exchange field. The three possible scenarios
discussed in the text: 
(a) $q < k_{\mathrm F}^\downarrow < k_{\mathrm F}^\uparrow$,
(b) $k_{\mathrm F}^\downarrow < k_{\mathrm F}^\uparrow < q$,
(c) $k_{\mathrm F}^\downarrow < q < k_{\mathrm F}^\uparrow$.}\label{fig35}
\end{center}
\end{figure}

Assume now that the Fermi surface of a typical magnetic metal simply
consists of two spheres with different radii for the different spins. This oversimplified
scenario is that of a free-electron model with internal exchange field
\begin{equation}
E^\sigma=\frac{\hbar^2k^2}{2m}+\sigma \Delta/2
\;{,}
\end{equation}
with $\sigma=-1$ ($\sigma=+1$) for majority (minority) spins and $\Delta$ the 
exchange energy. The spin-dependent Fermi wave-vectors are then respectively
$k_{\mathrm F}^\uparrow=\sqrt{2m(E_{\mathrm F}+\Delta/2)}/\hbar$ and
$k_{\mathrm F}^\downarrow=\sqrt{2m(E_{\mathrm F}-\Delta/2)}/\hbar$.

The transport through an interface between such a magnetic metal and the
nanotube is determined by the overlap between the corresponding Fermi surfaces.
Three possible scenarios are possible. 
First the Fermi-wave vector of the carbon nanotube is smaller than both 
$k_{\mathrm F}^\uparrow$ and $k_{\mathrm F}^\downarrow$
(see figure \ref{fig35}a). In this case in the
magnetic metal there is always a $k$-vector that matches the 
Fermi-wave vector of the nanotube for both spins. Therefore both spins can be
injected into the tube and the total resistance will be small and
spin-independent.

Secondly the Fermi-wave vector of the carbon nanotube is larger than both 
$k_{\mathrm F}^\uparrow$ and $k_{\mathrm F}^\downarrow$
(see figure \ref{fig35}b). In this case there are no available states in 
the metallic contact whose wave-vectors match exactly the Fermi
wave-vector of the carbon nanotube. Therefore in the zero-bias zero-temperature
limit the resistance is infinite. Nevertheless as one increases the temperature, 
phonon assisted transport starts to be possible.
Spin electrons can be scattered out of the Fermi surface into states with large 
longitudinal momentum. At temperature $T$ the fraction of electrons with energy 
above $E_{\mathrm F}$ is simply proportional to the Fermi distribution function. 
However, because of the exchange energy, spin-up electrons will
possess higher momentum than spin-down. Therefore one can find more
spin-up states with a longitudinal momentum matching the one of the nanotube than 
spin-down states. 
This gives a temperature-induced spin-dependent resistance. Hence one should 
expect that the increase of the temperature will decrease the
resistance for spin-up electrons, leaving unchanged that of spin-down electrons.

Finally if the Fermi wave-vector of the carbon nanotube is larger than
$k_{\mathrm F}^\downarrow$ but smaller than $k_{\mathrm F}^\uparrow$
(see figure \ref{fig35}c), only the majority electrons
can enter the nanotube and the system becomes fully spin-polarized. In this
situation a spin-valve structure made by magnetic contacts and carbon nanotube as
spacer is predicted to show an infinite GMR at zero temperature, similar to the case of
the half-metals \cite{ss_2}. The increase of the temperature will
produce a degradation of the polarization because also the spin-down electrons may
occupy high energy states with large longitudinal momentum. Both the spins can be
injected and the spin-polarization will depend on the number of occupied states with
longitudinal momentum matching the one of the nanotube.

Two important aspects must be pointed out. First all these considerations are
based on the assumption of perfectly crystalline systems. This may not be true
in reality and the effects of breaking the translational invariance must be
considered. From a qualitative point of view disorder will smear the Fermi
surface and eventually produce some states with large longitudinal momentum.
This will improve the conductance through the nanotube, even if its
spin-polarization will be in general dependent on the nature of disorder. 

Second, in contacts made from transition metals the simple parabolic band model
introduced here is largely non-realistic. The Fermi surface of magnetic metal can
comprise different manifolds with different orbital components and the degree
of polarization of a junction depends upon how the different manifolds couple
to the nanotube. In this case simple theories are only speculative and more
realistic bandstructure calculations are needed. These are rather problematic 
since the problem include the need of describing transition metal leads and 
a molecule comprising a large number of degrees of freedom.

For this reason most of the calculations to date have used simple tight-binding 
models without self-consistent procedures \cite{Mehrez,Kromp1,Kromp2}. These roughly 
agree on the possibility of large GMR ratios in transition metals contacted nanotubes,
although the actual values predicted are somehow affected by the different methods and the
contact geometry. 
In addition more sophisticated calculations based on DFT and non-equilibrium Green's 
function method for simpler carbon chains have been reported \cite{GuoGMR,PatiGMR}. 
These give interesting insights into the spin transport properties through organic materials
although they fail in providing a realistic description of magnetic transition metal contacts.
In the case of reference \cite{GuoGMR} the contacts are made from Al leads in which a
strong magnetic field is applied, while in the case of reference \cite{PatiGMR} the contacts
are Au wires and the spin polarization is obtained by inserting a Co atoms between the C
chains and the electrodes.

\subsection{Molecular devices}

Together with their indisputable spectacular properties carbon nanotubes as building blocks 
in nanoscale devices have an important drawback, namely their saturated orbitals
make the bonding with electrodes problematic. This is not only structurally weak
but also inefficient at transmitting electrons. For this reason molecules other
than nanotubes look more appealing for constructing electronic and therefore spin-
devices. Molecules offer an almost infinite range of HOMO-LUMO gaps and molecular orbital
types, and they can be functionalized in several ways in order to attach them to metallic
contacts. Therefore they appear as the ultimate systems for engineering spin-devices.
 
Recently there have been several investigations of transport in molecular spin-valve like 
devices. These include hot electron coherent spin transfer across molecular bridges 
\cite{Aws5}, spin-injection in $\pi$-conjugated molecules \cite{mol_GMR,GMR_poly} and 
organic tunneling junctions \cite{mol_TMR}. All these demonstrate convincingly the 
possibility of performing spin-physics in organic compounds, although problems connected 
with a generally poor reproducibility still need to be addressed.

The modeling of such devices is even more challenging than that involving nanotubes
since large biases can be applied and an accurate description of the electrostatic is also
needed. This adds up to the already discussed difficulty of describing magnetic transition
metal leads, and makes the problem a tough theoretical challenge. A few seminal
studies on transport through 1,4-benzene-dithiolate molecular spin-valves 
\cite{pati,kirk} have appeared. However they either overlook the electrostatic problem 
\cite{kirk}, or the leads are non-magnetic and the spin-polarization is obtained by a 
magnetic cluster added at the top of Au electrodes \cite{pati}. 

To the best of my knowledge the code  {\it{Smeagol}} \cite{smeagol1} is the only one to date able of 
calculating accurate $I$-$V$ characteristics of devices employing magnetic leads. Here
I briefly summarize the spin-transport properties of spin-valve made from Ni leads and
a molecule as spacer \cite{smeagol2}. As an example I will analyze two different molecules, 
respectively [8]-alkane-dithiolate (octane-dithiolate) and 1,4-[3]-phenyl-dithiolate (tricene-dithiolate). 
A schematic density of states and the charge density isosurfaces of the HOMO and LUMO 
states for the isolated molecules are presented in figures \ref{fig36} and \ref{fig37}. 
\begin{figure}[ht]
\begin{center}
\includegraphics[width=9.5cm]{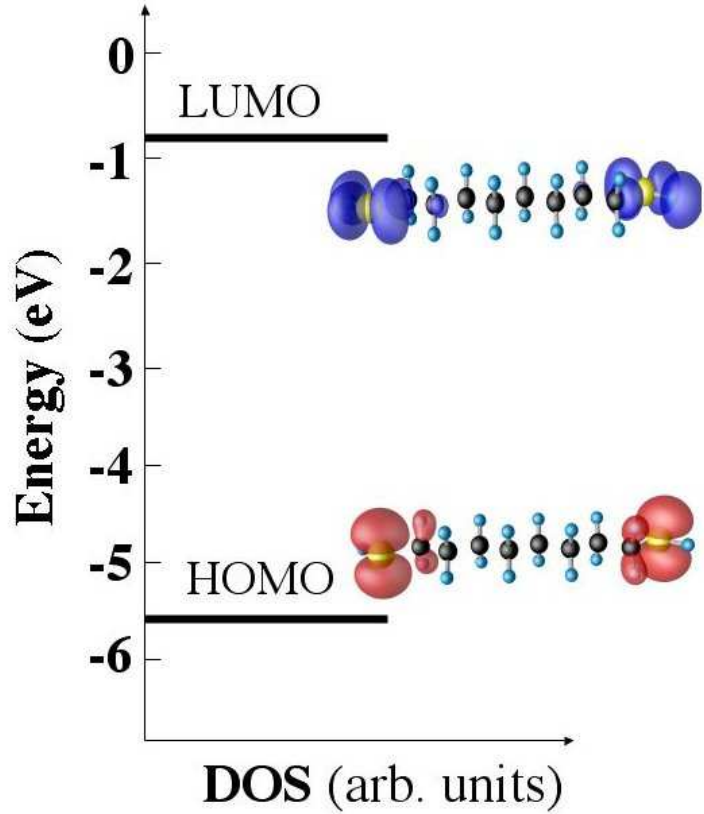}
\caption{[8]-alkane (octane) molecule:
DOS and charge density isosurface plots for the relevant molecular states of
the isolated molecule. $E_\mathrm{F}$ denotes the position of the Fermi
level for such an isolated molecule.}\label{fig36}
\end{center}
\end{figure}
\begin{figure}[ht]
\begin{center}
\includegraphics[width=9.5cm,clip=true,angle=0.0]{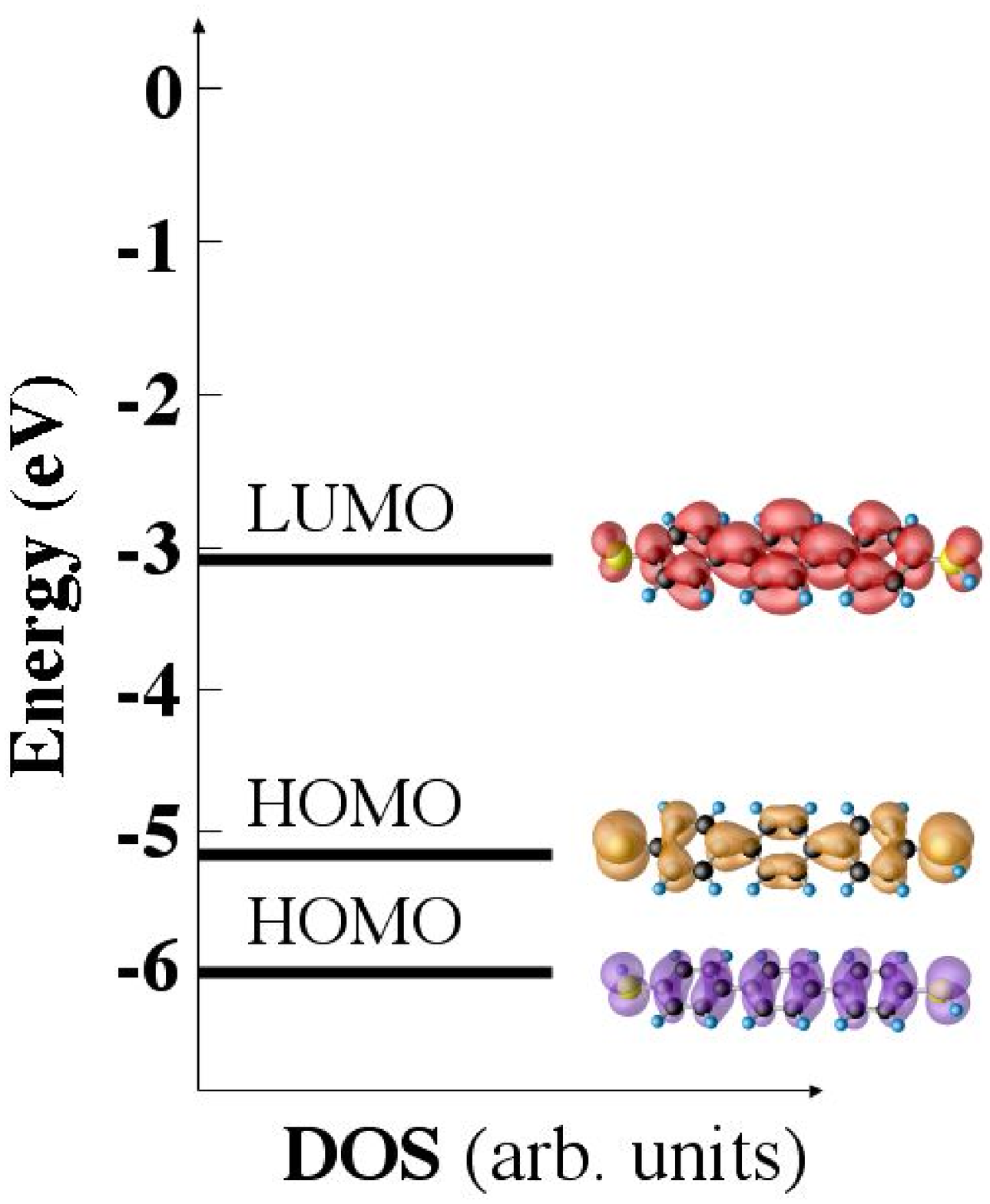}
\caption{1,4-[3]-phenyl (tricene) molecule:
DOS and charge density isosurface plots for the relevant molecular states of
the isolated molecule. $E_\mathrm{F}$ denotes the position of the Fermi
level for such an isolated molecule.}\label{fig37}
\end{center}
\end{figure}

The two molecules present two fundamental differences. First the HOMO-LUMO gap 
(this is the DFT-LDA gap, which might be substantially smaller than the actual gap) in
octane (5~eV) is about twice as big as the one of tricene (2.5~eV). Secondly in octane the charge
density of both the HOMO and the LUMO states has predominant amplitude around
the S atoms of the thiol group, with little density on the central carbon backbone. 
In contrast the HOMO and LUMO states of tricene are delocalized throughout the whole molecule 
with charge density concentrated both on the end-groups and on the central phenyl groups. 
This leads us to expect that the octane and the tricene will form respectively TMR and GMR 
devices, as actually found in our calculations \cite{smeagol2}. 

In the case of the 1,4-[n]-phenyl-dithiolate we find that the current does 
not scale sensibly with the number of phenyl groups in the molecule and the transport is 
essentially resonant through a molecular state. This picture is enforced by the fact that the
zero-bias transmission coefficient approach unity for energies close to the leads Fermi 
level. In addition from the study of the evolution of the orbital resolved density of states 
as a function of the distance between the thiol group and the electrodes we identify such a 
resonant state as the HOMO state of tricene. However it is worth mention that this appears 
rather broad and spin-splitted, because of the strong coupling with the $d$ orbitals of the 
leads \cite{smeagol2}. Therefore the Ni/tricene/Ni spin-valve behaves as an all-metal
spin-valve.

In contrast the zero-bias transmission coefficient of the Ni/octane/Ni junction presents
a sharp peak at $E_\mathrm{F}$ that scales exponentially with the number of alkane groups
$T\propto\mathrm{e}^{-\beta n}$ ($\beta\sim0.88$). This clearly demonstrates that the device is 
in a tunneling regime. It is interesting to note that such an exponent is similar to that found 
for the same molecules attached to gold (111) surfaces \cite{Sankey}. Also in this case the
coupling between the thiol groups and the electrodes is strongly, however because of
the highly localized nature of the HOMO and LUMO states, there in not a molecular
state extending through the entire structure close to the Fermi level. Hence the 
Ni/octane/Ni junction has the features of a TMR spin-valve.

In both cases the the spin-transport properties can be qualitatively
understood in terms of transport through a single molecular state (see figure \ref{fig38}).
Let $t^\uparrow(E)$ be the majority spin hopping integral from one
of the leads to the molecular state, and $t^\downarrow(E)$ the
same quantity for the minority spins. Then neglecting multiple
scattering from the contacts, the total transmission coefficients
of the entire spin-valve in the parallel state can be written
$T^{\uparrow\uparrow}(E)=(t^{\uparrow})^2$ and
$T^{\downarrow\downarrow}(E)=(t^{\downarrow})^2$ respectively for
the majority and minority spins. Similarly the transmission in the
anti-parallel configuration is
$T^{{\uparrow\downarrow}}(E)=T^{\downarrow\uparrow}(E)=
t^{\uparrow}t^{\downarrow}$. Thus $T^{{\uparrow\downarrow}}(E)$ is
 a convolution of the transmission coefficients for
the parallel case $T^{\uparrow\downarrow}\propto
\sqrt{{T^{\uparrow\uparrow}}T^{\downarrow\downarrow}}$.
\begin{figure}[ht]
\begin{center}
\includegraphics[width=14.0cm,clip=true,angle=0.0]{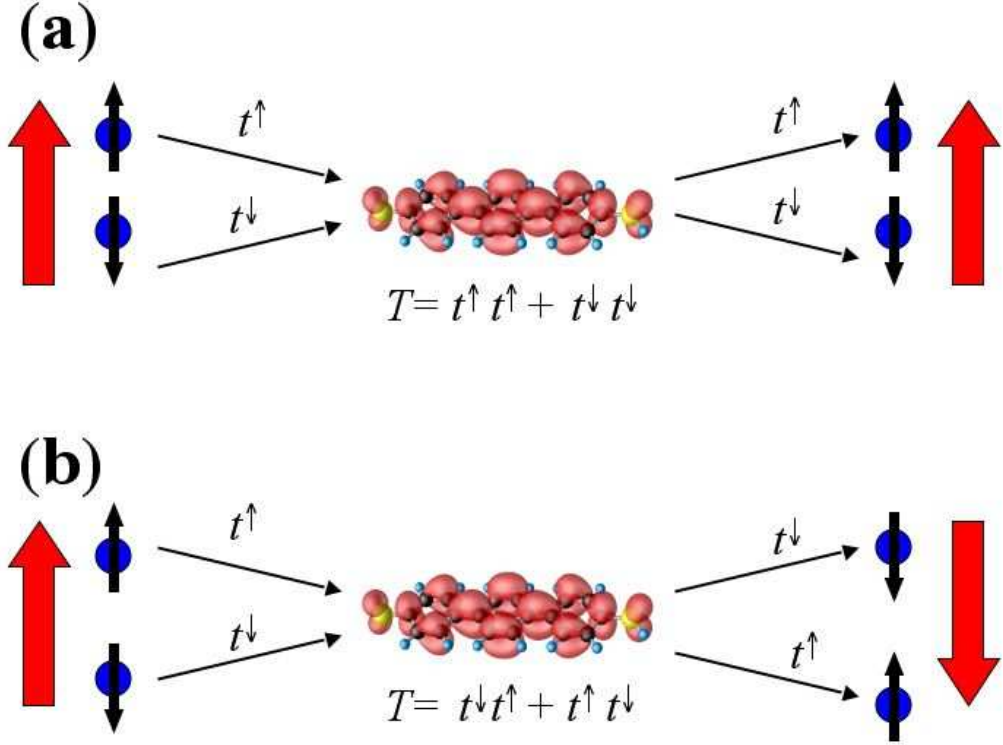}
\caption{Scheme of the spin-transport mechanism through a single molecular state. 
$t^\uparrow(E)$ ($t^\downarrow(E)$) is the majority (minority) spin hopping integral 
from one of the leads to the molecular state. Neglecting quantum interference, 
in the parallel case (a) the total transmission coefficient is simply 
$T=(t^{\uparrow})^2+(t^{\downarrow})^2$, while in the antiparallel (b) 
$T=2t^{\uparrow}t^{\downarrow}$. Note that if either $t^\uparrow$ or
$t^\downarrow$ vanishes, the current in the antiparallel configuration
will also vanish (infinite GMR).}\label{fig38}
\end{center}
\end{figure}

This qualitative argument gives us a tool for understanding the MR.
For the case in which only one spin couples to the molecular
state, the total transmission in the antiparallel case will be
identically zero since either $t^{\uparrow}$ or $t^{\downarrow}$
vanishes. This is the most desirable situation in real devices
since, in principle, an infinite $r_\mathrm{GMR}$ can be obtained.
Furthermore if for a particular energy window the transport is
through two distinct molecular states, which are respectively
coupled to the majority and minority spin only, then in this
window one will find $T^{\uparrow\uparrow}(E)\ne0$,
$T^{\downarrow\downarrow}(E)\ne0$ but
$T^{\uparrow\downarrow}(E)=0$.

The $I$-$V$ characteristics of both the molecules are strongly non-linear with the bias
and consequently also the GMR ratio suffers this non-linearity. In table \ref{Table7} I present a 
summary of our results, including the maximum and the minimum of the GMR ratio over the
bias range studied ($\pm2$~Volt) and the device resistance at 1~Volt for the parallel configuration
of the leads. These clearly demonstrate that molecular spin valves can yield large MR ratios. In the
case of TMR junction (octane) both the resistance and the TMR ratio are in the same range as 
in recent experiments on octane-based Ni spin-valves \cite{mol_TMR}. However
a direct comparison is difficult since the actual number of
molecules bridging the two electrodes is not known with precision, and a degradation of the
GMR signal due to spin-flip and electron-phonon scattering,
misalignment of the magnetization of the
contacts and current shortcut through highly conductive pin-holes,
can drastically reduce $r_\mathrm{GMR}$. 
\begin{table}[htbp]
\begin{center}
\begin{tabular}{cccc} \hline
\ Device \ & \  GMR$_\mathrm{max}$ (\%) \  & \ GMR$_\mathrm{min}$ (\%) \ &
 \ $R$ ($\Omega$) \ \\ \hline\hline
{\bf Ni/S-octane-S/Ni}      & 110   & 30 & 5.9$\times$10$^6$    \\ 
{\bf Ni/S-tricene-S/Ni}     & 660   & 51 & 8.2$\times$10$^4$     \\ 
{\bf Ni/Se-tricene-Se/Ni}   & 865   & 82 & 1.2$\times$10$^5$ \\ 
{\bf Ni/Te-tricene-Te/Ni}   & 870   & 124  & 2$\times$10$^5$  \\ \hline\hline
\end{tabular}
\caption{Summary table of the $I$-$V$ characteristics and GMR ratios for triacene
and octane based spin-valves. GMR$_\mathrm{max}$ (GMR$_\mathrm{min}$) is
the maximum (minimum) GMR ratio over the region of bias investigated and
$R (\Omega)$ is the resistance at 1~Volt for the parallel configuration of the 
leads.}
\label{Table7}
\end{center} 
\end{table}

In contrast in the case of metallic (tricene) junctions the GMR ratio can reach very high 
values exceeding a few hundreds percent. In addition in the table I have reported results
for the same triacene molecule, where now the anchoring groups are Se and Te atoms.
Since S, Se and Te belong to the same chemical group in the periodic table they present
a similar chemistry, and the main difference is simply a larger separation between such a 
group and the Ni contacts. This in fact increases with the atomic number of the anchor.
Such an increase reduces the overlap between the molecule and the electrodes and enhances 
the device resistance. However since this change in resistance depends on the spin
direction, being more pronounced for the minority spins, we find a general increase of the
GMR maximum when going from S to Se to Te. These findings have the important consequence
that the GMR signal can be tuned by an appropriate choice of the anchoring chemistry.

The field of molecular- spin-electronics is certainly in its early infancy, although it has already 
demonstrated part of its potential. Clearly more investigations are needed both theoretical
and experimental, in particular addressing the problems of stability, efficiency and device
production yield. However the future looks very promising, in particular since novel 
phenomena such as molecular spin rectification or spin inversion can be demonstrated.

\subsection{Magnetic proximity}

I wish to close this section with a brief discussion of a recently discovered phenomenon,
which is only indirectly connected to spin-transport, but that may open a novel avenue to
spin-manipulation at the nanoscale. This is the magnetic proximity effect. The basic idea is 
quite simple: there is always some charge transfer at the contact between a conducting 
molecule and a metal associated with the alignment of their respective chemical potentials. 
In a ferromagnet, the densities of states for spin up and spin down electrons at the Fermi 
level are different. Therefore one expects some degree of spin transfer to accompany the charge 
transfer. Roughly speaking, the transferred electrons are more abundant for one of the two spin directions leading to a spin imbalance in the electron transfer that accounts for a contact-induced
magnetic moment on the molecule. In the limit of a half metal, where either the majority or 
minority electrons are absent at the Fermi level, the spin transfer is equal to the charge transfer. 

A first indirect evidence of this effect was found recently by the group of J.M.D.~Coey,
who investigated the magnetic state of a carbon meteorite, and concluded that some unaccounted 
magnetization could be attributed to a tiny moment of the carbon atoms induced 
by the proximity with magnetic impurities \cite{meteo}. Subsequently Ferreira and Sanvito
derived a close system of equations for the induced magnetic moment and for the energetic
of a carbon nanotube on a magnetic surface \cite{ss_ferreira}. The calculation was based
on a simple tight-binding model with a reasonable choice of parameters for the coupling between
the nanotube and a cobalt surface. The calculations revealed that induced magnetic moments 
of the order of 0.1~$\mu_\mathrm{B}$ per carbon atom in contact with the surface can be achieved
at room temperature. This paved the way for a more controlled set of experiments in order to
demonstrate the magnetic proximity effect.

Experimentally the problem is to detect such a small spin transfer against the huge background 
coming from the ferromagnetic substrate. One possible strategy is to chose a
smooth thin metal film as a substrate and look for the stray field around the nanotube. 
A uniformly magnetized thin film creates no stray field whatever its direction of magnetization. 
In contrast a magnetized nanotube will produce a stray field, which will be directly detectable.
This was the basic idea behind the experiments from C\'espedes et al. \cite{CNT_prox}, who
measured induced magnetic moments in excess of 0.1$\mu_\mathrm{B}$ per carbon atom in contact,
in good agreement with the theoretical prediction. The experiment consists in taking AFM and MFM
images of nanotubes on various surfaces. The difference between the topographic (AFM)
and the magnetic images (MFM) is a direct measure of the stray field coming from the nanotube
and therefore provides evidence for the induced magnetic moment. 

In figure \ref{fig39} and \ref{fig40} I show the typical AFM and MFM pictures for nanotubes 
respectively on non-magnetic and magnetic substrates. When the substrate is not magnetic
the MFM-AFM image does not show the presence of the tube, which indicates that this latter
does not produce any stray field. In contrast the same picture for a nanotube on magnetite 
(see figure \ref{fig40}) unambiguously shows contrast and demonstrates the presence 
of an induced magnetic moment with a clear magnetic domain pattern. 
\begin{figure}[ht]
\begin{center}
\includegraphics[width=9.5cm,clip=true,angle=0.0]{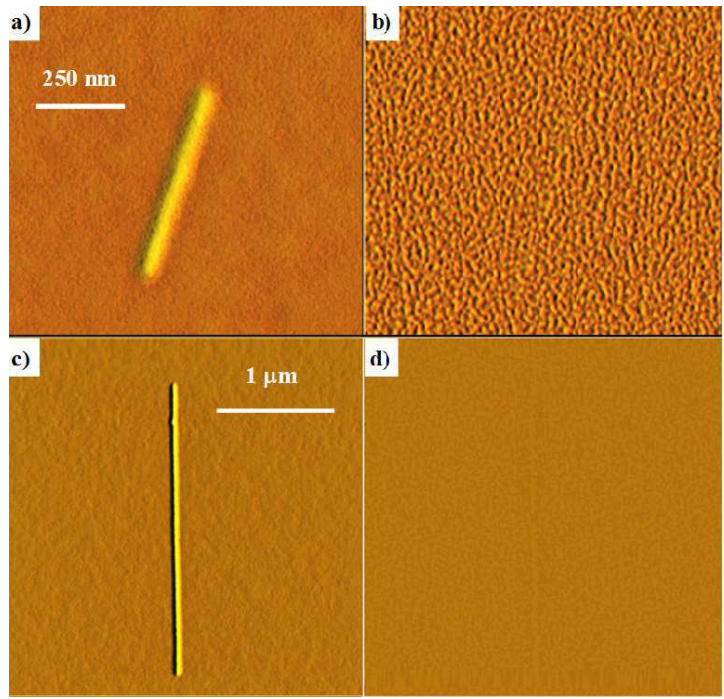}
\caption{AFM images of a carbon nanotube on copper a) and silicon c) substrate and their
corresponding MFM scans b) and d). The MFM images do not show any magnetic contrast
indicating no induced magnetic moment.}\label{fig39}
\end{center}
\end{figure}

In the another experiment, Mertins et al. \cite{Mertins} produced a multilayer of thin, alternating 
iron and carbon layers, of thickness 2.55 and 0.55 nm, respectively.  Then, they probed locally
the magnetic moment of the carbon by X-ray magneto-optical reflectivity of polarized synchrotron 
radiation. In this type of measurement the Fe and C absorption edges differ by about 500~eV
enabling one of establishing with precision whether the magnetic moment comes from C or not. 
With this method magnetic moment of the order of 0.05~$\mu_\mathrm{B}$ were found.
\begin{figure}[ht]
\begin{center}
\includegraphics[width=9.5cm,clip=true,angle=0.0]{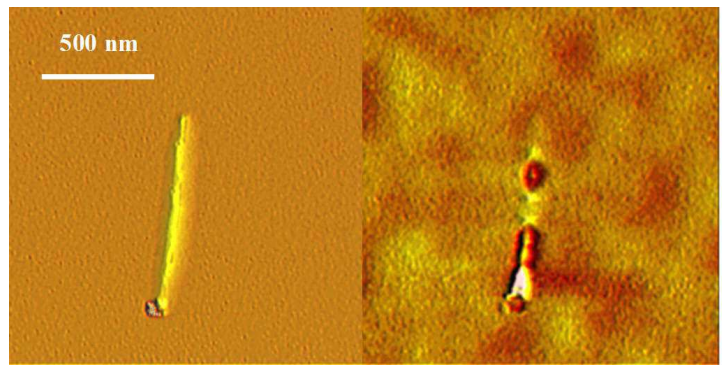}
\caption{AFM images of a carbon nanotube on cobalt substrate (left) and the
corresponding MFM scans (right). The MFM image shows magnetic contrast
indicating an induced magnetic moment.}\label{fig40}
\end{center}
\end{figure}

It is important to point out that the magnetic proximity effect does not imply intrinsic ferromagnetism
\cite{phyworld} and no spin aligning potential exists in carbon. This means that a magnetic moment 
is detected only when a second ferromagnetic material is present and when good contact is made. 
However this effect may have important applications in spintronics. For instance one can 
envision four terminal devices, where two non-magnetic contacts are used for charge injection
while two other magnetic electrodes are used for the spin-manipulation. These types of set-up
clearly add a novel dimension to spintronics and perhaps will help in overcoming the
spin-injection problem.

\setcounter{equation}{0}
\section{Conclusion}

A decade after the discovery of the GMR effect the future of spin-electronics in
nanoscale systems looks bright. This is mainly due to the improved understanding of
the spin-transport mechanisms and the better control over the device processing.
At the same time the possibility of conducting spin-transport measurements
in systems comprising a handful of atoms has opened completely new 
prospectives. We can envision in a near future new devices where the spin and 
molecular functionalities will be combined achieving a broad range of applications,
from biological sensors to tools for coherent quantum data processing.

From the theoretical side the last decade has also witnessed a rapid evolution of
computational methods for both electronic structure and quantum transport. Several
numerical implementations are currently available. The most advanced of them are
based on {\it ab initio} schemes and therefore do not depend on parameters obtained
from experiments. These open the way to a physics ``without compromises'', where the
numerical predictions must reproduce the experimental data, if the systems under
investigation are the same. For this reason {\it ab initio} transport schemes have
became an invaluable tools.

In this review I have discussed the two main philosophies behind quantum transport, 
namely the scattering theory and the non-equilibrium Green function method. Then
I have overviewed the numerical tools available to date and I have reviewed the main
successes of quantitative transport theory applied to spin problems. These include
the GMR and TMR effects, spin transport through atomic point contacts and through organic 
molecules. I deliberately have not addressed problems connected with inelastic effects
both of mechanical and magnetic origin, and with the need for more accurate and
flexible electronic structure methods. Here the field is rather vast and in continuous
evolution and it deserves a review on its own. Certainly the future of modeling
spin transport at the nanoscale is dawning.

\setcounter{equation}{0}
\section{Acknowledgment}

This work is supported by Science Foundation of Ireland under 
the grant SFI02/IN1/I175. It is my privilege to thank Alexander 
Reily Rocha, Cormac Toher, Maria Tsoneva and Nicola Jean
for careful reading the manuscript and Colin Lambert, Jaime Ferrer, 
Michael Coey, Nicola Spaldin, John Jefferson, Alessio Filippetti, 
Tchadvar Todorov and Kieron Burke for useful discussions.

\setcounter{equation}{0}
\setcounter{section}{0}

\renewcommand{\thesection}{\Alph{section}}
\renewcommand{\theequation}{\Alph{section}.\arabic{equation}}

\setcounter{equation}{0}
\section{Retarded Green function for a double infinite system}
\label{AppB}

In this appendix I present the explicit calculation leading to the 
equation (\ref{eq2.27}) for the Green's function of a double infinite system. The
starting point is the equation (\ref{eq2.26})
\begin{equation}
g_{zz^\prime}=\left\{
\begin{array}[4]{r}
\sum_{m=1}^M{\phi}^{m}e^{ik_m(z-z^\prime)}
{\mathsf w}^{m\dag}\;\;\;\;\;z\geq{z}^\prime \\
\\
\sum_{m=1}^M{\phi}^{\bar{m}}e^{i\bar{k}_m(z-z^\prime)}
{\mathsf w}_{\bar{m}\dag}
\;\;\;\;\;z\leq{z}^\prime \\
\end{array}
\right. \; ,
\label{eqA.1}
\end{equation}
with ${\mathsf w}^{m}$ and ${\mathsf w}^{\bar{m}}$ two vectors to be determined. 
The expression of equation (\ref{eqA.1}) to be a Green function must be continuous 
for $z=z^\prime$ and must satisfy the Green equation
\begin{equation}
[(E-H)g]_{zz^\prime}=\delta_{zz^\prime} \; . 
\label{eqA.2}
\end{equation}
The first condition yields immediately to the relation
\begin{equation}
\sum_{m=1}^M{\phi}^{m}{\mathsf w}^{m\dag}=
\sum_{m=1}^M{\phi}^{\bar{m}}{\mathsf w}^{\bar{m}\dag} \; ,
\label{eqA.3}
\end{equation}
while the second gives
\begin{equation}
\sum_{m=1}^M\left[(E-H_0){\phi}^{m}{\mathsf w}^{m\dag}+
H_1{\phi}^{m}e^{ik_m}{\mathsf w}^{m\dag}+
H_{-1}{\phi}^{\bar{m}}e^{-i\bar{k}_m}{\mathsf w}^{\bar{m}\dag}
\right]=1
\; .
\label{eqA.4}
\end{equation}
The task is now to re-write the vectors ${\mathsf w}$'s as a function of the known
vectors $\phi$'s and their dual $\tilde\phi$'s. 
First note that by adding and subtracting to the (\ref{eqA.4}) the expression
\begin{equation}
{\cal W}=\sum_{m=1}^M\left[
H_{-1}{\phi}^{m}e^{-i{k}_m}{\mathsf w}^{m\dag}
\right]
\; ,
\label{eqA.5}
\end{equation}
it is possible to re-write the (\ref{eqA.4}) in the following compact form
\begin{equation}
\sum_{m=1}^MH_{-1}\left[
{\phi}^{m}e^{-ik_m}{\mathsf w}^{m\dag}-
{\phi}^{\bar{m}}e^{-i\bar{k}_m}{\mathsf w}^{\bar{m}\dag}
\right]=-1
\; ,
\label{eqA.6}
\end{equation}
where the definition of ${\phi}^{k}$ of equation (\ref{eq2.21}) has been used.
Now consider the continuity equation (\ref{eqA.3}) and multiply 
the left-hand side by 
the dual vector ${\tilde{\phi}}^{n}$ and 
the right-hand side by ${\tilde{\phi}}^{{\bar n}}$. 
It yields respectively to two expressions that
relate ${\mathsf w}^{{k}}$ to ${\mathsf w}^{\bar{k}}$
\begin{equation}
\sum_{m=1}^M\left[
{\tilde{\phi}}^{n\dag}{\phi}^{\bar{m}}{\mathsf w}^{\bar{m\dag}}
\right]={\mathsf w}^{n\dag}
\; ,
\label{eqA.7}
\end{equation}
\begin{equation}
\sum_{m=1}^M\left[
{\tilde{\phi}}^{\bar{n}\dag}{\phi}^{{m}}{\mathsf w}^{m\dag}
\right]={\mathsf w}^{\bar{n}\dag}
\; .
\label{eqA.8}
\end{equation}
If now one substitutes the equations (\ref{eqA.7}) and (\ref{eqA.8}) into the
equation (\ref{eqA.6}) and uses the continuity equation (\ref{eqA.3}), the 
following fundamental relation is obtained
\begin{equation}
\sum_{m=1}^M\sum_{n=1}^M H_{-1}\left[
{{\phi}}_{m} e^{-i{k}_m}{\tilde{\phi}}^{m\dag}-
{{\phi}}^{\bar{m}} e^{-i\bar{k}_m}{\tilde{\phi}}^{\bar{m}\dag}
\right]{{\phi}}^{\bar{n}}{\mathsf w}^{\bar{n}\dag}=1
\; ,
\label{eqA.9}
\end{equation}
from which it follows immediately
\begin{equation}
\left[ \sum_{m=1}^M H_{-1}\left(
{\phi}^m e^{-ik_m}{\tilde{\phi}}^{m\dag}-
{{\phi}}^{\bar{m}} e^{-i\bar{k}_m}{\tilde{\phi}}^{\bar{m}\dag}
\right) \right] ^{-1}=
\sum_{n=1}^M {\phi}^{\bar{n}}{\mathsf w}^{\bar{n\dag}}=
\sum_{n=1}^M {{\phi}}^{n}{\mathsf w}^{n\dag}
\; .
\label{eqA.10}
\end{equation}
In the second equality of the equation (\ref{eqA.10}) I have used the continuity
equation (\ref{eqA.3}). Note that the equation (\ref{eqA.10}) expresses
explicitly the vectors ${\mathsf w}$'s in term of the known quantities $\phi$,
$\tilde{\phi}$ and $H_{-1}$. Therefore ${\mathsf w}$'s may be computed by simply
multiplying the (\ref{eqA.10}) for the correct dual vectors. By doing so I obtain
\begin{equation}
{\mathsf w}^{\bar{n}\dag}={\tilde{\phi}}^{\bar{n}\dag} {\cal V}^{-1}
\; ,
\label{eqA.11}
\end{equation}

\begin{equation}
{\mathsf w}^{n\dag}={\tilde{\phi}}^{n^\dag} {\cal V}^{-1}
\; ,
\label{eqA.12}
\end{equation}
with ${\cal V}$ the operator defined in section \ref{gsgf}
\begin{equation}
{\cal{V}}=\sum^{M}_{m=1}H_{-1}\left[
{\phi}^{m}e^{-ik_m}\tilde{{\phi}}^{m\dag}-
{\phi}^{\bar{m}}e^{-i\bar{k}_m}\tilde{{\phi}}^{\bar{m}\dag}
\right]\;{.}
\label{eqA.13}
\end{equation}
The equation (\ref{eqA.13}) concludes the demonstration. In fact by substituting
the expressions for ${\mathsf w}^{\bar{k}}$ and ${\mathsf w}^{k}$ into the
starting Ansatz (\ref{eqA.1}), one obtains the final expression for the
Green function of an infinite system
\begin{equation}
g_{zz^\prime}=\left\{
\begin{array}[4]{r}
\sum_{m=1}^M{\phi}_{m}e^{ik_m(z-z^\prime)}
\tilde{{\phi}}^{m\dag}{\cal{V}}^{-1}\;\;\;\;\;z\geq{z}^\prime \\
\\
\sum_{m=1}^M{\phi}_{\bar{m}}e^{i\bar{k}_m(z-z^\prime)}
\tilde{{\phi}}^{\bar{m}\dag}{\cal{V}}^{-1}
\;\;\;\;\;z\leq{z}^\prime \\
\end{array}
\right.\;{.}
\label{eqA.14}
\end{equation}

\setcounter{equation}{0}
\section{General current operator and rotation in the degenerate space}
\label{AppC}

I this appendix I will generalized the current operator introduced in section
\ref{Cur_orb_cur} and discuss the rotation needed to diagonalize it in the case in which the vectors 
$\phi^k$'s are degenerate.
Let us start by considering the current matrix at the position $z$. It can be
easily expressed as the time derivative of the density matrix at the same point
$z$
\begin{equation}
{\cal J}_z(t)=\frac{\partial}{\partial t}\psi_z(t)\psi_z(t)^\dagger
\; ,
\label{eqB.1}
\end{equation}
where $\psi_z(t)$ are the coefficients of the time-dependent wave-function at the 
position $z$ satisfying the time-dependent Schr\"odinger equation 
\begin{equation}
i\frac{\partial}{\partial t}\psi_z(t)=H_0 \psi_z(t)+ 
H_1 \psi_{z+1}(t)+ H_{-1}\psi_{z-1}(t)
\; .
\label{eqB.2}
\end{equation}
By explicitly evaluating the time derivative in equation (\ref{eqB.1}), and by
using the Schr\"odinger equation and its complex conjugate, the current matrix
can be written in the following transparent form
\begin{equation}
{\cal J}_z={\cal J}_{0}+{\cal J}_{z-1\rightarrow{z}}
+{\cal J}_{z+1\rightarrow{z}}
\;{,}
\label{eqB.3}
\end{equation}
where ${\cal J}_{0}$, ${\cal J}_{z-1\rightarrow{z}}$ and 
${\cal J}_{z+1\rightarrow{z}}$ are defined respectively as
\begin{equation}
{\cal J}_{0}=-i\left[H_0\psi_z\psi_z^\dagger-\psi_z\psi_z^\dagger H_0
\right]
\;{,}
\label{eqB.4}
\end{equation}
\begin{equation}
{\cal J}_{z+1\rightarrow{z}}=-i\left[H_{1}\psi_{z+1}\psi_z^\dagger-\psi_z
\psi_{z+1}^\dagger H_{-1}\right]
\;{.}
\label{eqB.5}
\end{equation}
\begin{equation}
{\cal J}_{z-1\rightarrow{z}}=-i\left[H_{-1}\psi_{z-1}\psi_{z}^\dagger-\psi_{z}
\psi_{z-1}^\dagger H_1\right]
\;{,}
\label{eqB.6}
\end{equation}
In the calculation of the relations above I simplified the time-dependent
component of the wave-function and expressed the current matrix by mean of the
column vectors introduced in section \ref{gsgf} through the time-independent Schr\"odinger
equation (\ref{eq2.19}). Note that ${\cal J}_0$ does depend only on the value of the
wave-function at the position $z$, while ${\cal J}_{z-1\rightarrow{z}}$ and 
${\cal J}_{z+1\rightarrow{z}}$ depend also respectively on its value at the position
$z-1$ and $z+1$. Note also that the expressions (\ref{eqB.5}) and (\ref{eqB.6}) are formally
identical to their one-dimensional counterparts (\ref{occ_lr}) and (\ref{occ_rl})
once the on-site energy $\epsilon_0$ and the hopping integral $\gamma_0$ are replaced
respectively by the intra- and inter-slice matrices $H_0$ and $H_1$.

Now evaluate the expectation value of the
current by taking the trace of the current matrix. It is easy to show that
\begin{equation}
J_z={\mathrm Tr}\;{\cal J}_z=J_{z-1\rightarrow{z}}+J_{z+1\rightarrow{z}}
\; ,
\label{eqB.7}
\end{equation}
with
\begin{equation}
J_{z+1\rightarrow{z}}=-i\left[\psi_z^\dagger H_{1}\psi_{z+1}-
\psi_{z+1}^\dagger H_{-1}\psi_z\right]
=-2\Im\left({\psi_{z+1}^\dagger H_{-1}\psi_z}\right)
\;{.}
\label{eqB.8}
\end{equation}
\begin{equation}
J_{z-1\rightarrow{z}}=-i\left[\psi_z^\dagger H_{-1}\psi_{z-1}-
\psi_{z-1}^\dagger H_{1}\psi_z
\right]
=2\Im\left({\psi_z^\dagger H_{-1}\psi_{z-1}}\right)
\;{,}
\label{eqB.9}
\end{equation}
In the calculation of the equations (\ref{eqB.8}) and (\ref{eqB.9}) I used the circular
property of the trace ${\mathrm Tr}\;AB={\mathrm Tr}\;BA$. Note that the
expectation value of ${\cal J}_0$ vanishes. As in the one dimensional case 
the relations (\ref{eqB.8}) and (\ref{eqB.9}) have a clear interpretation. 
${\cal J}_{z+1\rightarrow{z}}$ and ${\cal J}_{z-1\rightarrow{z}}$ represent the current 
matrices for electrons propagating respectively to the left (left-moving) and to the right 
(right-moving).
Note that in the case of a translational-invariant system $\psi_z$ can be
written in the Bloch form of equation (\ref{eq2.20})
\begin{equation}
\psi_z=n_{k}^{1/2}e^{ik{z}}\phi^{k}{\;}.
\label{eqB.10}
\end{equation}
If now one substitutes the (\ref{eqB.10}) into the (\ref{eqB.8}-\ref{eqB.9}) it
is easy to show that $J_z=0$ as expected from the translational invariance.
Moreover note again that local current conservation is achieved at the length
scale of a single slice, i.e. by taking the trace of the matrices ${\cal J}$. This is
equivalent to sum up all the orbital currents of the cell.

The final part of this appendix is dedicated to show that the states of the form
$\psi_z=n_{k}^{1/2}e^{ik{z}}\phi^{k}$ diagonalize the current. 
As anticipated in section \ref{gsgf} this is not strictly valid in the case of different $\phi^{k}$ 
corresponding to the same $k$. Nevertheless in such a case I will show that there is always
a rotation in the degenerate space that diagonalizes the current.

Consider for instance the right-moving current (all the following arguments can
be applied to the left-moving counterpart), and a Bloch state
\begin{equation}
\psi_z=\sum_m \alpha_m{e^{ik_mz}}{\phi}^{m}
\;{,}
\label{eqB.11}
\end{equation}
and calculate the expectation value of the current for such a state. 
It easy to show that this yields to the equation
\begin{equation}
{J}_{z-1\rightarrow{z}}=-i\sum_{m,n}\alpha_m\alpha_{n}^\star
\left[\phi^{n\dagger} H_{-1}\phi^{m} e^{-ik_m}-
\phi^{n^\dagger} H_{1}\phi^{m} e^{ik_{n}}
\right]=-i\sum_{m,n}\alpha_m\alpha_{n}^\star
\left[\phi^{n\dagger}(H_{-1}e^{-ik_m}-H_{1}e^{ik_{n}})
\phi_{n}\right]
\;{.}
\label{eqB.12}
\end{equation}
If one now assumes that the off-diagonal matrix elements vanish ($\phi_k$'s
diagonalize the current), then the states (\ref{eqB.11}) diagonalize the current and
curry unitary flux if the normalization constant is taken to be
\begin{equation}
\frac{1}{\nu_m^{1/2}}=\alpha_m=\frac{1}
{[-i\phi^{m\dagger}(H_{-1}e^{-ik_m}-H_1e^{ik_m})\phi^{m}]^{1/2}}
\;{,}
\label{eqB.13}
\end{equation}
with $\nu_m$ defining the group velocity.
Note that the states (\ref{eqB.11}) with the normalization constant
(\ref{eqB.13}) are the ones introduced in section \ref{gsgf}, which guarantee the
unitarity of the ${\cal S}$ matrix. 

The final step is to demonstrate that $\psi_z={e^{ikz}}{\phi}^{k}$ diagonalizes
the current. To achieve this, consider the Schr\"odinger equation evaluated on
such a Bloch state and its complex conjugate
\begin{equation}
\left(E-H_0\right){\phi}^{m}=
\left(H_1e^{ik_m}+H_{-1}e^{-ik_m}\right){\phi}^{m}
{\;},
\label{eqB.14}
\end{equation}
\begin{equation}
{\phi}^{m\dagger}\left(E-H_0\right)=
{\phi}^{m\dagger}\left(H_1e^{ik_m}+H_{-1}e^{-ik_m}\right)
{\;}.
\label{eqB.15}
\end{equation}
By multiplying the first equation by ${\phi}^{n\dagger}$ to the left and 
the second by ${\phi}^{n}$ to the right one obtains the relation
\begin{equation}
\phi^{n\dagger} H_{-1}\phi^{m}
e^{-ik_{m}}+\phi^{n\dagger} H_{1}\phi^{m}e^{ik_{m}}=
\phi^{n\dagger} H_{-1}\phi_{m}
e^{-ik_{n}}+\phi^{n\dagger} H_{1}\phi_{m}
e^{ik_{n}}
\;{.}
\label{eqB.16}
\end{equation}
The (\ref{eqB.16}) is identically satisfied if $k_{m}=k_{n}$,
also if $\phi^{m}\neq\phi^{n}$. This occurs
when one or more wave-vectors $k_{m}$ are degenerate 
(ie there are many $\phi^{m}$'s for the same $k_{m}$).
In the case this does not happen few algebraic manipulations yield to the relation
\begin{equation}
\phi^{n\dagger} H_{-1}\phi^{m}
e^{-ik_{n}}
=\phi^{n\dagger} H_{1}\phi^{m}e^{ik_{m}}
\;{.}
\label{eqB.17}
\end{equation}
The last equality shows the cancellation of the off-diagonal terms in the 
expression of the expectation value of the right-going current 
(\ref{eqB.12}). This means that in the case in which there is no degeneracy in
$k$, the states ${\phi}^{m}$ diagonalize the current. Nevertheless in
the case in which degeneracy is present one can perform a unitary rotation in
the degenerate space and construct a new basis $\varphi$ in which the current is
diagonal. To show explicitly how to obtain this rotation consider a set of
vectors ${\phi}_\mu^{k}\;$ $(\mu=1,...,N)\;$ corresponding to the same wave 
vector $k$ and construct the ``reduced'' $N\times N$ 
current matrix, whose matrix elements are
\begin{equation}
(J^{\mathrm R})_{\mu\nu}={\phi^{k}_{\mu}}^\dagger [i(H_{-1}e^{-ik}-H_1e^{ik})]
\phi^{k}_\nu
\;{.}
\label{eqB.18}
\end{equation}
Since $H_{-1}=H_{1}$ the ``reduced current'' is an hermitian matrix. Therefore
it has always a diagonal form. Moreover the transformation matrix $\cal U$ 
which diagonalizes $J^{\mathrm R}$, is a unitary matrix. If $\cal D$ is the 
diagonal form of $J^{\mathrm R}$ I can write such a unitary transformation as
\begin{equation}
{\cal D}={\cal U}^{-1}{J^{\mathrm R}}\:{\cal U}
\;{.}
\label{eqB.19}
\end{equation}
It follows that the transformation $\cal U$ also transforms the basis 
${\phi}_\mu^{k}$
into a ``rotated'' basis $\varphi_\mu^{k}$ which diagonalize the current (note 
that the ``reduced current'' is simply the total current introduced above 
calculated onto a subset of the total Hilbert Space). By evaluating
the equation (\ref{eqB.19}) the explicit definition of $\varphi_\mu^{k}$ is
obtained
\begin{equation}
\varphi_\mu^{k}=\sum_{\nu=1}^N \varphi_\nu^{k} {\cal U}_{\nu\mu}
\;{.}
\label{eqB.20}
\end{equation}
The equation (\ref{eqB.20}) completes the demonstration.

\setcounter{equation}{0}
\section{Green function-Wave function projector}
\label{AppD}

In this appendix I will show that the projector that maps the RGF
for the double infinite system on the corresponding wave-function
projects also the total RGF (system = scatterer + leads) on
the corresponding total-wave function. 
Consider the total Hamiltonian
\begin{equation}
H=H_{\mathrm{lead}}+H_{\mathrm{scat}}
\; ,
\label{eqC.1}
\end{equation}
where $H_{\mathrm{lead}}$ describes the leads and $H_{\mathrm{scat}}$ describes the scattering
region. The Schr\"odinger equation and the Green equation for the
leads (without any scattering region) are respectively
\begin{equation}
(E-H_{\mathrm{lead}})\psi_{\mathrm{lead}}=0
\; ,
\label{eqC.2}
\end{equation}
\begin{equation}
(E-H_{\mathrm{lead}})g_{\mathrm{lead}}={\cal I}
\; ,
\label{eqC.3}
\end{equation}
with $g_{\mathrm{lead}}$ the Green function, $E$ the energy and ${\cal I}$ the identity 
matrix. The corresponding equations for the whole system (scatterer + leads) 
are
\begin{equation}
[E-(H_{\mathrm{lead}}+H_{\mathrm{scat}})]\psi=0
\; ,
\label{eqC.4}
\end{equation}
\begin{equation}
[E-(H_{\mathrm{lead}}+H_{\mathrm{scat}})]G={\cal I}
\; .
\label{eqC.5}
\end{equation}
Furthermore $\psi$, $\psi_{\mathrm{lead}}$ and $g_{\mathrm{lead}}$, $G$ are related by the 
respective Dyson's equations
\begin{equation}
\psi=({\cal I}-g_{\mathrm{lead}}H_{\mathrm{scat}})^{-1}\psi_{\mathrm{lead}}
\; ,
\label{eqC.6}
\end{equation}
\begin{equation}
G=({\cal I}-g_{\mathrm{lead}}H_{\mathrm{scat}})^{-1}g_{\mathrm{lead}}
\; .
\label{eqC.7}
\end{equation}
Define now the projector $P$ in such a way that
\begin{equation}
\psi_{\mathrm{lead}}=g_{\mathrm{lead}}\cdot P
\; .
\label{eqC.8}
\end{equation}
If one now uses the Dyson equation for $\psi$ (\ref{eqC.6}) together with the
definition of $P$, it follows immediately
\begin{equation}
\psi=({\cal I}-g_{\mathrm{lead}}H_{\mathrm{scat}})^{-1}g_{\mathrm{lead}}\cdot P=G\cdot P
\; ,
\label{eqC.9}
\end{equation}
where I used the Dyson equation for $G$ (\ref{eqC.7}). The equation (\ref{eqC.9})
completes the demonstration and shows that $P$ also maps $G$ onto $\psi$.


\begin{thebibliography}{100}

\bibitem{Simonds}
J.~L. Simonds,
\newblock Physics Today {\bf 48}, 26 (1995).

\bibitem{baibich_gmr}
M.~N. Baibich et~al.,
\newblock Phys. Rev. Lett. {\bf 61}, 2472 (1988).

\bibitem{binasch_gmr}
G.~Binasch, P.~Gr\"unberg, F.~Saurenbach, and W.~Zinn,
\newblock Phys. Rev. B {\bf 39}, 4828 (1989).

\bibitem{exchange}
P.~Bruno,
\newblock Phys. Rev. B {\bf 52}, 411 (1995).

\bibitem{DMS1}
H.~Ohno,
\newblock Science {\bf 281}, 951 (1998).

\bibitem{DMS2}
H.~Ohno,
\newblock J. Magn. Magn. Mater {\bf 200}, 110 (1999).

\bibitem{Aws1}
J.~M. Kikkawa and D.~D. Awschalom,
\newblock Phys. Rev. Lett. {\bf 80}, 4113 (1998).

\bibitem{Aws2}
J.~M. Kikkawa and D.~D. Awschalom,
\newblock Nature {\bf 397}, 139 (1999).

\bibitem{Aws3}
I.~Malajovich, J.~M. Kikkawa, D.~D. Awschalom, J.~J. Berry, and N.~Samarth,
\newblock Phys. Rev. Lett. {\bf 84}, 1015 (2000).

\bibitem{Aws4}
M.~Poggio et~al.,
\newblock Phys. Rev. Lett. {\bf 91}, 207602 (207602).

\bibitem{spintronics_wolf}
S.~A. Wolf et~al.,
\newblock Science {\bf 294}, 1488 (2001).

\bibitem{spintronics_prinz1}
G.~Prinz,
\newblock Science {\bf 282}, 1660 (1998).

\bibitem{spintronics_prinz2}
G.~Prinz,
\newblock Phys. Today {\bf 48}, 58 (1995).

\bibitem{spin_computer}
D.~P.~D. Vincenzo,
\newblock Science {\bf 270}, 255 (1995).

\bibitem{moltronics}
C.~Joachim, J.~K. Gimzewski, and A.~Aviram,
\newblock Nature (London) {\bf 408}, 541 (2000).

\bibitem{mol_trans}
S.~J. Tans, A.~R.~M. Verschueren, and C.~Dekker,
\newblock Nature {\bf 393}, 49 (1998).

\bibitem{mol_ndr}
J.~Chen, M.~A. Reed, A.~M. Rawlett, and J.~M. Tour,
\newblock Science {\bf 286}, 1550 (1999).

\bibitem{mol_rec}
Z.~Yao, H.~W.~C. Postman, L.~Balents, and C.~Dekker,
\newblock Nature {\bf 402}, 273 (1999).

\bibitem{CNT_GMR}
K.~Tsukagoshi, B.~W. Alphenaar, and H.~Ago,
\newblock Nature {\bf 401}, 572 (1999).

\bibitem{CNT_prox}
O.~C\'espedes, M.~S. Ferreira, S.~Sanvito, M.~Kociak, and J.~M.~D. Coey,
\newblock J.Phys.: Condens Matter {\bf 16}, L155 (2004).

\bibitem{mol_GMR}
Z.~H. Xiong, D.~Wu, Z.~V. Vardeny, and J.~Shi,
\newblock Nature {\bf 427}, 821 (2004).

\bibitem{BMR1}
N.~Garc\'{\i}a, M.~Mu{\~n}oz, and Y.~W. Zhao,
\newblock Phys. Rev. Lett. {\bf 82}, 2923 (1999).

\bibitem{BMR2}
S.~Z. Hua and H.~D. Chopra,
\newblock Phys. Rev. B, Rapid Communications {\bf 67}, 060401 (2003).

\bibitem{BMR3}
J.~J. Versluijs, M.~A. Bari, and J.~M.~D. Coey,
\newblock Phys. Rev. Lett. {\bf 87}, 026601 (2001).

\bibitem{GMR_poly}
V.~Dediu, M.~Murgia, F.~C. Matacotta, C.~Taliani, and S.~Barbanera,
\newblock Solid State Communication {\bf 122}, 181 (2002).

\bibitem{Aws5}
M.~Ouyang and D.~D. Awschalom,
\newblock Science {\bf 301}, 1074 (2003).

\bibitem{valet_fert}
T.~Valet and A.~Fert,
\newblock Phys. Rev. B {\bf 48}, 7099 (1993).

\bibitem{thomasfermi1}
L.~H. Thomas,
\newblock Proc. Cambridge Phil. Soc. {\bf 23}, 542 (1927).

\bibitem{thomasfermi2}
E.~Fermi,
\newblock Z. Physik {\bf 48}, 73 (1928).

\bibitem{ashcroft}
N.~D. Ashcroft and N.~D. Mermin,
\newblock {\em Solid State Physics},
\newblock Int. Thompson Pub., 1976.

\bibitem{pettifor}
D.~G. Pettifor,
\newblock {\em Bonding and structure of molecules and solids},
\newblock Oxford University Press, Oxford, UK, 2002.

\bibitem{RKKY1}
M.~A. Ruderman and C.~Kittel,
\newblock Phys. Rev. {\bf 96}, 99 (1954).

\bibitem{RKKY2}
T.~Kasuya,
\newblock Prog. Theor. Phys. {\bf 16}, 58 (1956).

\bibitem{RKKY3}
K.~Yoshida,
\newblock Phys. Rev. {\bf 106}, 893 (1957).

\bibitem{nicola}
N.~A. Spaldin,
\newblock {\em Magnetic Materials: Fundamentals and device applications},
\newblock Cambridge University Press, Cambridge, UK, 2003.

\bibitem{brunodw}
P.~Bruno,
\newblock Phys. Rev. Lett. {\bf 83}, 2425 (1999).

\bibitem{kramer_loc}
B.~Kramer and A.~MacKinnon,
\newblock Rep. Prog. Phys. {\bf 56}, 1469 (1993).

\bibitem{in_mfp}
D.~Bozec et~al.,
\newblock Phys. Rev. Lett. {\bf 85}, 1314 (2000).

\bibitem{2current_mott}
N.~Mott,
\newblock Proc. Roy. Soc. A {\bf 153}, 699 (1936).

\bibitem{sdl_Co}
W.~C. Chiang, Q.~Yang, W.~P.~P. Jr., R.~Loloee, and J.~Bass,
\newblock J. Appl. Phys. {\bf 81}, 4570 (1997).

\bibitem{sdl_py}
S.~Steenwyk, S.~Y. Hsu, R.~Loloee, J.~Bass, and W.~P.~P. Jr.,
\newblock J. Mag. Mag. Mater. {\bf 170}, L1 (1997).

\bibitem{pauljim}
P.~Delaney and J.~Greer,
\newblock Phys. Rev. Lett. {\bf 93}, 036805 (2004).

\bibitem{HF_Ratner}
S.~N. Yalirakia, A.~E. Roitberg, C.~Gonzalez, V.~Mujica, and M.~A. Ratner,
\newblock J. Chem. Phys. {\bf 111}, 6997 (1999).

\bibitem{HKohn}
H.~Hohenberg and W.~Kohn,
\newblock Phys. Rev. {\bf 136}, B864 (1964).

\bibitem{KSham}
W.~Kohn and L.~Sham,
\newblock Phys. Rev. {\bf 140}, A1133 (1965).

\bibitem{pabloON}
P.~Ordej\'on,
\newblock Comp. Mat. Sci. {\bf 12}, 157 (1998).

\bibitem{TDDFT_Gross}
M.~A.~L. Marques and E.~K.~U. Gross,
\newblock {\em Time-Dependent Density Functional Theory},
\newblock in Lecture Notes in Physics: Vol 620, Springer-Verlag, Heidelberg,
  2003.

\bibitem{Runge_Gross}
E.~Runge and E.~K.~U. Gross,
\newblock Phys. Rev. Lett. {\bf 52}, 997 (1984).

\bibitem{baer}
R.~Baer, T.~Seideman, S.~Ilani, and D.~Neuhauser,
\newblock J. Chem. Phys. {\bf 120}, 3387 (2004).

\bibitem{burke_car}
R.~Gebauer and R.~Car,
\newblock Phys. Rev. B {\bf 70}, 125324 (2004).

\bibitem{sutton}
A.~P. Sutton,
\newblock {\em Electronic Structure of Materials},
\newblock Oxford University Press, Oxford, UK, 1996.

\bibitem{tbdft}
M.~Elstner et~al.,
\newblock Phys. Rev. B {\bf 58}, 7260 (1998).

\bibitem{Alex_tb}
A.~P. Horsfield and A.~M. Bratkovsky,
\newblock J. Phys.: Condens. Matter {\bf 12}, R1 (2000).

\bibitem{todorov_tdtb}
T.~N. Todorov,
\newblock J. Phys.: Cond. Matter {\bf 14}, 3049 (2002).

\bibitem{ss_1}
S.~Sanvito, J.~H. Jefferson, and C.~J. Lambert,
\newblock Phys. Rev. B {\bf 60}, 7385 (1999).

\bibitem{gehring}
G.~A. Gehring,
\newblock {\em An introduction to the theory of normal and ferromagnetic
  materials},
\newblock in ``Spin electronics'' Lecture Notes in Physics: Vol 569,
  Springer-Verlag, Heidelberg, 2001.

\bibitem{siesta}
J.~M. Soler et~al.,
\newblock J. Phys. Cond. Matter {\bf 14}, 2745 (2002).

\bibitem{Fermi_per}
T.-S. Choy,
\newblock Ph.D. Thesis, University of Florida (2001) ,
  http://www.phys.ufl.edu/fermisurface/.

\bibitem{slater_koster}
J.~C. Slater and G.~F. Koster,
\newblock Phys. Rev. {\bf 94}, 1498 (1954).

\bibitem{ss_2}
J.~M.~D. Coey and S.~Sanvito,
\newblock J. Phys.: Appl. Phys. {\bf 37}, 988 (2004).

\bibitem{Mazin_PRL}
I.~I. Mazin,
\newblock Phys. Rev. Lett. {\bf 83}, 1427 (1999).

\bibitem{BOEqs}
P.~B. Allen,
\newblock Phys. Rev. B {\bf 17}, 3725 (1978).

\bibitem{CrO}
S.~P. Lewis, P.~B. Allen, and T.~Sasaki,
\newblock Phys. Rev. B {\bf 55}, 10253 (1997).

\bibitem{LCMO}
B.~Nadgorny et~al.,
\newblock Phys. Rev. B {\bf 63}, 184433 (2001).

\bibitem{TMO}
D.~Singh,
\newblock Phys. Rev. B {\bf 55}, 313 (1997).

\bibitem{landauer_original1}
R.~Landauer,
\newblock Phil. Mag. {\bf 21}, 863 (1970).

\bibitem{landauer_original2}
R.~Landauer,
\newblock IBM J. Res. Develop. {\bf 1}, 233 (1957).

\bibitem{multi_channel}
M.~B\"uttiker, Y.~Imry, R.~Landauer, and S.~Pinhas,
\newblock Phys. Rev. B {\bf 31}, 6207 (1985).

\bibitem{sharvin}
Y.~V. Sharvin,
\newblock Zh. Eksp. Teor. Phys. {\bf 48}, 984 (1965).

\bibitem{sharvin2}
Y.~V. Sharvin,
\newblock Sov. Phys. JETP {\bf 21}, 655 (1965).

\bibitem{jullier}
M.~Julliere,
\newblock Phys. Lett. A {\bf 50}, 225 (1975).

\bibitem{tsymbal}
E.~Y. Tsymbal and D.~G. Pettifor,
\newblock J. Phys.: Condens. Matter {\bf 9}, L411 (1997).

\bibitem{Andreev}
A.~F. Andreev,
\newblock Sov. Phys. JETP {\bf 19}, 1228 (1964).

\bibitem{TedrovM}
R.~Meservey and P.~M. Tedrow,
\newblock Phys. Rep. {\bf 238}, 173 (1994).

\bibitem{Beenaker}
M.~J.~M. de~Jong and C.~Beenakker,
\newblock Phys. Rev. Lett. {\bf 74}, 1657 (1995).

\bibitem{ss_3}
F.~Taddei, S.~Sanvito, and C.~J. Lambert,
\newblock J. Low Temp. Phys. {\bf 124}, 305 (2001).

\bibitem{Schmit}
G.~Schmidt, D.~Ferrand, L.~W. Molenkamp, A.~T. Filip, and B.~J. van Wees,
\newblock Phys. Rev. B {\bf 62}, R4790 (2000).

\bibitem{Rashba1}
E.~I. Rashba,
\newblock Eur. Phys. J. B {\bf 29}, 513 (2002).

\bibitem{Rashba2}
V.~Y. Kravchenko and E.~I. Rashba,
\newblock Phys. Rev. B {\bf 67}, 121310 (2003).

\bibitem{Roukes}
H.~X. Tang, F.~G. Monzon, R.~Lifshitz, M.~C. Cross, and M.~L. Roukes,
\newblock Phys. Rev. B {\bf 61}, 4437 (2000).

\bibitem{Aws6}
Y.~Ohno et~al.,
\newblock Nature {\bf 402}, 790 (2001).

\bibitem{Molenkamp}
R.~Fiederling et~al.,
\newblock Nature {\bf 402}, 787 (1999).

\bibitem{AlexB_1}
A.~M. Bratkovsky,
\newblock Appl. Phys. Lett. {\bf 72}, 2334 (1998).

\bibitem{PR_Agrait}
N.~Agra{\"i}t, A.~L. Yeyati, and J.~M. van Ruitenbeek,
\newblock Phys. Rep. {\bf 377}, 81 (2003).

\bibitem{ssh}
A.~J. Heeger, S.~Kivelson, and J.~R. Schrieffer,
\newblock Rev. Mod. Phys. {\bf 60}, 782 (1988).

\bibitem{torque_exp}
S.~I. Kiselev et~al.,
\newblock Nature {\bf 425}, 380 (2003).

\bibitem{dw_motion_exp}
M.~Kla{\"u}i et~al.,
\newblock Appl. Phys. Lett. {\bf 83}, 105 (2003).

\bibitem{dw_motion_the}
L.~Berger,
\newblock J. Appl. Phys. {\bf 55}, 1954 (1984).

\bibitem{torque_the}
L.~Berger,
\newblock J. Appl. Phys. {\bf 71}, 2721 (1992).

\bibitem{tatara}
G.~Tatara and H.~Kohno,
\newblock Phys. Rev. Lett. {\bf 92}, 086601 (2004).

\bibitem{TTod_Max}
M.~D. Ventra, Y.-C. Chen, and T.~Todorov,
\newblock Phys. Rev. Lett. {\bf 92}, 176803 (2004).

\bibitem{papacon}
D.~Papaconstantopoulos,
\newblock {\em Handbook of the Band Structure of Elemental Solids},
\newblock Plenum, New York, 1986.

\bibitem{papa_web}
http://cst www.nrl.navy.mil/bind/.

\bibitem{But_IBM}
M.~B\"uttiker,
\newblock IBM J. Res. Develop. {\bf 32}, 317 (1988).

\bibitem{Kohn_transport}
A.~Kamenev and W.~Kohn,
\newblock Phys. Rev. B {\bf 63}, 155304 (2001).

\bibitem{wave_function_method1}
N.~D. Lang,
\newblock Phys. Rev. B {\bf 36}, 8173 (1987).

\bibitem{econbook}
E.~Economou,
\newblock {\em Green's Functions in Quantum Physics},
\newblock Springer-Verlag, New York, 1983.

\bibitem{Joann1}
D.~Lee and J.~D. Joannopoulos,
\newblock Phys. Rev. B {\bf 23}, 4988 (1981).

\bibitem{Joann2}
D.~Lee and J.~D. Joannopoulos,
\newblock Phys. Rev. B {\bf 23}, 4997 (1981).

\bibitem{smeagol1}
A.~R. Rocha et~al.,
\newblock in preparation , http://www.smeagol.tcd.ie.

\bibitem{smeagol2}
A.~R. Rocha et~al.,
\newblock Nature Materials  {\bf 4}, 335 (2005).
  http://www.nature.com/nmat/journal/vaop/ncurrent/abs/nmat1349.html.

\bibitem{sanvito}
S.~Sanvito, C.~J. Lambert, J.~H. Jefferson, and A.~M. Bratkovsky,
\newblock Phys. Rev. B {\bf 59}, 11936 (1999).

\bibitem{lhr93}
C.~Lambert, V.~Hui, and S.~Robinson,
\newblock J. Phys.: Condens. Matter {\bf 5}, 4187 (1993).

\bibitem{wexler}
G.~Wexler,
\newblock Proc. Phys. Soc. {\bf 89}, 927 (1966).

\bibitem{bauer1}
G.~E.~W. Bauer, A.~Brataas, K.~M. Schep, and P.~J. Kelly,
\newblock J. Appl. Phys. {\bf 75}, 6704 (1994).

\bibitem{kelly1}
K.~M. Schep, P.~J. Kelly, and G.~E.~W. Bauer,
\newblock Phys. Rev. Lett. {\bf 74}, 586 (1995).

\bibitem{kelly2}
K.~M. Schep, P.~J. Kelly, and G.~E.~W. Bauer,
\newblock Phys. Rev. B {\bf 57}, 8907 (1998).

\bibitem{ss_4}
S.~Sanvito and N.~A. Hill,
\newblock Phys. Rev. Lett. {\bf 87}, 267202 (2001).

\bibitem{ss_5}
S.~Sanvito,
\newblock Phys. Rev. B {\bf 68}, 054425 (2003).

\bibitem{lopez1}
M.~P. Lopez-Sancho, J.~M. Lopez-Sancho, and J.~Rubio,
\newblock J. Phys. F {\bf 14}, 1205 (1984).

\bibitem{lopez2}
M.~P. Lopez-Sancho, J.~M. Lopez-Sancho, and J.~Rubio,
\newblock J. Phys. F {\bf 15}, 851 (1985).

\bibitem{marcoBN}
M.~B. Nardelli,
\newblock Phys. Rev. B {\bf 60}, 7828 (1999).

\bibitem{mat1}
J.~Mathon,
\newblock J. Phys. Condens. Matter {\bf 1}, 2505 (1989).

\bibitem{mat2}
J.~Mathon,
\newblock Phys. Rev. B {\bf 55}, 960 (1997).

\bibitem{KKR1}
J.~Korringa,
\newblock Physics {\bf 13}, 392 (1947).

\bibitem{KKR2}
W.~Kohn and N.~Rostoker,
\newblock Phys. Rev. {\bf 94}, 1111 (1954).

\bibitem{LKKR1}
J.~M. MacLaren, S.~Crampin, D.~D. Vvedensky, and J.~B. Pendry,
\newblock Phys. Rev. B {\bf 40}, 12164 (1989).

\bibitem{LKKR2}
J.~M. MacLaren, X.~G. Zhang, W.~H. Butler, and X.~Wang,
\newblock Phys. Rev. B {\bf 59}, 5470 (1999).

\bibitem{LKKR3}
W.~Wildberger, R.~Zeller, and P.~H. Dederichs,
\newblock Phys. Rev. B {\bf 55}, 10074 (1997).

\bibitem{LKKR4}
P.~Mavropoulos, N.~Papanikolaou, and P.~H. Dederichs,
\newblock Phys. Rev. B {\bf 69}, 125104 (2004).

\bibitem{kubo}
R.~Kubo,
\newblock J. Phys. Soc. Jpn. {\bf 12}, 570 (1957).

\bibitem{wan1}
A.~Calzolari, N.~Marzari, I.~Souza, and M.~B. Nardelli,
\newblock Phys. Rev. B {\bf 69}, 035108 (2004).

\bibitem{marzari}
N.~Marzari and D.~Vanderbilt,
\newblock Phys. Rev. B {\bf 56}, 12847 (1997).

\bibitem{ramsak1}
A.~Ram\v{s}ak and T.~Rejec,
\newblock cond-mat/0310452 .

\bibitem{ramsak2}
T.~Rejec and A.~Ram\v{s}ak,
\newblock Phys. Rev. B {\bf 68}, 035342 (2003).

\bibitem{ramsak3}
T.~Rejec and A.~Ram\v{s}ak,
\newblock Phys. Rev. B {\bf 68}, 033306 (2003).

\bibitem{kieron}
K.~Burke, R.~Gaudoin, and F.~Evers,
\newblock preprint .

\bibitem{kieron2}
K.~Burke, R.~Car, and R.~Gebauer,
\newblock cond-mat/0410352 .

\bibitem{Upd1}
S.~K. Upadhyay, A.~Palanisami, R.~N. Louie, and R.~A. Buhrman,
\newblock Phys. Rev. Lett. {\bf 81}, 3247 (1998).

\bibitem{Upd2}
S.~K. Upadhyay, R.~N. Louie, and R.~A. Buhrman,
\newblock Appl. Phys. Lett. {\bf 74}, 3881 (1999).

\bibitem{Soul}
R.~J. Soulen et~al.,
\newblock Science {\bf 282}, 85 (1998).

\bibitem{BdG}
P.~G. de~Gennes,
\newblock {\em Superconductivity of Metals and Alloys},
\newblock Addison-Wesley, New York, 1989.

\bibitem{Lambert_Raimondi}
C.~J. Lambert and R.~Raimondi,
\newblock J. Phys.: Condens. Matter {\bf 10}, 901 (1998).

\bibitem{caroli}
C.~Caroli, R.~Combescot, P.~Nozieres, and D.~Saint-James,
\newblock J. Phys. C: Solid State Phys. {\bf 5}, 21 (1972).

\bibitem{ferrer}
J.~Ferrer, A.~Mart\'{\i}n-Rodero, and F.~Flores,
\newblock Phys. Rev. B {\bf 38}, 10113 (1988).

\bibitem{datta}
S.~Datta,
\newblock {\em Electronic Transport in Mesoscopic Systems},
\newblock Cambridge University Press, Cambridge, UK, 1995.

\bibitem{Haug}
H.~Haug and A.~P. Jauho,
\newblock {\em Quantum Kinetics in Transport and Optics of Semiconductors},
\newblock Springer, Berlin, 1996.

\bibitem{Xue}
Y.~Xue, S.~Datta, and M.~A. Ratner,
\newblock Chem. Phys. {\bf 281}, 151 (2002).

\bibitem{datta2}
S.~Datta,
\newblock Nanotechnology {\bf 15}, S433 (2004).

\bibitem{Janak_Williams}
J.~Janak and A.~Williams,
\newblock Phys. Rev. B {\bf 14}, 4199 (1976).

\bibitem{Wang_Callaway}
C.~Wang and J.~Callaway,
\newblock Phys. Rev. B {\bf 15}, 298 (1977).

\bibitem{Moruzzi_77}
V.~Moruzzi, A.~Williams, and J.~Janak,
\newblock Phys. Rev. B {\bf 15}, 2854 (1997).

\bibitem{vbh}
U.~V. Barth and L.~Hedin,
\newblock J. Phys. C {\bf 5}, 1629 (1972).

\bibitem{Schwarz_Mohn}
K.~Schwarz and P.~Mohn,
\newblock J. Phys. F {\bf 14}, L129 (1984).

\bibitem{Moruzzi_FSM}
V.~Moruzzi, P.~Marcus, K.~Schwarz, and P.~Mohn,
\newblock Phys. Rev. B {\bf 34}, 1784 (1986).

\bibitem{Fu_Ho}
C.-L. Fu and K.-M. Ho,
\newblock Phys. Rev. B {\bf 28}, 5480 (1983).

\bibitem{PBE96}
J.~Perdew, K.~Burke, and M.~Ernzerhof,
\newblock Phys. Rev. Lett. {\bf 77}, 3865 (1996).

\bibitem{David_Singhs_book}
D.~Singh,
\newblock {\em Planewaves, pseudopotentials and the LAPW method},
\newblock Kluwer Academic Publishers, 1994.

\bibitem{anis1}
V.~Anisimov, J.~Zaanen, and O.~Andersen,
\newblock Phys. Rev. B {\bf 44}, 943 (1991).

\bibitem{anis2}
V.~Anisimov, F.~Aryasetiawan, and A.~Liechtenstein,
\newblock J. Phys.:Condens. Matter {\bf 9}, 767 (1997).

\bibitem{pzu}
J.~Perdew and A.~Zunger,
\newblock Phys. Rev. B {\bf 23}, 5048 (1981).

\bibitem{PAO}
O.~Sankey and D.~Niklewski,
\newblock Phys. Rev. B {\bf 40}, 3979 (1989).

\bibitem{TM}
N.~Troullier and J.~Martins,
\newblock Phys. Rev. B {\bf 43}, 1993 (1991).

\bibitem{Lou82}
S.~Louie, S.~Froyen, and M.~Cohen,
\newblock Phys. Rev. B {\bf 26}, 1738 (1982).

\bibitem{mcdcal}
J.~Taylor, H.~Guo, and J.~Wang,
\newblock Phys. Rev. B {\bf 63}, 245407 (2001).

\bibitem{fireball}
P.~Ordej\'on, E.~Artacho, and J.~Soler,
\newblock Phys. Rev. B {\bf 53}, R10 (1996).

\bibitem{multigrid}
A.~Brandt,
\newblock Math. Comput. {\bf 31}, 333 (1977).

\bibitem{mcdcal1}
B.~Larade, J.~Taylor, Q.~Zheng, H.~M.~P. Pomorski, and H.~Guo,
\newblock Phys. Rev. B {\bf 64}, 195402 (2001).

\bibitem{mcdcal2}
C.~Roland, B.~Larade, J.~Taylor, and H.~Guo,
\newblock Phys. Rev. B {\bf 65}, R041401 (2002).

\bibitem{mcdcal3}
H.~Mehrez et~al.,
\newblock Phys. Rev. B {\bf 65}, 195419 (2002).

\bibitem{transiesta}
M.~Bradbyge, J.~Taylor, K.~Stokbro, J.-L. Mozos, and P.~Ordej\'on,
\newblock Phys. Rev. B {\bf 65}, 165401 (2002).

\bibitem{transiesta1}
M.~Brandbyge, K.~Stokbro, J.~Taylor, J.-L. Mozos, and P.~Ordej\'on,
\newblock Phys. Rev. B {\bf 67}, 193104 (2003).

\bibitem{gaussian98}
Gaussian 98  (Gaussian Inc., Pittsburgh, PA, 1998).

\bibitem{palacios1}
J.~J. Palacios, A.~J. P\'erez-Jim\'enez, E.~Louis, E.~SanFabi\'an, and J.~A.
  Verg\'es,
\newblock Phys. Rev. B {\bf 66}, 035322 (2002).

\bibitem{palacios2}
E.~Louis, J.~A. Verg\'es, J.~J. Palacios, A.~J. P\'erez-Jim\'enez, and
  E.~SanFabi\'an,
\newblock Phys. Rev. B {\bf 67}, 155321 (2003).

\bibitem{bethe}
J.~Joannopoulos and F.~Yndurain,
\newblock Phys. Rev. B {\bf 10}, 5164 (1974).

\bibitem{Lang2}
N.~D. Lang,
\newblock Phys. Rev. B {\bf 52}, 5335 (1995).

\bibitem{diVentra}
M.~D. Ventra, S.~T. Pantelides, and N.~D. Lang,
\newblock Phys. Rev. Lett. {\bf 84}, 979 (2000).

\bibitem{tbdft2}
T.~Frauenheim et~al.,
\newblock J. Phys.: Condens. Matter {\bf 14}, 3015 (2002).

\bibitem{dicarlo}
A.~Pecchia and A.~D. Carlo,
\newblock Rep. Prog. Phys. {\bf 67}, 1497 (2004).

\bibitem{parkin_ex}
S.~Parkin, N.~More, and K.~Roche,
\newblock Phys. Rev. Lett. {\bf 64}, 2304 (1990).

\bibitem{Bruno_ex}
P.~Bruno,
\newblock Phys. Rev. B {\bf 52}, 411 (1995).

\bibitem{MSU91}
W.~P. Jr. et~al.,
\newblock Phys. Rev. Lett. {\bf 66}, 3060 (1991).

\bibitem{Gijs93}
M.~Gijs, S.~Lenczowski, and J.~Giesbers,
\newblock Phys. Rev. Lett. {\bf 70}, 3343 (1993).

\bibitem{Gijs97}
M.~Gijs and G.~Bauer,
\newblock Adv. Phys. {\bf 46}, 285 (1997).

\bibitem{Ansermet98}
J.-P. Ansermet,
\newblock J.Phys.: Cond. Matter {\bf 10}, 6027 (1998).

\bibitem{mertig1}
P.~Zahn, J.~Binder, I.~Mertig, R.~Zeller, and P.~Dederichs,
\newblock Phys. Rev. Lett. {\bf 80}, 4309 (1998).

\bibitem{mertig2}
P.~Zahn, I.~Mertig, M.~Richter, and H.~Eschrig,
\newblock Phys. Rev. Lett. {\bf 75}, 2996 (1995).

\bibitem{tsy1}
E.~Tsymbal and D.~Pettifor,
\newblock Phys. Rev. B {\bf 54}, 15314 (1996).

\bibitem{mat3}
J.~Mathon,
\newblock Phys. Rev. B {\bf 54}, 55 (1996).

\bibitem{el3}
L.~Piraux, S.~Dubois, A.~Fert, and L.~Belliard,
\newblock Eur. Phys. J. B {\bf 4}, 413 (1998).

\bibitem{mertig3}
I.~Mertig, R.~Zeller, and P.~Dederichs,
\newblock Phys. Rev. B {\bf 47}, 16178 (1993).

\bibitem{mertig4}
V.~Stepanyuk, R.~Zeller, P.~Dederichs, and I.~Mertig,
\newblock Phys. Rev. B {\bf 49}, 5157 (1994).

\bibitem{leed99}
D.~Bozec, M.~Howson, B.~Hickey, S.~Shatz, and N.~Wiser,
\newblock J. Phys.: Condens. Matter {\bf 12}, 4263 (2000).

\bibitem{MSU97}
W.-C. Chiang, Q.~Yang, W.~P. Jr., R.~Loloee, and J.~Bass,
\newblock J. Appl. Phys. {\bf 81}, 4570 (1997).

\bibitem{Didier99}
D.~Bozec, M.~Walker, B.~Hickey, M.~Howson, and N.~Wiser,
\newblock Phys. Rev. B {\bf 60}, 3037 (1999).

\bibitem{ss_7}
S.~Sanvito, C.~Lambert, J.~Jefferson, and A.~Bratkovsky,
\newblock J. Phys.C: Condens. Matter {\bf 10}, L691 (1998).

\bibitem{ss_6}
S.~Sanvito, C.~Lambert, and J.~Jefferson,
\newblock Phys. Rev. B {\bf 61}, 14225 (2000).

\bibitem{And80}
P.~Anderson, D.~Thouless, E.~Abrahams, and D.~Fisher,
\newblock Phys. Rev. B {\bf 22}, 3519 (1980).

\bibitem{Been97}
C.~Beenakker,
\newblock Rev. Mod. Phys. {\bf 69}, 731 (1997).

\bibitem{Kram93}
B.~Kramer and A.~MacKinnon,
\newblock Rep. Prog. Phys. {\bf 56}, 1469 (1993).

\bibitem{Falko99}
V.~Fal'ko, C.~Lambert, and A.~Volkov,
\newblock JETP Letter {\bf 69}, 532 (1999).

\bibitem{Falko99b}
V.~Fal'ko, C.~Lambert, and A.~Volkov,
\newblock Phys. Rev. B {\bf 60}, 15394 (1999).

\bibitem{ss_8}
F.~Taddei, S.~Sanvito, and C.~Lambert,
\newblock Phys. Rev. Lett. {\bf 82}, 4938 (1999).

\bibitem{ss_9}
F.~Taddei, S.~Sanvito, and C.~Lambert,
\newblock Phys. Rev. B {\bf 63}, 12404 (2001).

\bibitem{BTK}
G.~Blonder, M.~Tinkham, and T.~Klapwijk,
\newblock Phys. Rev. B {\bf 25}, 4515 (1982).

\bibitem{KellySup}
K.~Xia, P.~Kelly, G.~Bauer, and I.~Turek,
\newblock Phys. Rev. Lett. {\bf 89}, 166603 (2002).

\bibitem{ss_10}
F.~Taddei, S.~Sanvito, and C.~Lambert,
\newblock J. Comput. Theor. Nanosci. , in press.

\bibitem{Julliere75}
M.~Julliere,
\newblock Phys. Lett. {\bf 50A}, 225 (1975).

\bibitem{Slon89}
J.~Slonczewski,
\newblock Phys. Rev. B {\bf 39}, 6995 (1989).

\bibitem{deTeresa}
J.~de~Teresa et~al.,
\newblock Science {\bf 286}, 507 (1999).

\bibitem{Mac99}
J.~MacLaren, X.-G. Zhang, W.~Butler, and X.~Wang,
\newblock Phys. Rev. B {\bf 59}, 5470 (1999).

\bibitem{Wang98}
K.~Wang, S.~Zhang, P.~Levy, L.~Szunyogh, and P.~Weinberger,
\newblock J. Magn. Magn. Mater. {\bf 189}, L131 (1998).

\bibitem{Tsy97t}
E.~Tsymbal and D.~Pettifor,
\newblock J. Phys.: Condens. Matter {\bf 9}, L411 (1997).

\bibitem{Math97t}
J.~Mathon,
\newblock Phys. Rev. B {\bf 56}, 11810 (1997).

\bibitem{wang98bis}
K.~Wang, P.~Levy. S.~Zhang, and L.~Szunyogh,
\newblock Phil. Mag. {\bf 83}, 1255 (2003).

\bibitem{But01}
W.~Butler, X.-G. Zhang, T.~Schulthess, and J.~MacLaren,
\newblock Phys. Rev. B {\bf 63}, 054416 (2001).

\bibitem{wenn02}
O.~Wunnicke et~al.,
\newblock Phys. Rev. B {\bf 65}, 064425 (2002).

\bibitem{TsyPet}
E.~Tsymbal and D.~Pettifor,
\newblock Phys. Rev. B {\bf 58}, 432 (1998).

\bibitem{olenik}
E.~Tsymbal, V.~Burlakov, and I.~Oleinik,
\newblock Phys. Rev. B {\bf 66}, 073201 (2002).

\bibitem{mertigtun}
P.~Zahn and I.~Mertig,
\newblock Phys. Rev. Lett. {\bf 75}, 2996 (1995).

\bibitem{mathontun}
J.~Mathon and A.~Umerski,
\newblock Phys. Rev. B {\bf 63}, 220403(R) (2001).

\bibitem{Alex97tris}
A.~Bratkovsky,
\newblock {\em in Science and Technology of Magnetic Oxides},
\newblock Eds.: M.Hundley, J.Nickel, R.Ramesh, Y.Tokura, MRS, Vol.494, 1998.

\bibitem{Alex97}
A.~Bratkovsky,
\newblock Phys. Rev. B {\bf 56}, 2344 (1997).

\bibitem{Alex97bis}
A.~Bratkovsky,
\newblock JETP Letter {\bf 65}, 453 (1997).

\bibitem{TMRexpBIAS}
J.~Zhang and R.~White,
\newblock J. Appl. Phys. {\bf 83}, 6512 (1998).

\bibitem{TMRbias}
C.~Zhang et~al.,
\newblock Phys. Rev. B {\bf 69}, 134406 (2004).

\bibitem{MgO}
W.~Wulfhekel et~al.,
\newblock Appl. Phys. Lett. {\bf 78}, 509 (2001).

\bibitem{cabrera}
G.~Cabrera and L.~Falicov,
\newblock Phys. Status Solidi B {\bf 61}, 539 (1974).

\bibitem{tatara1}
G.~Tatara and H.~Fukuyama,
\newblock Phys. Rev. Lett. {\bf 78}, 3773 (1997).

\bibitem{otani}
Y.~Otani, S.~Kim, K.~Fukamichi, O.~Kitakani, and Y.~Shimada,
\newblock IEEE Trans. Magn. {\bf 34}, 1096 (1998).

\bibitem{ruediger}
U.~Ruediger, S.~Z. J.~Yu, A.~D. Kent, and S.~S.~P. Parkin,
\newblock Phys. Rev. Lett. {\bf 80}, 5639 (1998).

\bibitem{gregg}
J.~F. Gregg et~al.,
\newblock Phys. Rev. Lett. {\bf 77}, 1580 (1996).

\bibitem{viret}
M.~Viret et~al.,
\newblock Phys. Rev. B {\bf 53}, 8464 (1996).

\bibitem{vanHoof}
J.~van Hoof, K.~Schep, A.~Brataas, G.~Bauer, and P.~Kelly,
\newblock Phys. Rev. Lett. {\bf 59}, 138 (1999).

\bibitem{kudDW}
J.~Kudrnovsky et~al.,
\newblock Phys. Rev. B {\bf 62}, 15084 (2000).

\bibitem{kudDW2}
J.~Kudrnovsky, V.~Drchal, I.~Turek, P.~Streda, and P.~Bruno,
\newblock Surf. Sci. {\bf 482-485}, 1107 (2001).

\bibitem{BMR4}
N.~Garcia, H.~Wang, H.~Cheng, and N.~D. Nikolic,
\newblock IEEE Trans. Magn. {\bf 39}, 2776 (2003).

\bibitem{viret1}
M.~Viret et~al.,
\newblock Phys. Rev. B {\bf 66}, 220401(R) (2002).

\bibitem{viret2}
M.~Montero et~al.,
\newblock Phys. Rev. B {\bf 70}, 184418 (2004).

\bibitem{egel1}
W.~Egelhoff et~al.,
\newblock J. Appl. Phys. {\bf 95}, 7554 (2004).

\bibitem{egel2}
J.~Mallett, E.~Svedberg, H.~Ettedgui, T.~Moffat, and W.~Egelhoff,
\newblock Phys. Rev. B {\bf 70}, 172406 (2004).

\bibitem{egel3}
S.~Chung, M.~Mu{\~n}oz, N.~Garc\'{i}a, W.~Egelhoff, and R.~Gomez,
\newblock J. Appl. Phys. {\bf 93}, 7939 (2003).

\bibitem{JEW}
J.~Wegrowe et~al.,
\newblock Phys. Rev. B {\bf 67}, 104418 (2003).

\bibitem{sheer}
E.~Scheer et~al.,
\newblock Nature {\bf 394}, 154 (1998).

\bibitem{agrait}
N.~Agrait, A.~Levy-Yeyati, and J.~van Ruitenbeek,
\newblock Phys. Rep. {\bf 377}, 81 (2003).

\bibitem{his1}
H.~Oshima and K.~Miyano,
\newblock Appl. Phys. Lett. {\bf 73}, 2203 (1998).

\bibitem{his2}
T.~Ono, Y.~Ooka, and H.~Miyajima,
\newblock Appl. Phys. Lett, {\bf 75}, 1622 (1999).

\bibitem{his3}
V.~Rodrigues, J.~Bettini, P.~Silva, and D.~Ugarte,
\newblock Phys. Rev. Lett. {\bf 91}, 096801 (2003).

\bibitem{his4}
C.~Untiedt, D.~Dekker, D.~Djukic, and J.~van Ruitenbeek,
\newblock Phys. Rev. B {\bf 69}, 081401 (2004).

\bibitem{dalcorso}
A.~Smogunov, A.~D. Corso, and E.~Tosatti,
\newblock Phys. Rev. B {\bf 70}, 045417 (2004).

\bibitem{mertigpc}
A.~Bagrets, N.~Papanikolaou, and I.~Mertig,
\newblock Phys. Rev. B {\bf 70}, 064410 (2004).

\bibitem{palaciospc}
J.~Palacios, D.~Jacob, and J.~Fern\'andez-Rossier,
\newblock cond-mat/0406249 .

\bibitem{butlerpc}
J.~Velev and W.~Butler,
\newblock Phys. Rev. B {\bf 69}, 094425 (2004).

\bibitem{papapc}
N.~Papanikolaou,
\newblock J. Phys.: Condens. Matter {\bf 15}, 5049 (2003).

\bibitem{oscar1}
O.~C\'espedes et~al.,
\newblock J. Magn. Magn. Mater. {\bf 242}, 492 (2002).

\bibitem{oscar2}
O.~C\'espedes et~al.,
\newblock J. Magn. Magn. Mater. {\bf 272-276}, 1571 (2004).

\bibitem{alexpc}
A.~R. Rocha and S.~Sanvito,
\newblock Phys. Rev. B {\bf 70}, 094406 (2004).

\bibitem{Dress}
R.~Saito, G.~Dresslhaus, and M.~Dresselhaus,
\newblock {\em Physical Properties of Carbon Nanotubes},
\newblock Imperial College Press, London, 1996.

\bibitem{Heer}
S.~Franck, P.~Poncharal, Z.~Wang, and W.~de~Heer,
\newblock Science {\bf 280}, 1744 (1998).

\bibitem{todorovCN}
C.~White and T.~Todorov,
\newblock Nature {\bf 393}, 240 (1998).

\bibitem{CNT1}
B.~Zhao, I.~M\"onch, T.~M\"uhl, H.~Vinzelberg, and C.~Schneider,
\newblock J. Appl. Phys. {\bf 91}, 7026 (2002).

\bibitem{CNT2}
B.~Zhao, I.~M\"onch, H.~Vinzelberg, T.~M\"uhl, and C.~Schneider,
\newblock Appl. Phys. Lett. {\bf 80}, 3144 (2002).

\bibitem{CNT4}
S.~Sahoo, T.~Kontos, C.~Sch\"onenberger, and C.~S\"urgers,
\newblock cond-mat/0411623 .

\bibitem{CNT5}
A.~Jensen, J.~Nygard, and J.~Borggreen,
\newblock in Toward the controllable quantum states, Proceedings of the
  International Symposium on Mesoscopic Superconductivity and Spintronics, H.
  Takayanagi and J. Nitta , 33 (2003).

\bibitem{Tersoff}
J.~Tersoff,
\newblock Appl. Phys. Lett. {\bf 74}, 2122 (1999).

\bibitem{MaxJTersoff}
P.~Delaney and M.~D. Ventra,
\newblock Appl. Phys. Lett. {\bf 75}, 4028 (1999).

\bibitem{Mehrez}
H.~Mehrez, J.~Taylor, H.~Guo, J.~Wang, and C.~Roland,
\newblock Phys. Rev. Lett. {\bf 84}, 2682 (2000).

\bibitem{Kromp1}
S.~Krompiewski, R.~Gutierrez, and G.~Cuniberti,
\newblock Phys. Rev. B {\bf 69}, 155423 (2004).

\bibitem{Kromp2}
S.~Krompiewski,
\newblock J. Phys.: Condens. Matter {\bf 16}, 2981 (2004).

\bibitem{GuoGMR}
Y.~Wei, Y.~Xu, J.~Wang, and H.~Guo,
\newblock Phys. Rev. B {\bf 70}, 193406 (2004).

\bibitem{PatiGMR}
R.~Pati et~al.,
\newblock Phys. Rev. B {\bf 68}, 014412 (2003).

\bibitem{mol_TMR}
J.~Petta, S.~Slater, and D.~Ralph,
\newblock Phys. Rev. Lett. {\bf 93}, 136601 (2004).

\bibitem{pati}
R.~Pati, L.~Sanapati, P.~Ajayan, and K.~Nayak,
\newblock Phys. Rev. B {\bf 68}, 100407(R) (2003).

\bibitem{kirk}
E.~Emberly and G.~Kirczenow,
\newblock Chem. Phys. {\bf 281}, 311 (2002).

\bibitem{Sankey}
J.~Tomfohr and O.~Sankey,
\newblock Phys. Rev. B {\bf 65}, 245105 (2002).

\bibitem{meteo}
J.~Coey, M.~Venkatesan, C.~Fitzgerald, A.~Douvalis, and I.~Sanders,
\newblock Nature {\bf 420}, 156 (2002).

\bibitem{ss_ferreira}
M.~Ferreira and S.~Sanvito,
\newblock Phys. Rev. B {\bf 69}, 035407 (2004).

\bibitem{Mertins}
H.~Mertins et~al.,
\newblock Europhys. Lett. {\bf 66}, 743 (2004).

\bibitem{phyworld}
J.~Coey and S.~Sanvito,
\newblock Physics World {\bf November}, 33 (2004).

\end{thebibliography}
\end{document}